\documentclass[longauth]{aa_gaia} 
\usepackage{graphicx}
\usepackage{txfonts}
\usepackage{longtable}

\newcommand\gaia{\textit{Gaia}}

\newcommand\tyctwo{\textit{Tycho}-2}
\newcommand\gdrone{\gaia~DR1}
\newcommand\gdrtwo{\gaia~DR2}
\newcommand\gdrthree{\gaia~DR3}

\newcommand{\kms}{{\mathrm{km\,s^{-1}}}}

\newcommand{\gbp}{{G_\mathrm{BP}}}
\newcommand{\grp}{{G_\mathrm{RP}}}
\newcommand{\grvs}{{G_\mathrm{RVS}}}
\newcommand{\teff}{{T_\mathrm{eff}}}
\providecommand{\masyr}{\,mas\,yr$^{-1}$}

\usepackage{color}
\usepackage{amstext}

\begin{document}

   \title{\gaia{} Data Release 2:\\ Mapping the Milky Way disc kinematics}


\author{
{\it Gaia} Collaboration
\and D.        ~Katz                          \inst{\ref{inst:0001}}
\and T.        ~Antoja                        \inst{\ref{inst:0002},\ref{inst:0003}}
\and M.        ~Romero-G\'{o}mez              \inst{\ref{inst:0003}}
\and R.        ~Drimmel                       \inst{\ref{inst:0005}}
\and C.        ~Reyl\'{e}                     \inst{\ref{inst:0006}}
\and G.M.      ~Seabroke                      \inst{\ref{inst:0007}}
\and C.        ~Soubiran                      \inst{\ref{inst:0008}}
\and C.        ~Babusiaux                     \inst{\ref{inst:0001},\ref{inst:0010}}
\and P.        ~Di Matteo                     \inst{\ref{inst:0001}}
\and F.        ~Figueras                      \inst{\ref{inst:0003}}
\and E.        ~Poggio                        \inst{\ref{inst:0013},\ref{inst:0005}}
\and A.C.      ~Robin                         \inst{\ref{inst:0006}}
\and D.W.      ~Evans                         \inst{\ref{inst:0016}}
\and A.G.A.    ~Brown                         \inst{\ref{inst:0017}}
\and A.        ~Vallenari                     \inst{\ref{inst:0018}}
\and T.        ~Prusti                        \inst{\ref{inst:0002}}
\and J.H.J.    ~de Bruijne                    \inst{\ref{inst:0002}}
\and C.A.L.    ~Bailer-Jones                  \inst{\ref{inst:0021}}
\and M.        ~Biermann                      \inst{\ref{inst:0022}}
\and L.        ~Eyer                          \inst{\ref{inst:0023}}
\and F.        ~Jansen                        \inst{\ref{inst:0024}}
\and C.        ~Jordi                         \inst{\ref{inst:0003}}
\and S.A.      ~Klioner                       \inst{\ref{inst:0026}}
\and U.        ~Lammers                       \inst{\ref{inst:0027}}
\and L.        ~Lindegren                     \inst{\ref{inst:0028}}
\and X.        ~Luri                          \inst{\ref{inst:0003}}
\and F.        ~Mignard                       \inst{\ref{inst:0030}}
\and C.        ~Panem                         \inst{\ref{inst:0031}}
\and D.        ~Pourbaix                      \inst{\ref{inst:0032},\ref{inst:0033}}
\and S.        ~Randich                       \inst{\ref{inst:0034}}
\and P.        ~Sartoretti                    \inst{\ref{inst:0001}}
\and H.I.      ~Siddiqui                      \inst{\ref{inst:0036}}
\and F.        ~van Leeuwen                   \inst{\ref{inst:0016}}
\and N.A.      ~Walton                        \inst{\ref{inst:0016}}
\and F.        ~Arenou                        \inst{\ref{inst:0001}}
\and U.        ~Bastian                       \inst{\ref{inst:0022}}
\and M.        ~Cropper                       \inst{\ref{inst:0007}}
\and M.G.      ~Lattanzi                      \inst{\ref{inst:0005}}
\and J.        ~Bakker                        \inst{\ref{inst:0027}}
\and C.        ~Cacciari                      \inst{\ref{inst:0044}}
\and J.        ~Casta\~{n}eda                 \inst{\ref{inst:0003}}
\and L.        ~Chaoul                        \inst{\ref{inst:0031}}
\and N.        ~Cheek                         \inst{\ref{inst:0047}}
\and F.        ~De Angeli                     \inst{\ref{inst:0016}}
\and C.        ~Fabricius                     \inst{\ref{inst:0003}}
\and R.        ~Guerra                        \inst{\ref{inst:0027}}
\and B.        ~Holl                          \inst{\ref{inst:0023}}
\and E.        ~Masana                        \inst{\ref{inst:0003}}
\and R.        ~Messineo                      \inst{\ref{inst:0053}}
\and N.        ~Mowlavi                       \inst{\ref{inst:0023}}
\and K.        ~Nienartowicz                  \inst{\ref{inst:0055}}
\and P.        ~Panuzzo                       \inst{\ref{inst:0001}}
\and J.        ~Portell                       \inst{\ref{inst:0003}}
\and M.        ~Riello                        \inst{\ref{inst:0016}}
\and P.        ~Tanga                         \inst{\ref{inst:0030}}
\and F.        ~Th\'{e}venin                  \inst{\ref{inst:0030}}
\and G.        ~Gracia-Abril                  \inst{\ref{inst:0061},\ref{inst:0022}}
\and G.        ~Comoretto                     \inst{\ref{inst:0036}}
\and M.        ~Garcia-Reinaldos              \inst{\ref{inst:0027}}
\and D.        ~Teyssier                      \inst{\ref{inst:0036}}
\and M.        ~Altmann                       \inst{\ref{inst:0022},\ref{inst:0067}}
\and R.        ~Andrae                        \inst{\ref{inst:0021}}
\and M.        ~Audard                        \inst{\ref{inst:0023}}
\and I.        ~Bellas-Velidis                \inst{\ref{inst:0070}}
\and K.        ~Benson                        \inst{\ref{inst:0007}}
\and J.        ~Berthier                      \inst{\ref{inst:0072}}
\and R.        ~Blomme                        \inst{\ref{inst:0073}}
\and P.        ~Burgess                       \inst{\ref{inst:0016}}
\and G.        ~Busso                         \inst{\ref{inst:0016}}
\and B.        ~Carry                         \inst{\ref{inst:0030},\ref{inst:0072}}
\and A.        ~Cellino                       \inst{\ref{inst:0005}}
\and G.        ~Clementini                    \inst{\ref{inst:0044}}
\and M.        ~Clotet                        \inst{\ref{inst:0003}}
\and O.        ~Creevey                       \inst{\ref{inst:0030}}
\and M.        ~Davidson                      \inst{\ref{inst:0083}}
\and J.        ~De Ridder                     \inst{\ref{inst:0084}}
\and L.        ~Delchambre                    \inst{\ref{inst:0085}}
\and A.        ~Dell'Oro                      \inst{\ref{inst:0034}}
\and C.        ~Ducourant                     \inst{\ref{inst:0008}}
\and J.        ~Fern\'{a}ndez-Hern\'{a}ndez   \inst{\ref{inst:0088}}
\and M.        ~Fouesneau                     \inst{\ref{inst:0021}}
\and Y.        ~Fr\'{e}mat                    \inst{\ref{inst:0073}}
\and L.        ~Galluccio                     \inst{\ref{inst:0030}}
\and M.        ~Garc\'{i}a-Torres             \inst{\ref{inst:0092}}
\and J.        ~Gonz\'{a}lez-N\'{u}\~{n}ez    \inst{\ref{inst:0047},\ref{inst:0094}}
\and J.J.      ~Gonz\'{a}lez-Vidal            \inst{\ref{inst:0003}}
\and E.        ~Gosset                        \inst{\ref{inst:0085},\ref{inst:0033}}
\and L.P.      ~Guy                           \inst{\ref{inst:0055},\ref{inst:0099}}
\and J.-L.     ~Halbwachs                     \inst{\ref{inst:0100}}
\and N.C.      ~Hambly                        \inst{\ref{inst:0083}}
\and D.L.      ~Harrison                      \inst{\ref{inst:0016},\ref{inst:0103}}
\and J.        ~Hern\'{a}ndez                 \inst{\ref{inst:0027}}
\and D.        ~Hestroffer                    \inst{\ref{inst:0072}}
\and S.T.      ~Hodgkin                       \inst{\ref{inst:0016}}
\and A.        ~Hutton                        \inst{\ref{inst:0107}}
\and G.        ~Jasniewicz                    \inst{\ref{inst:0108}}
\and A.        ~Jean-Antoine-Piccolo          \inst{\ref{inst:0031}}
\and S.        ~Jordan                        \inst{\ref{inst:0022}}
\and A.J.      ~Korn                          \inst{\ref{inst:0111}}
\and A.        ~Krone-Martins                 \inst{\ref{inst:0112}}
\and A.C.      ~Lanzafame                     \inst{\ref{inst:0113},\ref{inst:0114}}
\and T.        ~Lebzelter                     \inst{\ref{inst:0115}}
\and W.        ~L\"{ o}ffler                  \inst{\ref{inst:0022}}
\and M.        ~Manteiga                      \inst{\ref{inst:0117},\ref{inst:0118}}
\and P.M.      ~Marrese                       \inst{\ref{inst:0119},\ref{inst:0120}}
\and J.M.      ~Mart\'{i}n-Fleitas            \inst{\ref{inst:0107}}
\and A.        ~Moitinho                      \inst{\ref{inst:0112}}
\and A.        ~Mora                          \inst{\ref{inst:0107}}
\and K.        ~Muinonen                      \inst{\ref{inst:0124},\ref{inst:0125}}
\and J.        ~Osinde                        \inst{\ref{inst:0126}}
\and E.        ~Pancino                       \inst{\ref{inst:0034},\ref{inst:0120}}
\and T.        ~Pauwels                       \inst{\ref{inst:0073}}
\and J.-M.     ~Petit                         \inst{\ref{inst:0006}}
\and A.        ~Recio-Blanco                  \inst{\ref{inst:0030}}
\and P.J.      ~Richards                      \inst{\ref{inst:0132}}
\and L.        ~Rimoldini                     \inst{\ref{inst:0055}}
\and L.M.      ~Sarro                         \inst{\ref{inst:0134}}
\and C.        ~Siopis                        \inst{\ref{inst:0032}}
\and M.        ~Smith                         \inst{\ref{inst:0007}}
\and A.        ~Sozzetti                      \inst{\ref{inst:0005}}
\and M.        ~S\"{ u}veges                  \inst{\ref{inst:0021}}
\and J.        ~Torra                         \inst{\ref{inst:0003}}
\and W.        ~van Reeven                    \inst{\ref{inst:0107}}
\and U.        ~Abbas                         \inst{\ref{inst:0005}}
\and A.        ~Abreu Aramburu                \inst{\ref{inst:0142}}
\and S.        ~Accart                        \inst{\ref{inst:0143}}
\and C.        ~Aerts                         \inst{\ref{inst:0084},\ref{inst:0145}}
\and G.        ~Altavilla                     \inst{\ref{inst:0119},\ref{inst:0120},\ref{inst:0044}}
\and M.A.      ~\'{A}lvarez                   \inst{\ref{inst:0117}}
\and R.        ~Alvarez                       \inst{\ref{inst:0027}}
\and J.        ~Alves                         \inst{\ref{inst:0115}}
\and R.I.      ~Anderson                      \inst{\ref{inst:0152},\ref{inst:0023}}
\and A.H.      ~Andrei                        \inst{\ref{inst:0154},\ref{inst:0155},\ref{inst:0067}}
\and E.        ~Anglada Varela                \inst{\ref{inst:0088}}
\and E.        ~Antiche                       \inst{\ref{inst:0003}}
\and B.        ~Arcay                         \inst{\ref{inst:0117}}
\and T.L.      ~Astraatmadja                  \inst{\ref{inst:0021},\ref{inst:0161}}
\and N.        ~Bach                          \inst{\ref{inst:0107}}
\and S.G.      ~Baker                         \inst{\ref{inst:0007}}
\and L.        ~Balaguer-N\'{u}\~{n}ez        \inst{\ref{inst:0003}}
\and P.        ~Balm                          \inst{\ref{inst:0036}}
\and C.        ~Barache                       \inst{\ref{inst:0067}}
\and C.        ~Barata                        \inst{\ref{inst:0112}}
\and D.        ~Barbato                       \inst{\ref{inst:0013},\ref{inst:0005}}
\and F.        ~Barblan                       \inst{\ref{inst:0023}}
\and P.S.      ~Barklem                       \inst{\ref{inst:0111}}
\and D.        ~Barrado                       \inst{\ref{inst:0172}}
\and M.        ~Barros                        \inst{\ref{inst:0112}}
\and M.A.      ~Barstow                       \inst{\ref{inst:0174}}
\and S.        ~Bartholom\'{e} Mu\~{n}oz      \inst{\ref{inst:0003}}
\and J.-L.     ~Bassilana                     \inst{\ref{inst:0143}}
\and U.        ~Becciani                      \inst{\ref{inst:0114}}
\and M.        ~Bellazzini                    \inst{\ref{inst:0044}}
\and A.        ~Berihuete                     \inst{\ref{inst:0179}}
\and S.        ~Bertone                       \inst{\ref{inst:0005},\ref{inst:0067},\ref{inst:0182}}
\and L.        ~Bianchi                       \inst{\ref{inst:0183}}
\and O.        ~Bienaym\'{e}                  \inst{\ref{inst:0100}}
\and S.        ~Blanco-Cuaresma               \inst{\ref{inst:0023},\ref{inst:0008},\ref{inst:0187}}
\and T.        ~Boch                          \inst{\ref{inst:0100}}
\and C.        ~Boeche                        \inst{\ref{inst:0018}}
\and A.        ~Bombrun                       \inst{\ref{inst:0190}}
\and R.        ~Borrachero                    \inst{\ref{inst:0003}}
\and D.        ~Bossini                       \inst{\ref{inst:0018}}
\and S.        ~Bouquillon                    \inst{\ref{inst:0067}}
\and G.        ~Bourda                        \inst{\ref{inst:0008}}
\and A.        ~Bragaglia                     \inst{\ref{inst:0044}}
\and L.        ~Bramante                      \inst{\ref{inst:0053}}
\and M.A.      ~Breddels                      \inst{\ref{inst:0197}}
\and A.        ~Bressan                       \inst{\ref{inst:0198}}
\and N.        ~Brouillet                     \inst{\ref{inst:0008}}
\and T.        ~Br\"{ u}semeister             \inst{\ref{inst:0022}}
\and E.        ~Brugaletta                    \inst{\ref{inst:0114}}
\and B.        ~Bucciarelli                   \inst{\ref{inst:0005}}
\and A.        ~Burlacu                       \inst{\ref{inst:0031}}
\and D.        ~Busonero                      \inst{\ref{inst:0005}}
\and A.G.      ~Butkevich                     \inst{\ref{inst:0026}}
\and R.        ~Buzzi                         \inst{\ref{inst:0005}}
\and E.        ~Caffau                        \inst{\ref{inst:0001}}
\and R.        ~Cancelliere                   \inst{\ref{inst:0208}}
\and G.        ~Cannizzaro                    \inst{\ref{inst:0209},\ref{inst:0145}}
\and T.        ~Cantat-Gaudin                 \inst{\ref{inst:0018},\ref{inst:0003}}
\and R.        ~Carballo                      \inst{\ref{inst:0213}}
\and T.        ~Carlucci                      \inst{\ref{inst:0067}}
\and J.M.      ~Carrasco                      \inst{\ref{inst:0003}}
\and L.        ~Casamiquela                   \inst{\ref{inst:0003}}
\and M.        ~Castellani                    \inst{\ref{inst:0119}}
\and A.        ~Castro-Ginard                 \inst{\ref{inst:0003}}
\and P.        ~Charlot                       \inst{\ref{inst:0008}}
\and L.        ~Chemin                        \inst{\ref{inst:0220}}
\and A.        ~Chiavassa                     \inst{\ref{inst:0030}}
\and G.        ~Cocozza                       \inst{\ref{inst:0044}}
\and G.        ~Costigan                      \inst{\ref{inst:0017}}
\and S.        ~Cowell                        \inst{\ref{inst:0016}}
\and F.        ~Crifo                         \inst{\ref{inst:0001}}
\and M.        ~Crosta                        \inst{\ref{inst:0005}}
\and C.        ~Crowley                       \inst{\ref{inst:0190}}
\and J.        ~Cuypers$^\dagger$             \inst{\ref{inst:0073}}
\and C.        ~Dafonte                       \inst{\ref{inst:0117}}
\and Y.        ~Damerdji                      \inst{\ref{inst:0085},\ref{inst:0231}}
\and A.        ~Dapergolas                    \inst{\ref{inst:0070}}
\and P.        ~David                         \inst{\ref{inst:0072}}
\and M.        ~David                         \inst{\ref{inst:0234}}
\and P.        ~de Laverny                    \inst{\ref{inst:0030}}
\and F.        ~De Luise                      \inst{\ref{inst:0236}}
\and R.        ~De March                      \inst{\ref{inst:0053}}
\and R.        ~de Souza                      \inst{\ref{inst:0238}}
\and A.        ~de Torres                     \inst{\ref{inst:0190}}
\and J.        ~Debosscher                    \inst{\ref{inst:0084}}
\and E.        ~del Pozo                      \inst{\ref{inst:0107}}
\and M.        ~Delbo                         \inst{\ref{inst:0030}}
\and A.        ~Delgado                       \inst{\ref{inst:0016}}
\and H.E.      ~Delgado                       \inst{\ref{inst:0134}}
\and S.        ~Diakite                       \inst{\ref{inst:0006}}
\and C.        ~Diener                        \inst{\ref{inst:0016}}
\and E.        ~Distefano                     \inst{\ref{inst:0114}}
\and C.        ~Dolding                       \inst{\ref{inst:0007}}
\and P.        ~Drazinos                      \inst{\ref{inst:0249}}
\and J.        ~Dur\'{a}n                     \inst{\ref{inst:0126}}
\and B.        ~Edvardsson                    \inst{\ref{inst:0111}}
\and H.        ~Enke                          \inst{\ref{inst:0252}}
\and K.        ~Eriksson                      \inst{\ref{inst:0111}}
\and P.        ~Esquej                        \inst{\ref{inst:0254}}
\and G.        ~Eynard Bontemps               \inst{\ref{inst:0031}}
\and C.        ~Fabre                         \inst{\ref{inst:0256}}
\and M.        ~Fabrizio                      \inst{\ref{inst:0119},\ref{inst:0120}}
\and S.        ~Faigler                       \inst{\ref{inst:0259}}
\and A.J.      ~Falc\~{a}o                    \inst{\ref{inst:0260}}
\and M.        ~Farr\`{a}s Casas              \inst{\ref{inst:0003}}
\and L.        ~Federici                      \inst{\ref{inst:0044}}
\and G.        ~Fedorets                      \inst{\ref{inst:0124}}
\and P.        ~Fernique                      \inst{\ref{inst:0100}}
\and F.        ~Filippi                       \inst{\ref{inst:0053}}
\and K.        ~Findeisen                     \inst{\ref{inst:0001}}
\and A.        ~Fonti                         \inst{\ref{inst:0053}}
\and E.        ~Fraile                        \inst{\ref{inst:0254}}
\and M.        ~Fraser                        \inst{\ref{inst:0016},\ref{inst:0270}}
\and B.        ~Fr\'{e}zouls                  \inst{\ref{inst:0031}}
\and M.        ~Gai                           \inst{\ref{inst:0005}}
\and S.        ~Galleti                       \inst{\ref{inst:0044}}
\and D.        ~Garabato                      \inst{\ref{inst:0117}}
\and F.        ~Garc\'{i}a-Sedano             \inst{\ref{inst:0134}}
\and A.        ~Garofalo                      \inst{\ref{inst:0276},\ref{inst:0044}}
\and N.        ~Garralda                      \inst{\ref{inst:0003}}
\and A.        ~Gavel                         \inst{\ref{inst:0111}}
\and P.        ~Gavras                        \inst{\ref{inst:0001},\ref{inst:0070},\ref{inst:0249}}
\and J.        ~Gerssen                       \inst{\ref{inst:0252}}
\and R.        ~Geyer                         \inst{\ref{inst:0026}}
\and P.        ~Giacobbe                      \inst{\ref{inst:0005}}
\and G.        ~Gilmore                       \inst{\ref{inst:0016}}
\and S.        ~Girona                        \inst{\ref{inst:0287}}
\and G.        ~Giuffrida                     \inst{\ref{inst:0120},\ref{inst:0119}}
\and F.        ~Glass                         \inst{\ref{inst:0023}}
\and M.        ~Gomes                         \inst{\ref{inst:0112}}
\and M.        ~Granvik                       \inst{\ref{inst:0124},\ref{inst:0293}}
\and A.        ~Gueguen                       \inst{\ref{inst:0001},\ref{inst:0295}}
\and A.        ~Guerrier                      \inst{\ref{inst:0143}}
\and J.        ~Guiraud                       \inst{\ref{inst:0031}}
\and R.        ~Guti\'{e}rrez-S\'{a}nchez     \inst{\ref{inst:0036}}
\and R.        ~Haigron                       \inst{\ref{inst:0001}}
\and D.        ~Hatzidimitriou                \inst{\ref{inst:0249},\ref{inst:0070}}
\and M.        ~Hauser                        \inst{\ref{inst:0022},\ref{inst:0021}}
\and M.        ~Haywood                       \inst{\ref{inst:0001}}
\and U.        ~Heiter                        \inst{\ref{inst:0111}}
\and A.        ~Helmi                         \inst{\ref{inst:0197}}
\and J.        ~Heu                           \inst{\ref{inst:0001}}
\and T.        ~Hilger                        \inst{\ref{inst:0026}}
\and D.        ~Hobbs                         \inst{\ref{inst:0028}}
\and W.        ~Hofmann                       \inst{\ref{inst:0022}}
\and G.        ~Holland                       \inst{\ref{inst:0016}}
\and H.E.      ~Huckle                        \inst{\ref{inst:0007}}
\and A.        ~Hypki                         \inst{\ref{inst:0017},\ref{inst:0314}}
\and V.        ~Icardi                        \inst{\ref{inst:0053}}
\and K.        ~Jan{\ss}en                    \inst{\ref{inst:0252}}
\and G.        ~Jevardat de Fombelle          \inst{\ref{inst:0055}}
\and P.G.      ~Jonker                        \inst{\ref{inst:0209},\ref{inst:0145}}
\and \'{A}.L.  ~Juh\'{a}sz                    \inst{\ref{inst:0320},\ref{inst:0321}}
\and F.        ~Julbe                         \inst{\ref{inst:0003}}
\and A.        ~Karampelas                    \inst{\ref{inst:0249},\ref{inst:0324}}
\and A.        ~Kewley                        \inst{\ref{inst:0016}}
\and J.        ~Klar                          \inst{\ref{inst:0252}}
\and A.        ~Kochoska                      \inst{\ref{inst:0328},\ref{inst:0329}}
\and R.        ~Kohley                        \inst{\ref{inst:0027}}
\and K.        ~Kolenberg                     \inst{\ref{inst:0331},\ref{inst:0084},\ref{inst:0187}}
\and M.        ~Kontizas                      \inst{\ref{inst:0249}}
\and E.        ~Kontizas                      \inst{\ref{inst:0070}}
\and S.E.      ~Koposov                       \inst{\ref{inst:0016},\ref{inst:0337}}
\and G.        ~Kordopatis                    \inst{\ref{inst:0030}}
\and Z.        ~Kostrzewa-Rutkowska           \inst{\ref{inst:0209},\ref{inst:0145}}
\and P.        ~Koubsky                       \inst{\ref{inst:0341}}
\and S.        ~Lambert                       \inst{\ref{inst:0067}}
\and A.F.      ~Lanza                         \inst{\ref{inst:0114}}
\and Y.        ~Lasne                         \inst{\ref{inst:0143}}
\and J.-B.     ~Lavigne                       \inst{\ref{inst:0143}}
\and Y.        ~Le Fustec                     \inst{\ref{inst:0346}}
\and C.        ~Le Poncin-Lafitte             \inst{\ref{inst:0067}}
\and Y.        ~Lebreton                      \inst{\ref{inst:0001},\ref{inst:0349}}
\and S.        ~Leccia                        \inst{\ref{inst:0350}}
\and N.        ~Leclerc                       \inst{\ref{inst:0001}}
\and I.        ~Lecoeur-Taibi                 \inst{\ref{inst:0055}}
\and H.        ~Lenhardt                      \inst{\ref{inst:0022}}
\and F.        ~Leroux                        \inst{\ref{inst:0143}}
\and S.        ~Liao                          \inst{\ref{inst:0005},\ref{inst:0356},\ref{inst:0357}}
\and E.        ~Licata                        \inst{\ref{inst:0183}}
\and H.E.P.    ~Lindstr{\o}m                  \inst{\ref{inst:0359},\ref{inst:0360}}
\and T.A.      ~Lister                        \inst{\ref{inst:0361}}
\and E.        ~Livanou                       \inst{\ref{inst:0249}}
\and A.        ~Lobel                         \inst{\ref{inst:0073}}
\and M.        ~L\'{o}pez                     \inst{\ref{inst:0172}}
\and S.        ~Managau                       \inst{\ref{inst:0143}}
\and R.G.      ~Mann                          \inst{\ref{inst:0083}}
\and G.        ~Mantelet                      \inst{\ref{inst:0022}}
\and O.        ~Marchal                       \inst{\ref{inst:0001}}
\and J.M.      ~Marchant                      \inst{\ref{inst:0369}}
\and M.        ~Marconi                       \inst{\ref{inst:0350}}
\and S.        ~Marinoni                      \inst{\ref{inst:0119},\ref{inst:0120}}
\and G.        ~Marschalk\'{o}                \inst{\ref{inst:0320},\ref{inst:0374}}
\and D.J.      ~Marshall                      \inst{\ref{inst:0375}}
\and M.        ~Martino                       \inst{\ref{inst:0053}}
\and G.        ~Marton                        \inst{\ref{inst:0320}}
\and N.        ~Mary                          \inst{\ref{inst:0143}}
\and D.        ~Massari                       \inst{\ref{inst:0197}}
\and G.        ~Matijevi\v{c}                 \inst{\ref{inst:0252}}
\and T.        ~Mazeh                         \inst{\ref{inst:0259}}
\and P.J.      ~McMillan                      \inst{\ref{inst:0028}}
\and S.        ~Messina                       \inst{\ref{inst:0114}}
\and D.        ~Michalik                      \inst{\ref{inst:0028}}
\and N.R.      ~Millar                        \inst{\ref{inst:0016}}
\and D.        ~Molina                        \inst{\ref{inst:0003}}
\and R.        ~Molinaro                      \inst{\ref{inst:0350}}
\and L.        ~Moln\'{a}r                    \inst{\ref{inst:0320}}
\and P.        ~Montegriffo                   \inst{\ref{inst:0044}}
\and R.        ~Mor                           \inst{\ref{inst:0003}}
\and R.        ~Morbidelli                    \inst{\ref{inst:0005}}
\and T.        ~Morel                         \inst{\ref{inst:0085}}
\and D.        ~Morris                        \inst{\ref{inst:0083}}
\and A.F.      ~Mulone                        \inst{\ref{inst:0053}}
\and T.        ~Muraveva                      \inst{\ref{inst:0044}}
\and I.        ~Musella                       \inst{\ref{inst:0350}}
\and G.        ~Nelemans                      \inst{\ref{inst:0145},\ref{inst:0084}}
\and L.        ~Nicastro                      \inst{\ref{inst:0044}}
\and L.        ~Noval                         \inst{\ref{inst:0143}}
\and W.        ~O'Mullane                     \inst{\ref{inst:0027},\ref{inst:0099}}
\and C.        ~Ord\'{e}novic                 \inst{\ref{inst:0030}}
\and D.        ~Ord\'{o}\~{n}ez-Blanco        \inst{\ref{inst:0055}}
\and P.        ~Osborne                       \inst{\ref{inst:0016}}
\and C.        ~Pagani                        \inst{\ref{inst:0174}}
\and I.        ~Pagano                        \inst{\ref{inst:0114}}
\and F.        ~Pailler                       \inst{\ref{inst:0031}}
\and H.        ~Palacin                       \inst{\ref{inst:0143}}
\and L.        ~Palaversa                     \inst{\ref{inst:0016},\ref{inst:0023}}
\and A.        ~Panahi                        \inst{\ref{inst:0259}}
\and M.        ~Pawlak                        \inst{\ref{inst:0413},\ref{inst:0414}}
\and A.M.      ~Piersimoni                    \inst{\ref{inst:0236}}
\and F.-X.     ~Pineau                        \inst{\ref{inst:0100}}
\and E.        ~Plachy                        \inst{\ref{inst:0320}}
\and G.        ~Plum                          \inst{\ref{inst:0001}}
\and E.        ~Poujoulet                     \inst{\ref{inst:0419}}
\and A.        ~Pr\v{s}a                      \inst{\ref{inst:0329}}
\and L.        ~Pulone                        \inst{\ref{inst:0119}}
\and E.        ~Racero                        \inst{\ref{inst:0047}}
\and S.        ~Ragaini                       \inst{\ref{inst:0044}}
\and N.        ~Rambaux                       \inst{\ref{inst:0072}}
\and M.        ~Ramos-Lerate                  \inst{\ref{inst:0425}}
\and S.        ~Regibo                        \inst{\ref{inst:0084}}
\and F.        ~Riclet                        \inst{\ref{inst:0031}}
\and V.        ~Ripepi                        \inst{\ref{inst:0350}}
\and A.        ~Riva                          \inst{\ref{inst:0005}}
\and A.        ~Rivard                        \inst{\ref{inst:0143}}
\and G.        ~Rixon                         \inst{\ref{inst:0016}}
\and T.        ~Roegiers                      \inst{\ref{inst:0432}}
\and M.        ~Roelens                       \inst{\ref{inst:0023}}
\and N.        ~Rowell                        \inst{\ref{inst:0083}}
\and F.        ~Royer                         \inst{\ref{inst:0001}}
\and L.        ~Ruiz-Dern                     \inst{\ref{inst:0001}}
\and G.        ~Sadowski                      \inst{\ref{inst:0032}}
\and T.        ~Sagrist\`{a} Sell\'{e}s       \inst{\ref{inst:0022}}
\and J.        ~Sahlmann                      \inst{\ref{inst:0027},\ref{inst:0440}}
\and J.        ~Salgado                       \inst{\ref{inst:0441}}
\and E.        ~Salguero                      \inst{\ref{inst:0088}}
\and N.        ~Sanna                         \inst{\ref{inst:0034}}
\and T.        ~Santana-Ros                   \inst{\ref{inst:0314}}
\and M.        ~Sarasso                       \inst{\ref{inst:0005}}
\and H.        ~Savietto                      \inst{\ref{inst:0446}}
\and M.        ~Schultheis                    \inst{\ref{inst:0030}}
\and E.        ~Sciacca                       \inst{\ref{inst:0114}}
\and M.        ~Segol                         \inst{\ref{inst:0449}}
\and J.C.      ~Segovia                       \inst{\ref{inst:0047}}
\and D.        ~S\'{e}gransan                 \inst{\ref{inst:0023}}
\and I-C.      ~Shih                          \inst{\ref{inst:0001}}
\and L.        ~Siltala                       \inst{\ref{inst:0124},\ref{inst:0454}}
\and A.F.      ~Silva                         \inst{\ref{inst:0112}}
\and R.L.      ~Smart                         \inst{\ref{inst:0005}}
\and K.W.      ~Smith                         \inst{\ref{inst:0021}}
\and E.        ~Solano                        \inst{\ref{inst:0172},\ref{inst:0459}}
\and F.        ~Solitro                       \inst{\ref{inst:0053}}
\and R.        ~Sordo                         \inst{\ref{inst:0018}}
\and S.        ~Soria Nieto                   \inst{\ref{inst:0003}}
\and J.        ~Souchay                       \inst{\ref{inst:0067}}
\and A.        ~Spagna                        \inst{\ref{inst:0005}}
\and F.        ~Spoto                         \inst{\ref{inst:0030},\ref{inst:0072}}
\and U.        ~Stampa                        \inst{\ref{inst:0022}}
\and I.A.      ~Steele                        \inst{\ref{inst:0369}}
\and H.        ~Steidelm\"{ u}ller            \inst{\ref{inst:0026}}
\and C.A.      ~Stephenson                    \inst{\ref{inst:0036}}
\and H.        ~Stoev                         \inst{\ref{inst:0471}}
\and F.F.      ~Suess                         \inst{\ref{inst:0016}}
\and J.        ~Surdej                        \inst{\ref{inst:0085}}
\and L.        ~Szabados                      \inst{\ref{inst:0320}}
\and E.        ~Szegedi-Elek                  \inst{\ref{inst:0320}}
\and D.        ~Tapiador                      \inst{\ref{inst:0476},\ref{inst:0477}}
\and F.        ~Taris                         \inst{\ref{inst:0067}}
\and G.        ~Tauran                        \inst{\ref{inst:0143}}
\and M.B.      ~Taylor                        \inst{\ref{inst:0480}}
\and R.        ~Teixeira                      \inst{\ref{inst:0238}}
\and D.        ~Terrett                       \inst{\ref{inst:0132}}
\and P.        ~Teyssandier                   \inst{\ref{inst:0067}}
\and W.        ~Thuillot                      \inst{\ref{inst:0072}}
\and A.        ~Titarenko                     \inst{\ref{inst:0030}}
\and F.        ~Torra Clotet                  \inst{\ref{inst:0486}}
\and C.        ~Turon                         \inst{\ref{inst:0001}}
\and A.        ~Ulla                          \inst{\ref{inst:0488}}
\and E.        ~Utrilla                       \inst{\ref{inst:0107}}
\and S.        ~Uzzi                          \inst{\ref{inst:0053}}
\and M.        ~Vaillant                      \inst{\ref{inst:0143}}
\and G.        ~Valentini                     \inst{\ref{inst:0236}}
\and V.        ~Valette                       \inst{\ref{inst:0031}}
\and A.        ~van Elteren                   \inst{\ref{inst:0017}}
\and E.        ~Van Hemelryck                 \inst{\ref{inst:0073}}
\and M.        ~van Leeuwen                   \inst{\ref{inst:0016}}
\and M.        ~Vaschetto                     \inst{\ref{inst:0053}}
\and A.        ~Vecchiato                     \inst{\ref{inst:0005}}
\and J.        ~Veljanoski                    \inst{\ref{inst:0197}}
\and Y.        ~Viala                         \inst{\ref{inst:0001}}
\and D.        ~Vicente                       \inst{\ref{inst:0287}}
\and S.        ~Vogt                          \inst{\ref{inst:0432}}
\and C.        ~von Essen                     \inst{\ref{inst:0503}}
\and H.        ~Voss                          \inst{\ref{inst:0003}}
\and V.        ~Votruba                       \inst{\ref{inst:0341}}
\and S.        ~Voutsinas                     \inst{\ref{inst:0083}}
\and G.        ~Walmsley                      \inst{\ref{inst:0031}}
\and M.        ~Weiler                        \inst{\ref{inst:0003}}
\and O.        ~Wertz                         \inst{\ref{inst:0509}}
\and T.        ~Wevers                        \inst{\ref{inst:0016},\ref{inst:0145}}
\and \L{}.     ~Wyrzykowski                   \inst{\ref{inst:0016},\ref{inst:0413}}
\and A.        ~Yoldas                        \inst{\ref{inst:0016}}
\and M.        ~\v{Z}erjal                    \inst{\ref{inst:0328},\ref{inst:0516}}
\and H.        ~Ziaeepour                     \inst{\ref{inst:0006}}
\and J.        ~Zorec                         \inst{\ref{inst:0518}}
\and S.        ~Zschocke                      \inst{\ref{inst:0026}}
\and S.        ~Zucker                        \inst{\ref{inst:0520}}
\and C.        ~Zurbach                       \inst{\ref{inst:0108}}
\and T.        ~Zwitter                       \inst{\ref{inst:0328}}
}
\institute{
     GEPI, Observatoire de Paris, Universit\'{e} PSL, CNRS, 5 Place Jules Janssen, 92190 Meudon, France\relax                                                                                                \label{inst:0001}
\and Science Support Office, Directorate of Science, European Space Research and Technology Centre (ESA/ESTEC), Keplerlaan 1, 2201AZ, Noordwijk, The Netherlands\relax                                       \label{inst:0002}
\and Institut de Ci\`{e}ncies del Cosmos, Universitat  de  Barcelona  (IEEC-UB), Mart\'{i} i  Franqu\`{e}s  1, 08028 Barcelona, Spain\relax                                                                  \label{inst:0003}
\and INAF - Osservatorio Astrofisico di Torino, via Osservatorio 20, 10025 Pino Torinese (TO), Italy\relax                                                                                                   \label{inst:0005}
\and Institut UTINAM UMR6213, CNRS, OSU THETA Franche-Comt\'{e} Bourgogne, Universit\'{e} Bourgogne Franche-Comt\'{e}, 25000 Besan\c{c}on, France\relax                                                      \label{inst:0006}
\and Mullard Space Science Laboratory, University College London, Holmbury St Mary, Dorking, Surrey RH5 6NT, United Kingdom\relax                                                                            \label{inst:0007}
\and Laboratoire d'astrophysique de Bordeaux, Univ. Bordeaux, CNRS, B18N, all{\'e}e Geoffroy Saint-Hilaire, 33615 Pessac, France\relax                                                                       \label{inst:0008}
\and Univ. Grenoble Alpes, CNRS, IPAG, 38000 Grenoble, France\relax                                                                                                                                          \label{inst:0010}
\and Universit\`{a} di Torino, Dipartimento di Fisica, via Pietro Giuria 1, 10125 Torino, Italy\relax                                                                                                        \label{inst:0013}
\and Institute of Astronomy, University of Cambridge, Madingley Road, Cambridge CB3 0HA, United Kingdom\relax                                                                                                \label{inst:0016}
\and Leiden Observatory, Leiden University, Niels Bohrweg 2, 2333 CA Leiden, The Netherlands\relax                                                                                                           \label{inst:0017}
\and INAF - Osservatorio astronomico di Padova, Vicolo Osservatorio 5, 35122 Padova, Italy\relax                                                                                                             \label{inst:0018}
\and Max Planck Institute for Astronomy, K\"{ o}nigstuhl 17, 69117 Heidelberg, Germany\relax                                                                                                                 \label{inst:0021}
\and Astronomisches Rechen-Institut, Zentrum f\"{ u}r Astronomie der Universit\"{ a}t Heidelberg, M\"{ o}nchhofstr. 12-14, 69120 Heidelberg, Germany\relax                                                   \label{inst:0022}
\and Department of Astronomy, University of Geneva, Chemin des Maillettes 51, 1290 Versoix, Switzerland\relax                                                                                                \label{inst:0023}
\and Mission Operations Division, Operations Department, Directorate of Science, European Space Research and Technology Centre (ESA/ESTEC), Keplerlaan 1, 2201 AZ, Noordwijk, The Netherlands\relax          \label{inst:0024}
\and Lohrmann Observatory, Technische Universit\"{ a}t Dresden, Mommsenstra{\ss}e 13, 01062 Dresden, Germany\relax                                                                                           \label{inst:0026}
\and European Space Astronomy Centre (ESA/ESAC), Camino bajo del Castillo, s/n, Urbanizacion Villafranca del Castillo, Villanueva de la Ca\~{n}ada, 28692 Madrid, Spain\relax                                \label{inst:0027}
\and Lund Observatory, Department of Astronomy and Theoretical Physics, Lund University, Box 43, 22100 Lund, Sweden\relax                                                                                    \label{inst:0028}
\and Universit\'{e} C\^{o}te d'Azur, Observatoire de la C\^{o}te d'Azur, CNRS, Laboratoire Lagrange, Bd de l'Observatoire, CS 34229, 06304 Nice Cedex 4, France\relax                                        \label{inst:0030}
\and CNES Centre Spatial de Toulouse, 18 avenue Edouard Belin, 31401 Toulouse Cedex 9, France\relax                                                                                                          \label{inst:0031}
\and Institut d'Astronomie et d'Astrophysique, Universit\'{e} Libre de Bruxelles CP 226, Boulevard du Triomphe, 1050 Brussels, Belgium\relax                                                                 \label{inst:0032}
\and F.R.S.-FNRS, Rue d'Egmont 5, 1000 Brussels, Belgium\relax                                                                                                                                               \label{inst:0033}
\and INAF - Osservatorio Astrofisico di Arcetri, Largo Enrico Fermi 5, 50125 Firenze, Italy\relax                                                                                                            \label{inst:0034}
\and Telespazio Vega UK Ltd for ESA/ESAC, Camino bajo del Castillo, s/n, Urbanizacion Villafranca del Castillo, Villanueva de la Ca\~{n}ada, 28692 Madrid, Spain\relax                                       \label{inst:0036}
\and INAF - Osservatorio di Astrofisica e Scienza dello Spazio di Bologna, via Piero Gobetti 93/3, 40129 Bologna, Italy\relax                                                                                \label{inst:0044}
\and Serco Gesti\'{o}n de Negocios for ESA/ESAC, Camino bajo del Castillo, s/n, Urbanizacion Villafranca del Castillo, Villanueva de la Ca\~{n}ada, 28692 Madrid, Spain\relax                                \label{inst:0047}
\and ALTEC S.p.a, Corso Marche, 79,10146 Torino, Italy\relax                                                                                                                                                 \label{inst:0053}
\and Department of Astronomy, University of Geneva, Chemin d'Ecogia 16, 1290 Versoix, Switzerland\relax                                                                                                      \label{inst:0055}
\and Gaia DPAC Project Office, ESAC, Camino bajo del Castillo, s/n, Urbanizacion Villafranca del Castillo, Villanueva de la Ca\~{n}ada, 28692 Madrid, Spain\relax                                            \label{inst:0061}
\and SYRTE, Observatoire de Paris, Universit\'{e} PSL, CNRS, Sorbonne Universit\'{e}, LNE, 61 avenue de l’Observatoire 75014 Paris, France\relax                                                         \label{inst:0067}
\and National Observatory of Athens, I. Metaxa and Vas. Pavlou, Palaia Penteli, 15236 Athens, Greece\relax                                                                                                   \label{inst:0070}
\and IMCCE, Observatoire de Paris, Universit\'{e} PSL, CNRS, Sorbonne Universit\'{e}, Univ. Lille, 77 av. Denfert-Rochereau, 75014 Paris, France\relax                                                     \label{inst:0072}
\and Royal Observatory of Belgium, Ringlaan 3, 1180 Brussels, Belgium\relax                                                                                                                                  \label{inst:0073}
\and Institute for Astronomy, University of Edinburgh, Royal Observatory, Blackford Hill, Edinburgh EH9 3HJ, United Kingdom\relax                                                                            \label{inst:0083}
\and Instituut voor Sterrenkunde, KU Leuven, Celestijnenlaan 200D, 3001 Leuven, Belgium\relax                                                                                                                \label{inst:0084}
\and Institut d'Astrophysique et de G\'{e}ophysique, Universit\'{e} de Li\`{e}ge, 19c, All\'{e}e du 6 Ao\^{u}t, B-4000 Li\`{e}ge, Belgium\relax                                                              \label{inst:0085}
\and ATG Europe for ESA/ESAC, Camino bajo del Castillo, s/n, Urbanizacion Villafranca del Castillo, Villanueva de la Ca\~{n}ada, 28692 Madrid, Spain\relax                                                   \label{inst:0088}
\and \'{A}rea de Lenguajes y Sistemas Inform\'{a}ticos, Universidad Pablo de Olavide, Ctra. de Utrera, km 1. 41013, Sevilla, Spain\relax                                                                     \label{inst:0092}
\and ETSE Telecomunicaci\'{o}n, Universidade de Vigo, Campus Lagoas-Marcosende, 36310 Vigo, Galicia, Spain\relax                                                                                             \label{inst:0094}
\and Large Synoptic Survey Telescope, 950 N. Cherry Avenue, Tucson, AZ 85719, USA\relax                                                                                                                      \label{inst:0099}
\and Observatoire Astronomique de Strasbourg, Universit\'{e} de Strasbourg, CNRS, UMR 7550, 11 rue de l'Universit\'{e}, 67000 Strasbourg, France\relax                                                       \label{inst:0100}
\and Kavli Institute for Cosmology, University of Cambridge, Madingley Road, Cambride CB3 0HA, United Kingdom\relax                                                                                          \label{inst:0103}
\and Aurora Technology for ESA/ESAC, Camino bajo del Castillo, s/n, Urbanizacion Villafranca del Castillo, Villanueva de la Ca\~{n}ada, 28692 Madrid, Spain\relax                                            \label{inst:0107}
\and Laboratoire Univers et Particules de Montpellier, Universit\'{e} Montpellier, Place Eug\`{e}ne Bataillon, CC72, 34095 Montpellier Cedex 05, France\relax                                                \label{inst:0108}
\and Department of Physics and Astronomy, Division of Astronomy and Space Physics, Uppsala University, Box 516, 75120 Uppsala, Sweden\relax                                                                  \label{inst:0111}
\and CENTRA, Universidade de Lisboa, FCUL, Campo Grande, Edif. C8, 1749-016 Lisboa, Portugal\relax                                                                                                           \label{inst:0112}
\and Universit\`{a} di Catania, Dipartimento di Fisica e Astronomia, Sezione Astrofisica, Via S. Sofia 78, 95123 Catania, Italy\relax                                                                        \label{inst:0113}
\and INAF - Osservatorio Astrofisico di Catania, via S. Sofia 78, 95123 Catania, Italy\relax                                                                                                                 \label{inst:0114}
\and University of Vienna, Department of Astrophysics, T\"{ u}rkenschanzstra{\ss}e 17, A1180 Vienna, Austria\relax                                                                                           \label{inst:0115}
\and CITIC – Department of Computer Science, University of A Coru\~{n}a, Campus de Elvi\~{n}a S/N, 15071, A Coru\~{n}a, Spain\relax                                                                        \label{inst:0117}
\and CITIC – Astronomy and Astrophysics, University of A Coru\~{n}a, Campus de Elvi\~{n}a S/N, 15071, A Coru\~{n}a, Spain\relax                                                                            \label{inst:0118}
\and INAF - Osservatorio Astronomico di Roma, Via di Frascati 33, 00078 Monte Porzio Catone (Roma), Italy\relax                                                                                              \label{inst:0119}
\and Space Science Data Center - ASI, Via del Politecnico SNC, 00133 Roma, Italy\relax                                                                                                                       \label{inst:0120}
\and University of Helsinki, Department of Physics, P.O. Box 64, FI-00014 Helsinki, Finland\relax                                                                                                            \label{inst:0124}
\and Finnish Geospatial Research Institute FGI, Geodeetinrinne 2, FI-02430 Masala, Finland\relax                                                                                                             \label{inst:0125}
\and Isdefe for ESA/ESAC, Camino bajo del Castillo, s/n, Urbanizacion Villafranca del Castillo, Villanueva de la Ca\~{n}ada, 28692 Madrid, Spain\relax                                                       \label{inst:0126}
\and STFC, Rutherford Appleton Laboratory, Harwell, Didcot, OX11 0QX, United Kingdom\relax                                                                                                                   \label{inst:0132}
\and Dpto. de Inteligencia Artificial, UNED, c/ Juan del Rosal 16, 28040 Madrid, Spain\relax                                                                                                                 \label{inst:0134}
\and Elecnor Deimos Space for ESA/ESAC, Camino bajo del Castillo, s/n, Urbanizacion Villafranca del Castillo, Villanueva de la Ca\~{n}ada, 28692 Madrid, Spain\relax                                         \label{inst:0142}
\and Thales Services for CNES Centre Spatial de Toulouse, 18 avenue Edouard Belin, 31401 Toulouse Cedex 9, France\relax                                                                                      \label{inst:0143}
\and Department of Astrophysics/IMAPP, Radboud University, P.O.Box 9010, 6500 GL Nijmegen, The Netherlands\relax                                                                                             \label{inst:0145}
\and European Southern Observatory, Karl-Schwarzschild-Str. 2, 85748 Garching, Germany\relax                                                                                                                 \label{inst:0152}
\and ON/MCTI-BR, Rua Gal. Jos\'{e} Cristino 77, Rio de Janeiro, CEP 20921-400, RJ,  Brazil\relax                                                                                                             \label{inst:0154}
\and OV/UFRJ-BR, Ladeira Pedro Ant\^{o}nio 43, Rio de Janeiro, CEP 20080-090, RJ, Brazil\relax                                                                                                               \label{inst:0155}
\and Department of Terrestrial Magnetism, Carnegie Institution for Science, 5241 Broad Branch Road, NW, Washington, DC 20015-1305, USA\relax                                                                 \label{inst:0161}
\and Departamento de Astrof\'{i}sica, Centro de Astrobiolog\'{i}a (CSIC-INTA), ESA-ESAC. Camino Bajo del Castillo s/n. 28692 Villanueva de la Ca\~{n}ada, Madrid, Spain\relax                                \label{inst:0172}
\and Leicester Institute of Space and Earth Observation and Department of Physics and Astronomy, University of Leicester, University Road, Leicester LE1 7RH, United Kingdom\relax                           \label{inst:0174}
\and Departamento de Estad\'{i}stica, Universidad de C\'{a}diz, Calle Rep\'{u}blica \'{A}rabe Saharawi s/n. 11510, Puerto Real, C\'{a}diz, Spain\relax                                                       \label{inst:0179}
\and Astronomical Institute Bern University, Sidlerstrasse 5, 3012 Bern, Switzerland (present address)\relax                                                                                                 \label{inst:0182}
\and EURIX S.r.l., Corso Vittorio Emanuele II 61, 10128, Torino, Italy\relax                                                                                                                                 \label{inst:0183}
\and Harvard-Smithsonian Center for Astrophysics, 60 Garden Street, Cambridge MA 02138, USA\relax                                                                                                            \label{inst:0187}
\and HE Space Operations BV for ESA/ESAC, Camino bajo del Castillo, s/n, Urbanizacion Villafranca del Castillo, Villanueva de la Ca\~{n}ada, 28692 Madrid, Spain\relax                                       \label{inst:0190}
\and Kapteyn Astronomical Institute, University of Groningen, Landleven 12, 9747 AD Groningen, The Netherlands\relax                                                                                         \label{inst:0197}
\and SISSA - Scuola Internazionale Superiore di Studi Avanzati, via Bonomea 265, 34136 Trieste, Italy\relax                                                                                                  \label{inst:0198}
\and University of Turin, Department of Computer Sciences, Corso Svizzera 185, 10149 Torino, Italy\relax                                                                                                     \label{inst:0208}
\and SRON, Netherlands Institute for Space Research, Sorbonnelaan 2, 3584CA, Utrecht, The Netherlands\relax                                                                                                  \label{inst:0209}
\and Dpto. de Matem\'{a}tica Aplicada y Ciencias de la Computaci\'{o}n, Univ. de Cantabria, ETS Ingenieros de Caminos, Canales y Puertos, Avda. de los Castros s/n, 39005 Santander, Spain\relax             \label{inst:0213}
\and Unidad de Astronom\'ia, Universidad de Antofagasta, Avenida Angamos 601, Antofagasta 1270300, Chile\relax                                                                                               \label{inst:0220}
\and CRAAG - Centre de Recherche en Astronomie, Astrophysique et G\'{e}ophysique, Route de l'Observatoire Bp 63 Bouzareah 16340 Algiers, Algeria\relax                                                       \label{inst:0231}
\and University of Antwerp, Onderzoeksgroep Toegepaste Wiskunde, Middelheimlaan 1, 2020 Antwerp, Belgium\relax                                                                                               \label{inst:0234}
\and INAF - Osservatorio Astronomico d'Abruzzo, Via Mentore Maggini, 64100 Teramo, Italy\relax                                                                                                               \label{inst:0236}
\and Instituto de Astronomia, Geof\`{i}sica e Ci\^{e}ncias Atmosf\'{e}ricas, Universidade de S\~{a}o Paulo, Rua do Mat\~{a}o, 1226, Cidade Universitaria, 05508-900 S\~{a}o Paulo, SP, Brazil\relax          \label{inst:0238}
\and Department of Astrophysics, Astronomy and Mechanics, National and Kapodistrian University of Athens, Panepistimiopolis, Zografos, 15783 Athens, Greece\relax                                            \label{inst:0249}
\and Leibniz Institute for Astrophysics Potsdam (AIP), An der Sternwarte 16, 14482 Potsdam, Germany\relax                                                                                                    \label{inst:0252}
\and RHEA for ESA/ESAC, Camino bajo del Castillo, s/n, Urbanizacion Villafranca del Castillo, Villanueva de la Ca\~{n}ada, 28692 Madrid, Spain\relax                                                         \label{inst:0254}
\and ATOS for CNES Centre Spatial de Toulouse, 18 avenue Edouard Belin, 31401 Toulouse Cedex 9, France\relax                                                                                                 \label{inst:0256}
\and School of Physics and Astronomy, Tel Aviv University, Tel Aviv 6997801, Israel\relax                                                                                                                    \label{inst:0259}
\and UNINOVA - CTS, Campus FCT-UNL, Monte da Caparica, 2829-516 Caparica, Portugal\relax                                                                                                                     \label{inst:0260}
\and School of Physics, O'Brien Centre for Science North, University College Dublin, Belfield, Dublin 4, Ireland\relax                                                                                       \label{inst:0270}
\and Dipartimento di Fisica e Astronomia, Universit\`{a} di Bologna, Via Piero Gobetti 93/2, 40129 Bologna, Italy\relax                                                                                      \label{inst:0276}
\and Barcelona Supercomputing Center - Centro Nacional de Supercomputaci\'{o}n, c/ Jordi Girona 29, Ed. Nexus II, 08034 Barcelona, Spain\relax                                                               \label{inst:0287}
\and Department of Computer Science, Electrical and Space Engineering, Lule\aa{} University of Technology, Box 848, S-981 28 Kiruna, Sweden\relax                                                            \label{inst:0293}
\and Max Planck Institute for Extraterrestrial Physics, High Energy Group, Gie{\ss}enbachstra{\ss}e, 85741 Garching, Germany\relax                                                                           \label{inst:0295}
\and Astronomical Observatory Institute, Faculty of Physics, Adam Mickiewicz University, S{\l}oneczna 36, 60-286 Pozna{\'n}, Poland\relax                                                                    \label{inst:0314}
\and Konkoly Observatory, Research Centre for Astronomy and Earth Sciences, Hungarian Academy of Sciences, Konkoly Thege Mikl\'{o}s \'{u}t 15-17, 1121 Budapest, Hungary\relax                               \label{inst:0320}
\and E\"{ o}tv\"{ o}s Lor\'and University, Egyetem t\'{e}r 1-3, H-1053 Budapest, Hungary\relax                                                                                                               \label{inst:0321}
\and American Community Schools of Athens, 129 Aghias Paraskevis Ave. \& Kazantzaki Street, Halandri, 15234 Athens, Greece\relax                                                                             \label{inst:0324}
\and Faculty of Mathematics and Physics, University of Ljubljana, Jadranska ulica 19, 1000 Ljubljana, Slovenia\relax                                                                                         \label{inst:0328}
\and Villanova University, Department of Astrophysics and Planetary Science, 800 E Lancaster Avenue, Villanova PA 19085, USA\relax                                                                           \label{inst:0329}
\and Physics Department, University of Antwerp, Groenenborgerlaan 171, 2020 Antwerp, Belgium\relax                                                                                                           \label{inst:0331}
\and McWilliams Center for Cosmology, Department of Physics, Carnegie Mellon University, 5000 Forbes Avenue, Pittsburgh, PA 15213, USA\relax                                                                 \label{inst:0337}
\and Astronomical Institute, Academy of Sciences of the Czech Republic, Fri\v{c}ova 298, 25165 Ond\v{r}ejov, Czech Republic\relax                                                                            \label{inst:0341}
\and Telespazio for CNES Centre Spatial de Toulouse, 18 avenue Edouard Belin, 31401 Toulouse Cedex 9, France\relax                                                                                           \label{inst:0346}
\and Institut de Physique de Rennes, Universit{\'e} de Rennes 1, 35042 Rennes, France\relax                                                                                                                  \label{inst:0349}
\and INAF - Osservatorio Astronomico di Capodimonte, Via Moiariello 16, 80131, Napoli, Italy\relax                                                                                                           \label{inst:0350}
\and Shanghai Astronomical Observatory, Chinese Academy of Sciences, 80 Nandan Rd, 200030 Shanghai, China\relax                                                                                              \label{inst:0356}
\and School of Astronomy and Space Science, University of Chinese Academy of Sciences, Beijing 100049, China\relax                                                                                           \label{inst:0357}
\and Niels Bohr Institute, University of Copenhagen, Juliane Maries Vej 30, 2100 Copenhagen {\O}, Denmark\relax                                                                                              \label{inst:0359}
\and DXC Technology, Retortvej 8, 2500 Valby, Denmark\relax                                                                                                                                                  \label{inst:0360}
\and Las Cumbres Observatory, 6740 Cortona Drive Suite 102, Goleta, CA 93117, USA\relax                                                                                                                      \label{inst:0361}
\and Astrophysics Research Institute, Liverpool John Moores University, 146 Brownlow Hill, Liverpool L3 5RF, United Kingdom\relax                                                                            \label{inst:0369}
\and Baja Observatory of University of Szeged, Szegedi \'{u}t III/70, 6500 Baja, Hungary\relax                                                                                                               \label{inst:0374}
\and Laboratoire AIM, IRFU/Service d'Astrophysique - CEA/DSM - CNRS - Universit\'{e} Paris Diderot, B\^{a}t 709, CEA-Saclay, 91191 Gif-sur-Yvette Cedex, France\relax                                        \label{inst:0375}
\and Warsaw University Observatory, Al. Ujazdowskie 4, 00-478 Warszawa, Poland\relax                                                                                                                         \label{inst:0413}
\and Institute of Theoretical Physics, Faculty of Mathematics and Physics, Charles University in Prague, Czech Republic\relax                                                                                \label{inst:0414}
\and AKKA for CNES Centre Spatial de Toulouse, 18 avenue Edouard Belin, 31401 Toulouse Cedex 9, France\relax                                                                                                 \label{inst:0419}
\and Vitrociset Belgium for ESA/ESAC, Camino bajo del Castillo, s/n, Urbanizacion Villafranca del Castillo, Villanueva de la Ca\~{n}ada, 28692 Madrid, Spain\relax                                           \label{inst:0425}
\and HE Space Operations BV for ESA/ESTEC, Keplerlaan 1, 2201AZ, Noordwijk, The Netherlands\relax                                                                                                            \label{inst:0432}
\and Space Telescope Science Institute, 3700 San Martin Drive, Baltimore, MD 21218, USA\relax                                                                                                                \label{inst:0440}
\and QUASAR Science Resources for ESA/ESAC, Camino bajo del Castillo, s/n, Urbanizacion Villafranca del Castillo, Villanueva de la Ca\~{n}ada, 28692 Madrid, Spain\relax                                     \label{inst:0441}
\and Fork Research, Rua do Cruzado Osberno, Lt. 1, 9 esq., Lisboa, Portugal\relax                                                                                                                            \label{inst:0446}
\and APAVE SUDEUROPE SAS for CNES Centre Spatial de Toulouse, 18 avenue Edouard Belin, 31401 Toulouse Cedex 9, France\relax                                                                                  \label{inst:0449}
\and Nordic Optical Telescope, Rambla Jos\'{e} Ana Fern\'{a}ndez P\'{e}rez 7, 38711 Bre\~{n}a Baja, Spain\relax                                                                                              \label{inst:0454}
\and Spanish Virtual Observatory\relax                                                                                                                                                                       \label{inst:0459}
\and Fundaci\'{o}n Galileo Galilei - INAF, Rambla Jos\'{e} Ana Fern\'{a}ndez P\'{e}rez 7, 38712 Bre\~{n}a Baja, Santa Cruz de Tenerife, Spain\relax                                                          \label{inst:0471}
\and INSA for ESA/ESAC, Camino bajo del Castillo, s/n, Urbanizacion Villafranca del Castillo, Villanueva de la Ca\~{n}ada, 28692 Madrid, Spain\relax                                                         \label{inst:0476}
\and Dpto. Arquitectura de Computadores y Autom\'{a}tica, Facultad de Inform\'{a}tica, Universidad Complutense de Madrid, C/ Prof. Jos\'{e} Garc\'{i}a Santesmases s/n, 28040 Madrid, Spain\relax            \label{inst:0477}
\and H H Wills Physics Laboratory, University of Bristol, Tyndall Avenue, Bristol BS8 1TL, United Kingdom\relax                                                                                              \label{inst:0480}
\and Institut d'Estudis Espacials de Catalunya (IEEC), Gran Capita 2-4, 08034 Barcelona, Spain\relax                                                                                                         \label{inst:0486}
\and Applied Physics Department, Universidade de Vigo, 36310 Vigo, Spain\relax                                                                                                                               \label{inst:0488}
\and Stellar Astrophysics Centre, Aarhus University, Department of Physics and Astronomy, 120 Ny Munkegade, Building 1520, DK-8000 Aarhus C, Denmark\relax                                                   \label{inst:0503}
\and Argelander-Institut f\"{ ur} Astronomie, Universit\"{ a}t Bonn,  Auf dem H\"{ u}gel 71, 53121 Bonn, Germany\relax                                                                                       \label{inst:0509}
\and Research School of Astronomy and Astrophysics, Australian National University, Canberra, ACT 2611 Australia\relax                                                                                       \label{inst:0516}
\and Sorbonne Universit\'{e}s, UPMC Univ. Paris 6 et CNRS, UMR 7095, Institut d'Astrophysique de Paris, 98 bis bd. Arago, 75014 Paris, France\relax                                                          \label{inst:0518}
\and Department of Geosciences, Tel Aviv University, Tel Aviv 6997801, Israel\relax                                                                                                                          \label{inst:0520}
}

   \date{Received ; accepted }


  \abstract
   {The second Gaia data release (\gdrtwo) contains high-precision positions, parallaxes, and proper motions for 1.3 billion sources as well as line-of-sight velocities for 7.2 million stars brighter than $\grvs = 12$~mag. Both samples provide a full sky coverage.}
   {To illustrate the potential of \gdrtwo{}, we provide a first look at the kinematics of the Milky Way disc, within a radius of several kiloparsecs around the Sun.}
   {We benefit for the first time from a sample of {6.4} million F-G-K stars with full 6D phase-space coordinates, precise parallaxes ($\sigma_\varpi / \varpi \leq 20\%$), and precise Galactic cylindrical velocities (median uncertainties of 0.9-1.4 $\kms$ and {20}\% of the stars with uncertainties smaller than 1~$\kms$ on all three components). From this sample, we extracted a sub-sample of {3.2} million giant stars to map the velocity field of the Galactic disc from $\sim$5~kpc to $\sim$13~kpc from the Galactic centre and up to 2~kpc above and below the plane. We also study the distribution of 0.3 million solar neighbourhood stars ($r < 200$~pc){,} with median velocity uncertainties of {0.4}~$\kms${,} in velocity space and use the full sample to examine how the over-densities evolve in more distant regions.}
   {\gdrtwo{} allows us to draw 3D maps of the Galactocentric median velocities and velocity dispersions with unprecedented accuracy, precision, and spatial resolution. The maps show the complexity and richness of the velocity field of the galactic disc. We observe streaming motions in all the components of the velocities as well as patterns in the velocity dispersions. For example, we confirm the previously reported negative and positive galactocentric radial velocity gradients in the inner and outer disc, respectively. Here, we see them as part of a \textup{non-axisymmetric} kinematic oscillation, and we map its azimuthal and vertical behaviour. We also witness a new global arrangement of stars in the velocity plane of the solar neighbourhood and in distant regions in which stars are organised in thin substructures with the shape of circular arches that are oriented approximately along the horizontal direction in the $U-V$ plane.
   Moreover, in distant regions, we see  variations in the velocity substructures more clearly than ever
before, in particular, variations in the velocity of the Hercules stream.}
   {\gdrtwo{} provides the largest existing full 6D phase-space coordinates catalogue. It also vastly increases the number of available distances and transverse velocities with respect to \gdrone{}. \gdrtwo{} offers a great wealth of information on the Milky Way {and reveals clear non-axisymmetric kinematic signatures within the Galactic disc, for instance}. It is now up to the astronomical community to explore its full potential.}


   \maketitle


\section{Introduction}\label{intro}
Our position in the disc of the Milky Way does not allow us to capture the global picture of our galaxy easily. Mapping its 3D structure requires large and precise astrometric catalogues. The second \gaia{} data release (\gaia~DR2, \citet{DR2-DPACP-36}) contains positions and parallaxes for 1.3 billions sources down to magnitude $G \sim 21$~mag, which multiplies by a huge factor the number of stars for which a distance can be derived with respect to \gaia~DR1. Not only does \gaia~DR2 provide the 3D location of a very large sample of stars in the Galaxy, it also contains full velocity information (proper motions and line-of-sight velocity) for 7.2~million stars brighter than $\grvs = 12$~mag, and
transverse velocity for an unprecedentedly large number of stars. This paper belongs to a series of six \gdrtwo{} performance verification papers that are meant to demonstrate the quality of the catalogue through a basic examination of some of the key science cases of the \gaia{}  mission. In this paper, we report a first look at the kinematic properties of the Milky Way disc as pictured by the second \gaia{} data release.

\gdrtwo{} contains unprecedented information about the Galaxy, {which should allow us to infer} its current structure, its equilibrium state,
its evolution, modes of mass growth over time, dark matter distribution {(and perhaps nature),}  to cite a few of the
questions of modern Galactic astrophysics. As an example, it has been known for several decades that the
Galactic disc contains large-scale non-axisymmetric features, including a central boxy/peanut-shaped bar \citep{okuda77, maihara78, weiland94, dwek95, binney97, babusiaux05, lopez05, rattenbury07, cao13}
and its possible in-plane extension \citep{hammersley00, benjamin05, cabrera07, wegg15}, a warp \citep{burke57, kerr57, westerhout57, weaver74, djorgovski89, evans98, gyuk99, Drimmel2001, lopez02, momany06, robin08, reyle09, Amores2017}, and spiral arms \citep{GG1976, TC1993, drimmel00, bissantz02, Churchwell2009, vallee14, Reid2014, hachisuka15, hou15}. However, full knowledge of these asymmetric
structures, that is, of their spatial extent, pattern speeds, and number (in case of spiral arms) is still lacking.
Since asymmetries constitute the driver of the secular evolution in galaxy discs  {(see e.g. \citealt{minchev12, fouvry15, halle15, aumer17} and   \citealt{kormendy13}, for a review)} by redistributing angular
momentum between the inner and outer disc and between its baryonic and dark matter content \citep{debattista00, bournaud02, athanassoula03, martinez06, combes11}, quantifying their
characteristics is fundamental for understanding to what extent the Milky Way has "simply" evolved secularly in the last
$\sim$9 Gyr \citep{hammer07, martig14}, or whether some more complex evolutionary scenarios need to be invoked. 

Non-axisymmetric features manifest themselves not only in configuration spaces, but also in kinematic spaces, where they leave specific signatures related to their spatial extension, rotation speed around the Galaxy centre, and growth rate {\citep{SiebertEtAl2012, FaureEtAl2014, MonariEtAl2014, Debattista2014, bovy15, grand15, MonariEtAl2016b, Grand2016,  Antoja2016, pasetto16}}.
Many  studies prior to \gaia{} \citep{eggen58, eggen96, Chereul1999, Dehnen1998a, Famaey2005, Antoja2008, Gomez2012b}, and especially since the epoch of the {\it Hipparcos} satellite
\citep{perryman97}, have studied the kinematics of stars in the solar neighborhood and have shown that the stellar velocity and phase-space distributions
are not smooth, but rather clumpy.  Several hypotheses were able
to explain the nature of this clumpy distribution, suggesting that they might be  remnants of stellar clusters \citep{eggen96},  substructures related to
orbital effects of the bar and/or the spiral arms
\citep[e.g.][]{Dehnen2000, Desimone2004, Quillen2005,  Chakrabarty2007, Antoja2009}, remnants of accreted systems
from the halo \citep{Helmi1999b, Villalobos2009, gomez10, refiorentin15, jeanbaptiste17}, or substructures induced in the stellar disc by external perturbations
\citep{Quillen2009, Minchev2009,  Gomez2012a, jeanbaptiste17}. Despite all this theoretical and observational work, it is still
an open issue how we
can distinguish between the different types of substructures.
 With RAVE \citep{Steinmetz2006}, LAMOST \citep{liu14} combined with TGAS \citep{GaiaBrown2016, Lindegren2016}, and APOGEE-2 South \citep{majewski16, Majewski2017}, \citet{Antoja2012, Antoja2014}, \citet{Monari2017} and \citet{Hunt2018} concluded that at least one of these substructures, the Hercules stream, evolves with Galactic radius, consistently with the effects of the Outer Lindlblad Resonance of the bar. However, other studies have suggested a pattern speed for the Milky Way bar that is slower than previous estimates, placing this resonance well outside the solar radius \citep{liu12, portail17, perez17}. 
To understand  the role of the stellar bar, it
is necessary both to map the kinematics of disc stars in the Galaxy over a larger spatial extent and to increase the
statistics (the number of stars with full 3D kinematic information) out to a few kpc from
the Sun. Extending the spatial scale of kinematic studies to larger regions of the Galactic disc is also essential for quantifying the amplitude of velocity gradients, detection of which is now limited to a region of a few kiloparsec  around the Sun (see  \citet{SiebertEtAl2011, CarrilloEtAl2017, LiuEtAl2017}),  and constrain their origin.

In addition to secular evolutionary processes, a disc galaxy like ours is expected to have experienced several
accretion events in its recent and early past  \citep{bullock05, delucia08, stewart08, cooper10, font11, brook12, martig12, pillepich15, deason16, rodriguez16}. While some of these accretions  are currently being caught in the act,
like {for} the Sagittarius dwarf galaxy  \citep{ibata94} and the Magellanic Clouds \citep{mathewson74, nidever10, Donghia2016a}, we need to find the vestiges of ancient accretion events to understand the evolution of our Galaxy
and how its mass growth has proceeded over time. 
Events that took place in the far past are expected to have induced a thickening of the early Galactic disc, first by
increasing the in-plane and vertical velocity dispersion of stars \citep{quinn93, walker96, villalobos08, Villalobos2009, zolotov09, purcell10, dimatteo11, qu11, font11, mccarthy12, cooper15, welker17}, and second by
agitating the gaseous disc from which new stars are born, generating early stellar populations with higher initial
velocity dispersions than those currently being formed \citep{brook04, brook07, forbes12, bird13, stinson13}. These complementary modes of formation of the Galactic disc can be imprinted on kinematics-age and kinematics-abundance relations \citep{stromberg46, spitzer51, Nordstrom2004, seabroke07, holmberg07, holmberg09, bovy12, haywood13, sharma14, bovy16, martig16, ness16, mackereth17, Robin2017}, and distinguishing between them requires full 3D kinematic information
for several million stars, in order to be able to separate the contribution of accreted from in-situ populations,
and to constrain impulsive signatures that are typical of accretions \citep{minchev14} versus a more quiescent cooling of the
Galactic disc over time.
Accretion events that took place in the more recent past of our Galaxy can also generate ripples and rings in a
galactic disc  \citep{Gomez2012a}, as well as in the inner stellar halo \citep{jeanbaptiste17}. 
{Such vertical perturbations of the disc are further complicated by the effect of spiral arms \citep{Donghia2016b, MonariEtAl2016b}, which together with the effect of accretion events might explain vertical wave modes as observed in SEGUE and RAVE \citep{Widrow2012, Williams2013, CarrilloEtAl2017}, as well as in-plane velocity anisotropy \citep{SiebertEtAl2012, MonariEtAl2016b}}.
Mapping the kinematics out to several kiloparsec from the Sun is crucial for understanding whether signs of these recent and ongoing accretion events are visible in the Galactic disc, to ultimately understand to what extent the Galaxy can be represented as a system in dynamical equilibrium \citep{hafner00, Dehnen1998b}, at least in its inner regions, or to recover the nature
of the perturber and the time of its accretion instead from the characteristics and strength of these ringing modes \citep{Gomez2012a}.

Signatures of interactions and gravitational disturbances of satellite galaxies can also affect the outer disc beyond the solar radius, in regions where the stellar surface density drops and the disc is more fragile to external
perturbations. Several works have discussed the possibility that the Galactic warp may be generated by the
interaction with the Magellanic Clouds \citep{burke57, weinberg06} or Sagittarius  \citep{bailin03}, while other scenarios suggest that a warped
structure in a galaxy disc may be generated by a dark matter halo distribution that is off-centred or tilted with respect to
the baryonic one \citep{bailin03b}, by bending instabilities  \citep{revaz04} in the disc, or by misaligned infall of material \citep{ostriker89, quinn92}. These scenarios
predict either long-lived, transient, or repeatedly excited structures, and it is clear that to understand the origin
of the Galactic warp, we need to understand its dynamical nature, since, for example, a long-lived warp would leave a
specific signature in the kinematics of stars in the outer disc \citep{abedi14, poggio17}.

In the coming years, the astronomical community will work towards answering these great questions about the Galaxy with the help of \gaia{} data. In this paper, we provide a first exploration of the kinematic properties of the Milky Way disc that already reveals novel results, shows the far-reaching possibilities of the data, and predicts their high future impact. The paper starts by a description of the \gdrtwo{} data that are used in this analysis (Sect.~2). Details are given about calculating distances, velocities, and their uncertainties, as well as about  the different data selections. In Sect.~3 we start by exploring the velocity components in 3D, their medians and dispersions, by searching for global trends as a function of position, distance from the Galactic centre, and  height above the plane. This analysis for the first time presents full 3D kinematic maps of the Galaxy up to several kiloparsec from the Sun. In Sect.~4 we zoom into the solar neighborhood and revisit its velocity distribution by searching for kinematic substructures at small scales with unprecedented accuracy, and also by showing how they evolve with spatial position. The full-sky coverage of \gaia{} overcomes limitations in angular coverage of earlier studies. Finally, in Sect.~5, we present the main conclusions of this work.


\section{Data}\label{data}

In this section, we describe and characterise briefly the \gdrtwo\ data that we used. We start with an overview of the content of DR2. Secondly, we detail how the distances, velocities, and their uncertainties are calculated. Next, we explain how we built a dereddened HR diagram to select different stellar populations, followed by details on the different data samples that are used throughout the paper, and details on their main characteristics. Finally, the last two subsections briefly characterise important aspects of the samples, such as the correlations between variables, and the anisotropy of the samples.

\subsection{DR2 data overview}
\gdrtwo\ provides astrometric parameters (positions, parallaxes, and proper motions) for 1.3 billion sources. The median
uncertainty for the bright sources (G$<$14~mag) is 0.03 mas for the parallax and 0.07\masyr for the proper motions. The
reference frame is aligned with the International Celestial Reference System (ICRS)
and non-rotating with respect to the quasars to within 0.1\masyr. The
systematics are below 0.1~mas and the parallax zeropoint uncertainty is small, about {0.03~mas}. Significant spatial correlations
between the astrometric parameters are also observed. For more details about the astrometric content of \gdrtwo, see
\citet{DR2-DPACP-51, DR2-DPACP-39} and references therein.

The photometric content of \gdrtwo\ consists of weighted-mean fluxes and their uncertainties for three passbands, $G$, $\gbp$
, and $\grp$. All sources have $G$ photometry, but only about 1.4 out of the 1.7 billion sources have both
$\gbp$ and $\grp$ photometry.
The sources without colour information mainly lie in crowded regions where the larger
windows for the BP and RP photometers have a higher chance of overlap between sources and make the photometry unreliable.
The processing for future data releases will include deblending algorithms that will increase the number of sources with colour information.
The precision at $G=12$, the magnitude most relevant for this kinematic study, is around 1 mmag or better for all
three passbands. However, there are systematics in the data at the 10 mmag level. For more details about the
photometric content of \gdrtwo, see \citet{DR2-DPACP-40} and references therein.

To facilitate the selection of specific types of stars, we also used the extinction $A_G$ and color excess $E(\gbp -
\grp)$ provided in \gdrtwo{}, whose estimation was described in \cite{DR2-DPACP-43}.  However, the accuracy of the astrophysical parameters, derived from \gaia\ data alone, is degenerate for some parts of the Hertzsprung-Russel (HR) diagram, especially for high extinction
values. To assist in the sample selections, we therefore also made use of 2MASS photometry of the \gaia{}
sources, specifically, of the \gaia/2MASS cross-match provided within GACS for \gdrtwo{} (see \citealt{DR2-DPACP-41}). Details of how the 2MASS photometry was used are described below in Sect.~\ref{photometry} and
\ref{dataselection}.

A novelty of \gdrtwo{} with respect to \gdrone{} is that it contains line-of-sight velocities\footnote{We use the term line-of-sight velocity for the Doppler-shift measured from the spectra and radial velocity for the Galactocentric velocity component $V_R$ defined in Sect.~\ref{velocities}.} for 7.2 million stars brighter than $\grvs = 12$~mag that were observed with the {\it \textup{Radial Velocity Spectrometer}} \citep{DR2-DPACP-46}. The stars are distributed throughout the full celestial sphere. This release contains line-of-sight velocities for stars with effective temperatures in the range $\sim [3550, 6900]$~K. Cooler and hotter stars will be published in future \gaia{} releases.
The precision of \gdrtwo{} line-of-sight velocities is at the $\kms$ level. At the bright end, the precision is of the order of 0.2 to 0.3~$\kms$. At the faint end, it is of the order of 1.4~$\kms$ for $\teff$ = 5000~K stars and $\sim$ {3.7} $\kms$ at $\teff$ = 6500~K. For more details about the Gaia spectroscopic processing pipeline and the \gdrtwo{} line-of-sight velocities, see \citet{DR2-DPACP-47} and \citet{DR2-DPACP-54} and references therein.

The global validation of \gdrtwo{} is described in~\citet{DR2-DPACP-39} and references therein.

\subsection{Calculation of distances, velocities, and uncertainties}\label{velocities}

In order to map the stars in position and velocity space, we must derive distances from the \gaia\ astrometry. For this purpose, we have selected only stars with $\varpi/\varepsilon_\varpi>5$ and adopted $1/\varpi$ as our distance estimate. It is well-known that the inverse of the parallax is biased when the uncertainty in parallax is significant \citep{Brown1997,Arenou1999,DR2-DPACP-38}. To quantify the distance bias introduced when using $1/\varpi$
as a distance estimator and a cut at 20\% relative uncertainty in parallax, we used the simulations described in Sect. \ref{dataselection}. We established that inverting the parallax leads to unbiased distances out to about 1.5 kpc, with overestimates of the order
of 17\% at 3 kpc. We therefore have to bear in mind that the distance bias in the extremes of our main sample is
non-negligible.

Note that this cut in relative uncertainty in parallax results in a cut in apparent magnitude, and other minor selection effects might be caused by this. However, after tests with our set of simulations, we  concluded that this cut does not introduce relevant artefacts in the kinematics. Alternatively, Bayesian methods might be used to infer distances from parallaxes instead of selecting stars with small relative uncertainty \citep[e.g.][]{BailerJones2015}. However, this is more complex in the sense that they require fixing a prior, and even the simplest sensible prior involves numerical solutions for most estimators and for all the confidence intervals. In this exploratory study, we chose to select small uncertainty in parallax since it is simpler and serves the
purposes of our work well.

\gaia\ provides the  five-parameter astrometric solution\footnote{Proper motion in right ascension $\mu_{\alpha}^*
\equiv\mu_{\alpha}\cos\delta$ of the source in ICRS at the reference epoch. This is the projection of the proper motion
vector in the direction of increasing right ascension. } and line-of-sight velocities, $(\alpha, \delta, \varpi,
\mu_{\alpha}^*,\mu_{\delta},V_{los})$, together with their associated uncertainties and correlations
between the astrometric quantities. From these observables and the derived distances, we computed heliocentric and
Galactic Cartesian and cylindrical positions and velocities. For the Cartesian heliocentric velocities, we took the usual
convention of $U$, $V,$ and $W$ oriented towards the Galactic centre, the direction of Galactic rotation, and the north
Galactic pole, respectively. The Galactic cylindrical coordinates are ($R$, $\phi$, $Z$, $V_R$, $V_\phi$, $V_Z$) with
$\phi$ in the direction of Galactic rotation and with an origin at the line Sun-Galactic centre. The Cartesian Galactic
coordinates are oriented such that the Sun is located at the X negative axis.  {For these transformations, we needed to adopt a height of the Sun above the plane. We used the value given by \citet{Chen2001} of $27$ pc, although other values can be $14\pm 4$ pc from COBE/DIRBE \citep{binney97} or $15.3^{+2.24}_{-2.16}$ from \gdrone{} \citep{Widmark2017}.}
We also adopted the distance of the Sun to the Galactic centre $R_{\odot}$ of $8.34$ kpc and the circular velocity at the solar radius of $V_\text{c}=240$ $\kms$ from \citet{Reid2014}. We took
the peculiar velocity of the Sun with respect of the local standard of rest from {\citet{Schonrich2010},} that is,
$(U_{\odot},V_{\odot},W_{\odot})=(11.1, 12.24, 7.25)$ $\kms$. The resulting value of
$(V_\text{c} + V_{\odot})/R_{\odot}$ is 30.2 $\kms$\,kpc$^{-1}$, which is compatible with the value
from the reflex motion of Sgr A* of \citet{Reid2004}. In these coordinate transformations, we propagated the full
covariance matrix. This means that we have the correlations between uncertainties in Cartesian and cylindrical
coordinates at our disposal.

\subsection{Intrinsic colour computation}\label{photometry}

\begin{figure}[h!]
\centering
\includegraphics[width=1.\hsize]{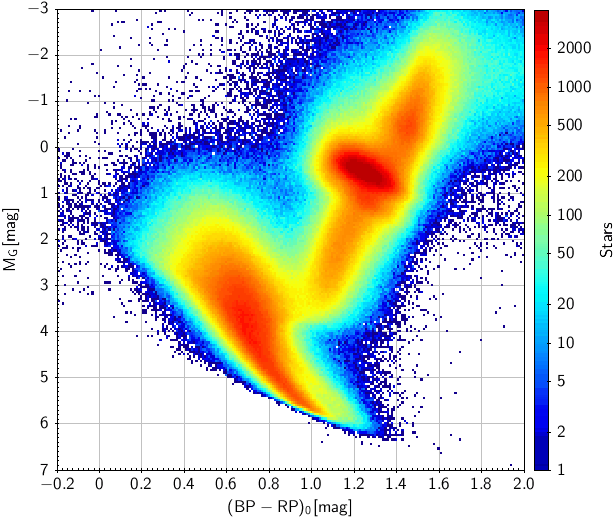}
\caption{De-reddened HR diagram for the main sample with 2MASS photometry and the number of stars per bin of 0.01mag $\times$ 0.05mag.} \label{fig1}
\end{figure}

To select stars in the HR diagram, we have used cuts in absolute magnitude and intrinsic colours. For this an extinction correction needed to be applied, in particular for distant giants and hot stars. While first extinction estimates by \gaia{} consortium have been made using the \gaia{} integrated bands alone, the addition of the 2MASS colours strongly helps to break the $\teff$-extinction degeneracy \citep{DR2-DPACP-43}. We used here the \gdrtwo\ provided cross-match with 2MASS \citep{DR2-DPACP-41}. We selected 2MASS stars with photometric quality flag AAA and photometric uncertainties lower than 0.05~mag. We used the same \gaia\ photometric cuts as in \cite{DR2-DPACP-31}: photometric uncertainties smaller than 5\% for $\gbp$ and $\grp$ and 2\% for $G,$ and a selection on the $\gbp$/$\grp$ excess flux factor based on the star colour.
To derive intrinsic colour-colour relations, we selected low-extinction intrinsically bright stars as in \cite{DR2-DPACP-31}, for example, using the 3D extinction map of \cite{Capitanio2017}\footnote{\url{http://stilism.obspm.fr/}} and the \gdrtwo\ distances, to select stars with $E(B-V)<$0.015 and $M_G<2.5$. For each photometric band $X=\gbp$,$\grp$,$J$,$H$, we built a fifth-order polynomial relation to model $(G-X)_0$ as a function of $(G-K_s)_0$. We used the extinction coefficient models described in \cite{Danielski18}, {computed using the nominal passbands}. We pre-selected intrinsically bright stars using the 2MASS $Ks$ magnitude, which is less strongly affected by extinction:
\begin{equation}
 K_s + 5 + 5 \log_{10} \left( \frac{\varpi + \varepsilon_\varpi}{1000} \right) < 4,
\end{equation}
where the astrometry is given in milliarseconds. Then the extinction $A_0$ and the intrinsic colour $(G-K_s)_0$ were determined for
each star through a maximum likelihood estimator (MLE).
This takes into account the photometric uncertainties, the intrinsic scatter around the intrinsic colour-colour relation (which is between 0.01 and 0.03~mag),
and the validity intervals of these relations as well as the positivity of the estimated extinction.
A chi-square test was performed to verify the validity of the resulting parameters, removing stars with a p-value limit lower than 0.05.
We also removed stars for which the MLE did not converge and those with an error on $(G-K_s)_0$ larger than 0.5~mag.
In total, we obtained extinction corrections for 90\% of the sample. Figure  \ref{fig1} shows the de-reddened HR diagram.

\subsection{Data selection}\label{dataselection}

As discussed above in Sect. \ref{velocities}, we selected sources with $\varpi/\varepsilon_\varpi>5$. This cut selects
stars with positive parallaxes and a relative parallax uncertainty smaller than 20\%. After this cut, we further selected
several samples that we use in the different sections of this study.

\begin{figure}[h!]
\centering
\includegraphics[width=0.9\hsize]{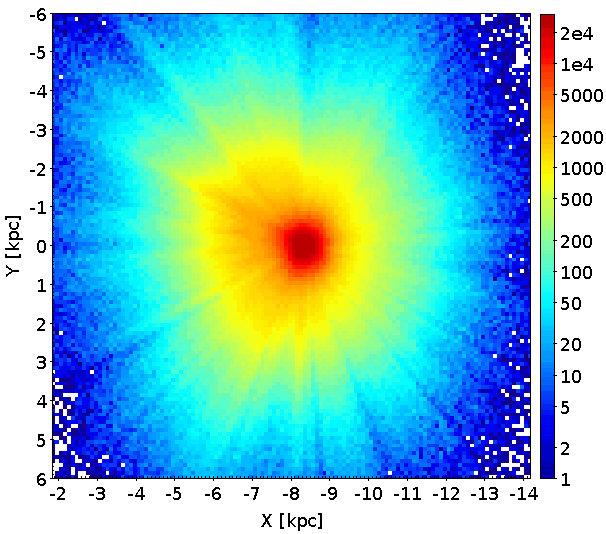}
\includegraphics[width=0.9\hsize]{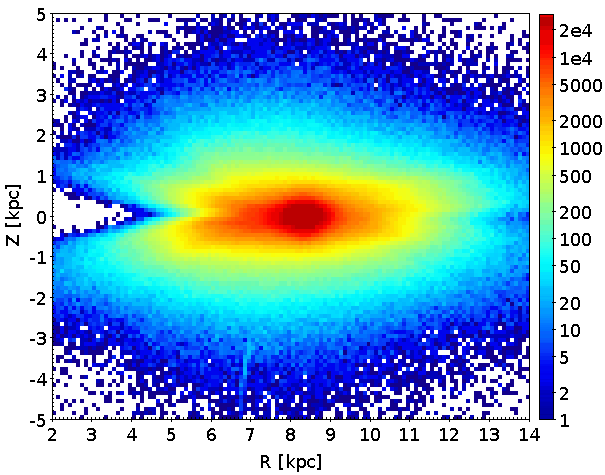}
\caption{{\em Top:} Surface density in the $(X,Y)$ plane {for the stars in the main sample that have available extinction-corrected photometry} (number of stars per bin of 100pc $\times$ 100pc). {\em Bottom:} Same for the $(R,Z)$ plane.}
\label{figdens}
\end{figure}

\begin{figure}
\centering
\includegraphics[width=0.9\hsize]{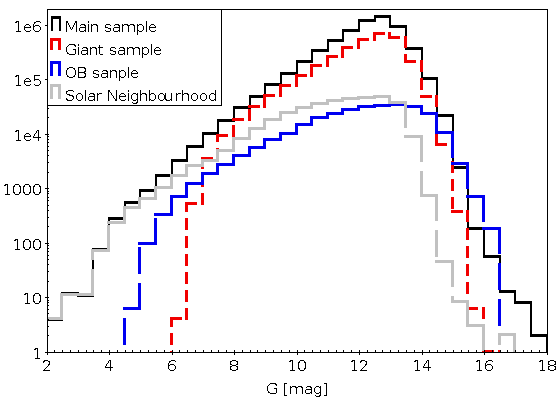}
\includegraphics[width=0.9\hsize]{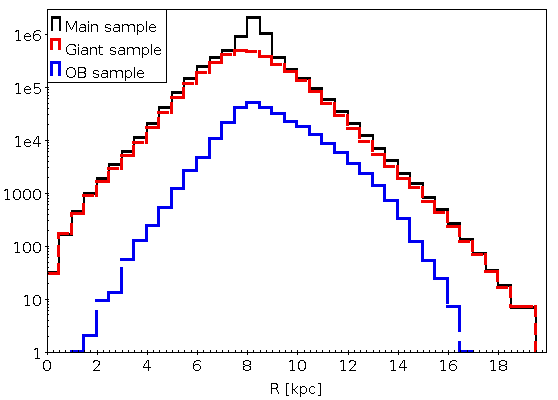}
\caption{{\em Top:} Histogram of the $G$ apparent magnitude for the four working samples. {\em Bottom:}  Histogram of the
Galactic radius for the main, giant, and OB samples. Stars in the solar neighbourhood sample are located at $d$ < 200 pc
(see Sect.~2).}
\label{fighist}
\end{figure}

\begin{figure}
\centering
\includegraphics[width=0.9\hsize]{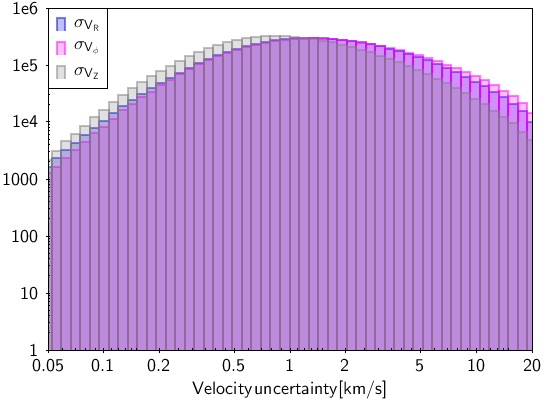}

\includegraphics[width=0.9\hsize]{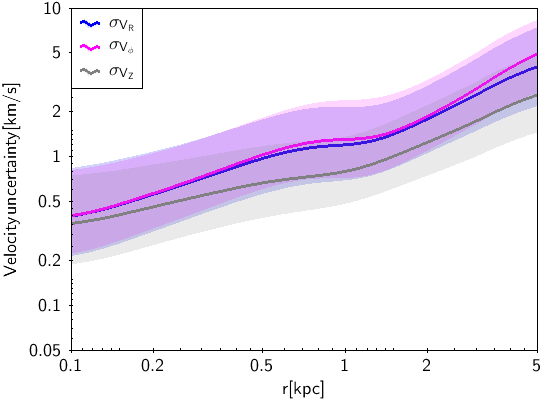}
\caption{{\em Top:} Histogram of velocity uncertainties in the Galactic cylindrical reference system  ($V_R$, $V_\phi$,$V_Z$) for the main sample. {\em Bottom:}  Median uncertainties in velocity as a function of the heliocentric distance for the main sample. The 25\% and 75\% quartiles are shown as colour-shaded areas.}
\label{figvelerrors}
\end{figure}

\paragraph{1. Main sample.} This sample consists of the {6\,376\,803} sources with an available  five-parameter astrometric
solution, line-of-sight velocities, and $\varpi/\varepsilon_\varpi > 5$. The intrinsic magnitudes and colours were calculated using Gaia and 2MASS photometry, as explained in Sect. \ref{photometry}. In the top and bottom panels of
Fig.~\ref{figdens}, we show the surface density per bins of $100$pc $\times 100$pc in $(X,Y)$ and $(R,Z)$ planes,
respectively, while in top and bottom panels of Fig.~\ref{fighist}, we show the $G$ apparent magnitude and the Galactic
radius distribution of the main sample (black lines) and the remaining working samples. For these stars, we computed
the full 6D phase space coordinates as detailed in Sect.~\ref{velocities}. The top panel of Fig.~\ref{figvelerrors}
shows the distribution of uncertainties in Galactic cylindrical velocities of the main sample. The median uncertainties are $(\varepsilon_{V_{r}}, \varepsilon_{V_{\phi}},\varepsilon_{V_z})=(1.4,1.4,0.9)$ $\kms$ 
, and
{$20\%$} of the stars have an uncertainty in all velocity components that is smaller than 1~$\kms$. The distributions in $\varepsilon_{V_{r}}$ and $\varepsilon_{V_{\phi}}$ are similar and
differ from the distribution for $\varepsilon_{V_{Z}}$, which is more precise. The reason is that most of the stars are located in the
Galactic plane: for these stars, the main contribution to the vertical velocity comes from the astrometric quantities,
which for this sample have smaller uncertainties than does the line-of-sight velocity. The uncertainties as a function of
distance are shown in the bottom panel. They seem to increase approximately linearly in this log-log plot. The median velocity uncertainty is below 1~$\kms$ at distances closer than 0.5 kpc, and below 2~$\kms$ at distances closer
than 2 kpc. In addition, uncertainties larger than 10~$\kms$ are only reached at distances larger than 5 kpc.

The main sample supersedes any previous full 6D phase-space sample in terms of quantity and precision of the data. For
instance, the main sample is about 12 times larger in number of stars than a sample made from UCAC proper motions \citep{Zacharias2013} and RAVE line-of-sight velocities and derived spectro-photometric distances \citep{Kunder2017}. Thus, the statistics enable studying the
Galaxy kinematics in more details and at much larger distances than before. At the faint end, the precision of the RAVE line-of-sight velocities is comparable to that of the RVS. However, with \gdrtwo,{} the precisions as a function of distance in the derived distances and in the proper motions are about two and more than $\text{ten}$ times better, respectively. This combination means that the precision in Galactocentric cylindrical velocities
of the main sample is approximately 5-7 times better. As an example, we show the Toomre diagram of
the main sample in Fig. 5.

\begin{figure}
\centering
\includegraphics[width=0.95\hsize]{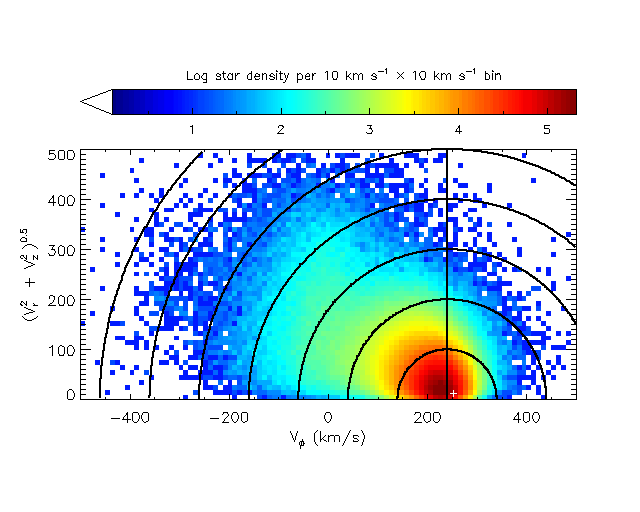}
\caption{Toomre diagram of the main sample.  The vertical line crosses the LSR at ($V_R$, $V_\phi$,$V_Z$) = (0, 240, 0) $\kms$.  The white dot is the peculiar velocity of the Sun: ($V_R$, $V_\phi$,$V_Z$) = (11.10, 252.24, 7.25) $\kms$.  The concentric circles show the total peculiar velocity, centred on the LSR.  The traditional use of the Toomre diagram to classify stars into stellar populations is complicated by the great range of the Galactic radius of the sample (Fig.~\ref{fighist}) and the possibility that both the mean $V_\phi$ of the thin disc and the $V_\phi$ lag between the thin and thick disc may vary with Galactic radius.  Nevertheless, it shows that the sample is dominated by the thin disc.  In the solar neighbourhood, the thin disc has an azimuthal velocity close to the LSR, and the thick disc lags behind by a few tens of $\kms$.}
\label{figtoomre}
\end{figure}

\paragraph{2. Giant sample.} This is a sub-selection of the main sample that includes only giant stars selected on their absolute magnitude in $G$ band $M_G<3.9$ and intrinsic colour $(\gbp - \grp)_0>0.95$.
The intrinsic magnitudes and colours were calculated by using Gaia and 2MASS photometry, as explained in Sect.
\ref{photometry}. This sample contains {3\,153\,160} sources. As noted in Fig.~\ref{fighist}, about half of the stars in the main sample are (red) giants, which are the main contribution at distances larger than 1 kpc from the Sun. That is why this sample is used in Sect.~\ref{maps} to analyse the large-scale kinematic maps in the Galactic disc. As
expected, {78}\% of the (red) giant sample is located within 3 kpc of the Sun. Nonetheless, the inner regions, that is, areas towards the Galactic centre with Galactic radius between 3-5 kpc, are still well sampled with more than {500 000} stars (see the bottom panel of Fig.~\ref{fighist}). Furthermore, in
the outskirts of the galactic disc, our red giant sample contains more than 10 000 stars at $R>13$~kpc, thus
reaching a significant number of stars.  Nonetheless, most of these stars belong to the tip of the red giant branch,
and their stellar evolutionary stage is therefore different from the red clump sources, most of which are located at about
$\pm2$ kpc from the Sun. The median uncertainties are
{$(\varepsilon_{V_{r}}, \varepsilon_{V_{\phi}},\varepsilon_{V_z})=(1.6,1.7,1.2)$}
$\kms{}$ , and $13\%$ of the stars have an uncertainty in all velocity components that is smaller than $1$ $\kms{}$.

\paragraph{3. Solar neighbourhood sample.} This is a sub-selection of the main sample with stars located within
200 pc of the Sun, that is, with $\varpi > 5$ mas. This comprises
{366\,182}
 stars with a median velocity uncertainty of
$(\varepsilon_U, \varepsilon_V, \varepsilon_W)=($
0.4, 0.4, 0.4
$)\,\kms$ and with {78}\% of stars having uncertainties smaller than $1\,\kms$
in all components.

\begin{figure*}[h!]
\centering
\includegraphics[width=0.32\hsize]{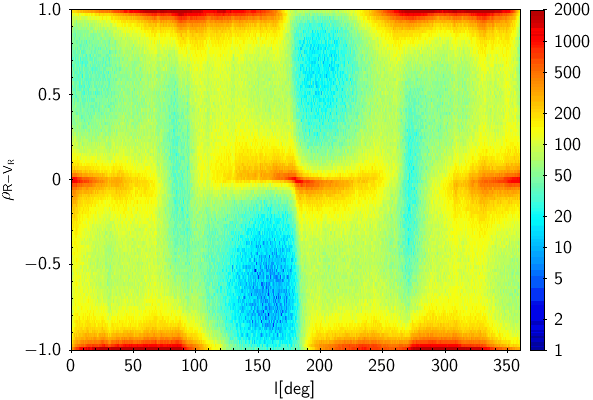}
\includegraphics[width=0.32\hsize]{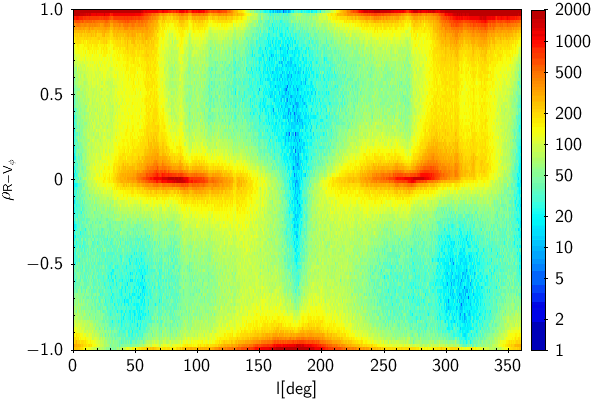}
\includegraphics[width=0.32\hsize]{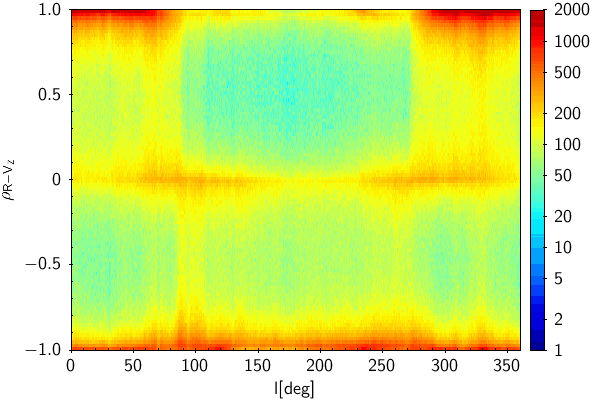}
\caption{Correlations in the main {sample} between the components of the Galactic velocity and the Galactic radius ($R$) as a function of longitude.  The colour scale indicates the number of stars per bins of [0.5,0.02].}\label{figcorrelations}
\end{figure*}

\paragraph{4. OB sample.}
This is the selection of OB stars used in Sect.~\ref{maps} to map the median vertical velocity of young stellar populations. This sample is different from those described above in that it
is not constrained to sources with available
line-of-sight velocities. However, the additional challenge is identifying young, intrinsically blue stars near the
Galactic plane that are significantly reddened.

An initial list of OB star candidates in DR2 was found using the following criteria:
\begin{eqnarray}
\varpi/\varepsilon_\varpi > 5 \\
(G_{BP} - G_{RP})_0 = (G_{BP} - G_{RP}) - E(G_{BP} - G_{RP}) < 0 \\
M_G = G + 5\log \varpi + 5 - A_G < 2,
\end{eqnarray}
where $A_G$ and $E(G_{BP} - G_{RP})$ are the extinction and colour excesses provided in \gdrtwo{} (see \citealt{DR2-DPACP-43}), and $\varpi$ is expressed in mas.
To ensure that our sample indeed consists of young stars rather than giants or red clump stars with erroneous
extinctions, a further selection was made using the 2MASS photometry that also satisfies the following conditions:
\begin{eqnarray}
J-H & < & 0.14(G-K_s) + 0.02 \\
J-K_s & < & 0.23(G-K_s).
\end{eqnarray}
These colour-colour selection criteria were adopted from those
described by Poggio et al (in prep) and are based on the observed 2MASS colours of
spectroscopically \textup{bona fide }OB stars from the \tyctwo{} stars found in \gdrone{} and the {\tyctwo{} spectral type} catalogue
\citep{wright03}. In addition, the photometric quality conditions $\varepsilon_{J,H,K_s} < 0.05$ and 2MASS photometric flag
equal to AAA were applied to avoid sources with problematic photometry. These selections yielded 285\,699 stars
whose 2MASS/\gaia{} colours and astrometry are consistent with our sources being OB stars. However, given the relatively
large uncertainties on the individual extinction parameters, our sample is likely to also contain a significant number
of upper main-sequence A stars. Nevertheless, such stars, being young, still serve our purpose here. The apparent magnitude and galactocentric radial distribution is shown in Fig.~\ref{fighist}.

\subsection{Simulation of Red Clump disc stars}
In order to analyse the effect of errors and biases throughout the different sections of this study, we used the simulation of \gaia{} data provided in \citet{RomeroGomez2015}. This is a test-particle simulation of Red Clump disc stars that evolved in a barred galactic potential. We only kept stars with $G\le13$ {from} the entire simulation to mimic the magnitude distributions
of our main sample. This led to a simulation of one million Red Clump disc stars with astrometric and line-of-sight
velocity uncertainties that matched those of \gdrtwo{}. We rescaled the end-of-mission astrometric uncertainty prescribed on the Gaia Science Performance webpage (see also \citealt{deBruijne2015}) to the \gdrtwo{} uncertainty for 22 months of mission\footnote{\url{http://www.rssd.esa.int/doc_fetch.php?id=359232}}, and for the bright stars, we included a multiplying factor of 3.6 to match the distribution of the uncertainty as a function of $G$ magnitude observed in the \gdrtwo{} data. The line-of-sight velocity uncertainties were also rescaled to match the uncertainty for the Red Clump-type of stars observed in our \gaia{} sample.

\begin{figure}[h!]
\centering
\includegraphics[width=0.9\hsize]{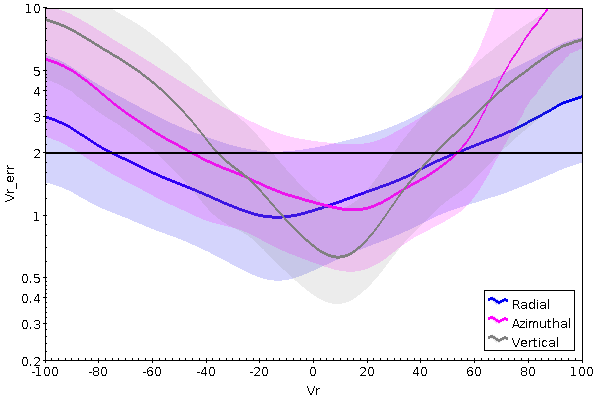}
\caption{Median of the uncertainty in the three Galactic cylindrical velocity components (radial $V_R$, azimuthal $V_\phi-V_c$ , and vertical $V_z$) as a function of the corresponding velocity
components for the main sample. The colour-shaded areas show the 25\% and 75\% quartiles. The horizontal dashed black line indicates the bias that would be introduced if a cut of 2 $\kms{}$ were performed.}
\label{figcorrelationseVV}
\end{figure}

\subsection{Correlations between astrometric and derived quantities}

\begin{figure}[h!]
\centering
\includegraphics[width=0.9\hsize]{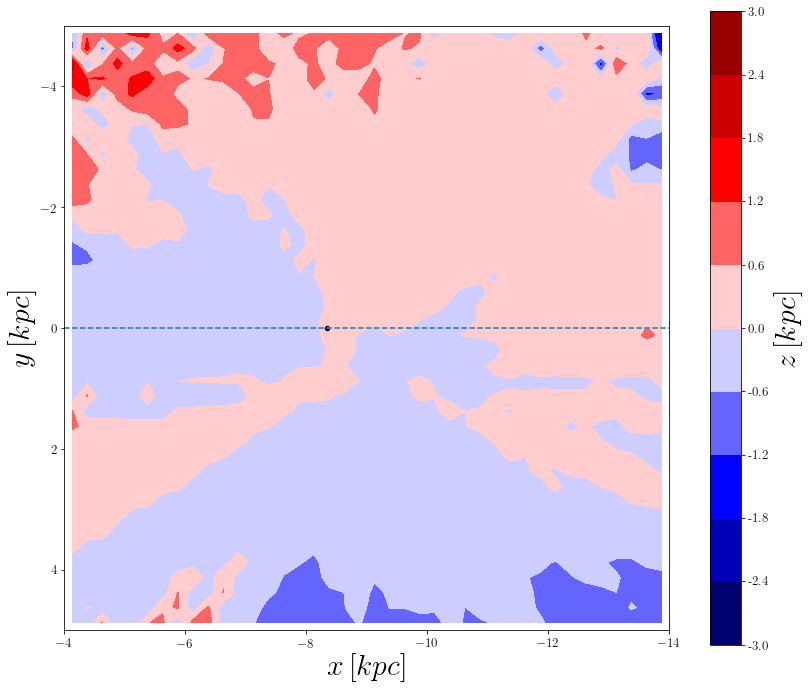}
\includegraphics[width=0.9\hsize]{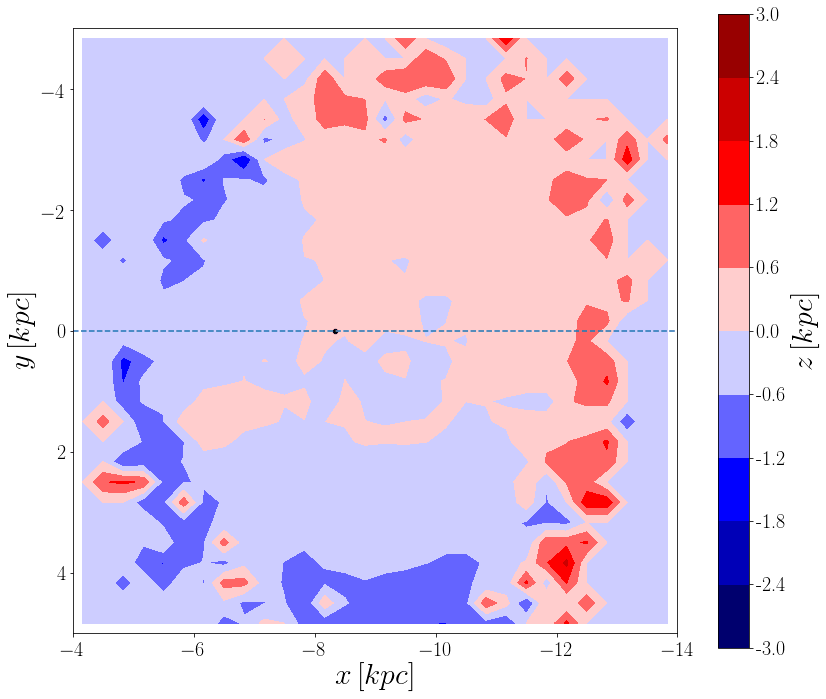}
\caption{Median vertical position $<Z>$ on the $XY$ plane. {\it Top:} Giant sample. {\it Bottom:} Simulations of Red Clump stars with G<13 (see text). The black dot marks the position of the Sun.}
\label{figmeanZ}
\end{figure}

\begin{figure*}
\centering
\includegraphics[width=0.32\hsize]{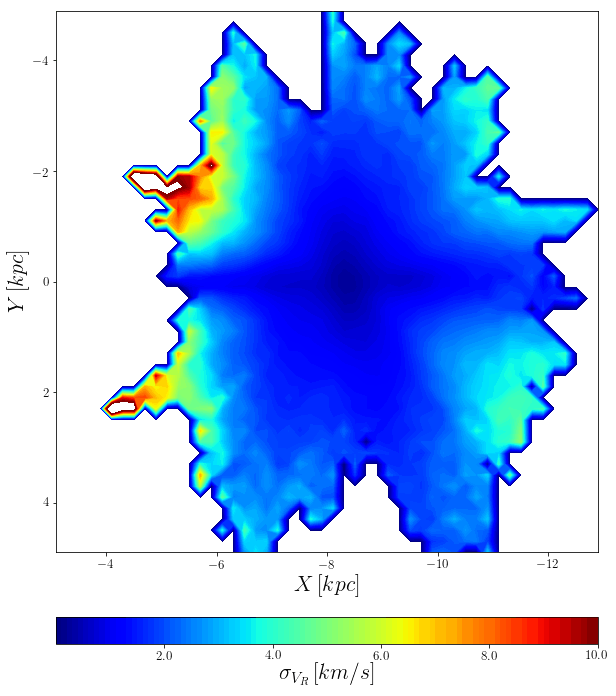}
\includegraphics[width=0.32\hsize]{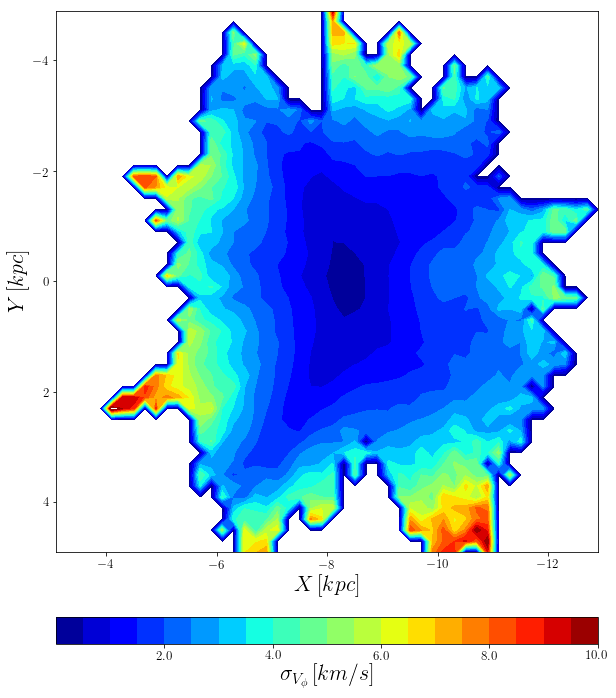}
\includegraphics[width=0.32\hsize]{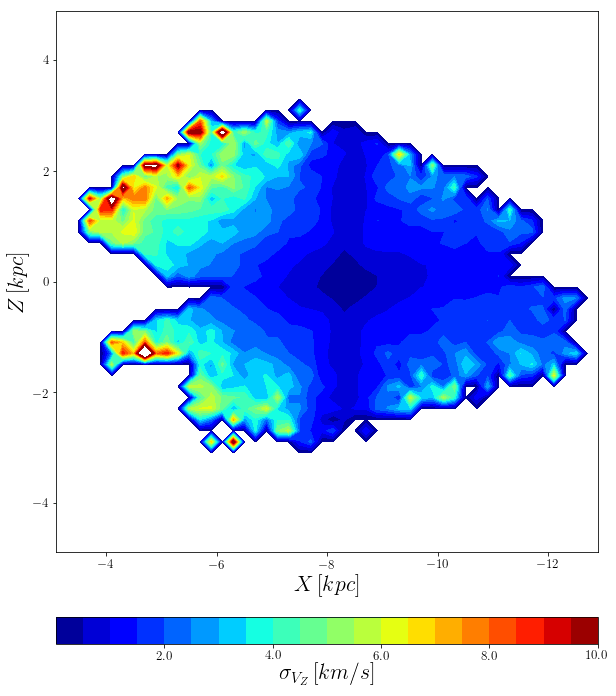}
\caption{Median uncertainty in the Galactic velocity components for the giant sample as a function of disc position. {\it Left:} Radial velocity
uncertainties  $\sigma_{V_{R}}$ in the XY plane. {\it Middle:} Azimuthal velocity uncertainties $\sigma_{V_{\phi}}$  in the XY plane. {\it Right:} Vertical velocity uncertainties $\sigma_{V_{Z}}$ in the XZ plane. In the first two panels, only stars with $|Z| < 200 $ pc
are considered. In the right panel, only stars with $|Y|<200$pc were taken.}
\label{figvelerrorsXYZ}
\end{figure*}

In Fig.~\ref{figcorrelations} we show for the main sample the correlation coefficient between the Galactic radius and the different components of the Galactic velocity as a function of the Galactic longitude. Most of the stars are concentrated in regions of correlations near unity, which are positive or negative depending on the Galactic longitude. This behaviour is mainly due to a geometric effect and not to especially strong correlations between the \gaia{} observables. The stars with correlation coefficients near to 1 in these panels do not have strong correlations between the \gaia{} observables. We note that the median absolute correlations of this sample are 
{$\rho_{\pi-\mu_{\alpha}}=-0.03$, $\rho_{\pi-\mu_{delta}}=0.01$ and $\rho_{\mu_{\alpha}-\mu_{\delta}}=0.01,$ and for 89$\%$} 
of the stars, all three correlations are weaker than 0.4. The behaviour in these panels arises because both the Galactic radius and the velocities are dependent on the heliocentric distance, which in our study we take as the inverted parallax. In this sense, any uncertainty in distance would translate into a proportional uncertainty in $R$ and ($V_{R}, V_{\phi}, {V_{Z}}$), its sign depending on the position in the Galaxy. Therefore, the uncertainties in radius and velocities are highly correlated. 

While the correlations on the observables might bias some derived quantities, this will only happen in the limit of large uncertainties and depending on the problem under study. We also note that if the errors on the astrometric basic parameters are random, as expected, these high correlations do not necessarily translate into a bias, meaning that this is not equivalent to having a systematic error. However, we emphasize that correlations are important in the uncertainty propagation and should not be neglected.

In our data selection we did not perform any cut in velocity uncertainty. Figure~\ref{figcorrelationseVV} shows the uncertainty in velocity as a function of velocity for the three Galactic components. Since the velocities and their uncertainties are correlated, removing stars with large uncertainties, such as
those above the dashed black line at 2~$\kms$, entails the removal of the stars with higher velocities. This can cause large biases on derived quantities such as the velocity dispersion, and we have checked that even the mean velocities as a function of Galactic radius or height above the plane appear to be highly biased (with differences of up to 20~$\kms{}$) when performing these data selections (see  Appendix \ref{app:}).

\subsection{Magnitude limit and asymmetric extinction}
Even though \gaia{} is unique in covering the whole sky, the effects of the scanning law, extinction, and other complex aspects of the completeness of the data (see \citealt{DR2-DPACP-39} and \citealt{DR2-DPACP-54})
complicate the selection function. As a consequence, the properties of the main sample depend strongly on the direction. To show one example, the average vertical position $Z$ in the $X$-$Y$ plane of the giant sample is displayed in
Fig. 8 (top). The median vertical position is a strong function of Galactic longitude, which is clearly affected by the extinction in our Galaxy, which is highly non-uniform. The values of median $Z$ are higher than 600 pc at distances beyond 3 kpc. In the bottom panel of Fig.~\ref{figmeanZ}, the same quantity is shown for the simulation of Red Clump stars described above. In this simulation, the 3D extinction model of \citet{Drimmel2001} was used. Similar trends are shown between the \gaia{} data and the simulation. To reduce the bias on the median $Z$ as a function of Galactic radius significantly, in Sect.~\ref{maps}, we divide the disc into layers of 400~pc height when it is observed face-on.

On similar lines, the uncertainties on the derived quantities also depend strongly on the position in the Galaxy in a complex way that is greatly related to extinction. Figure~\ref{figvelerrorsXYZ} shows the median velocity uncertainties as a function of position in configuration space. While the uncertainties globally increase as a function of distance from the Sun, as expected, this increase depends on {the} direction because it is affected by interstellar extinction. For instance, some blue spikes appear in these panels in lines of sight with lower extinction, while in other directions, the uncertainty achieves high values at close distances to the Sun. However, we note that the median velocity uncertainties are very small compared to other previous catalogues: they are of the order of 6-10~$\kms{}$ only at the extremes of the sample. We also emphasise that given the large number of stars, the uncertainties on the median velocities in a given Galactic position are much smaller than these median (individual) velocity uncertainties showed here. For instance, median velocities at 1 and 1.5 kpc have unprecedented precisions of 0.5 and 1~$\kms{}$, respectively (see colour-shaded areas in Figs.~\ref{fig:rvrmed}, \ref{fig:rvphimed}, and \ref{fig:rvzmed}).

%

\section{Mapping the disc median velocities and velocity dispersions}\label{maps}
Non-axisymmetric structures (e.g. bar and spiral arms) and external perturbers (e.g. the Sagittarius dwarf galaxy, the Magellanic Clouds, and dark matter sub-halos) are expected to disturb the Milky Way velocity field. In the past decade and thanks to large spectroscopic surveys and proper motion catalogues, RAVE~\citep{Steinmetz2006, Kunder2017}, SEGUE~\citep{yanny09}, APOGEE~\citep{Majewski2017, Abolfathi2017}, LAMOST~\citep{Cui2012, Zhao2012}, Tycho-2~\citep{Hog2000}, PPMX-L~\citep{Roeser2008, Roeser2010}, UCAC~\citep{Zacharias2004, Zacharias2010, Zacharias2013, Zacharias2017}, SPM4~\citep{Girard2011}, and Gaia-TGAS~\citep{GaiaBrown2016, Lindegren2016} streaming motions and velocity waves have been shown on a kiloparsec scale around the Sun~\citep{SiebertEtAl2011, Widrow2012, CarlinEtAl2013, Carlin2014, Williams2013, Pearl2017, CarrilloEtAl2017, Tian2017, LiuEtAl2017}. In this section, we take advantage of the large data volume, full sky coverage, accuracy, and precision of \gdrtwo{} {to re-examine these kinematic features at higher accuracy than ever before.} We study the kinematics of the sample of giant stars (described in Sect.~\ref{dataselection}), and map the medians ($\tilde{V}_R, \tilde{V}_\phi, \tilde{V}_Z$) and the dispersions ($\sigma_{V_R}, \sigma_{V_\phi}, \sigma_{V_Z}$) of the Galactocentric velocities as a function of the location in the Galaxy ($X, Y, R, \phi, Z$).

\begin{figure*}[]
\centering
\includegraphics[clip=true, trim = 10mm 40mm 20mm 40mm, width=0.41 \hsize]{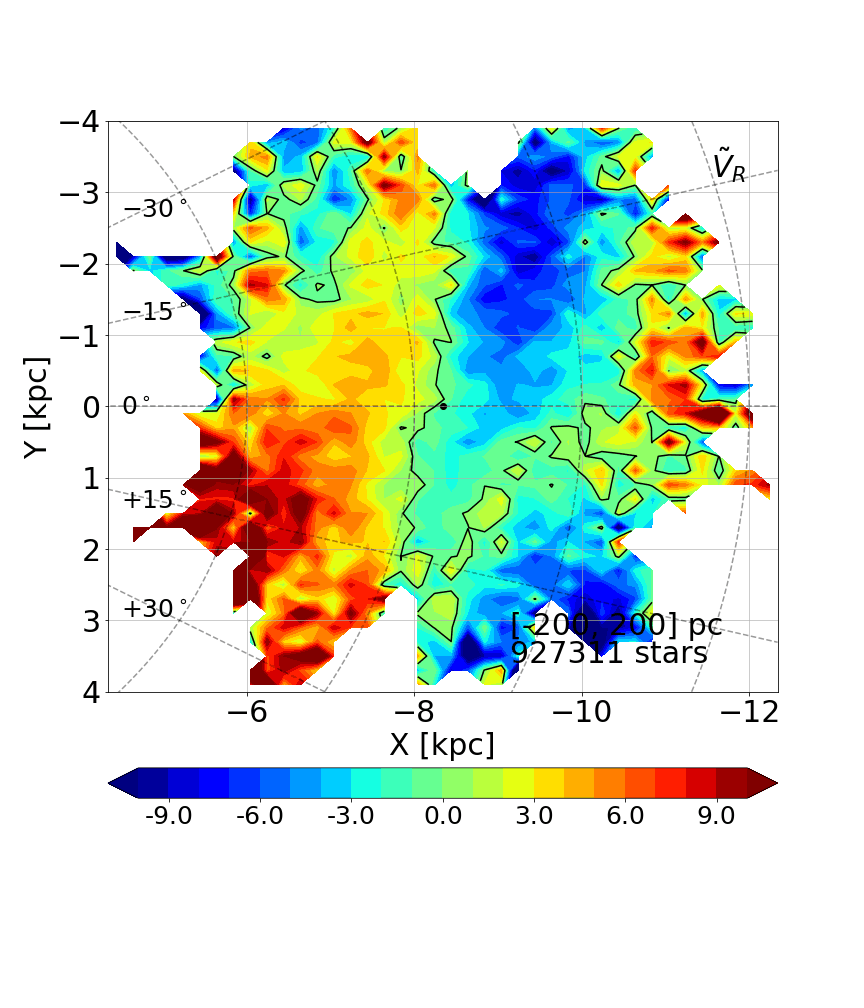}
\includegraphics[clip=true, trim = 10mm 40mm 20mm 40mm, width=0.41 \hsize]{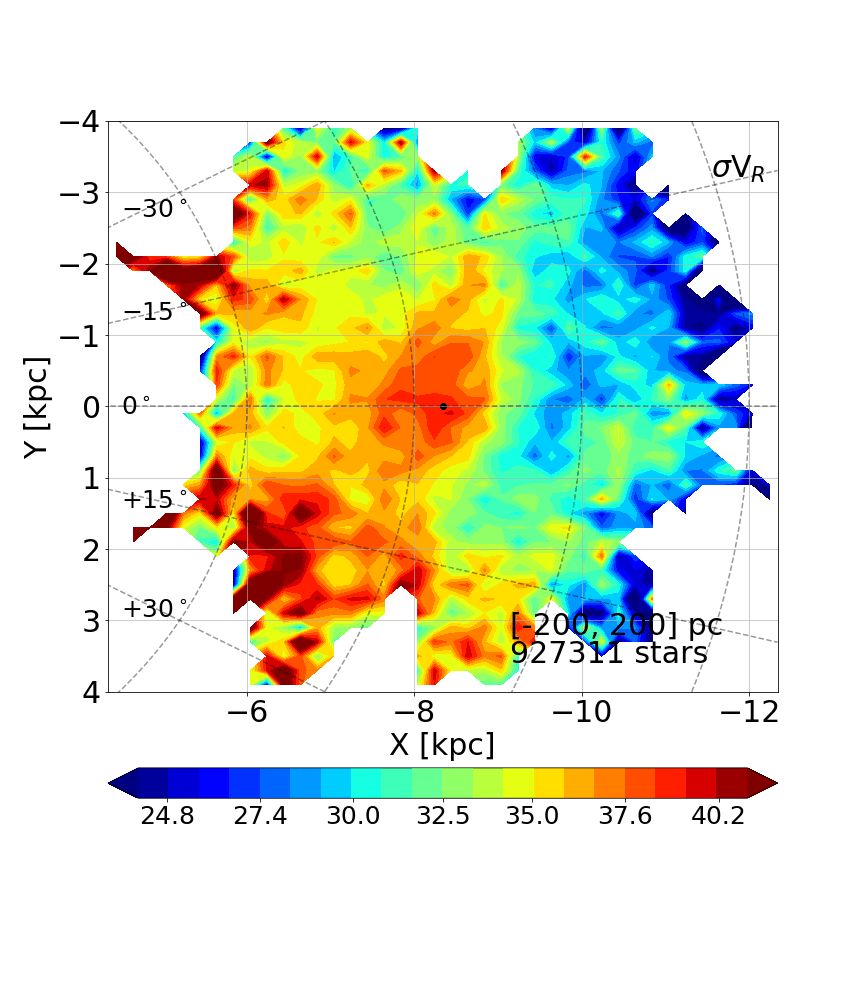}
\includegraphics[clip=true, trim = 10mm 40mm 20mm 40mm, width=0.41 \hsize]{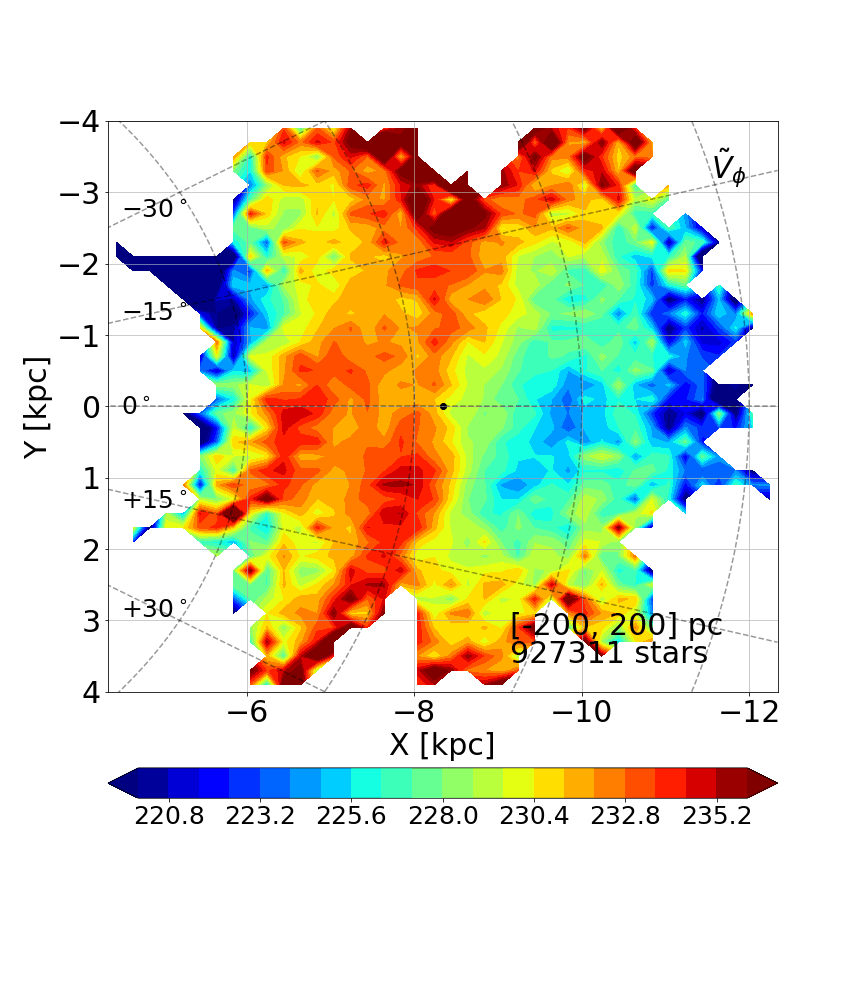}
\includegraphics[clip=true, trim = 10mm 40mm 20mm 40mm, width=0.41 \hsize]{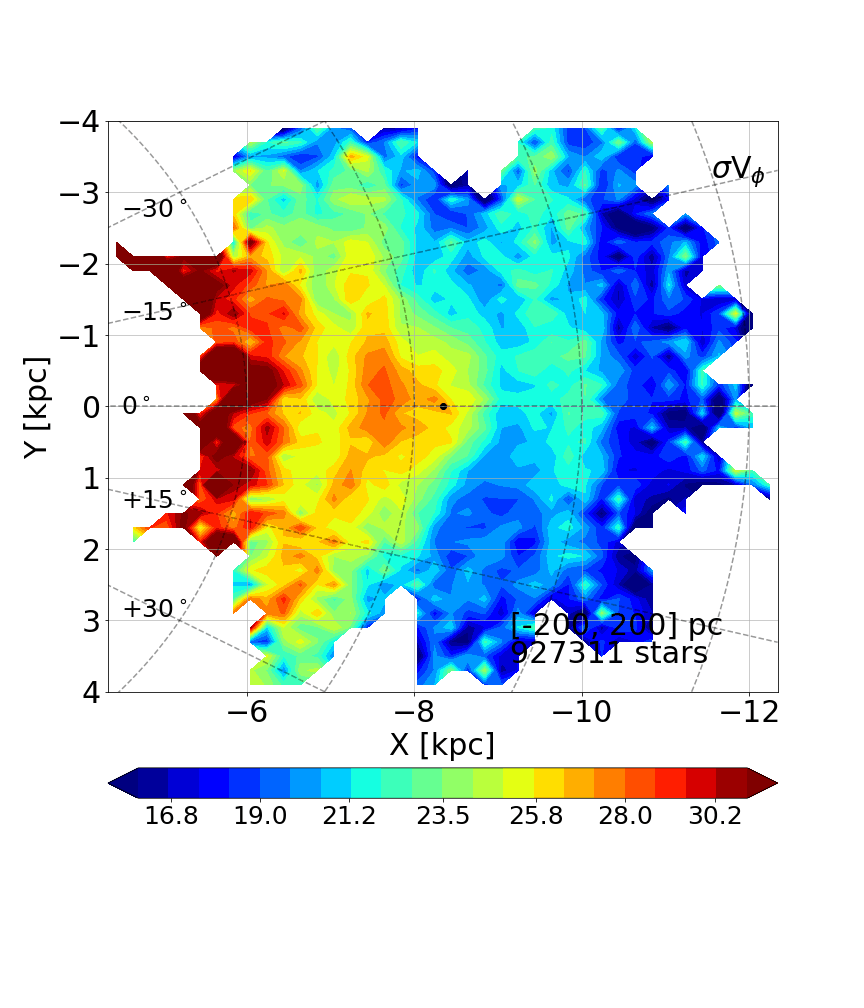}
\includegraphics[clip=true, trim = 10mm 40mm 20mm 40mm, width=0.41 \hsize]{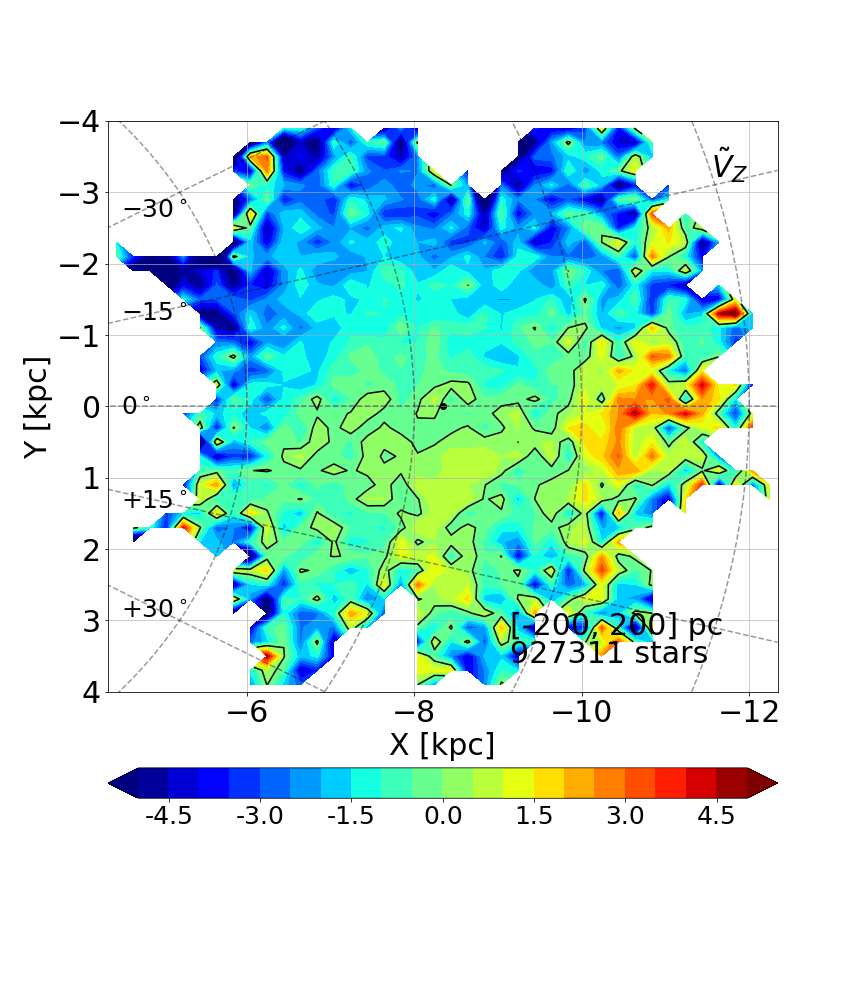}
\includegraphics[clip=true, trim = 10mm 40mm 20mm 40mm, width=0.41 \hsize]{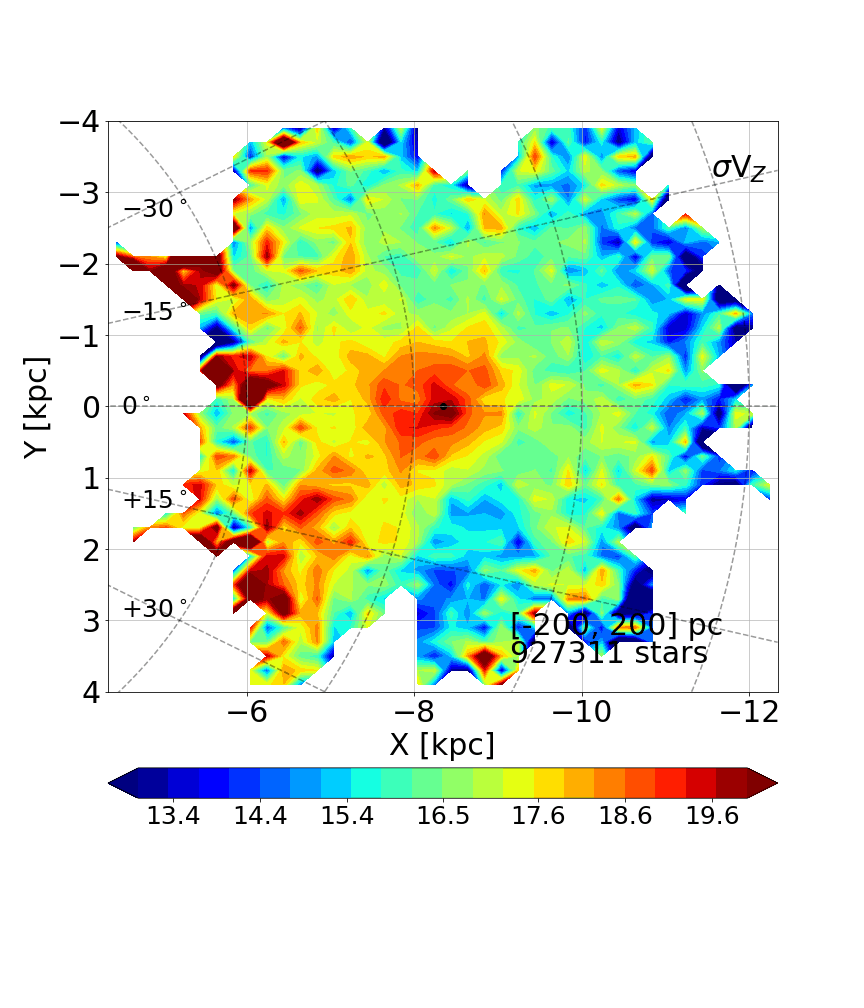}
\caption{Face-on views of the kinematics of the disc mid-plane ([-200, 200]~pc), derived using the giant sample: left, from top to bottom, median velocity maps $\tilde{V}_R, \tilde{V}_\phi, \tilde{V}_Z$ (in $\kms$), and right, from top to bottom, velocity dispersion maps $\sigma_{V_R}, \sigma_{V_\phi}, \sigma_{V_Z}$ (in $\kms$). The azimuths increase clockwise. They are labelled from $-30$ to $+30$~degrees, on the left of the maps. The Sun is represented by a black dot at $X=-8.34$~kpc and $Y=0$~kpc. The Galactic centre is located on the left side. The Milky Way rotates clockwise. The iso-velocity contours $\tilde{V}_R = 0$ and $\tilde{V}_Z = 0$~$\kms$ are pointed out as black lines. The numbers of stars used to produce the maps are given in the lower right corners.\label{fig:faceon}}
\end{figure*}

\begin{figure*}[]
\centering
\includegraphics[clip=true, trim = 10mm 40mm 20mm 40mm, width=0.41 \hsize]{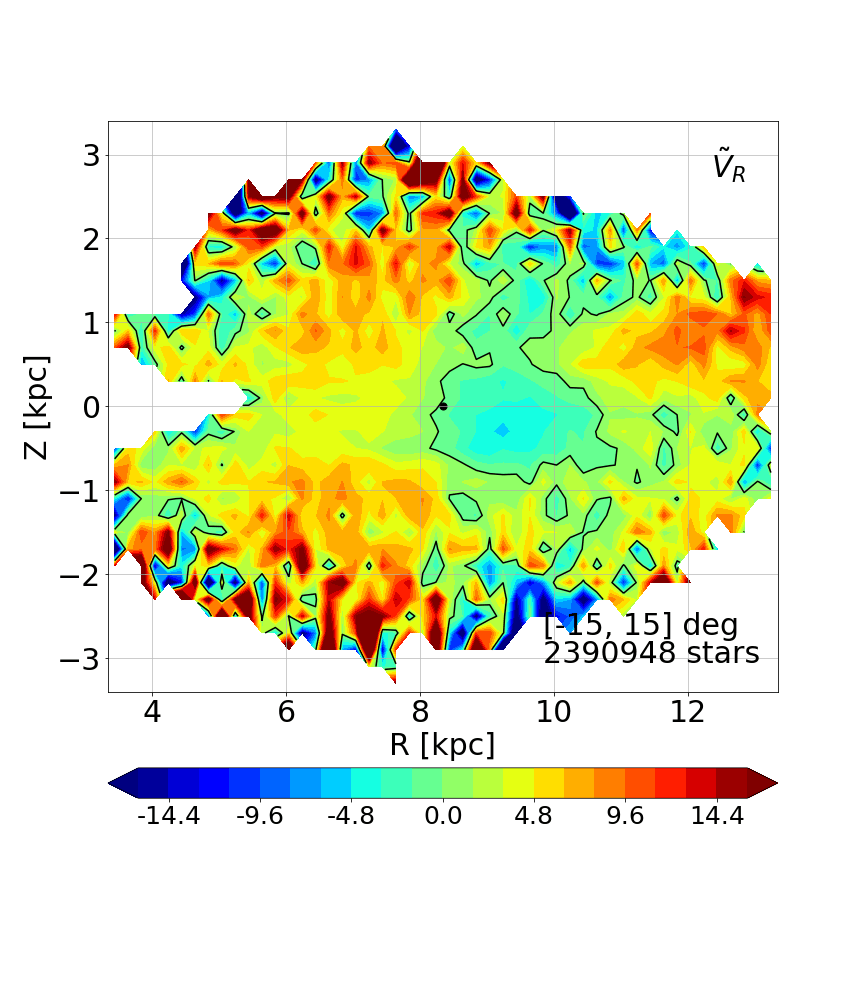}
\includegraphics[clip=true, trim = 10mm 40mm 20mm 40mm, width=0.41 \hsize]{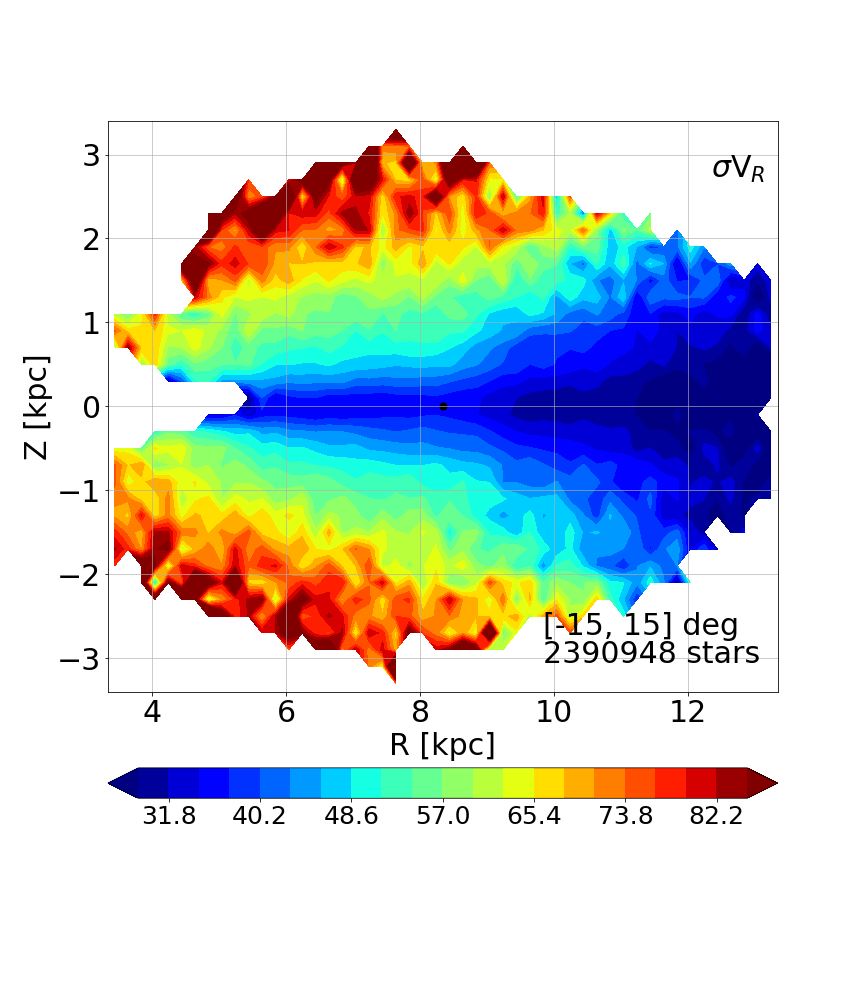}
\includegraphics[clip=true, trim = 10mm 40mm 20mm 40mm, width=0.41 \hsize]{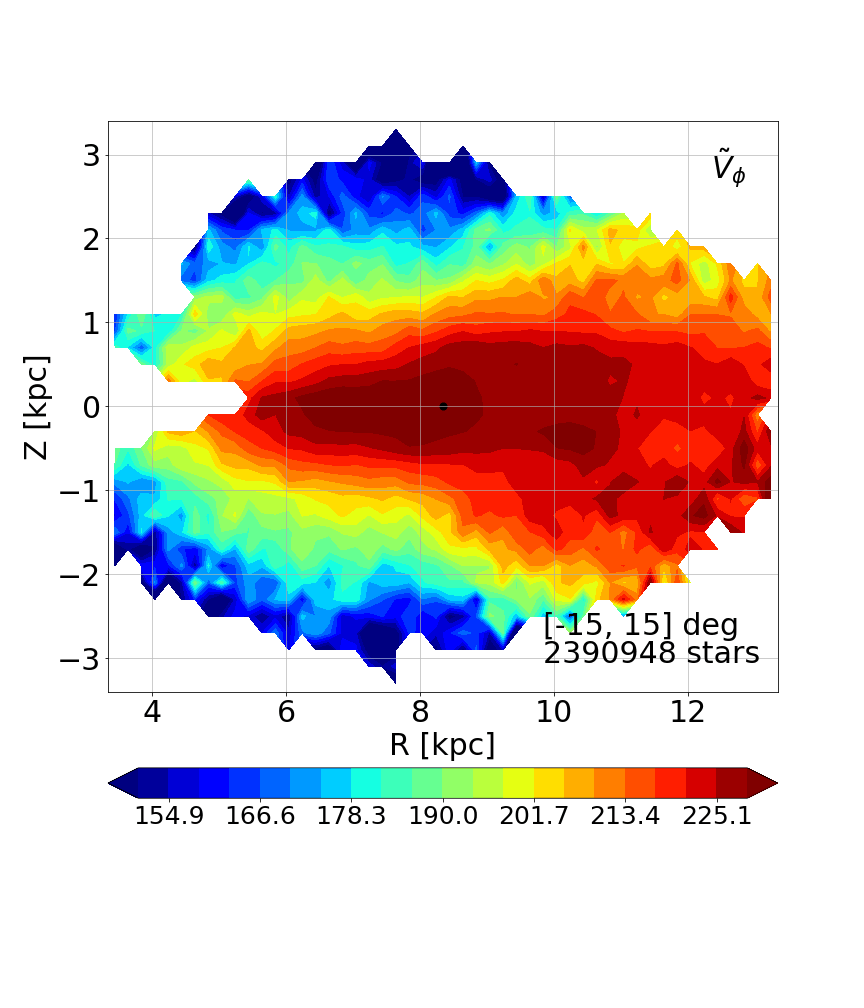}
\includegraphics[clip=true, trim = 10mm 40mm 20mm 40mm, width=0.41 \hsize]{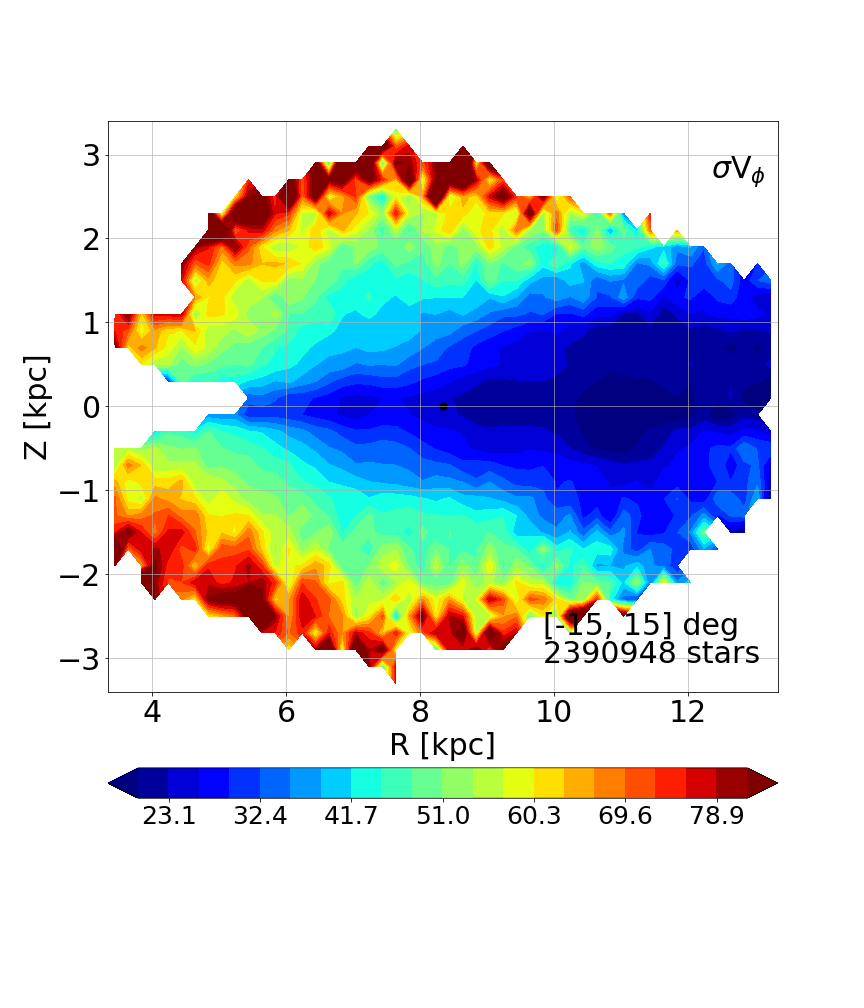}
\includegraphics[clip=true, trim = 10mm 40mm 20mm 40mm, width=0.41 \hsize]{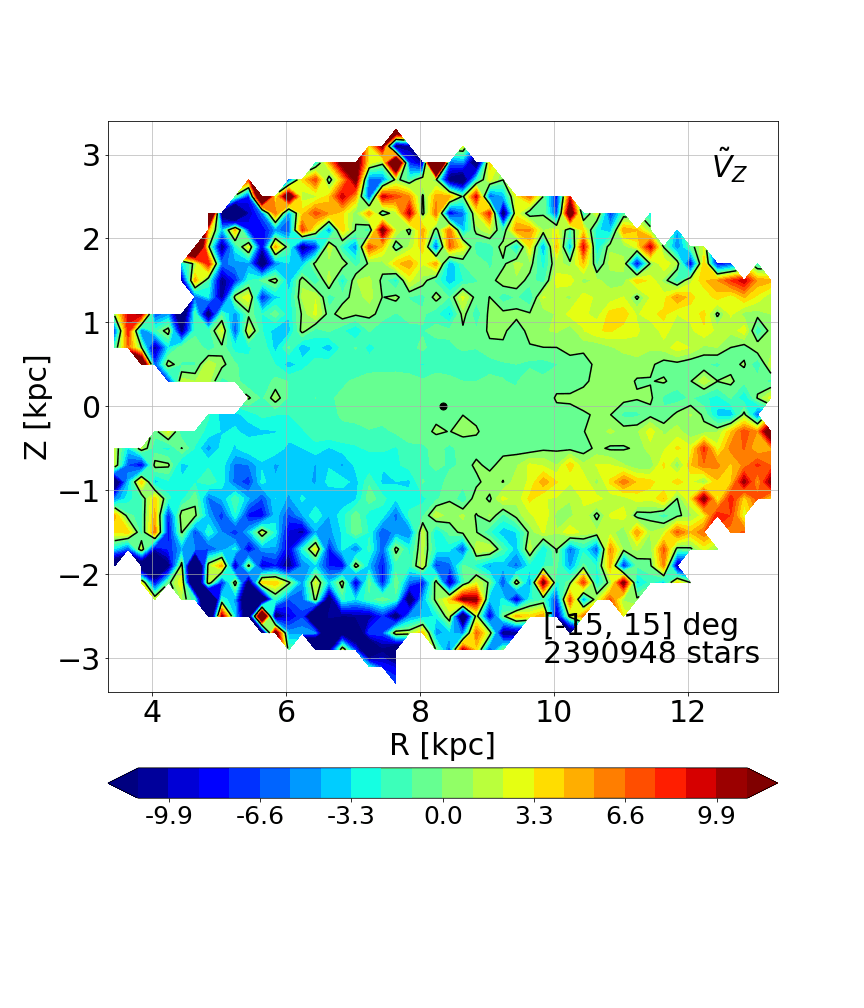}
\includegraphics[clip=true, trim = 10mm 40mm 20mm 40mm, width=0.41 \hsize]{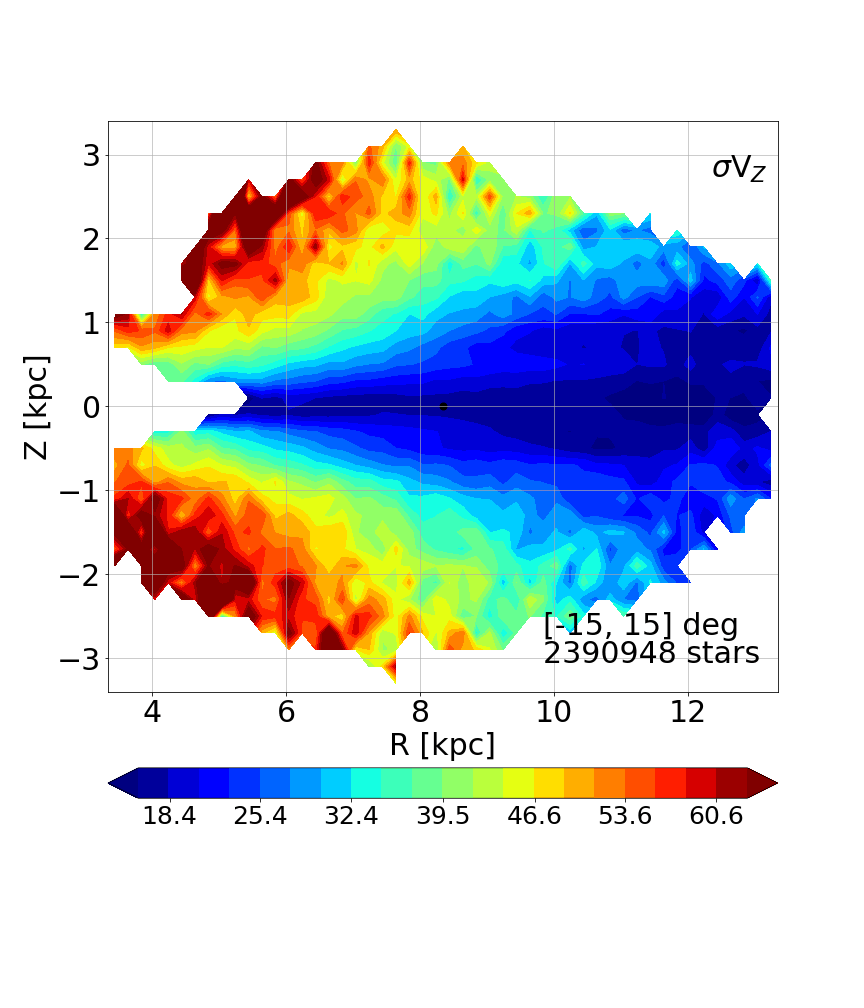}
\caption{Edge-on views of the kinematics of the disc for the azimuth range $\phi \in [-15, 15]$~degrees, derived using the giant sample: left, from top to bottom, median velocity maps $\tilde{V}_R, \tilde{V}_\phi, \tilde{V}_Z$ (in $\kms$), and right, from top to bottom, velocity dispersion maps $\sigma_{V_R}, \sigma_{V_\phi}, \sigma_{V_Z}$ (in $\kms$). The Sun is represented by a black dot at $R=8.34$~kpc and $Z=0$~kpc. The Galactic centre is located on the left side. The iso-velocity contours $\tilde{V}_R = 0$ and $\tilde{V}_Z = 0$~$\kms$ are pointed out as black lines. The numbers of stars used to produce the maps are given in the lower right corners.\label{fig:edgeon}}
\end{figure*}

\subsection{Method\label{sec:methodo}}
Four projections were used to study the kinematics (median velocities and velocity dispersions) of the giant sample.

\begin{enumerate}
\item \emph{\textup{Galactocentric Cartesian $XY$-Maps (face-on view).}} The sample was first divided vertically into layers of 400~pc height. The central layer was centred on the Galactic mid-plane and therefore contained stars with $Z$-coordinates in the range $[-200, 200]$~pc. The mosaic of $\tilde{V}_Z$ maps (Fig.~\ref{fig:xyvzmed}), presented in Appendix~\ref{app:maps}, is the exception. In order to determine possible vertical breathing modes, the layers were chosen symmetric with respect to the mid-plane. Each layer was then divided into $XY$-cells of 200~pc by 200~pc.
\item \emph{\textup{Galactocentric cylindrical $RZ$-maps (edge-on view).}} The sample was split into azimuth slices that were then divided into $RZ$-cells of 200~pc by 200~pc.
\item \emph{\textup{Galactocentric cylindrical radial projections.}} The sample was first split vertically into layers of 400~pc height and into two azimuth slices, $\phi = [-30, 0]$ and $[0, 30]$~deg,
respectively. The layers were centred on the Galactic mid-plane, except when we studied the median vertical velocity, for which specific attention was given to the possible north-south asymmetries. In this specific case, the giant sample was split into six layers, three above and three below the mid-plane. Each sub-sample was then divided into $R$-cells of 400~pc.
\item \emph{\textup{Galactocentric cylindrical vertical projections.}} The sample was first split into four azimuth slices of 15~degrees each and into three ranges in Galactic radius: $[6, 8]$, $[8, 10],$ and $[10, 12]$~kpc. Each sub-sample was then divided into $Z$-cells of 200~pc. This projection was used only to study the median vertical velocity, $\tilde{V}_Z$.\\
\end{enumerate}

When the cells were sufficiently populated, the medians ($\tilde{V}_i$, $i \in \{R, \phi, Z\}$) and the dispersions ($\sigma_{V_i}$, $i \in \{R, \phi, Z\}$)  of the velocities and their associated uncertainties were derived\footnote{according to formulae~\ref{eq:eMedLow} to \ref{eq:eDispUpp} (see Appendix~\ref{app:veloc}).}. A minimum of 30 stars per cell was required to compute the moments of the velocities in the $XY$-maps and $RZ$-maps. The minimum was 50 stars for the radial projections. Each face-on or edge-on map had its own colour range dynamics in order to heighten the contrast between the spatial structures within the map. {Conversely}, the different layers and azimuth slices shared the same scale in
the $R$-projections in order to facilitate the comparison.

The maps are (roughly) centred on the Sun $(X, Y)$ or $(R, Z)$ position, and the Galactic centre is located on the left side. In the face-on maps, the Milky Way rotates clockwise.

Figures~\ref{fig:faceon} and \ref{fig:edgeon} present the face-on and edge-on views of the median velocities and velocity dispersions for the mid-plane layer. For clarity, the full mosaics of face-on and edge-on maps, which offer vertical and azimuthal tomographic views of the disc kinematics, are presented in Appendix~\ref{app:maps}.

{To quantify and visualise the respective contributions of bending and breathing modes, we also map the bending and breathing velocities (Fig.~\ref{fig:xyvbendbreath}). We calculated them as the half-sum (mean) and half-difference of the median vertical velocities in symmetric layers with respect to the Galactic mid-plane:
\begin{equation}
V_{bending}(X,Y) = 0.5\ [\tilde{V}_Z((X,Y), L) + \tilde{V}_Z((X,Y), -L)]
\label{eq:bend}
\end{equation}
and
\begin{equation}
V_{breathing}(X,Y) = 0.5\ [\tilde{V}_Z((X,Y), L) - \tilde{V}_Z((X,Y), -L)]
\label{eq:breath}
,\end{equation}
where $\tilde{V}_Z((X,Y), L)$ is the median vertical velocity in the cell $(X,Y)$ and in the horizontal layer $L$. Layer $L$ was chosen to lie in the north Galactic hemisphere, and layer $-L$ is the symmetric layer in the south Galactic hemisphere. Formulae~\ref{eq:bend} and \ref{eq:breath} are similar to those defined by \citet{WidrowEtAl2014}, except that we calculated the half-difference for the breathing
velocity, while they used the full difference.}

%
%
%

\subsection{Radial velocity}
\begin{figure}[h!]
\centering
\includegraphics[width= \hsize]{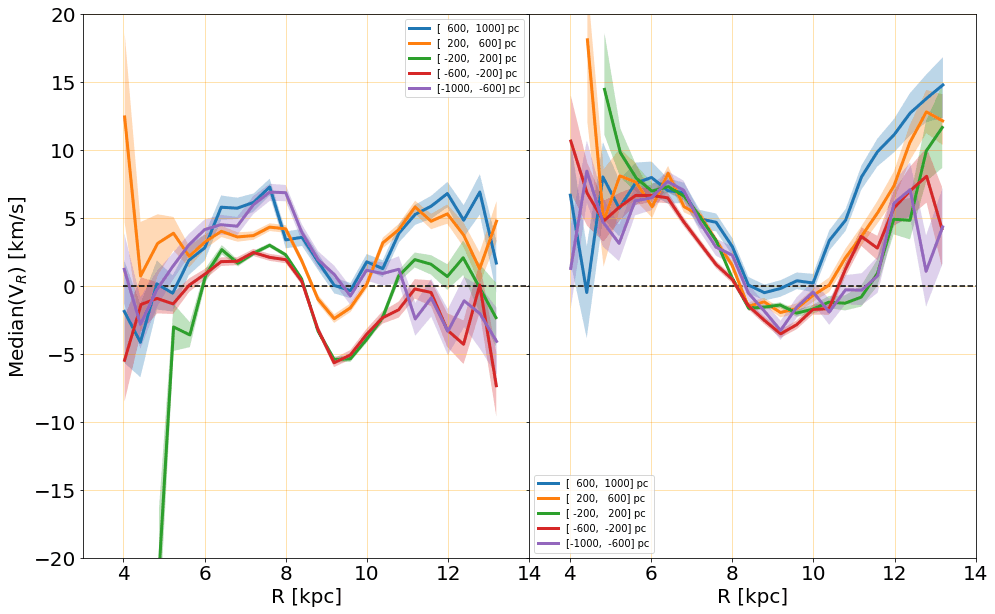}
\caption{Median radial velocities, $\tilde{V}_R$, of the giant sample as a function of Galactic radius for two azimuth slices: $[-30, 0]$~degrees (left), and $[0, +30]$~deg (right). The curves correspond to different $Z$ layers, defined in the legend. The shaded areas represent the $\pm 1-\sigma$ uncertainties on the median radial velocities.}
\label{fig:rvrmed}
\end{figure}

Figure~\ref{fig:rvrmed} shows the median radial velocity, $\tilde{V}_R$, as a function of Galactic radius for negative (left) and positive (right) azimuths and for different $Z$ layers (the different curves). The median radial velocity has a U-shape, with a minimum at about 9~kpc. Around this minimum and within a broad layer below and above the mid-plane, the median radial velocity is negative, meaning that more stars move inwards than outwards. At a distance from the minimum of 1 to 2~kpc, the median radial velocity becomes positive, meaning that more stars move outwards than inwards. At negative azimuths, the median radial velocity reaches maxima {at} around 6.5-7.5~kpc and 11-{13}~kpc and then decreases again. More than a U-shape, at negative azimuths, the median radial velocity seems to present oscillations. At positive azimuths, the signal is {partially} different. A maximum may be indicated at around 12-{13 kpc for Z in $[-1000, -200]$~pc, while in the other layers, the median radial velocity seems to continue to increase with Galactic radius}, but the data there are too noisy to conclude. In the inward direction, the radial velocity shows {a plateau starting at around} 7~kpc. {Farther inward, the rising of the green and orange curves at $R \leq 5-6$~kpc should be considered with caution, as it is significant only at the $\sim 1\sigma$ level.}

The vertical behaviour of the radial velocity oscillation varies with azimuth. At negative azimuths, the median radial velocity is minimum in the Galactic mid-plane and increases with distance to the plane. At positive azimuths, the median radial velocity shows {a much smaller} vertical gradient.

Seen face-on (upper left panel of Fig.~\ref{fig:faceon} and Fig.~\ref{fig:xyvrmed}), the negative radial velocities (blue-green pattern) have a semi-circular geometry with a small pitch angle that does not seem to present vertical variations.

Using RAVE data, \citet{SiebertEtAl2011} measured a negative radial velocity gradient from about 2~kpc inward of the Sun to about 1~kpc outward. This gradient was confirmed and further studied by several groups \citep{Williams2013, CarrilloEtAl2017}. \citet{CarrilloEtAl2017} also observed the onset of a positive gradient beyond the solar radius. Using samples of LAMOST giants, \citet{Tian2017} and \citet{LiuEtAl2017} also measured positive radial velocity and line-of-sight velocity gradients in the
direction of the galactic anti-centre, which flatten at around 2~kpc beyond the Sun. \citet{CarlinEtAl2013, Carlin2014} studied the motions of F-type stars observed with LAMOST in the direction of the Galactic anti-centre. They observed an inward mean motion of the stars in the Galactic plane and an inversion of the sense of the mean motion at a distance from the plane, in particular at $Z \lesssim -0.8$~kpc.

The negative and positive gradients revealed by previous studies are well visible in \gdrtwo\ data as part of oscillation(s) on
a kiloparsec scale. The full-sky coverage and large statistics of the \gdrtwo\ catalogue allows us to map the oscillation in 3D and to observe its semi-circular geometry, with a small pitch angle. At negative azimuth and around $R = 9$~kpc, the sign of the median radial velocity changes, that is, it is negative for $|Z| \lesssim 0.6-{0.8}$~kpc and positive at larger distances from the plane {(see Fig.~\ref{fig:rzvrmed} and \ref{fig:rvrmed})}, which is qualitatively in agreement with the observations of \citet{Carlin2014}. It should be noted that the vertical variation of $\tilde{V}_R$ is relatively modest, of the order of 5-{10}~$\kms$. Therefore a small change in the radial velocity zeropoint and in particular in the peculiar radial velocity of the Sun can modify the position of the inversion of the radial mean motion. {Different methods can indeed lead to estimates of the solar peculiar radial velocity with respect to the LSR that differ by a few $\kms$: that is, $U_\odot = 11.1\ \kms$ \citep{Schonrich2010}
and $U_\odot = 14.0\ \kms$ \citep{Schonrich2012}.}

%
%
%
%

\subsection{Azimuthal velocity}
\begin{figure}[h!]
\centering
\includegraphics[width= \hsize]{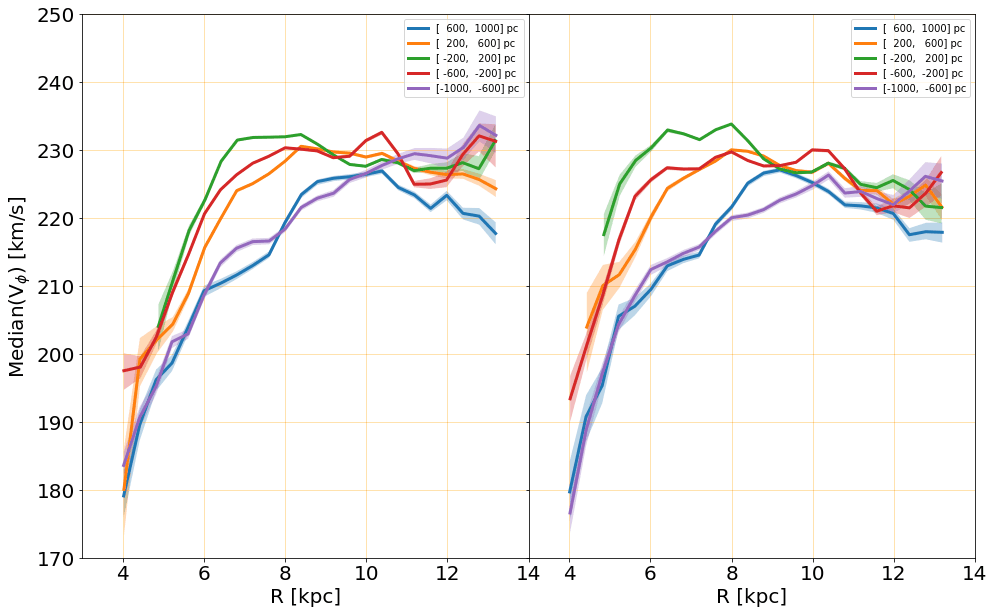}
\caption{Same as Fig.~\ref{fig:rvrmed} for the median azimuthal velocity, $\tilde{V}_\phi$.}
\label{fig:rvphimed}
\end{figure}

Figure~\ref{fig:rvphimed} shows the Milky Way stellar median rotation profiles from 4 to {13.2}~kpc from the Galactic centre. In the inner part of the Galaxy, the median azimuthal velocity presents a steep positive gradient with Galactic radius before it reaches a maximum at around 230~$\kms$ (a few $\kms$ below\footnote{This is expected for a mix of stars with different asymmetric drifts.} the value adopted in this study for the LSR: i.e. 240~$\kms$). When the maximum is reached, the azimuthal velocity presents a \emph{\textup{relatively}} flat profile, with variations of a few $\kms$ with Galactic radius. The asymmetric drift is expected to play a major role in the increase of the median velocity for increasing radius. At inner radii, the velocity dispersion in the radial velocity is larger (see Sect~\ref{sec:veldisp} and Fig~\ref{fig:vrdisp}), and the asymmetric drift correction is proportional to this dispersion squared. A detailed correction for the asymmetric drift is beyond the scope of this study, but only when this is completed can we assess whether the gradient in the azimuthal velocity is related to a gradient in the potential, to the effects of the non-axisymmetric perturbations such as the Galactic bar, and/or {to the increasing weight of the thin disc with respect to the $\alpha$-element-rich thick disc (the former presenting a greater radial scale length and a faster rotation than the latter, see \citealt{Bovy2012a, Robin2014}).}

The rotation profiles reach their maximum at shorter radius in the mid-plane than at larger distances from the plane: $R \sim 6-7$~kpc for $Z$ in $[-200, 200]$~pc, $R \sim 8$~kpc for $|Z|$ in $[200, 600]$~pc, and $R \sim 9-11$~kpc for $|Z|$ in $[600, 1000]$~pc. The comparison of the two panels of Figure~\ref{fig:rvphimed} and the comparison of the red and orange curves, on the one hand, and of blue and purple curves, on the other hand, show that the rotation profiles are relatively symmetric in azimuth and with respect to the mid-plane. At $R = 12$~kpc, most curves {are contained within a narrow range of median $V_\phi$.} The decrease with radius of the vertical gradient in azimuthal velocity is also visible in the edge-on maps (Fig.~\ref{fig:edgeon} middle left panel and Fig.~\ref{fig:rzvphimed}) as an outward flaring of the iso-velocity contours. This can be explained by an increase in asymmetric drift with $Z$. This change can be due to the different relative proportion of the thick and thin disc and/or of the different mean populations (young versus old), and to the variation in radial force in the galactic disc. \citet{Bienayme2015} have developed a dynamically self-consistent Staeckel potential using the mass distribution of the Besan\c{c}on Galaxy model. They showed the variation in asymmetric drift as a function of Galactocentric radius and distance to the Galactic plane. While the variations with $R$ are mild between 5 and 10~kpc (less than 20\%), the effect in $Z$ is very noticeable for the thin and thick discs
both. These variations are shown in \citet{Robin2017} and compared with the kinematics in \gdrone{}. The lag, typically of 5 to 20~$\kms$ in the Galactic plane, can be increased by 50 to 100\% at 1~kpc from the plane.

In addition to the large-scale variations, the median azimuthal velocity shows small-amplitude (a few $\kms$) variations with galactic radius, with maxima at $R \sim 6.5$~kpc (for $\phi > 0$ and $Z$ in $[-600, 200]$~pc), $R \sim 8$~kpc (for $\phi > 0$ and $Z$ in $[-600, 600]$~pc), and $R \sim 10$~kpc ($Z$ in $[-600, -200]$~pc). In the face-on maps (Figure~\ref{fig:xyvphimed}), in which the colour range dynamics was reduced to heighten the contrast between velocity features, these maxima are visible as red circular arcs. Super-imposed on this large-scale variation, the azimuthal velocity also shows arc-shaped oscillations with small amplitude on a kiloparsec scale.

%
%
%
%

\begin{figure}[h!]
\centering
\includegraphics[width= \hsize]{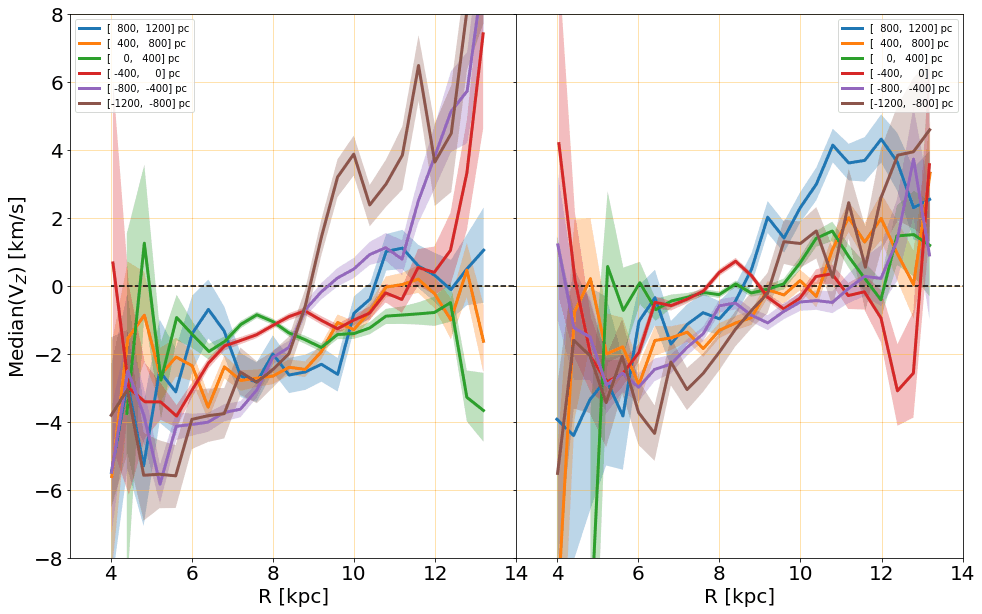}
\caption{Same as Fig.~\ref{fig:rvrmed} for the median vertical velocity, $\tilde{V}_Z$. The disc has been divided into six layers (the six curves), three above and three below the mid-plane.}
\label{fig:rvzmed}
\end{figure}

\subsection{Vertical velocity\label{vertvel}}
\begin{figure}[h!]
\centering
\includegraphics[width= \hsize]{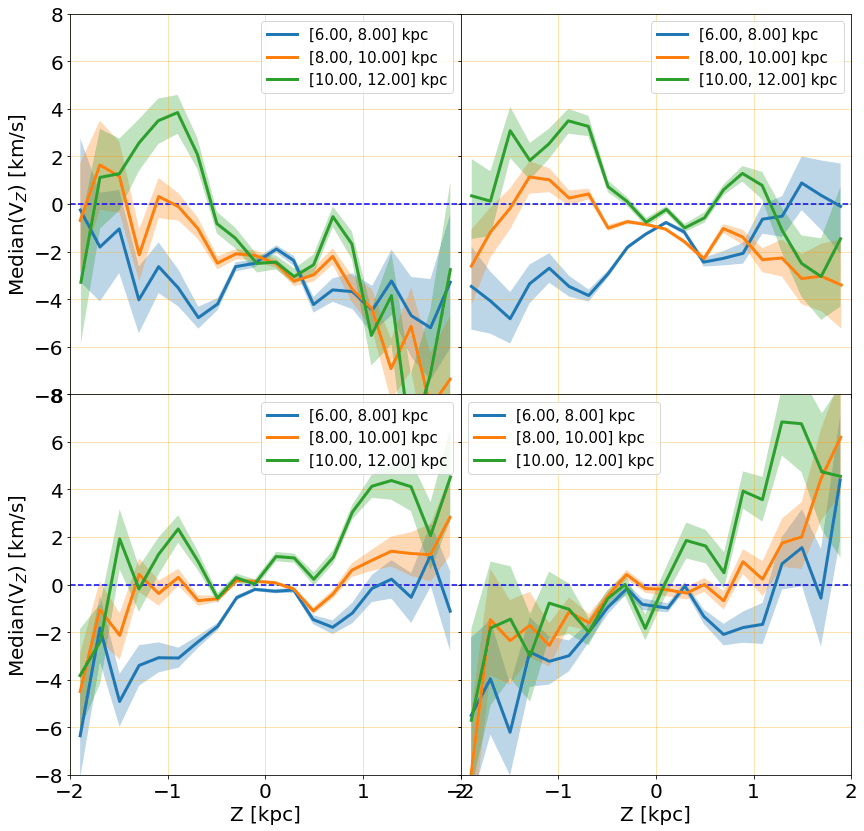}
\caption{Median vertical velocities, $\tilde{V}_Z$, of the giant sample as a function of height, $Z$, for four azimuth slices: $[-30, -15]$~degrees (upper left), $[-15, 0]$~degrees (upper right), $[0, +15]$~degrees (lower left), and $[+15, +30]$~degrees (lower right). The curves correspond to different ranges of galactic radius: $R \in [6, 8]$~kpc (blue), $[8,10]$~kpc (orange), and $[10, 12]$~kpc (green). The shaded areas represent the $\pm 1-\sigma$ uncertainties on the median vertical velocities.}
\label{fig:zvzmed}
\end{figure}

Figure~\ref{fig:rvzmed} shows a global increase in median vertical velocity, from {the inner to the outer disc}, but with complex vertical and azimuthal dependencies. The face-on (Fig.~\ref{fig:faceon} lower left panel and Fig.~\ref{fig:xyvzmed}) and edge-on maps (Fig.~\ref{fig:edgeon} lower left panel and Fig.~\ref{fig:rzvzmed}) show kiloparsec large, negative (green to blue) and positive (light green to red) velocity features, with an elaborate 3D geometry. Figure~\ref{fig:zvzmed} presents the vertical projection of $\tilde{V}_Z$ as a function of height $Z$ for different azimuth slices and ranges in Galactic radius. In the outer disc ($R > 10$~kpc), the positive velocity feature appears inclined with respect to the Galactic plane, that is, it is located below the mid-plane at $\phi \lesssim -15$~deg, extending over {most of} the width of the plane for $\phi$ in $\sim[-15, +15]$~deg and located mainly above the plane for $\phi \gtrsim 15$~degrees. Still in the outer disc and for $\phi \in [-15, +15]$~degrees, the median vertical velocity is mildly symmetric with respect to the mid-plane, with a minimum at around $Z = 0$~kpc and maxima at around $|Z| = 0.8-1.2$~kpc. {Globally, in the outer disc and at $\phi < -15$~deg, the vertical velocity shows a negative gradient with Z. The gradient flattens, but is still negative for $\phi \in [-15, 0]$~deg. It becomes positive for $\phi \in [0, +15]$~deg and steepens for
$\phi > +15$~deg. In addition to this evolving gradient, the vertical velocity shows two local maxima at around $|Z| = 0.8-1.2$~kpc.} In the inner disc, $R \in [6, 8]$~kpc and for $\phi > -15$~degrees, the vertical velocity shows a global increase with $Z$.

{Figure~\ref{fig:xyvbendbreath} shows the face-on maps of the bending and breathing velocities (defined in Sect.~\ref{sec:methodo}) for three groups of symmetric layers with respect to the Galactic mid-plane, from top to bottom: $[0, 400]$ and $[-400, 0]$~pc, $[400, 800]$ and $[-800, -400]$~pc, and $[800, 1200]$ and $[-1200, -800]$~pc. The bending velocity is negative (i.e. oriented towards the south Galactic pole) at negative azimuth for $|Z| \in [0, 400]$~pc and in the inner disc at larger distance from the mid-plane. It is positive in the outer disc. Close to the Galactic mid-plane, the signal is weak and localised. It becomes stronger and spatially more extended with greater distance from the mid-plane. The absolute value of the breathing velocity is mostly lower than 1~$\kms$ for $|Z| < 800$~pc. In the range $|Z| \in [800, 1200]$~pc, the breathing velocity is partly positive in the first, second, and fourth quadrants, and it is negative in the third.}

Using SEGUE spectra, \citet{Widrow2012} studied the vertical variations in mean vertical velocity, $\bar{V}_Z$, of a sample of high Galactic latitude ($|b| \in [54, 68]$~degrees) outer disc stars (Galactic longitude $l \in [100, 160]$~degrees). The mean vertical velocities they measured show a vertical asymmetry, with $\bar{V}_Z < 0$~$\kms$ below  $\sim0.5$~kpc and positive above. The mean vertical velocity also presents some oscillations. In the following year, \citet{Williams2013} studied the velocity field in an area of about 2~kpc around the Sun. Their ($R, Z$) maps show inversions of the sense of the mean vertical motion of the stars along the $Z$ axis, producing zones of compression and zones of rarefactions of the stars. Recently, \citet{CarrilloEtAl2017} compared the velocity field derived with different proper motion catalogues and found great differences in particular in the median vertical velocity, $\tilde{V}_Z$, maps. With the $\gdrone$ TGAS catalogue, they observed a breathing mode (a median motion of the stars away from the plane) in the inner disc and a downward bending {beyond the Sun,} over a distance of about 1~kpc.

The complex radial, vertical, and azimuthal dependencies of the vertical velocity make a comparison of samples with different selection functions difficult. The stars selected at positive azimuth and less than 2 kiloparsecs beyond the Sun (orange curves in the lower panels of Fig.~\ref{fig:zvzmed}) have some intersect with the sample of \citet{Widrow2012}. Although not identical, the vertical velocity profiles look compatible. In the inner disc {and $\phi > -15$~degrees}, we observe an increase in vertical velocity, with $Z$ having similarities with the vertical profile of the RAVE-TGAS sample\footnote{and distances from \citet{Astraatmadja2016}} of \citet{CarrilloEtAl2017}, but with smaller amplitudes at large $Z$ and a less pronounced symmetry with respect to the mid-plane (our inner disc $\tilde{V}_Z$ are mostly negative).
{It should be noted that because the median $V_Z$ values are relatively modest, a small change in the vertical velocity zeropoint can modify the position of the inversion of the vertical motion.}

%
%
%
%

\subsection{Radial, azimuthal, and vertical velocity dispersions\label{sec:veldisp}}
\begin{figure}[h!]
\centering
\includegraphics[width= \hsize]{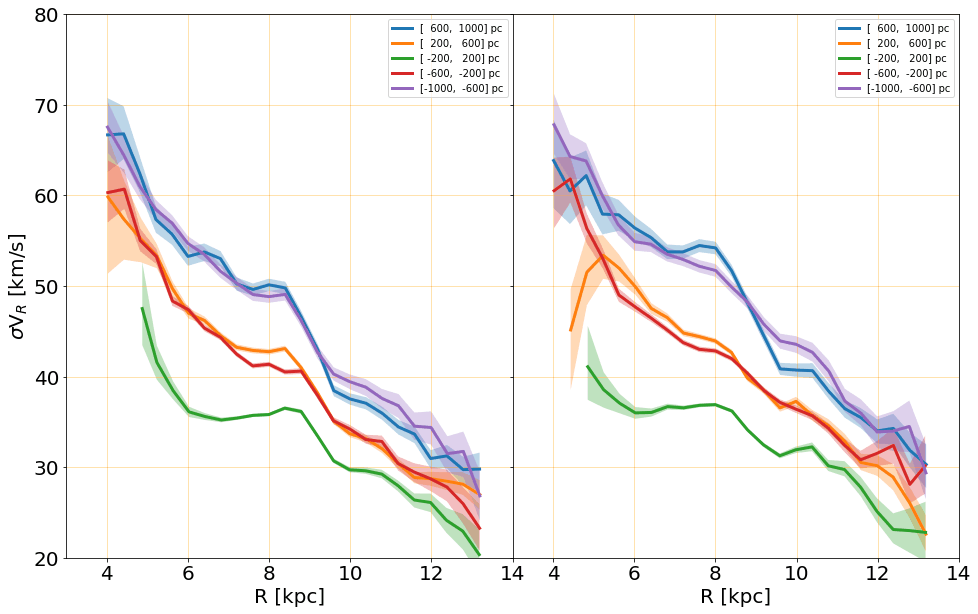}
\caption{Same as Fig.~\ref{fig:rvrmed} for the radial velocity dispersion, $\sigma_{V_R}$.}
\label{fig:vrdisp}
\end{figure}

\begin{figure}[h!]
\centering
\includegraphics[width= \hsize]{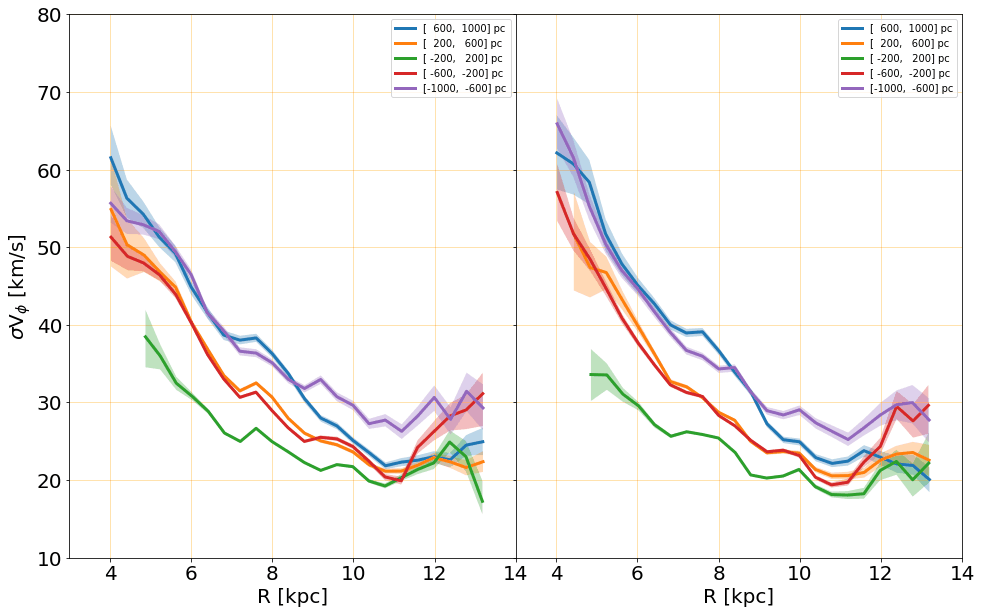}
\caption{Same as Fig.~\ref{fig:rvrmed} for the azimuthal velocity dispersion, $\sigma_{V_\phi}$.}
\label{fig:vphidisp}
\end{figure}

\begin{figure}[h!]
\centering
\includegraphics[width= \hsize]{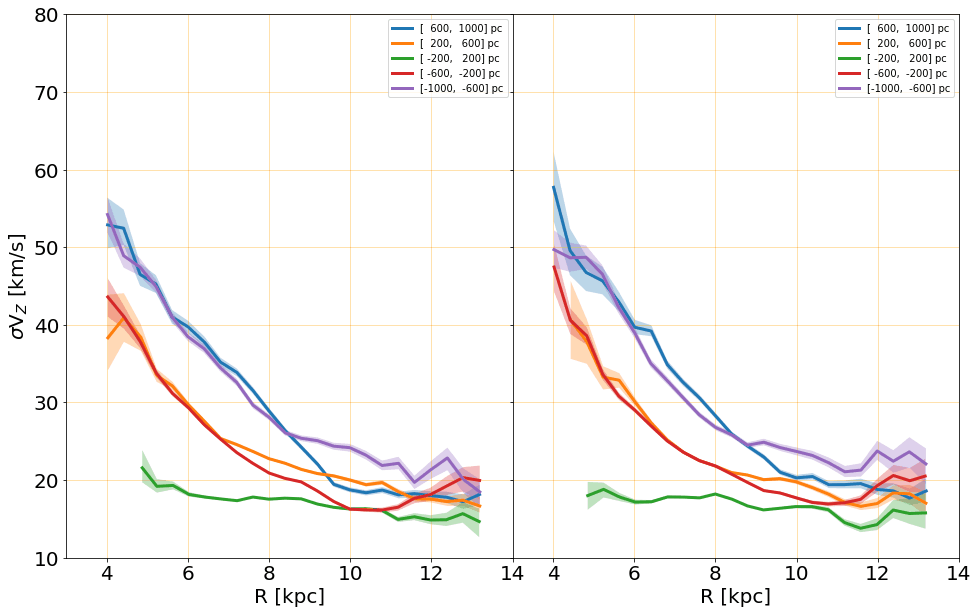}
\caption{Same as Fig.~\ref{fig:rvrmed} for the vertical velocity dispersion, $\sigma_{V_Z}$.}
\label{fig:vzdisp}
\end{figure}

Figures~\ref{fig:vrdisp}, \ref{fig:vphidisp}, and \ref{fig:vzdisp} show the dispersions of the three galactocentric components of the velocities, $\sigma_{V_R}$, $\sigma_{V_\phi}$ , and $\sigma_{V_Z}$, as a function of galactic radius for negative (left) and positive (right) azimuths and for different $Z$ layers (the different curves). The three velocity dispersions decrease with increasing radius. The gradient is significantly stronger at intermediate and large $Z$ than in the mid-plane, with the vertical velocity dispersion $\sigma_{V_Z}$ showing almost no gradient in the $Z$ layer $[-200, 200]$~pc. The dispersions are very symmetric with respect to the Galactic mid-plane, with the curves of symmetric layers showing very similar behaviours, including some kiloparsec-scale bumps/oscillations. 

As shown on the right side of Figure~\ref{fig:edgeon}, the iso-velocity dispersions flare outwards. Two effects can act together to produce these flares. On the one hand, there is a radial evolution in the relative proportion of the short-scale length thick disc and the colder longer-scale length thin disc. On the other hand, with increasing outward distance, the vertical component of the gravitational force weakens, and for the same velocity, a star can reach larger distances from the mid-plane. 

The velocity dispersions, in particular $\sigma_{V_R}$ and $\sigma_{V_\phi}$ , show small-amplitude fluctuations that extend on a kiloparsec scale both radially and vertically. The face-on view of the disc (Fig.~\ref{fig:faceon}) shows that these hot features have a semi-circular geometry that extends at least 20 to 30 degrees in azimuth.

%
%
%
%

\subsection{Discussion}
The Milky~Way is not an axisymmetric system at equilibrium. {In the past few years (less than a decade)}, asymmetric motions \citep{CasettiEtAl2011}, gradients \citep{SiebertEtAl2011}, and wave patterns \citep{Widrow2012} have been detected in the velocity field and were studied in increasingly more detail \citep{Williams2013, CarlinEtAl2013, SunEtAl2015, CarrilloEtAl2017, Pearl2017, Tian2017, LiuEtAl2017, BabaEtAl2017}. The second \gaia{} data release now offers a full-sky 3Dview of the complex Milky Way velocity pattern. It shows streaming motions in all three velocity components as well as small-amplitude fluctuations in the velocity dispersions.

Streaming motions might be produced by internal mechanisms {(e.g. response of the stars to the bar and/or spiral structure) or by external perturbers (e.g. satellite accretion(s), impact of low-mass dark matter halos), or by combinations of both.} It is beyond the scope of this paper to model the observations in
detail. Below, we review some of the results of the already rich literature and discuss them with regard to the \gdrtwo{} maps. 

\citet{SiebertEtAl2012} compared different models with two and four long-lived spiral arms. They successfully reproduced the gradient they had found the year before \citep{SiebertEtAl2011} with a two-arm model and a specific location of the Sun, near the inner ultra-harmonic 4:1 resonance. \citet{MonariEtAl2014} showed that the bar can also induce a negative radial velocity gradient. \citet{FaureEtAl2014} studied the stellar velocity response to stable spiral perturbations. Within corotation (located in their model at 12~kpc), the model produces inward motions within the arms and outward motions between the arms. The model also induces vertical breathing modes, with stars moving away from the plane at the outer edges of the arms and towards the plane at the inner edges (still within corotation). \citet{Debattista2014} also obtained breathing modes, with compression where the stars enter the spiral arms, and expansion where they exit. {\citet{MonariEtAl2016a} developed an analytical model, based on phase-space distribution functions, to study the perturbations induced by a spiral potential. The model predicts breathing modes.} \citet{Grand2016} used cosmological simulations to study the large-scale motions induced by the spiral arms in a Milky~Way-like galaxy. The simulation shows radially outwards and azimuthally backwards motions on the trailing edge of the arms, while on the leading edge, the effect is reversed: the streaming motion is oriented inwards and forwards (see also~\citealt{Antoja2016}). \citet{MonariEtAl2016b} studied the combined influence of the bar and two quasi-static spiral arms. The model produces horizontal (i.e. radial and azimuthal) streaming motion dominated by the influence of the bar and vertical breathing modes with spiral arms shape, but with the bar heightening the amplitude of the modes and shifting their locations. The vertical waves produced by internal mechanism models are usually breathing modes. Using N-body simulations, however, \citet{ChequersWidrow2017} recently showed that even in isolated Milky~Way-like galaxies, random noise in the distributions of halo and bulge stars can produce long-lived bending waves in the disc that are observable beyond the solar circle.

\begin{figure}[h!]
\centering
\includegraphics[width=\hsize]{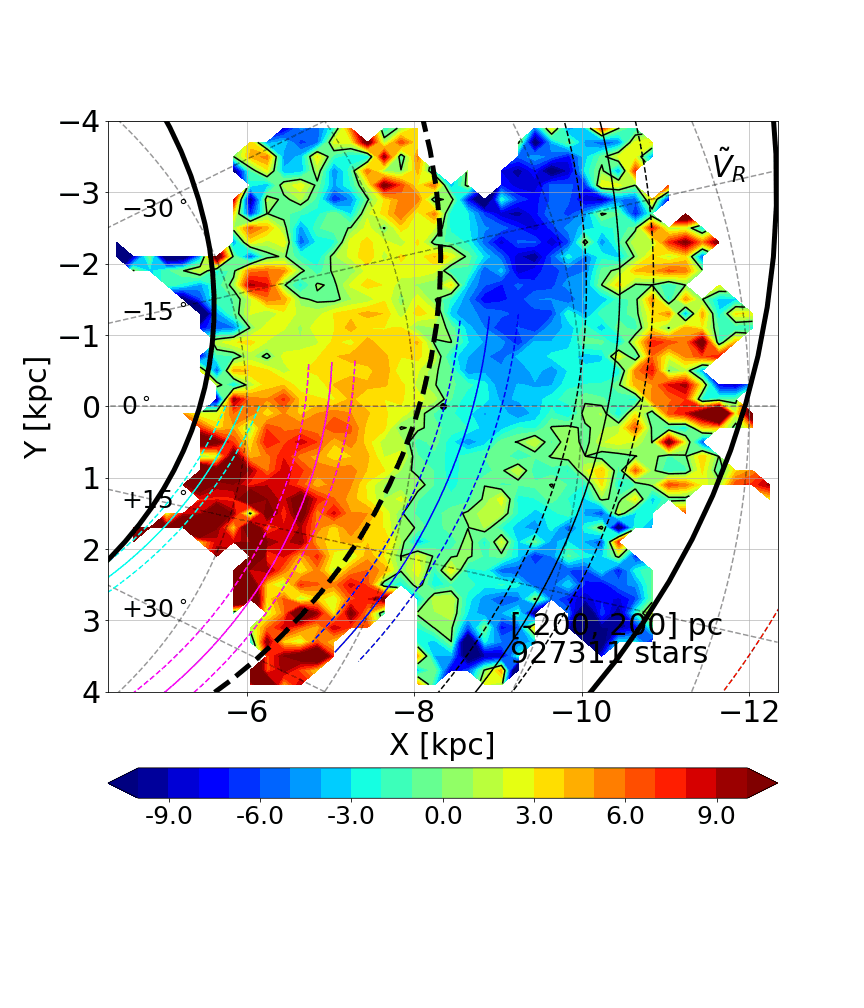}
\caption{Face-on map of the median radial velocity (in $\kms$) for the mid-plane layer ($[-200, +200]$~pc), derived using the giant sample. The two-arm model of \citet{drimmel00}, adjusted on near infra-red data, is over-plotted as thick black lines. The thick dashed line highlights the locus of minimum density between the two arms. The spiral arms model of \citet{Reid2014} is also over-plotted, i.e. from the inner to the outer disc: Scutum (cyan), Sagittarius (magenta), Local Arm (blue), Perseus (black), and the Outer Arm (red).}
\label{fig:xyvrmedarms}
\end{figure}

\begin{figure}[h!]
\centering
\includegraphics[width=\hsize]{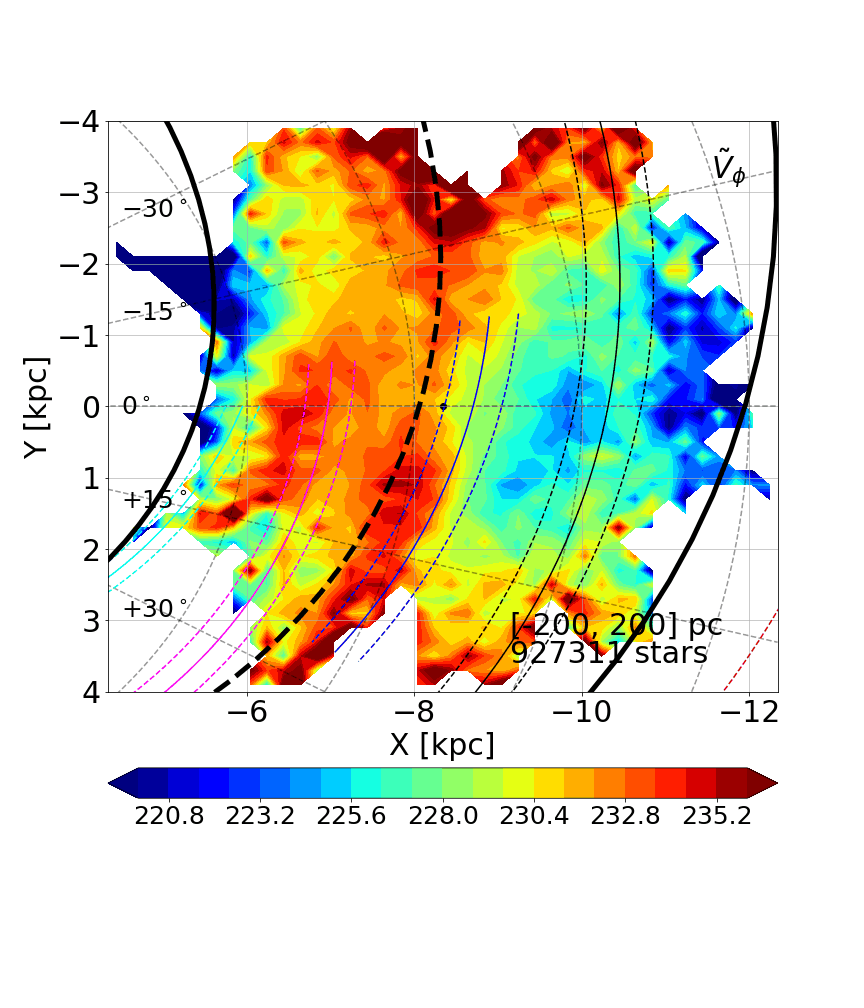}
\caption{Same as Fig.~\ref{fig:xyvrmedarms} for the median azimuthal velocity, $\tilde{V}_\phi$.}
\label{fig:xyvphimedarms}
\end{figure}

Figures~\ref{fig:xyvrmedarms} and \ref{fig:xyvphimedarms} show the face-on maps of the median radial and azimuthal velocities, respectively, for the mid-plane layer ($[-200, +200]$~pc). Two models of spiral arms are over-plotted. The two-arm model of \citet{drimmel00}, derived from near infra-red data, is represented by thick black lines, and the locus of the minimum density between the two arms is shown by the thick dashed line.
The spiral arms model of \citet{Reid2014} is represented with thin colour-coded lines (see caption of Fig.~\ref{fig:xyvrmedarms}). \citet{Reid2014} used masers as tracers of the spiral arms. It should be noted that these masers are associated with massive stars that are much younger than the giant stars whose kinematics is mapped in this section. The Local Arm shows some coincidence with the ridge of negative median radial velocities, and its trailing edge is close to the boundary between positive and negative $\tilde{V}_R$. This might even be fortuitous as the Local Arm is usually considered a weak structure~\citep{Churchwell2009}. The locus of minimum density between the two near-infrared arms also matches the boundary between positive and negative median radial velocities. The locus also correspond mildly with the semi-circular faster azimuthal velocity pattern (yellow-red arc in Fig.~\ref{fig:xyvphimedarms}). Dynamical models of bar and/or spiral arms predict streaming motions and changes in sign of the median radial velocity. It is therefore tempting to see a link between the radial velocity oscillation and the near-infrared arms. \citet{SiebertEtAl2012} indeed reproduced the negative radial gradient with a two-arm model. On the other hand, it should also be noted that \citet{LiuEtAl2017} obtained a radial oscillation by adjusting the positive radial gradient with a bar model. The mapping in 3D of $\tilde{V}_R$ and $\tilde{V}_\phi$ brings new constraints to the models.

Vertical to the Galactic disc, we expect the kinematics to reflect the large-scale warp. If the Milky Way warp is a long-lived structure, then we expect an associated kinematic signature towards the Galactic anti-centre in the vertical velocities. Figure~\ref{fig:faceon} (lower left plot) indeed seems to exhibit a systematic vertical velocity of about 2-3~$\kms$ at $R=10-11$~kpc in the direction of the anti-centre. However, this signal is weaker than expected from current empirical descriptions of the stellar warp, which assume the warp to be stable and non-precessing, and might indicate that the warp is instead an unstable transient feature.

\begin{figure}[h!]
\centering
\includegraphics[width=1.1\hsize]{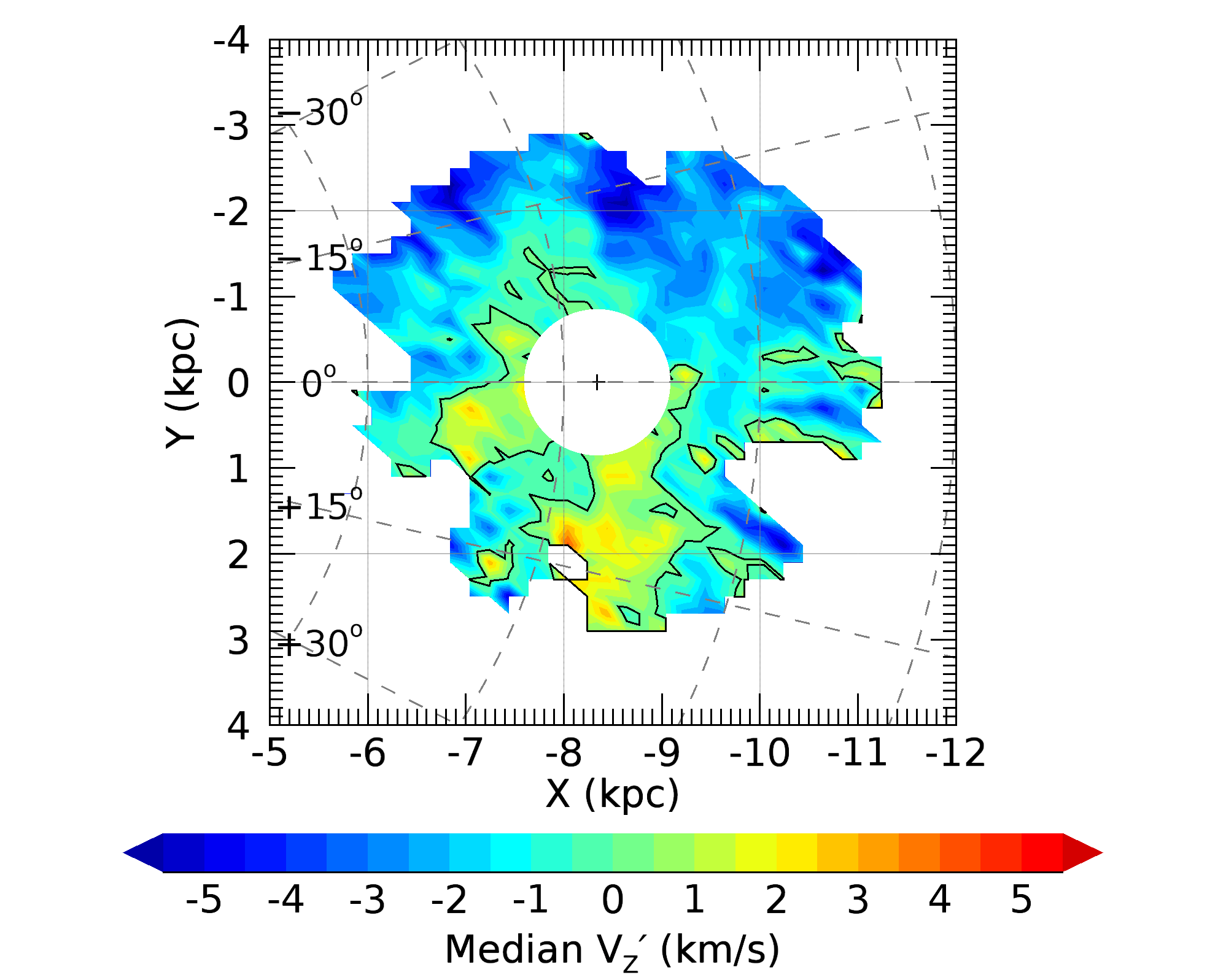}
\caption{Median vertical velocity $V^\prime_Z$ of the OB sample in the disc midplane ($Z = [-200,200]$pc) for the 200\,098 stars with $|Z| < 200$pc and $|b|< 15^\circ$. Orientation and coordinates are the same as in Fig.~\ref{fig:faceon}. The black cross shows the position of the Sun, and the white area around the Sun masks the region where the height of a 200~pc cell has a Galactic latitude $b>15^\circ$. The iso-velocity contours $V^\prime_Z$ = 0 km s$^-1$ are shown as black lines.}
\label{fig:vz_obstars}
\end{figure}

It is worth comparing the $\tilde{V}_Z$ map of the giants (Fig. \ref{fig:faceon}, lower left panel) to an equivalent map for the young OB sample. Line-of-sight velocities are not available for this sample (see Sect.~2.1 and \citet{DR2-DPACP-54} for details), so that we cannot directly calculate $V_Z$ for each star. However, at low Galactic latitudes, we can estimate a $V'_Z$ since most of the vertical motion is seen in the proper motions perpendicular to the Galactic plane. The vertical velocity
\begin{equation}
V_Z = \frac{4.74 \mu_b}{\varpi \cos b} + W_\odot + (S-S_\odot) \tan b,
\label{vzprime}
\end{equation}
where $S_\odot = U_\odot \cos l + V_\odot \sin l$, and similarly for $S$, which contains both differential Galactic rotation and the own peculiar motion of the star parallel to the Galactic plane, $S_*$. Neglecting the latter and assuming a flat rotation curve, we estimate $V^\prime_Z$ by taking $S \approx V_{LSR} (R_\odot/R - 1) \sin l$ in the above equation. Using stars in the OB sample within 200~pc of the Galactic plane, we map the median $V^\prime_Z$, shown in Fig. \ref{fig:vz_obstars} (our approximation means that we have effectively ignored a $\langle S_* \tan b \rangle$ for each cell). The resulting map is distinctly different from the map for the giants, and show no signs of a warp signature. We recall that our sample traces the motions of the gas from which these stars have recently been born. The lack of any warp signature here therefore again suggests that the warp is an unstable transient feature. 

External perturbations by satellites or dark-matter haloes, for
instance, can also disturb the disc structure and velocity field, and may in particular induce vertical waves \citep{WidrowEtAl2014, FeldmannSpolyar2015}. Simulating the interaction between a satellite and a galactic disc, \citet{WidrowEtAl2014} showed that the response of the disc depends, in particular, on the relative vertical velocity of the satellite. A satellite approaching "slowly" will primarily bend the disc, while a fast satellite will produce breathing and higher order modes. The Sagittarius dwarf galaxy has been interacting with the Milky Way for several Gyr. It is therefore a natural suspect for the perturbations observed in the velocities as well as in the outer disc. Simulations of the accretion of Sagittarius by the Milky~Way were indeed able to reproduce the local vertical velocity pattern measured by \citet{Widrow2012} \citep{GomezEtAl2013, LaporteEtAl2017}, but also large-scale outer-disc features such as the Monoceros ring \citep{LaporteEtAl2017}. The Large Magellanic Cloud can also induce vertical modes in the local disc, but with significantly smaller amplitudes than Sagittarius \citep{LaporteEtAl2017, LaporteEtAl2018}. External perturbers do not only modify the vertical velocities, but also the horizontal ones. The effect has in particular been studied in the velocity plane \citep{Minchev2009, Gomez2012a} and is discussed in the next section.

The vertical velocity field (Sect.~\ref{vertvel}) is quite complex, with different behaviour in the inner and outer disc and radial, azimuthal, and vertical dependencies. As previously observed by \citet{CarrilloEtAl2017}, it cannot be described by a single bending, breathing, or higher mode. We likely witness a superposition of modes that may be of several different origins.

%

\section{Kinematic substructure}\label{secuv}

In this section, we revisit the substructure in the velocity plane of the solar neighbourhood and explore it also in the velocity distribution of distant regions from the Sun. We focus on the in-plane velocities because they have been demonstrated to show most of the substructure. We remark that a detailed study of the kinematic substructure present in local and distant neighbourhoods is beyond the scope of the present study. Here we present only a first exploratory look and focus on the quality of the Gaia data, the aspects that allow us to scientifically verify the data, and the highlights of our findings. 

\subsection{Kinematic substructure in the solar neighbourhood}

\begin{figure*}\centering
\includegraphics[width=0.8\hsize]{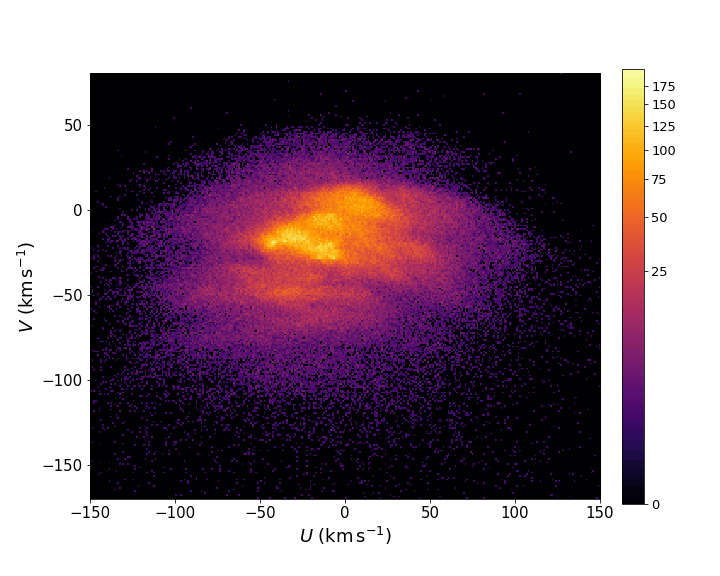}
\caption{Velocity plane of the stars in the solar neighbourhood. We show a 2D histogram of the velocity with a bin of $1\,\kms$, thus, the colour scale indicates the number of stars per ($\,\kms$)$^2$.}\label{figuv}

\end{figure*}

As explained in Sec. \ref{dataselection}, we built our solar neighbourhood sample by selecting stars closer than 200 pc to the Sun. The number of stars ({366\,182}) is an order of magnitude larger than local samples such as Geneva-Copenhagen Survey (GCS, \citealt{Nordstrom2004}) or RAVE (\citealt{Steinmetz2006}). Moreover, the uncertainties in velocity of this sample, which {are smaller than 1
$\kms{}$ for about} 80\% of the sample, allow us to probe substructure scales that are significantly smaller than ever
before. 

Figure \ref{figuv} shows a 2D histogram with a bin size of $1\,\kms$ of the in-plane Cartesian heliocentric velocities $U$ and $V$ of the solar neighbourhood sample. This velocity distribution is highly structured. We observe many nearly horizontal arch-like structures that have never been seen before. Even the dynamical stream of Hercules, located at negative $U$ velocities and $V\simeq-50\,\kms$  , now appears to be split into at least two of these branches (at $V\simeq-38\,\kms$ and $V\simeq-50\,\kms$) and perhaps the stream at $V\simeq-70\,\kms$ is also associated with the same structure. These arches appear for the whole range of $V$. We
note, for instance, the new low-density arch at $V\simeq 40\,\kms$. Some of them are not centred on $U$ and others are inclined in $V$.  Additionally, there is a clear under-density of stars also with an arched shape that extends from $(U,V)\simeq(-100, -25)$ to $(U,V)\simeq(75, -65)\,\kms$ immediately above the Hercules stream, which separates the velocity plane in two. This gap is not as horizontal as the over-dense arches.

Fig. \ref{figuv}  also shows more strongly rounded structures with sizes of about $10\,\kms$, especially at the centre of the distribution, which correspond to previously known moving groups and dynamical streams. We also see small high-density clumps that might be associated with known open clusters in the vicinity
of the Sun.  For instance, the yellow clumps at $(U,V)=(-42,-19)$ and  $(U,V)=(-7, -28)\,\kms$ correspond to the Hyades and Pleiades clusters, respectively. All of these substructures, that is, the medium and small structures, appear to be embedded in the larger arched substructures.

\begin{figure}
\centering
\includegraphics[width=0.8\hsize]{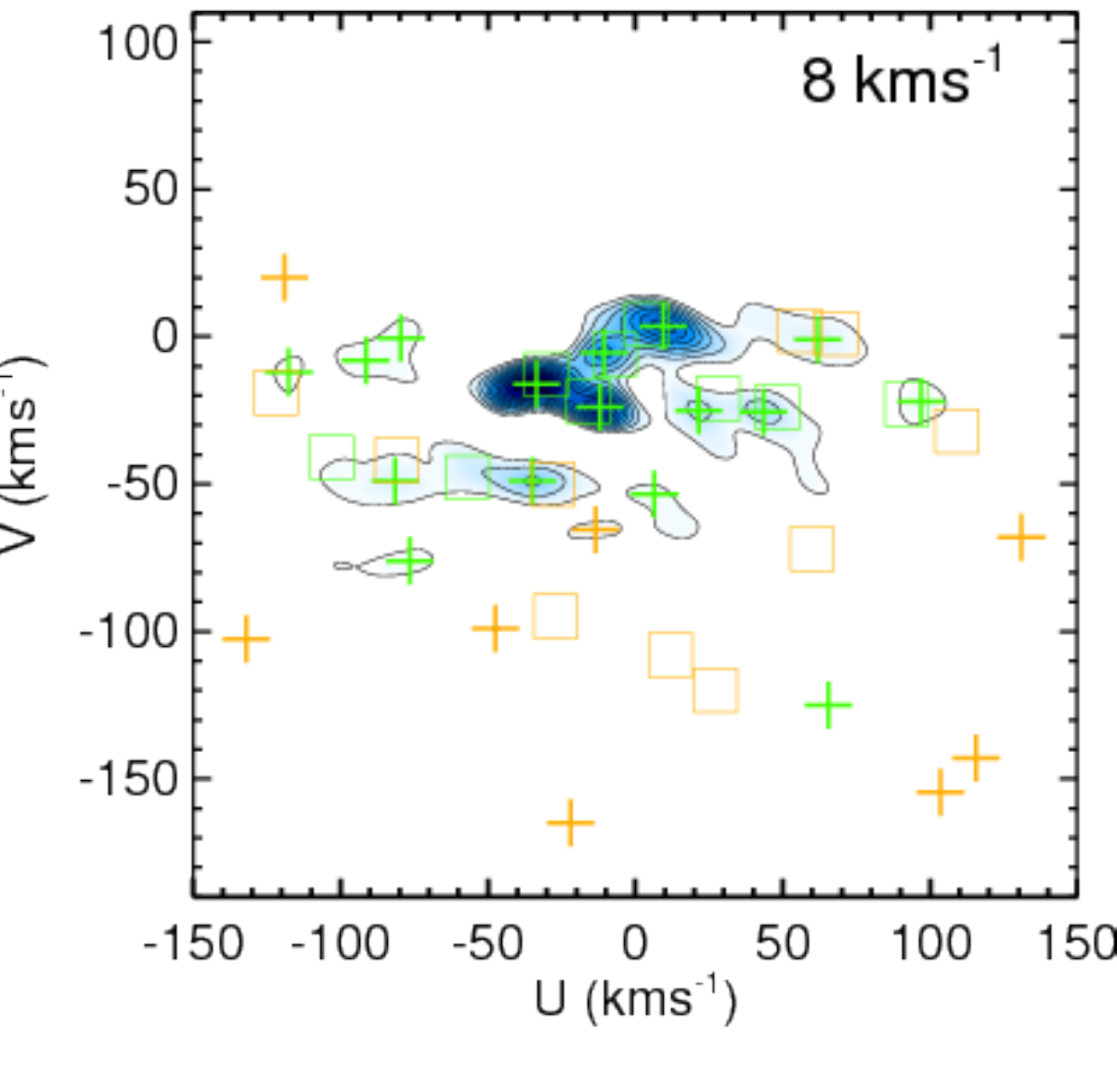}

\includegraphics[width=1.\hsize]{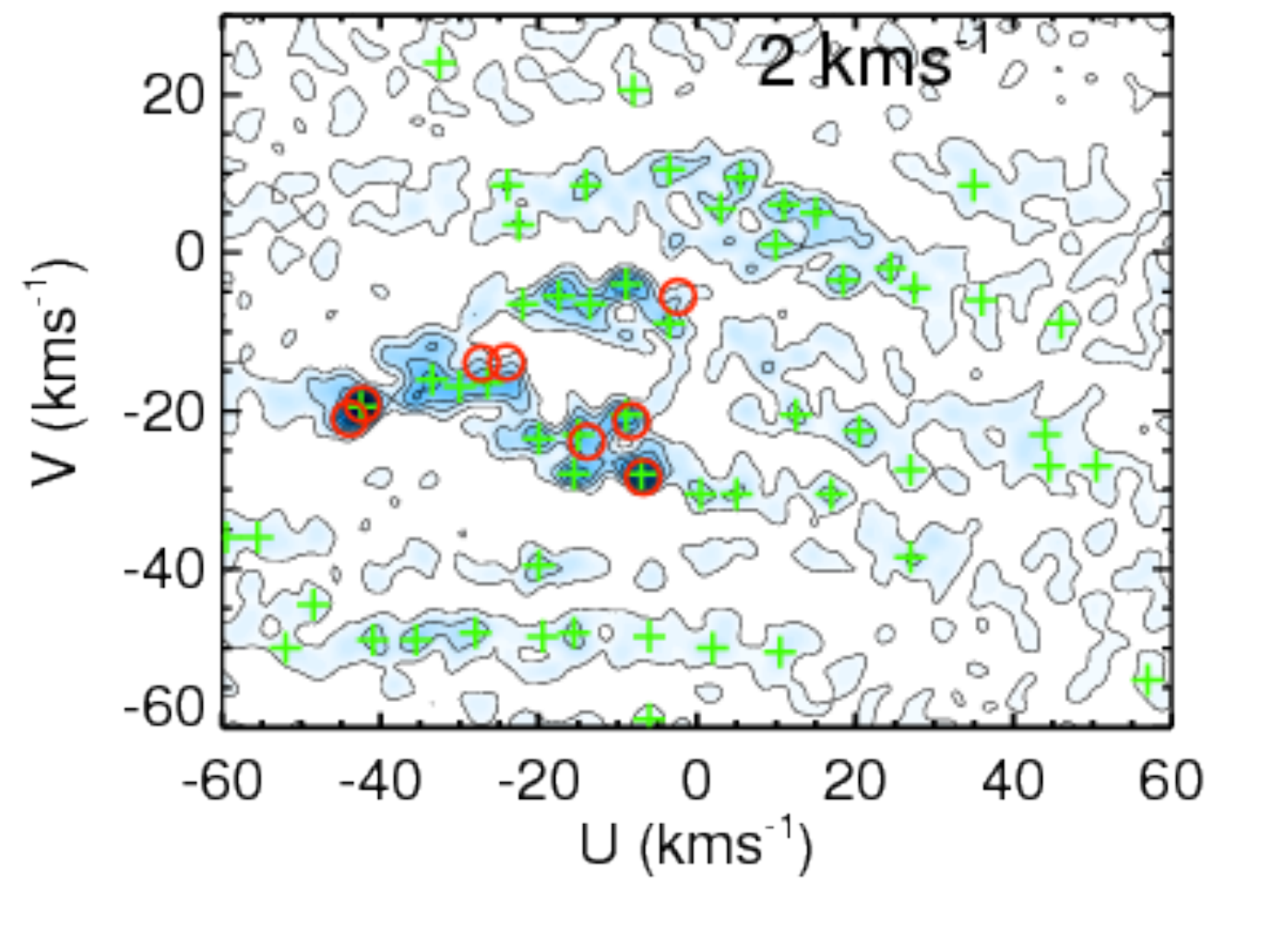}

\caption{Wavelet transform coefficients of the velocity plane in the solar neighbourhood showing the substructures at different scales. {\it Top:} Scales of around $8\,\kms$. 
The green and orange symbols mark peaks that are $>3\sigma$ and $2-3\sigma$ significant, respectively, for DR2 (crosses) or for the RAVE sample (squares) from the study of \citet{Antoja2012} (their table 2).
 {\it Bottom:} Scales of around $2\,\kms$. The green crosses mark peaks that are $>3\sigma$ significant. The red circles mark peaks for the velocity positions of  nearby open clusters that
are located at distances closer than 200 pc (see text). This panel shows a zoom-in on the central parts of the velocity plane.}\label{figuvwavelet}
\end{figure}

To study the substructure at different scales, we have performed the wavelet transform of the velocity plane of our neighbourhood sample. This is a mathematical transform that decomposes an image into the basis of the so-called wavelet function of different sizes. In practice, this transform yields a series of planes containing the substructures at different scales, and it has been extensively used to analyse the velocity substructure \citep[e.g.][]{Antoja2008, Chereul1999}. A thorough description of this technique can be found in \citet{Starck2002}.
 
In Fig.\ref{figuvwavelet} (top) we show the wavelet transform at scales of $8\,\kms$. This figure looks very similar to the wavelet transform applied to previous local sample \citep[e.g.][]{Antoja2011}, with several rounded structures organised in a larger arched structure. The green and orange crosses mark peaks that are $>3\sigma$ and $2-3\sigma$ significant, respectively. Most of the peaks detected at this scale are already known moving groups and dynamical streams. For comparison, we superpose the peaks detected with RAVE in \citet{Antoja2011} with squares following the same colour code. The most prominent structures are the moving groups of the Hyades ($(U,V)=(-33,-16)\,\kms$) and Pleiades ($(U,V)=(-11,-24)\,\kms$), which must not be confused with the open clusters with the same name, and the Sirius moving group $(U,V)=(10,3)\,\kms$. There are some groups that now have a larger significance than before, such as the two peaks inside the known Hercules stream (two green crosses at $V\simeq-50\,\kms$). Moreover, new significant groups appear, such as the substructures at $(U,V)\simeq(-90,0)\,\kms$. 

The kinematic substructure of smaller size can also be explored {thanks to} the precision of our data. We show it here in Fig. \ref{figuvwavelet} (bottom) with the wavelet transform at scales of $2\,\kms$. The plot shows a zoom-in into the central parts of the velocity plane. The figure again reveals prominent arched over-densities of stars, each one with a different inclination. We distinguish here at least five of these arches. Each of them shows internal smaller and rounder substructures, some of which mark the peaks of nearby clusters. 

The red  circles in this plot mark the velocity positions of eight nearby well-defined and fairly rich open clusters that
are located at distances closer than 200 pc (Hyades, ComaBer, Pleiades, Praesepe, alphaPer, IC2391, IC2602, and NGC2451A). Their heliocentric velocities have been computed from mean parameters provided in \cite{DR2-DPACP-31}. The Hyades and Pleiades clusters 
have the largest number of stars in the sample, thus showing high-intensity peaks in the wavelet transform at these small scales. Other clusters do no show a particular over-density in the wavelet space because they might be composed of a smaller number of stars in the local sample. 
Almost all clusters fall inside the larger substructure that
is formed by the arches, and seven out of eight lie in the same arch. 

\subsection{Kinematic substructure in distant regions }

To study the distant velocity distributions, we have selected stars from the main sample that are located in the Galactic disc with a cut of $|Z|<400$ pc and focused on a ring around the Sun between distances on the Galactic plane of 0.5 and 1 kpc. We furthermore partitioned this ring into sectors of $45\deg$ centred at the longitude of $l=[0,45,90,135,180,225,270,315]\deg$. The number of stars in these regions ranges from 
{90831 to 170001.}
 We then used Galactic cylindrical velocities since they are in a more natural reference system for these distant regions. We plot $-V_R$ instead of $V_R$ to obtain the same orientation as $U$ (i.e. velocity towards the Galactic centre). The median velocity uncertainties in the whole ring are of $(\epsilon_{V_{R}}, \epsilon_{V_{\phi}},\epsilon_{V_Z}) = ({1.2, 1.3, 0.6})$~$\kms$, and $80\%$ of the stars have uncertainties in all velocity components smaller than 
{3.}
$\kms$.

\begin{figure*}\centering
\includegraphics[width=0.9\hsize]{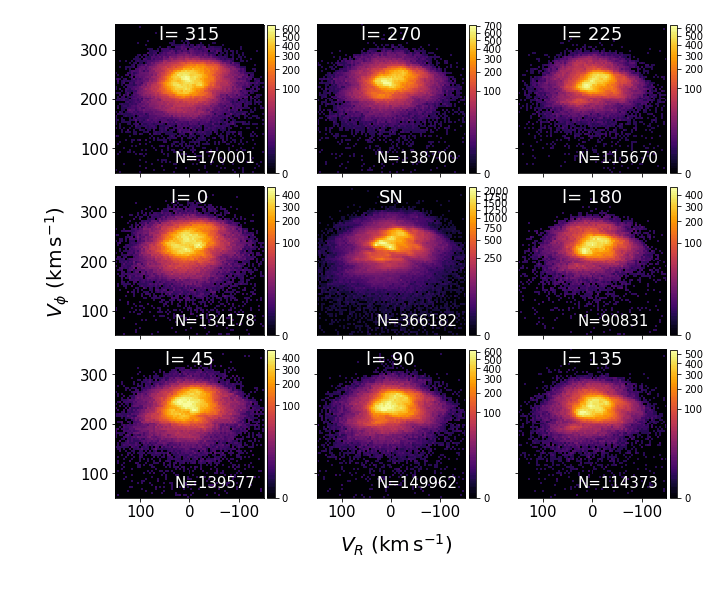}
\caption{Velocity plane of the stars in several distant regions.  The eight outer panels correspond to heliocentric sectors in the Milky Way disc selected from a ring between 0.5 and 1 kpc from the Sun, in a layer of $|Z|<400$ pc and $45\deg$ of amplitude centred at the Galactic longitudes indicated in each panel. The panels are 2D histograms with a bin size of $2\,\kms$, thus, the colour scale indicates the number of stars per 4 ($\,\kms$)$^2$. The middle panel shows the solar neighbourhood sample (defined in Sec. \ref{dataselection}) for comparison (the same as in Fig. \ref{figuv}, but with a larger bin size). The number of stars is indicated in each panel. We {have reversed the horizontal axis $V_R$} to obtain the same orientation as $U$ (i.e. velocity towards the Galactic centre).}\label{figuvdistant}
\end{figure*}

The 2D histograms of these eight distant regions are shown in Fig. \ref{figuvdistant}. In contrast to Fig. \ref{figuv}, we now use a bin of $2\,\kms$ to account for the larger uncertainty in these samples.
The panels are oriented such that the Galactic centre would be located at the left-hand side of the figure. For comparison, we include the distribution of the solar neighbourhood in the middle panel with the same bin size. 

Substructure is ubiquitous in these panels. Although we note a loss of definition in the substructures compared to the local volume because of the larger uncertainties, we still see clearly arched structure. Additionally, we note that the velocity structures change from region to region, {with greater changes in Galactic radius than in azimuth (i.e. greater changes for the different columns than for the different rows).} 
For example, the gap that separates the Hercules stream clearly moves from smaller to larger $V_\phi$  from the outer regions (three rightmost panels)  through the regions at the solar Galactic radius (three middle panels) to the inner regions (three leftmost panels). Effectively, the gap moves from $V_{\phi}\simeq 200\,\kms$ to  $V_{\phi}\simeq 240\,\kms$ {over a distance range of 2 kpc.} Although these findings are consistent with previous studies \citep{Antoja2014,Monari2017}, the resolution is unprecedented and we did not have to treat the data with any sophisticated method, as was required in previous work.

In most of the panels, but especially at the rightmost ones, this resolution allows us for the first time to see a structure below the  Hercules stream that is separated by a secondary gap. This gap is located at around  $V{_\phi}\simeq 200\,\kms$ in the three panels on the right, but lies at a different velocity in the other panels. In addition, the uppermost arch (at the highest $V_\phi$) that is observed in the solar volume is located at even larger $V{_\phi}$ in the inner Galactic regions (three left panels) and at smaller $V{_\phi}$ in the outer quadrants.

\subsection{Discussion}

We report a new global arrangement of stars in the velocity plane of the solar neighbourhood in which stars are organised in arches with nearly constant heliocentric velocity $V$ that extends to a wide range in $U$. This discovery is made possible by the higher
accuracy and greater number of stars in the Gaia local sample,
which exceeds the accuracy and number of stars of previous data
sets. 

Similar kinematic arches have appeared in several simulations, for example in the models by \citet{Dehnen2000} and \citet{Fux2001}, who both studied the effects of the Galactic bar on the local velocity distribution. The velocity arches appeared in their simulations, which use backwards integration, and the number of arches depended on the total integration time. 
\citet{Fux2001} related these arches to phase-mixing. Similar to what we observe here, their arches are not perfectly centred on $U$ or inclined in $V$. Similarly, \citet{Antoja2009} explained that arches also appeared in forward integrations in axisymmetric potentials as a result of a non-relaxed population and that these arches were not centred on $U$ when the bar was added to the potential. 

{Arched kinematic substructures have also been reported in simulations of spiral arm effects in \citet{Quillen2005} and \citet{Antoja2011}, and structures elongated in $U$  were observed
in the simulations of \citet{Desimone2004}, with transient episodes of spiral structure, although in the latter case they did not appear as strongly curved as in the {\gaia} data. The scenario
of transient spiral arms with the superposition of multiple waves with different frequencies is independently supported by several
other pieces of evidence, especially from simulations (see e.g.
\citealt{Sellwood2014}). Thus, the possibility that spiral arms or the possible transient multiple waves generate one or several of these arches should be explored in future work as well, in addition to the possibility that effects of phase-mixing might also play a role because these modes
are transient.}

Clear arches are also observed in the models by \citet{Minchev2009} and have been followed by \citet{Gomez2012a} in a phenomenon
that they dubbed "ringing", which these authors attribute to the impact of a satellite on the Galaxy disc. Using orbital integrations, a semi-analytical method, and N-body simulations, these authors found that as a response to the perturbation, the disc experiences a series of waves of constant energy that propagate through the disc and manifest themselves as arches in the $U-V$ plane. All these authors explained that these arches can give information on the time of the perturbation, the orbital parameters of the perturber, and its mass. 
It seems therefore that the arches are evidence of ongoing phase-mixing in the disc, but this might be attributed to several causes such as a perturbing satellite, the formation of the bar, or other changes in the Galactic potential to which the stellar disc is still adapting.

\citet{Minchev2009} searched for observational signatures of this scenario and focused on thick-disc samples, arguing that these stars do not experience additional perturbations by molecular clouds or spiral arms and that these arches might therefore be better distinguished.  These authors suggested that some of the previously known streams might be signatures of a minor merger. However, because of the low number of stars 
and the  velocity resolution at that time, these structures did not have a clear arched shape, and as the authors admit, were not significant enough. Recently,  \citet{Gomez2012b} found a few significant peaks in the energy distribution of the \citet{Schuster2006}  and \citet{Lee2011} catalogues that might well be linked to ringing and that showed rough similarities with those produced by a model of the Sagittarius dwarf galaxy.

For the first time, the arched substructures are clearly observed
here, and surprisingly, they have appeared in a sample that is
dominated by thin disc stars, since no selection on metallicity has been made here. We note their different inclinations and ranges in $U$, which must contain information on the mechanism
that causes them.
In any case, a detailed comparison with the different models is required to ascertain which exactly are the more plausible perturbing agents that may have caused a non-equilibrium state in the disc.

 At larger scales, the velocity distribution that \textit{Gaia} has measured shows rounded structures that are perfectly consistent with previous studies, and that might be of dynamical origin, related to  the spiral arms and bar gravitational potential, as suggested originally by \citet{Kalnajs1991} and later by many others (see Sec. \ref{intro}).
 It is also worth noting that the open clusters considered here, which are more prominent inside the solar neighbourhood, are all located on top of the arches in the velocity plane and thus seem to participate in a common dynamics. We note, however, that this is not a complete sample of clusters and that further study is required to extract definitive conclusions.
If it is confirmed that most of the clusters follow this arched organisation, the ages of the clusters might give us information on the timescales of the agent that created these arches. The clusters studied here are quite young, which might indicate that the causing agent(s) have been acting until quite recently.

The \textit{Gaia} data up to distances of 1 kpc in the Galactic plane show that the arches also exist in more distant regions. The quality of the data will certainly allow us to quantify the changes in substructure with position. We already observe movement
in {the} velocity of the gap between the Hercules stream and the remaining distribution as a function of radius, specifically at lower $V_\phi$ for larger Galactic radius. This was first discovered by \citet{Antoja2014} with RAVE data, based on which the authors demonstrated that this change is compatible with the orbital effects of the Galactic bar near the outer Lindblad resonance. This finding was also confirmed later on by \citet{Monari2017} using TGAS and LAMOST. However, while in \citet{Antoja2014}  a special projection of the velocities and complex analysis tools were required, here the gap and its variation with position is unambiguous by a simple inspection of data.  We also note that the range in Galactic radius that can be explored is larger in the present study (2 kpc) than in these previous studies ($\lesssim 1$ kpc). \citet{Hunt2018} also showed some variations in Hercules as a function of distance with APOGEE-2 South data, but only for a  specific range of Galactic radius, because their data were limited to a single line of sight. The thorough study of this velocity variation with {\it Gaia} data will  lead to a better determination of the bar properties, such as pattern speed and orientation angle.

%

\section{Conclusions}
Taking advantage of the full-sky coverage, precise distances, and velocities of \gdrtwo{} for more than {three} million giant stars, we have mapped the velocity field of the galaxy
in 3D over a large portion of the disc ($5 \leq R \leq 13$~kpc, $-30^\circ \leq \phi \leq +30^\circ$ , and $-2 \leq Z \leq +2$~kpc). The picture of the Milky Way disc kinematics drawn by \gdrtwo{} is both rich and complex. Streaming motions are observed in all three velocity components, and $\tilde{V}_Z$ likely shows a superposition of modes. The velocity dispersions also show some small-amplitude perturbations superimposed on a large-scale radial decrease that is quite symmetric in azimuth {and} with respect to the Galactic mid-plane.

This is also the first time that the velocity precision is high
enough and the statistics is large enough to resolve the small scales of the velocity plane of the solar neighbourhood. We find that stars clearly appear to be organised in kinematic arches that are oriented approximately along the horizontal direction in the $U-V$ plane. The origin of these arches is probably related to the non-equilibrium of the Galactic disc, which could have been induced by several causes or by a mix of them. For instance, similar arches have been seen in simulations of a satellite impact on the disc. These arches are also observed in the \gdrtwo{} data of more distant regions, and we also see the change in some of the previously known kinematic groups as a function of position in the disc in a manner consistent with models of the orbital effects of the Galactic bar. The velocity plane of the local neighbourhood and surroundings certainly appears to be as complex as ever, with a variety of structures of many different scales and shapes. Beyond these substructures, which are signatures  of current and past events of the formation and evolution of the Milky Way, this first inspection of the \gdrtwo{} data clearly promises exciting discoveries and significant advances in this field in the coming years.

\gdrtwo{} is now available to the astronomical community so that a variety of questions about the Milky Way can be addressed that
far exceed the specific examples visited in this paper. In particular, \gdrtwo{} provides transverse velocities for several hundred million stars. Although it was beyond the scope of this paper to use them, there is no doubt that they have much tell about the Galaxy kinematics and dynamics. Moreover, the joint use of \gdrtwo{} and large ground-based spectroscopic surveys will allow combining the precise \gdrtwo{} distances and velocities with detailed chemistry. We can expect this great wealth of information to trigger an intense activity in the galactic community in the years to come. Following \gdrtwo{}, \gdrthree{} is scheduled
to be released in a few years\footnote{\url{https://www.cosmos.esa.int/web/gaia/release}}. The many promises of the next release include a further improvement in the precision and accuracy of the astrometric, photometric, and spectroscopic data, an increase by a factor 5 to 10 of the stars with full velocities, detection and characterisation of multiple systems, and chemistry for millions of stars.

\section*{Acknowledgements\label{sec:acknowl}}

We are grateful to the anonymous referee for their constructive report that improved the quality of the manuscript, and in particular, for the suggestion to produce Fig.~\ref{fig:xyvbendbreath}.
This work presents results from the European Space Agency (ESA) space mission \gaia. \gaia\ data are being processed by the \gaia\ Data Processing and Analysis Consortium (DPAC). Funding for the DPAC is provided by national institutions, in particular the institutions participating in the \gaia\ MultiLateral Agreement (MLA). The \gaia\ mission website is \url{https://www.cosmos.esa.int/gaia}. The \gaia\ archive website is \url{https://archives.esac.esa.int/gaia}.

The \gaia\ mission and data processing have financially been supported by, in alphabetical order by country:
\begin{itemize}
\item the Algerian Centre de Recherche en Astronomie, Astrophysique et G\'{e}ophysique of Bouzareah Observatory;
\item the Austrian Fonds zur F\"{o}rderung der wissenschaftlichen Forschung (FWF) Hertha Firnberg Programme through grants T359, P20046, and P23737;
\item the BELgian federal Science Policy Office (BELSPO) through various PROgramme de D\'eveloppement d'Exp\'eriences scientifiques (PRODEX) grants and the Polish Academy of Sciences - Fonds Wetenschappelijk Onderzoek through grant VS.091.16N;
\item the Brazil-France exchange programmes Funda\c{c}\~{a}o de Amparo \`{a} Pesquisa do Estado de S\~{a}o Paulo (FAPESP) and Coordena\c{c}\~{a}o de Aperfeicoamento de Pessoal de N\'{\i}vel Superior (CAPES) - Comit\'{e} Fran\c{c}ais d'Evaluation de la Coop\'{e}ration Universitaire et Scientifique avec le Br\'{e}sil (COFECUB);
\item the Chilean Direcci\'{o}n de Gesti\'{o}n de la Investigaci\'{o}n (DGI) at the University of Antofagasta and the Comit\'e Mixto ESO-Chile;
\item the National Science Foundation of China (NSFC) through grants 11573054 and 11703065;  
\item the Czech-Republic Ministry of Education, Youth, and Sports through grant LG 15010, the Czech Space Office through ESA PECS contract 98058, and Charles University Prague through grant PRIMUS/SCI/17;    
\item the Danish Ministry of Science;
\item the Estonian Ministry of Education and Research through grant IUT40-1;
\item the European Commission’s Sixth Framework Programme through the European Leadership in Space Astrometry (\url{https://www.cosmos.esa.int/web/gaia/elsa-rtn-programme}{ELSA}) Marie Curie Research Training Network (MRTN-CT-2006-033481), through Marie Curie project PIOF-GA-2009-255267 (Space AsteroSeismology \& RR Lyrae stars, SAS-RRL), and through a Marie Curie Transfer-of-Knowledge (ToK) fellowship (MTKD-CT-2004-014188); the European Commission's Seventh Framework Programme through grant FP7-606740 (FP7-SPACE-2013-1) for the \gaia\ European Network for Improved data User Services (\url{http://genius-euproject.eu/}{GENIUS}) and through grant 264895 for the \gaia\ Research for European Astronomy Training (\url{https://www.cosmos.esa.int/web/gaia/great-programme}{GREAT-ITN}) network;
\item the European Research Council (ERC) through grants 320360 and 647208 and through the European Union’s Horizon 2020 research and innovation programme through grants 670519 (Mixing and Angular Momentum tranSport of massIvE stars -- MAMSIE) and 687378 (Small Bodies: Near and Far);\item the European Science Foundation (ESF), in the framework of the \gaia\ Research for European Astronomy Training Research Network Programme (\url{https://www.cosmos.esa.int/web/gaia/great-programme}{GREAT-ESF});
\item the European Space Agency (ESA) in the framework of the \gaia\ project, through the Plan for European Cooperating States (PECS) programme through grants for Slovenia, through contracts C98090 and 4000106398/12/NL/KML for Hungary, and through contract 4000115263/15/NL/IB for Germany;
\item the European Union (EU) through a European Regional Development Fund (ERDF) for Galicia, Spain;    
\item the Academy of Finland and the Magnus Ehrnrooth Foundation;
\item the French Centre National de la Recherche Scientifique (CNRS) through action 'D\'efi MASTODONS', the Centre National d'Etudes Spatiales (CNES), the L'Agence Nationale de la Recherche (ANR) 'Investissements d'avenir' Initiatives D’EXcellence (IDEX) programme Paris Sciences et Lettres (PSL$\ast$) through grant ANR-10-IDEX-0001-02, the ANR 'D\'{e}fi de tous les savoirs' (DS10) programme through grant ANR-15-CE31-0007 for project 'Modelling the Milky Way in the Gaia era' (MOD4Gaia), the R\'egion Aquitaine, the Universit\'e de Bordeaux, and the Utinam Institute of the Universit\'e de Franche-Comt\'e, supported by the R\'egion de Franche-Comt\'e and the Institut des Sciences de l'Univers (INSU);
\item the German Aerospace Agency (Deutsches Zentrum f\"{u}r Luft- und Raumfahrt e.V., DLR) through grants 50QG0501, 50QG0601, 50QG0602, 50QG0701, 50QG0901, 50QG1001, 50QG1101, 50QG1401, 50QG1402, 50QG1403, and 50QG1404 and the Centre for Information Services and High Performance Computing (ZIH) at the Technische Universit\"{a}t (TU) Dresden for generous allocations of computer time;
\item the Hungarian Academy of Sciences through the Lend\"ulet Programme LP2014-17 and the J\'anos Bolyai Research Scholarship (L.~Moln\'ar and E.~Plachy) and the Hungarian National Research, Development, and Innovation Office through grants NKFIH K-115709, PD-116175, and PD-121203;
\item the Science Foundation Ireland (SFI) through a Royal Society - SFI University Research Fellowship (M.~Fraser);
\item the Israel Science Foundation (ISF) through grant 848/16;
\item the Agenzia Spaziale Italiana (ASI) through contracts I/037/08/0, I/058/10/0, 2014-025-R.0, and 2014-025-R.1.2015 to the Italian Istituto Nazionale di Astrofisica (INAF), contract 2014-049-R.0/1/2 to INAF dedicated to the Space Science Data Centre (SSDC, formerly known as the ASI Sciece Data Centre, ASDC), and contracts I/008/10/0, 2013/030/I.0, 2013-030-I.0.1-2015, and 2016-17-I.0 to the Aerospace Logistics Technology Engineering Company (ALTEC S.p.A.), and INAF;
\item the Netherlands Organisation for Scientific Research (NWO) through grant NWO-M-614.061.414 and through a VICI grant (A.~Helmi) and the Netherlands Research School for Astronomy (NOVA);
\item the Polish National Science Centre through HARMONIA grant 2015/18/M/ST9/00544 and ETIUDA grants 2016/20/S/ST9/00162 and 2016/20/T/ST9/00170;\item the Portugese Funda\c{c}\~ao para a Ci\^{e}ncia e a Tecnologia (FCT) through grant SFRH/BPD/74697/2010; the Strategic Programmes UID/FIS/00099/2013 for CENTRA and UID/EEA/00066/2013 for UNINOVA;
\item the Slovenian Research Agency through grant P1-0188;
\item the Spanish Ministry of Economy (MINECO/FEDER, UE) through grants ESP2014-55996-C2-1-R, ESP2014-55996-C2-2-R, ESP2016-80079-C2-1-R, and ESP2016-80079-C2-2-R, the Spanish Ministerio de Econom\'{\i}a, Industria y Competitividad through grant AyA2014-55216, the Spanish Ministerio de Educaci\'{o}n, Cultura y Deporte (MECD) through grant FPU16/03827, the Institute of Cosmos Sciences University of Barcelona (ICCUB, Unidad de Excelencia 'Mar\'{\i}a de Maeztu') through grant MDM-2014-0369, the Xunta de Galicia and the Centros Singulares de Investigaci\'{o}n de Galicia for the period 2016-2019 through the Centro de Investigaci\'{o}n en Tecnolog\'{\i}as de la Informaci\'{o}n y las Comunicaciones (CITIC), the Red Espa\~{n}ola de Supercomputaci\'{o}n (RES) computer resources at MareNostrum, and the Barcelona Supercomputing Centre - Centro Nacional de Supercomputaci\'{o}n (BSC-CNS) through activities AECT-2016-1-0006, AECT-2016-2-0013, AECT-2016-3-0011, and AECT-2017-1-0020;
\item the Swedish National Space Board (SNSB/Rymdstyrelsen);
\item the Swiss State Secretariat for Education, Research, and Innovation through the ESA PRODEX programme, the Mesures d’Accompagnement, the Swiss Activit\'es Nationales Compl\'ementaires, and the Swiss National Science Foundation;
\item the United Kingdom Rutherford Appleton Laboratory, the United Kingdom Science and Technology Facilities Council (STFC) through grant ST/L006553/1, the United Kingdom Space Agency (UKSA) through grant ST/N000641/1 and ST/N001117/1, as well as a Particle Physics and Astronomy Research Council Grant PP/C503703/1.
\end{itemize}

%
%

\bibliographystyle{aa}
\bibliography{Gaia-DR2-MWmap}

\begin{appendix}
\section{Velocity moments and uncertainties\label{app:veloc}}
The lower and upper 1$\sigma$ uncertainties on the estimations of the median velocities are calculated as
\begin{equation}
\sigma_{\tilde{V_i}}^{low} = \sqrt{\pi\over 2 N} (\tilde{V_i} - Per(V_i, 15.85))
\label{eq:eMedLow}
\end{equation}
\begin{equation}
\sigma_{\tilde{V_i}}^{upp} = \sqrt{\pi\over 2 N} (Per(V_i, 84.15) - \tilde{V_i})
\label{eq:eMedUpp}
,\end{equation}
respectively, where N is the number of stars in the cell, $V_i$ is one of the three Galactocentric components of the velocity vector $V_R$, $V_\phi$ , or $V_Z$, and $\tilde{V_i}$ is the median of the distribution of $V_i$ , and $Per(V_i, 15.85)$ and $Per(V_i, 84.15)$ are the 15.85$^{th}$ and 84.15$^{th}$ percentiles of the distribution of $V_i$, respectively.\\

The dispersions of the velocities are calculated as
\begin{equation}
\sigma_{V_i} = {Per(V_i, 84.15) - Per(V_i, 15.85) \over 2}
\label{eq:disp}
,\end{equation}
using the same notation as above.\\

The lower and upper 1$\sigma$ uncertainties on the estimation of the velocity dispersions are calculated as
\begin{equation}
\sigma_{\sigma_{V_i}}^{low} = {\sqrt{2 \pi} \over e^{-0.5}} \sqrt{0.1585 \times 0.683 \over N} (\tilde{V_i} - Per(V_i, 15.85))
\label{eq:eDispLow}
\end{equation}

\begin{equation}
\sigma_{\sigma_{V_i}}^{upp} = {\sqrt{2 \pi} \over e^{-0.5}} \sqrt{0.1585 \times 0.683 \over N} (Per(V_i, 84.15) - \tilde{V_i})
\label{eq:eDispUpp}
,\end{equation}
respectively, using the same notations as in equations~\ref{eq:eMedLow} and~\ref{eq:eMedUpp}.

\section{Biases induced by selections on the velocity uncertainties\label{app:}}
\begin{figure*}[h!]
\centering
\includegraphics[width=0.30 \hsize]{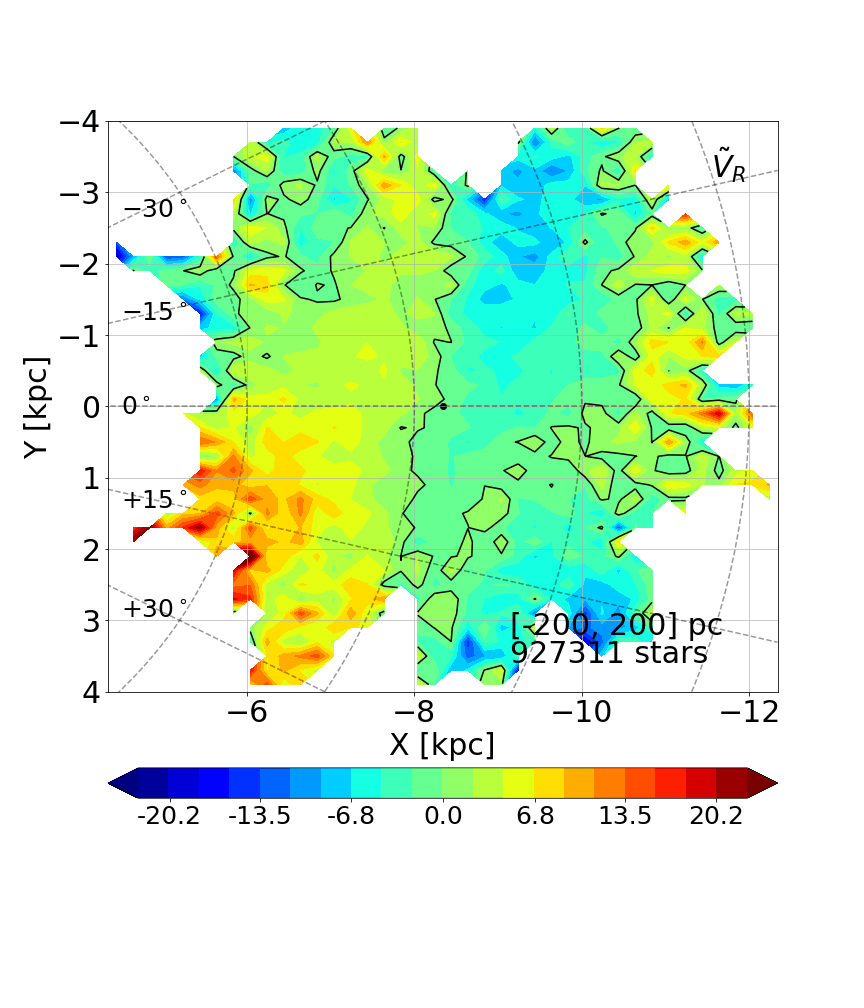}
\includegraphics[width=0.30 \hsize]{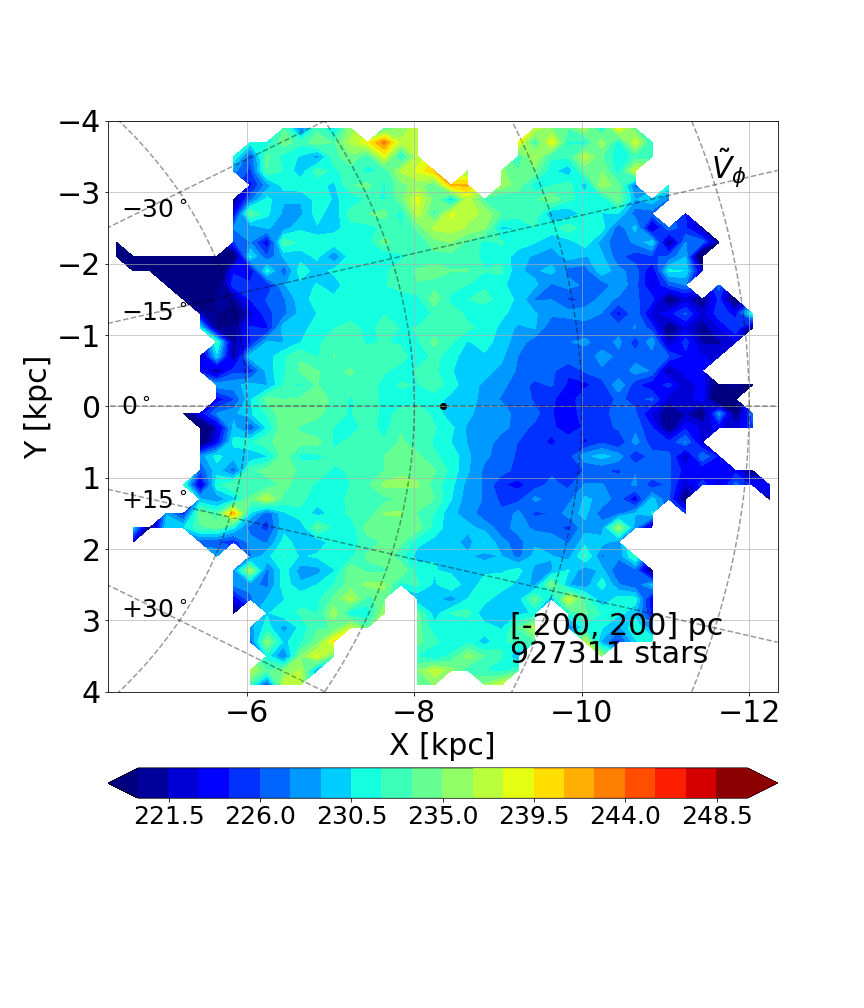}
\includegraphics[width=0.30 \hsize]{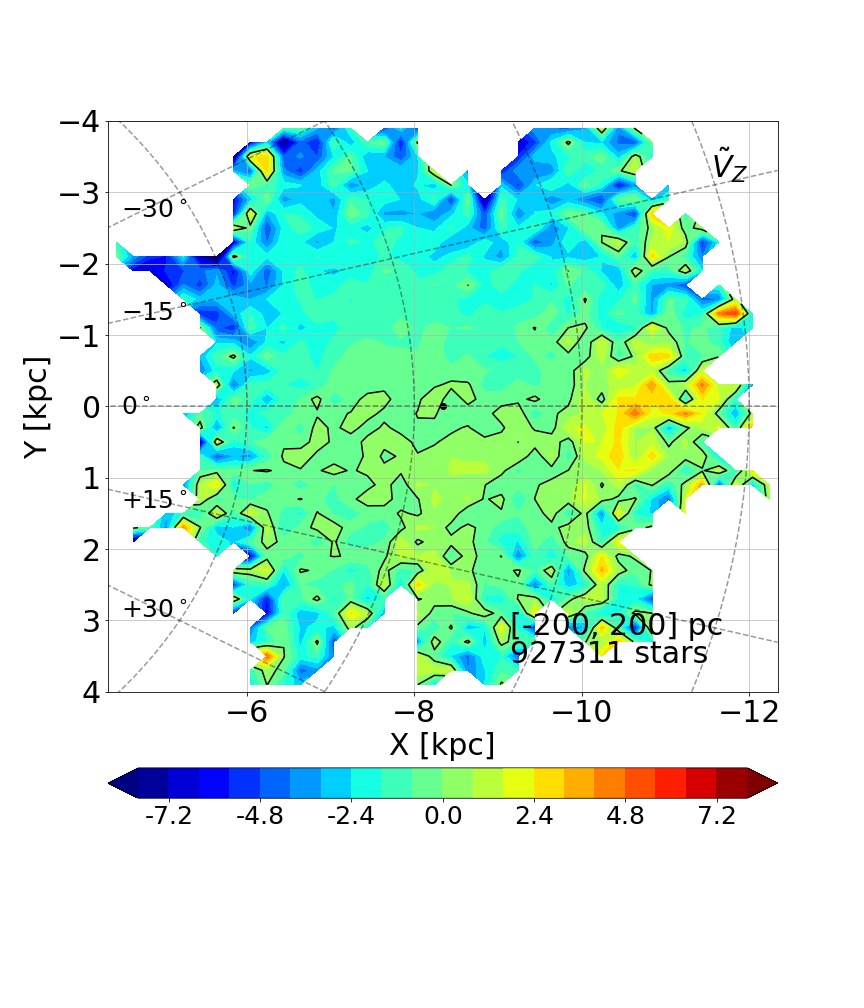}
\caption{Face-on view of the kinematic of the disc mid-plane ($[-200, +200]$~pc): Median radial velocity (left), median azimuthal velocity (centre), and median vertical velocity (right).\label{FigAll20}}
\end{figure*}

\begin{figure*}[h!]
\centering
\includegraphics[width=0.3 \hsize]{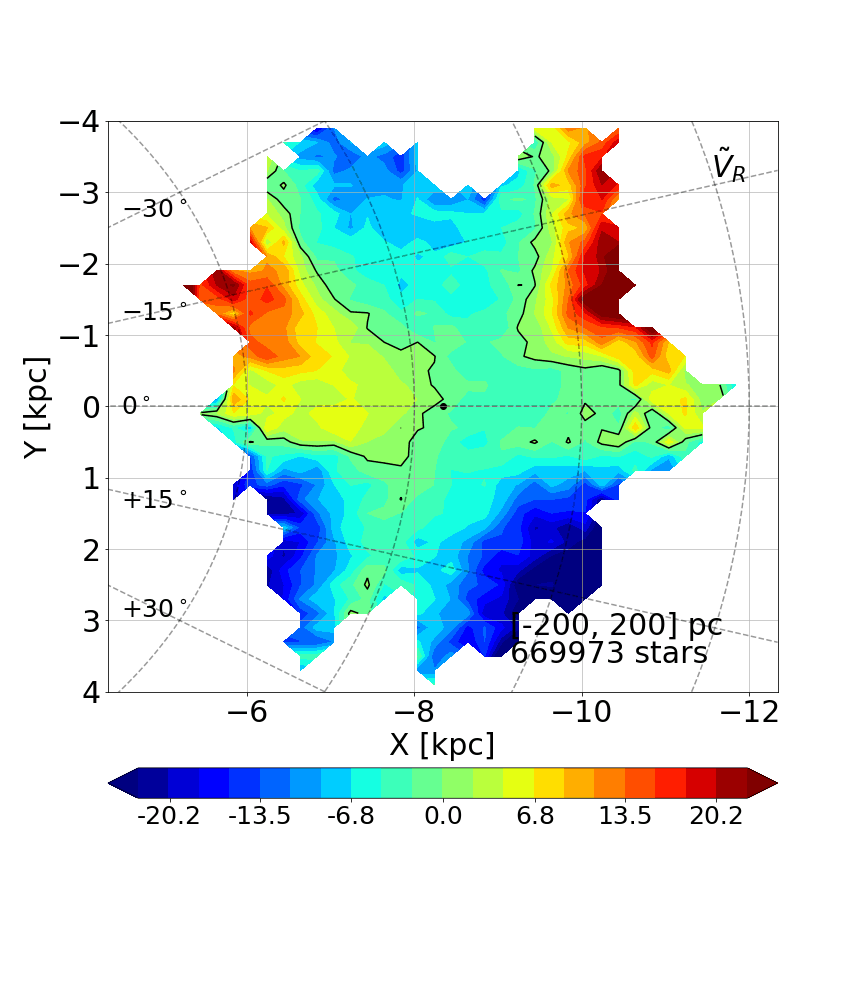}
\includegraphics[width=0.3 \hsize]{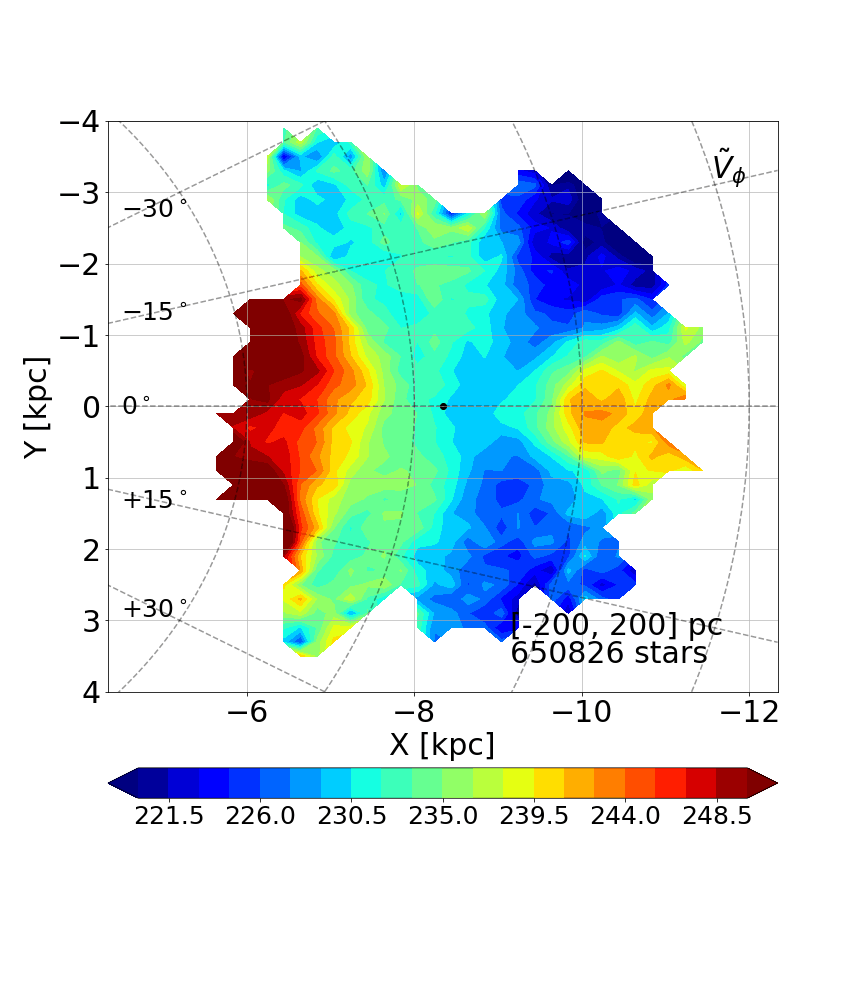}
\includegraphics[width=0.3 \hsize]{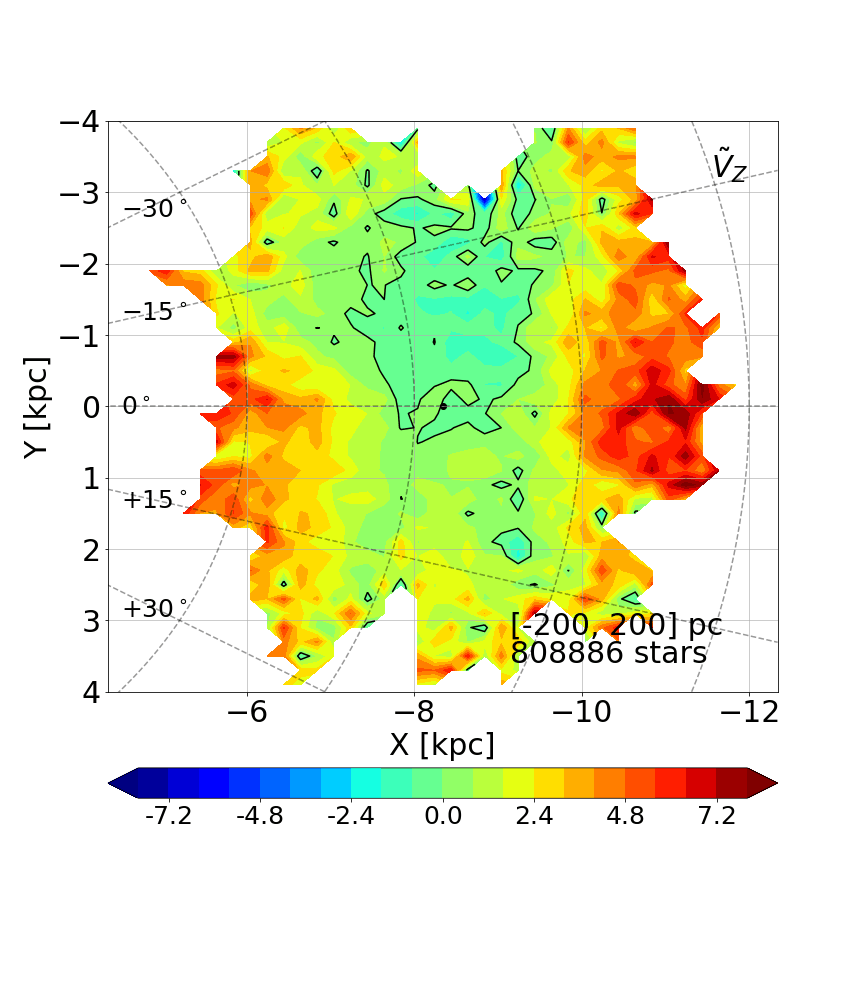}
\caption{Same as Fig.~\ref{FigAll20}, but selecting the stars with uncertainties smaller than or equal to 2~$\kms$ on the relevant component:  $\sigma_{V_R} \leq 2 \kms$ (left), $\sigma_{V_\phi} \leq 2 \kms$ (centre), and $\sigma_{V_Z} \leq 2 \kms$ (right).\label{FigAll02}}
\end{figure*}

Figures \ref{FigAll20} and  \ref{FigAll02} show the median cylindrical velocities in the face-on view of the disc plane for the main sample and for an additional selection of stars with velocity errors in each component smaller than $2\kms$, respectively. It is evident from the comparison of all panels in the two figures that this cut in velocity error introduces strong biases in the median velocity field. The differences in median velocity can be of up to $20\ \kms$ , and the global appearance of the velocity field changes substantially. The biases result from the correlations between velocities and velocity uncertainties. Selecting on the velocity uncertainties modifies the shape of the velocity distributions and therefore biases the measures of the moments of these distributions.

\section{Velocity maps \label{app:maps}}
\begin{figure*}[]
\centering
\includegraphics[clip=true, trim = 10mm 40mm 20mm 40mm, width=0.3 \hsize]{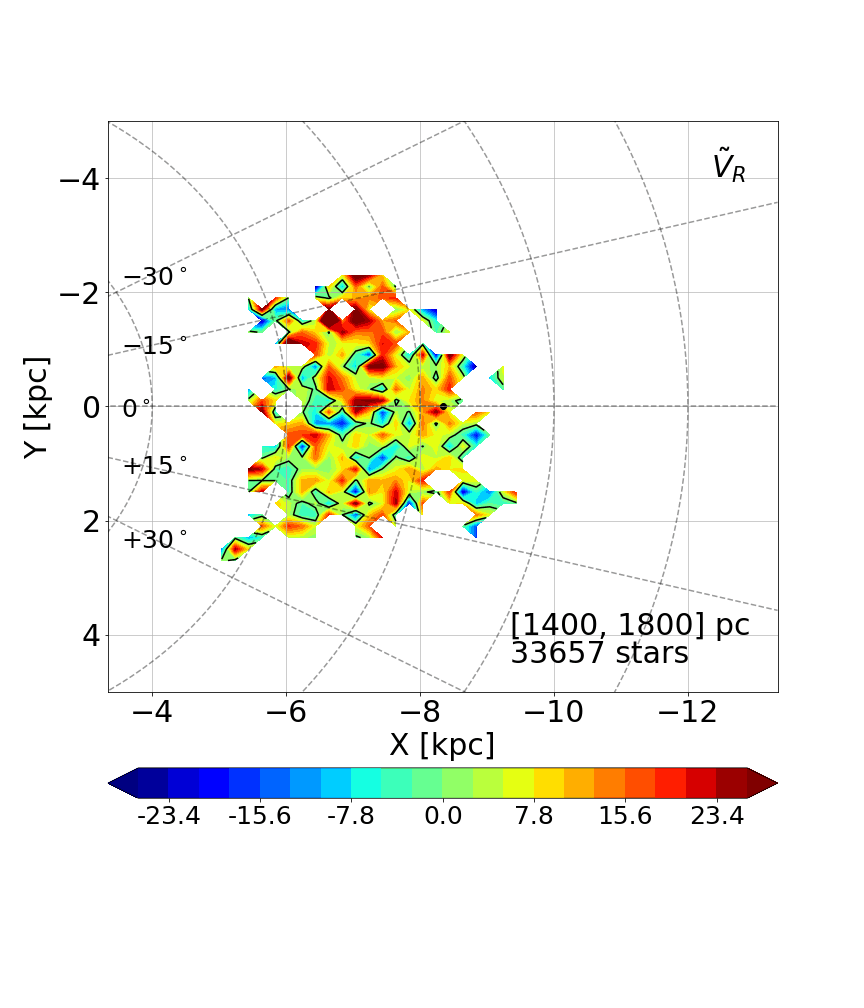}
\includegraphics[clip=true, trim = 10mm 40mm 20mm 40mm, width=0.3 \hsize]{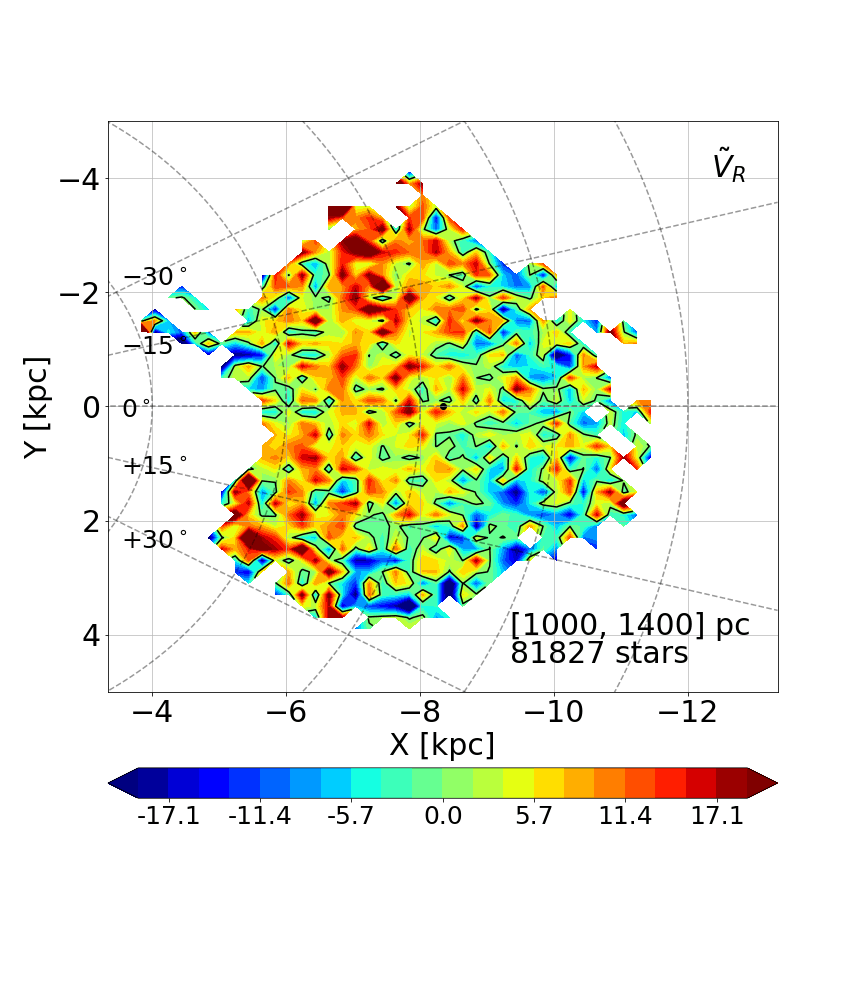}
\includegraphics[clip=true, trim = 10mm 40mm 20mm 40mm, width=0.3 \hsize]{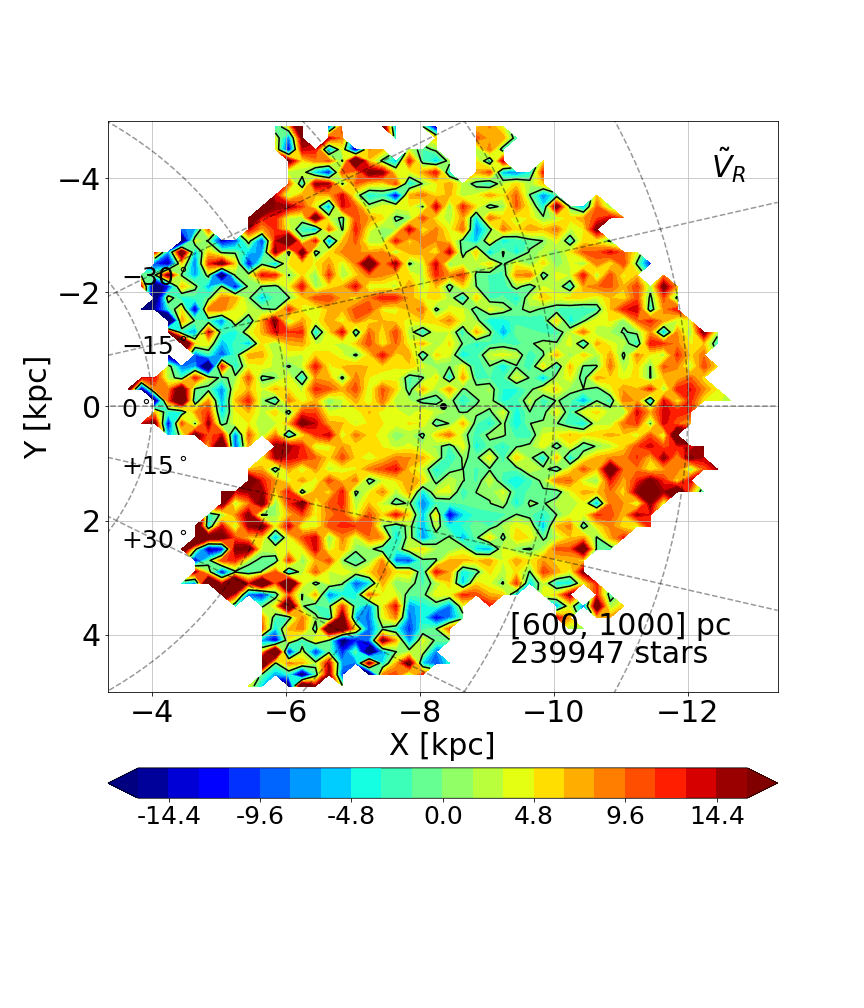}
\includegraphics[clip=true, trim = 10mm 40mm 20mm 40mm, width=0.3 \hsize]{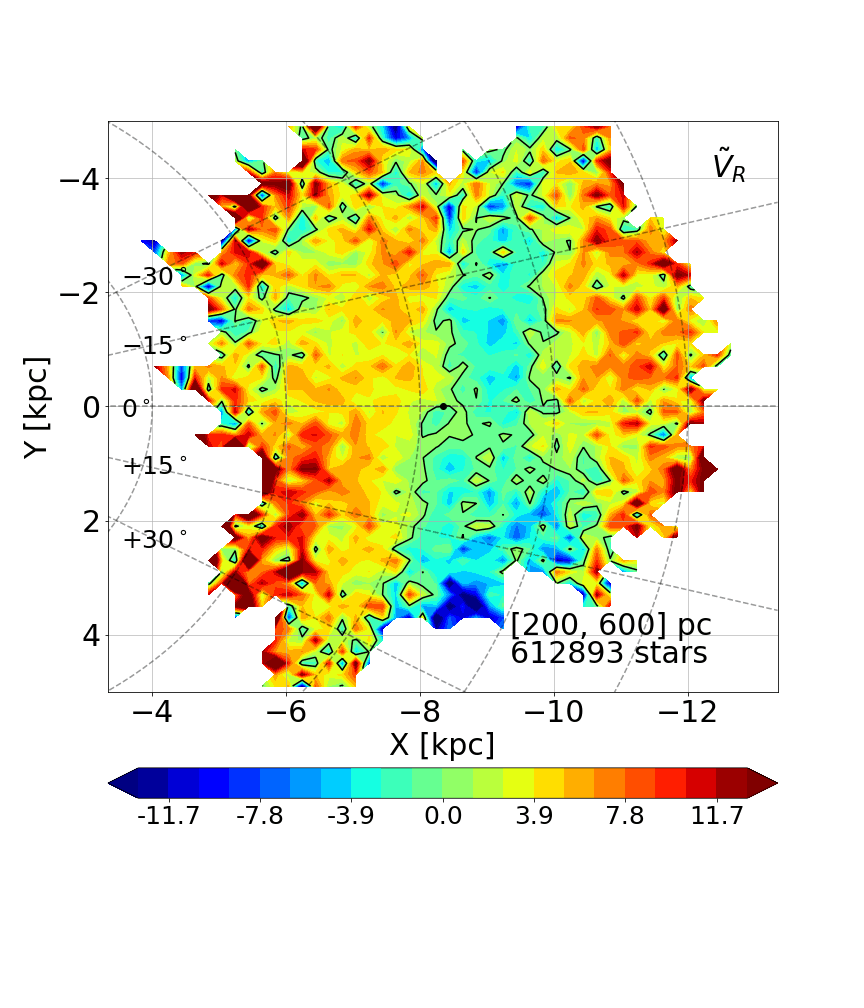}
\includegraphics[clip=true, trim = 10mm 40mm 20mm 40mm, width=0.3 \hsize]{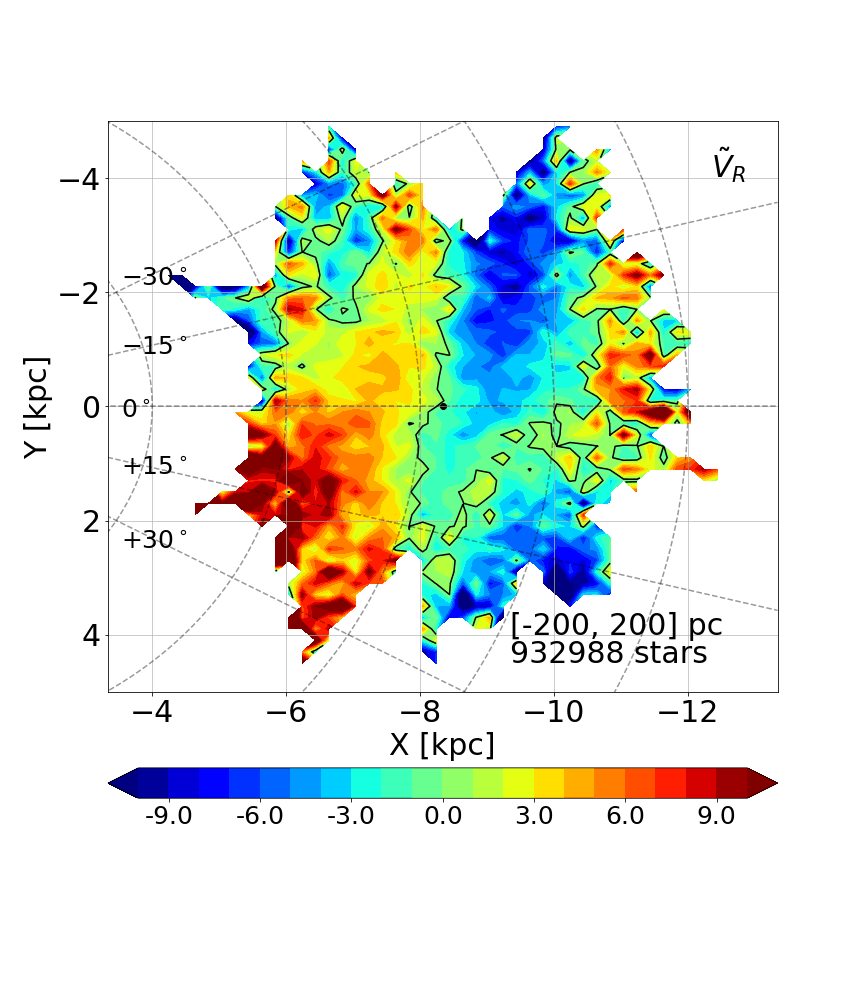}
\includegraphics[clip=true, trim = 10mm 40mm 20mm 40mm, width=0.3 \hsize]{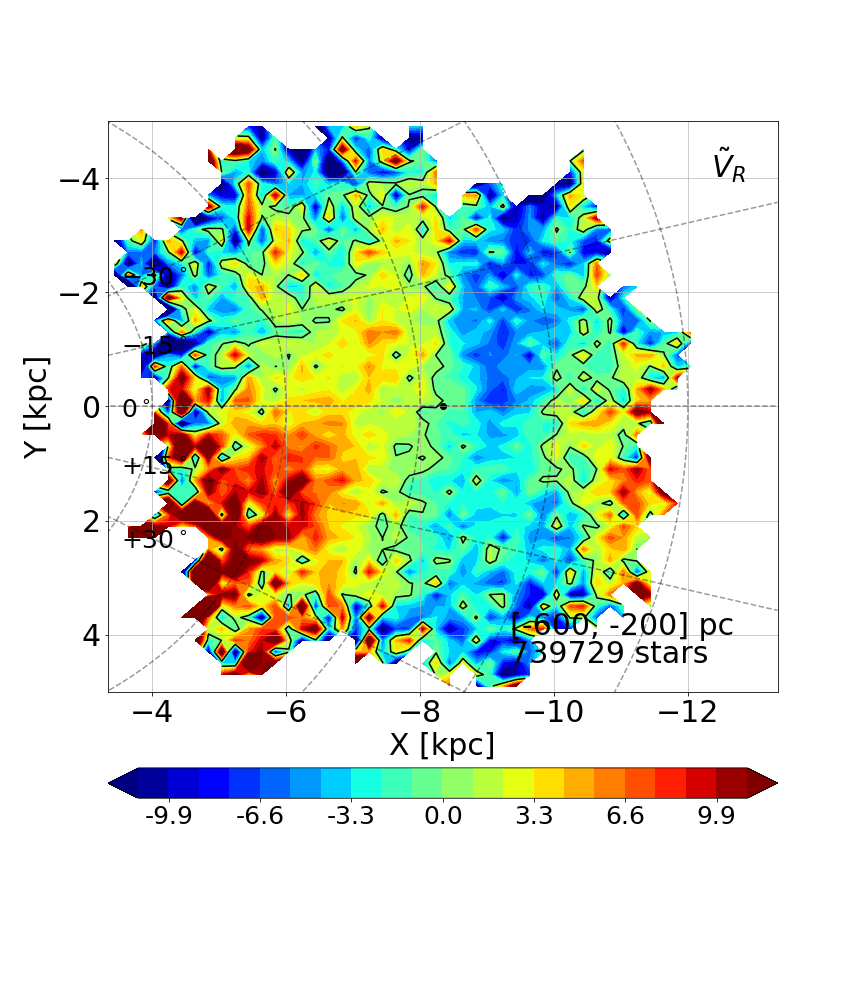}
\includegraphics[clip=true, trim = 10mm 40mm 20mm 40mm, width=0.3 \hsize]{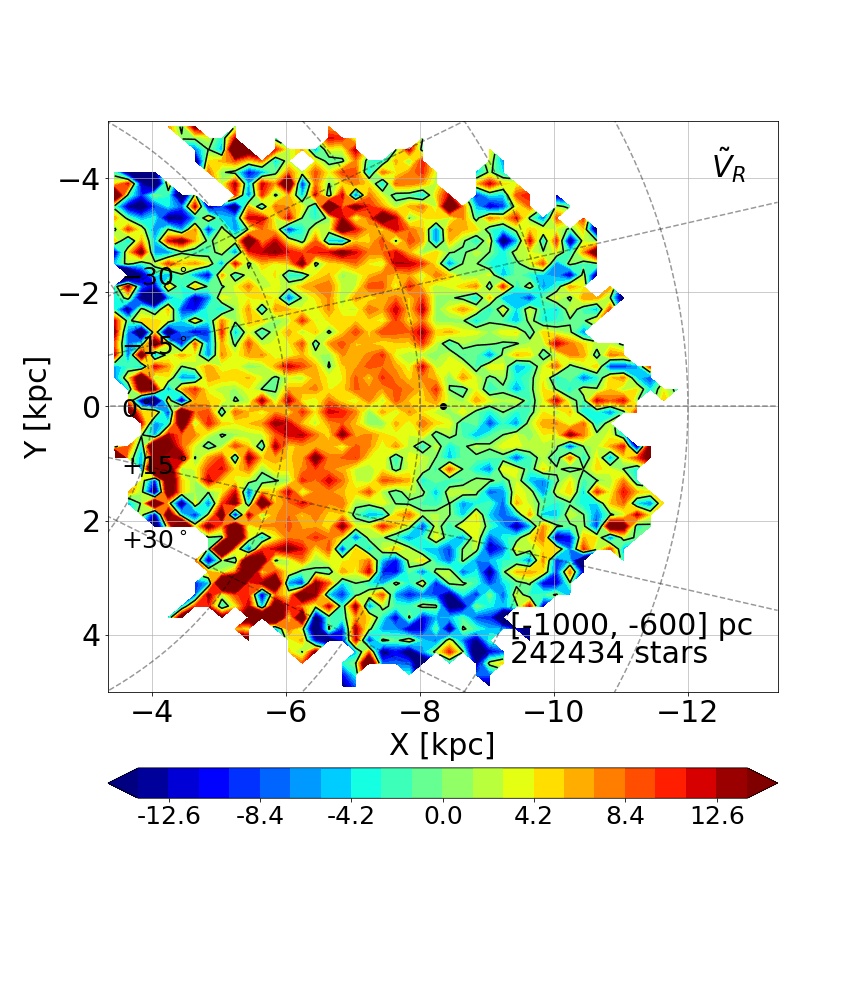}
\includegraphics[clip=true, trim = 10mm 40mm 20mm 40mm, width=0.3 \hsize]{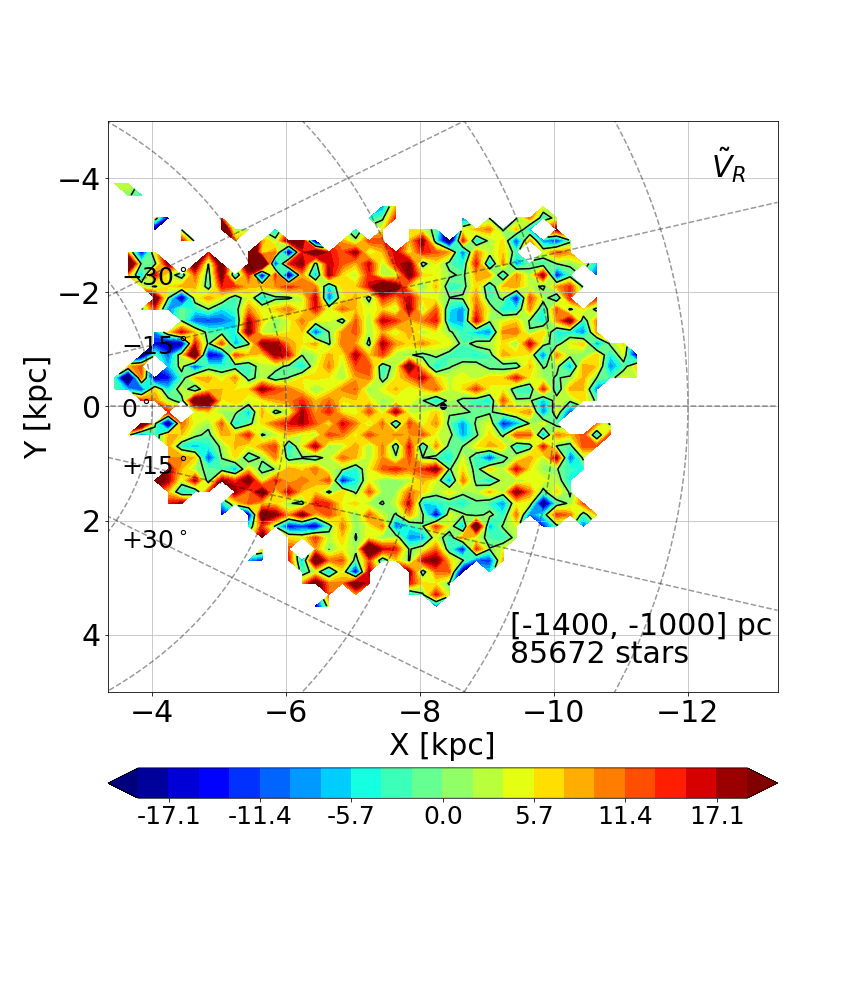}
\includegraphics[clip=true, trim = 10mm 40mm 20mm 40mm, width=0.3 \hsize]{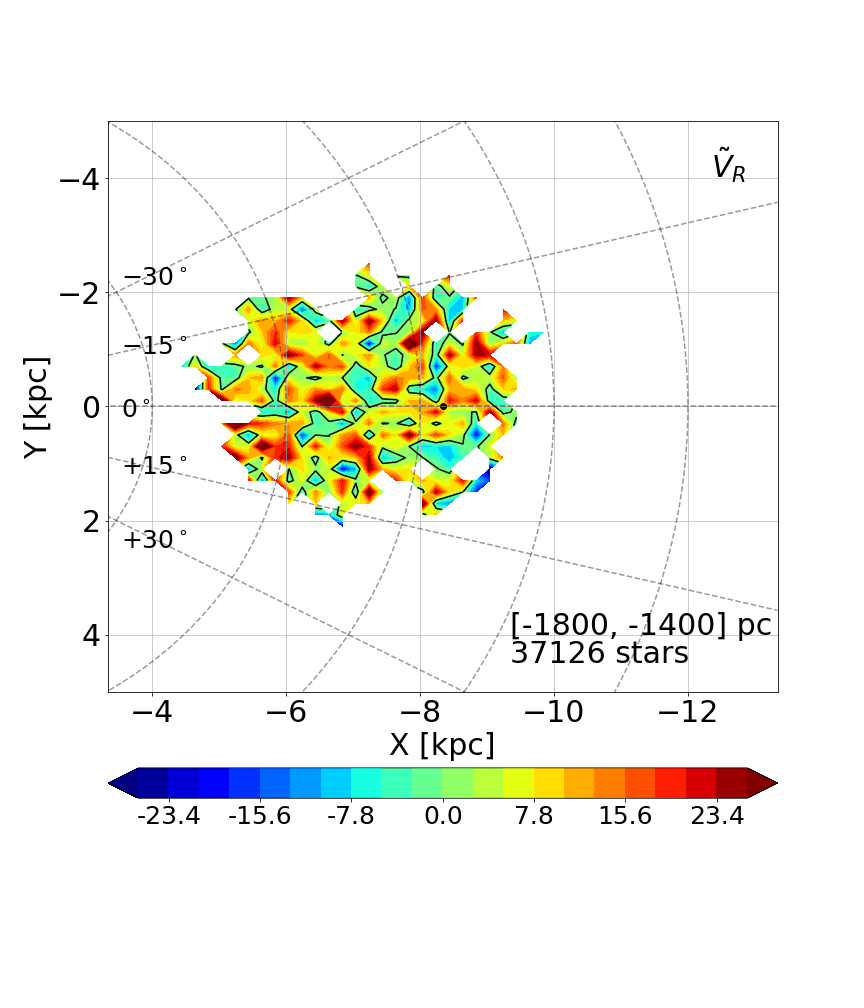}
\caption{Mosaic of face-on maps of the median radial velocity, $\tilde{V}_R$, of the giant sample. Each map corresponds to a $Z$ layer of 400~pc height, from $[+1400, +1800]$~pc (top left) to $[-1800, -1400]$~pc (bottom right). In each map, the azimuths increase clockwise. They are labelled from $-30$ to $+30$~degrees on the left of the maps. The Sun is represented by a black dot at $X=-8.34$~kpc and $Y=0$~kpc. The Galactic centre is located on the left side. The Milky Way rotates clockwise. The iso-velocity contours $\tilde{V}_R = 0$~$\kms$ are pointed out as black lines. The numbers of stars used to produce the maps are given in the lower left corners.}
\label{fig:xyvrmed}
\end{figure*}

\begin{figure*}[]
\centering
\includegraphics[clip=true, trim = 5mm 40mm 10mm 40mm, width=0.45 \hsize]{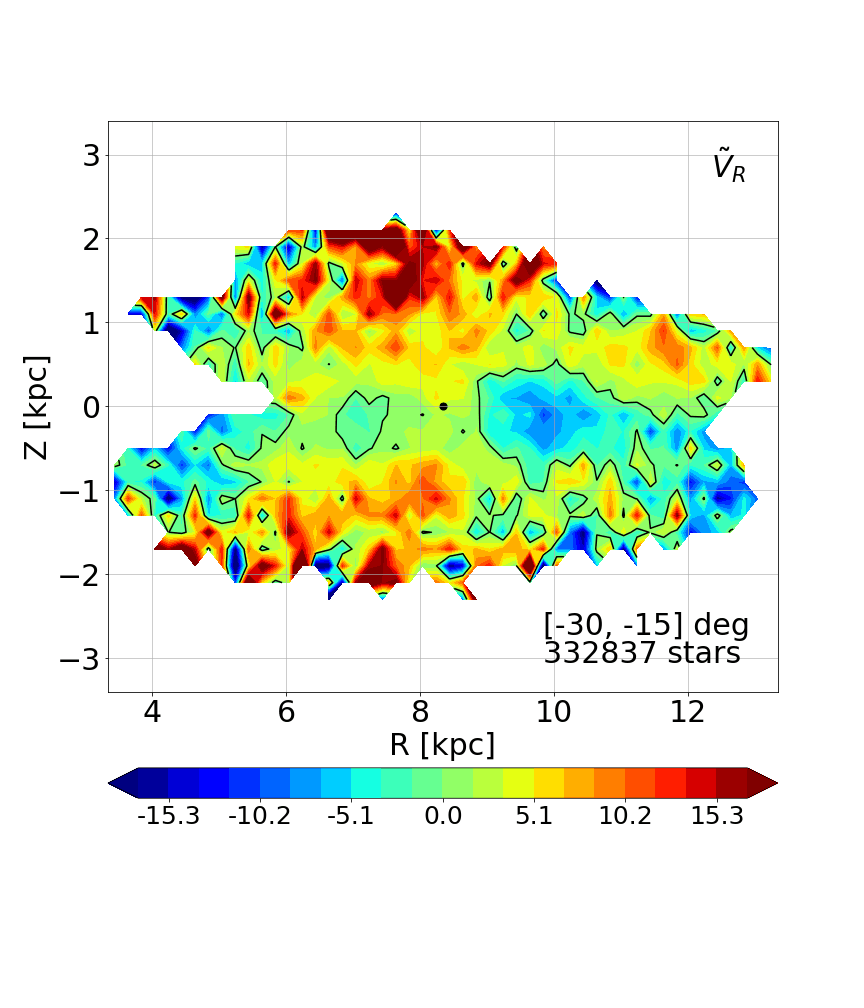}
\includegraphics[clip=true, trim = 5mm 40mm 10mm 40mm, width=0.45 \hsize]{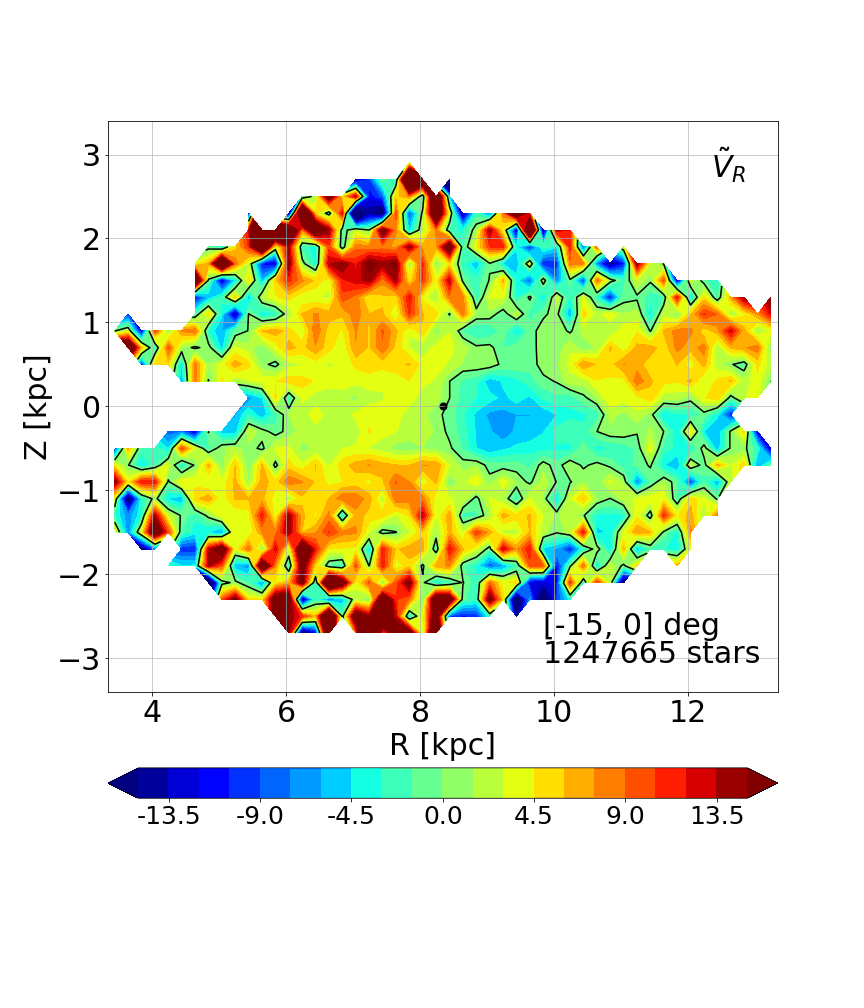}
\includegraphics[clip=true, trim = 5mm 40mm 10mm 40mm, width=0.45 \hsize]{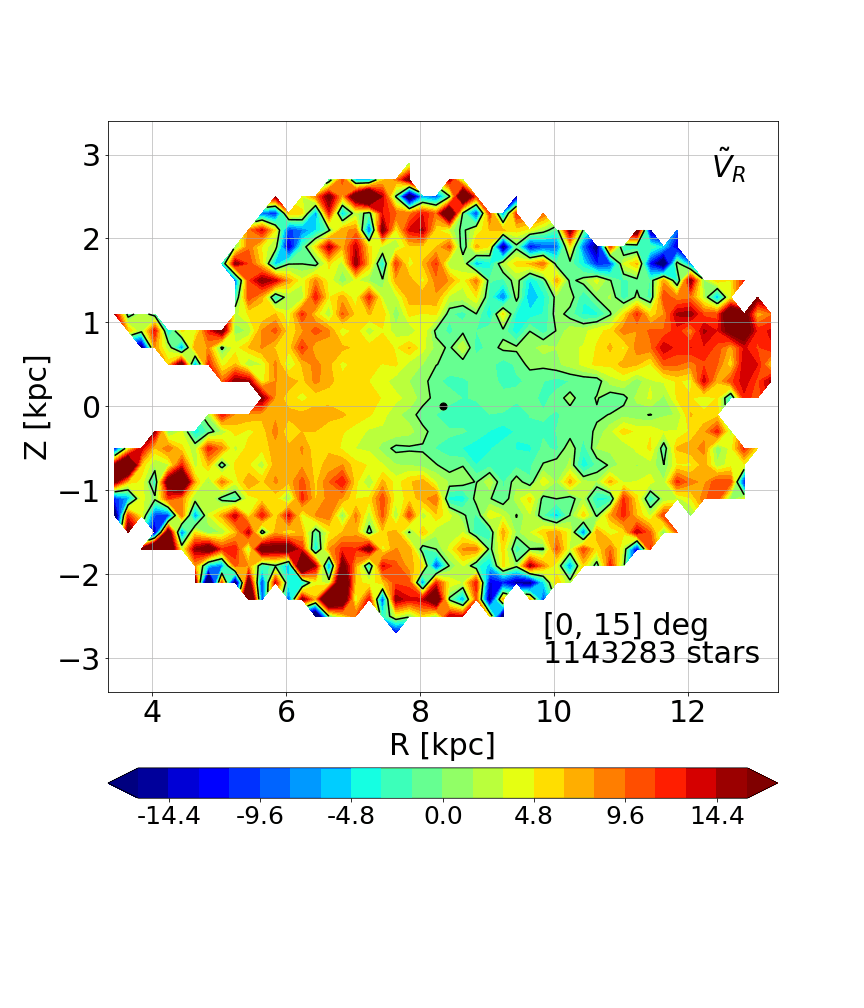}
\includegraphics[clip=true, trim = 5mm 40mm 10mm 40mm, width=0.45 \hsize]{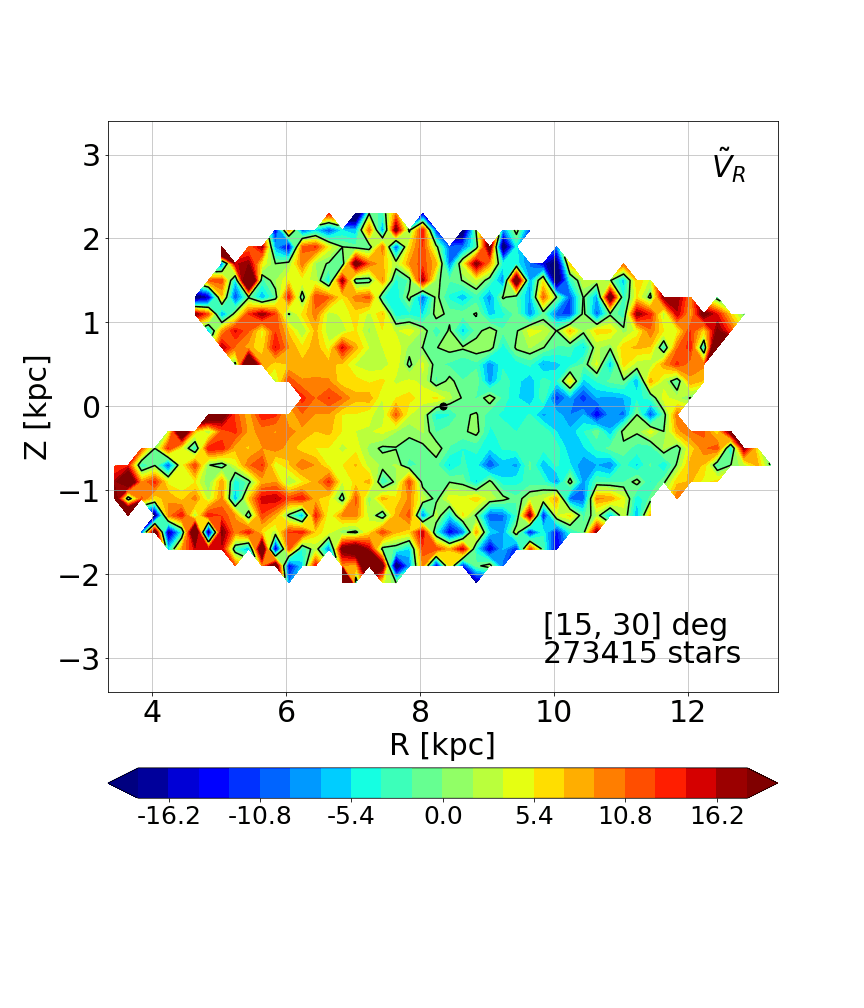}
\caption{Mosaic of edge-on maps of the median radial velocity, $\tilde{V}_R$, of the giant sample. Each map corresponds to a slice of 15~degrees in azimuth: $[-30, -15]$
(top left), $[-15, 0]$ (top right), $[0, +15]$ (bottom left), and $[+15, +30]$~degrees (bottom right). The Sun is represented by a black dot at $X=-8.34$~kpc and $Y=0$~kpc. The Galactic centre is located on the left side. The iso-velocity contours $\tilde{V}_R = 0$~$\kms$ are pointed out as black lines. The numbers of stars used to produce the maps are given in the lower left corners.}
\label{fig:rzvrmed}
\end{figure*}

\begin{figure*}[]
\centering
\includegraphics[clip=true, trim = 10mm 40mm 20mm 40mm, width=0.3 \hsize]{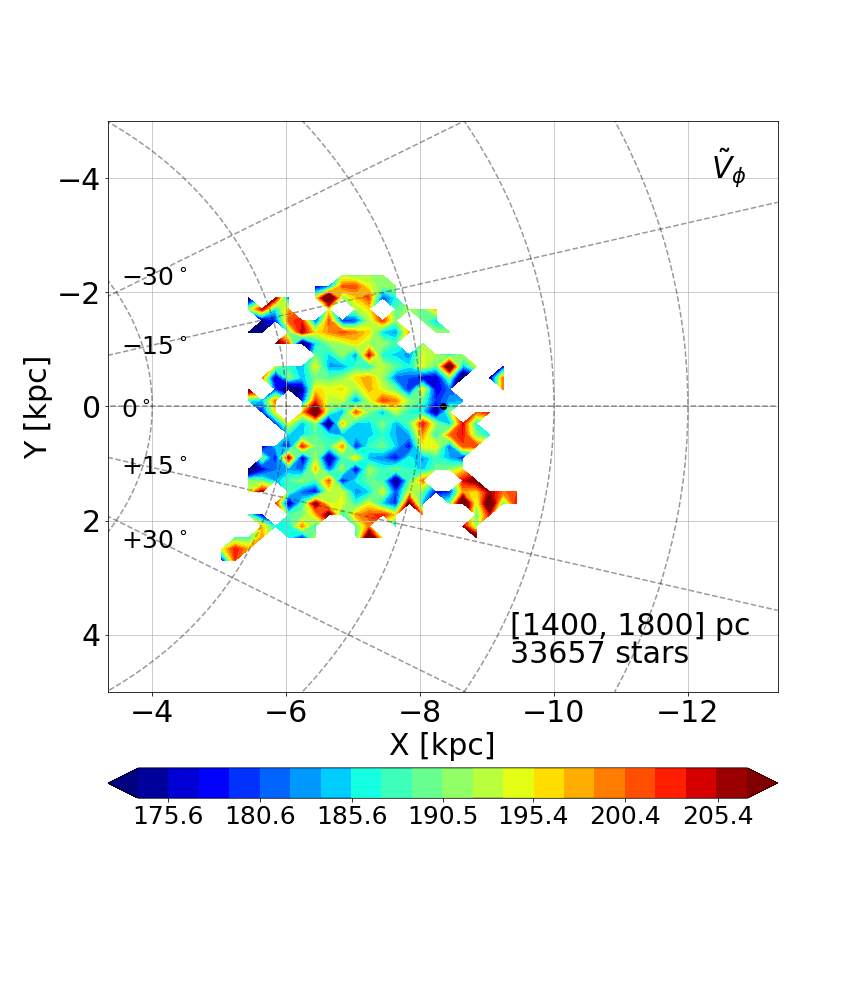}
\includegraphics[clip=true, trim = 10mm 40mm 20mm 40mm, width=0.3 \hsize]{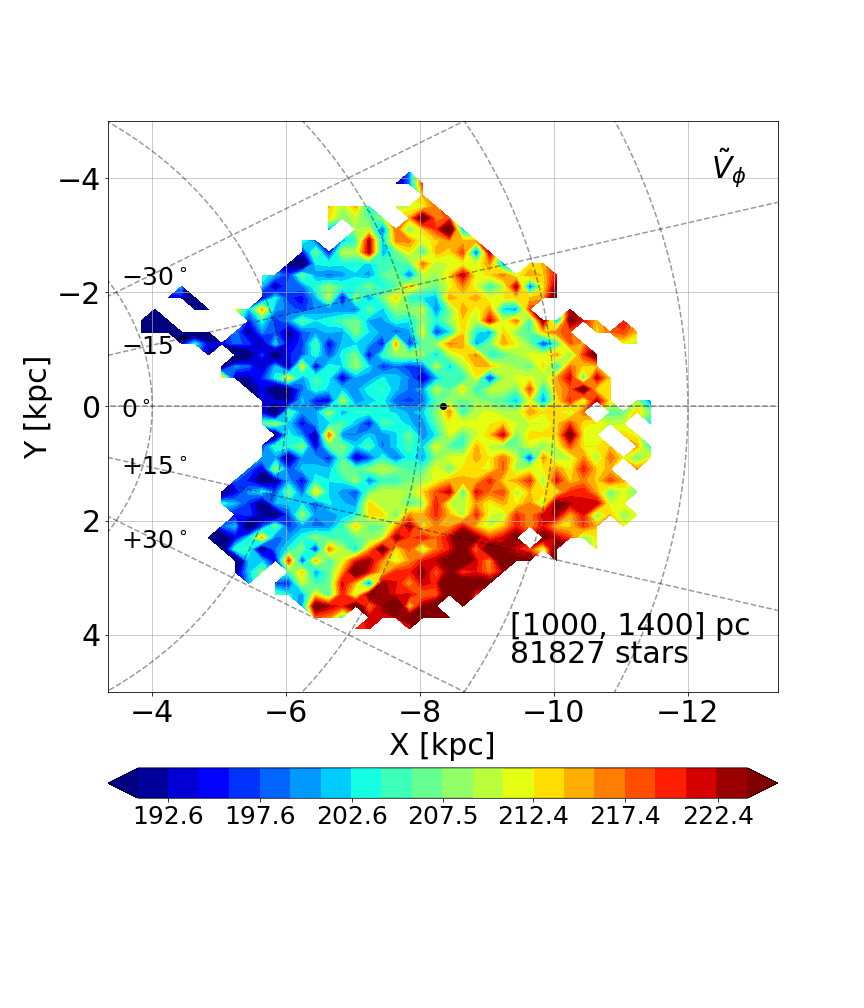}
\includegraphics[clip=true, trim = 10mm 40mm 20mm 40mm, width=0.3 \hsize]{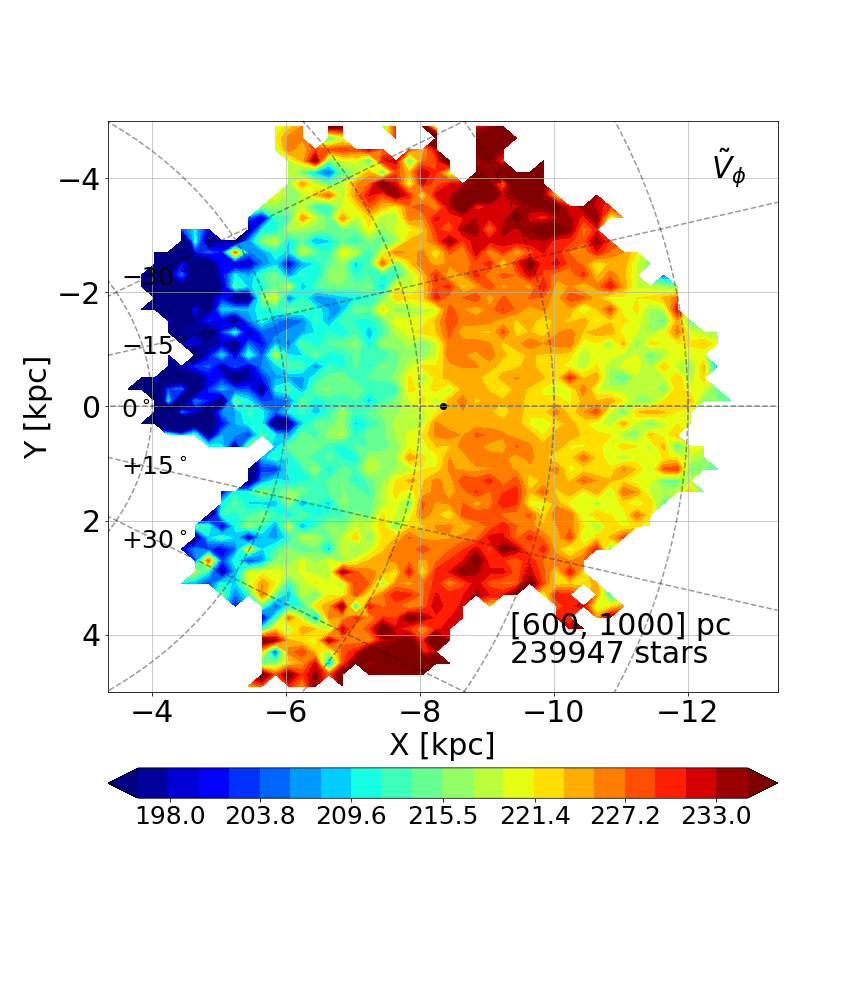}
\includegraphics[clip=true, trim = 10mm 40mm 20mm 40mm, width=0.3 \hsize]{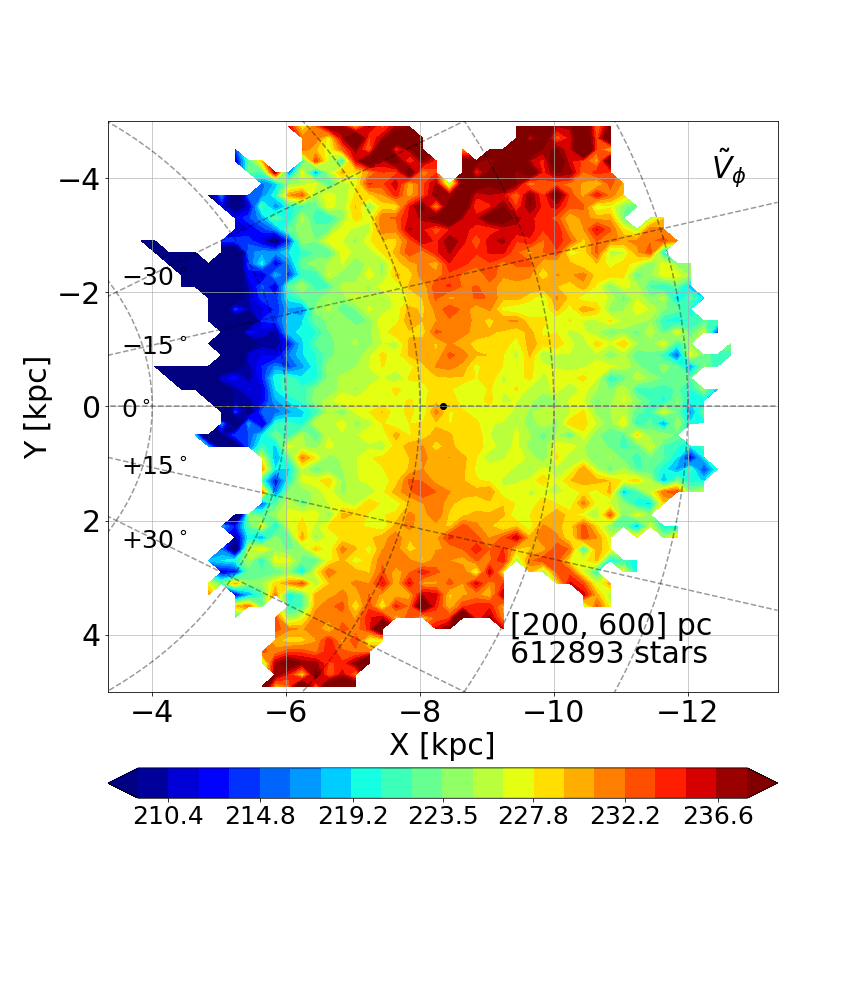}
\includegraphics[clip=true, trim = 10mm 40mm 20mm 40mm, width=0.3 \hsize]{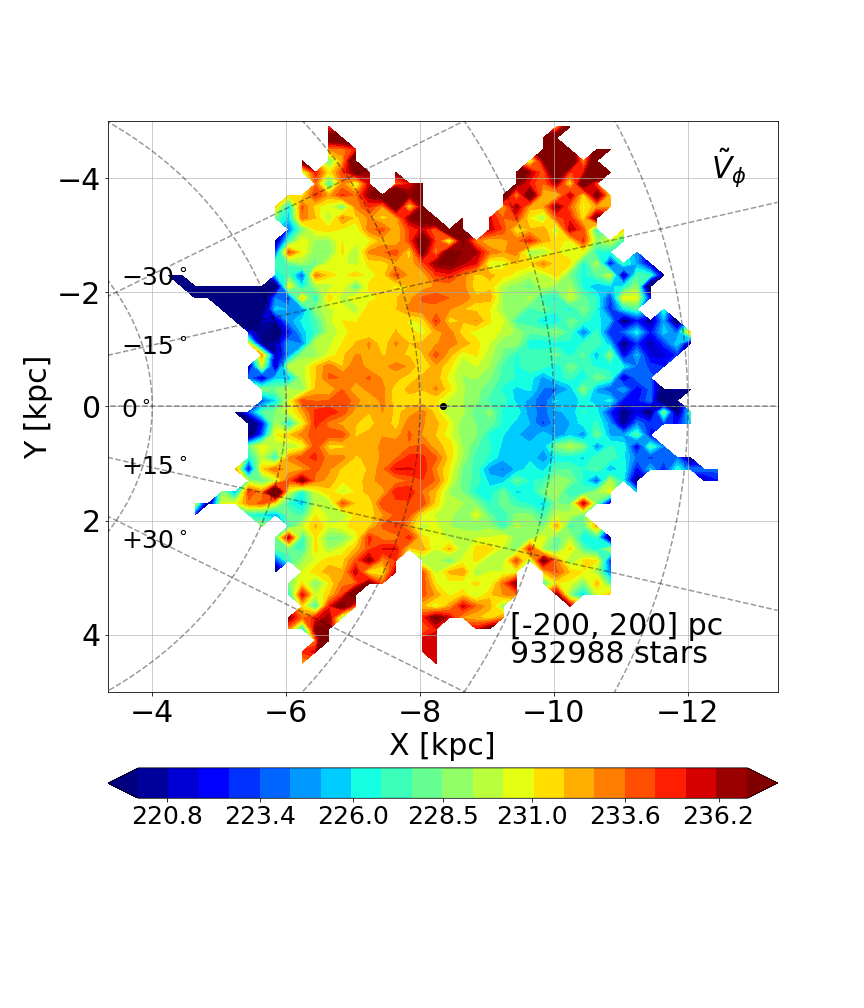}
\includegraphics[clip=true, trim = 10mm 40mm 20mm 40mm, width=0.3 \hsize]{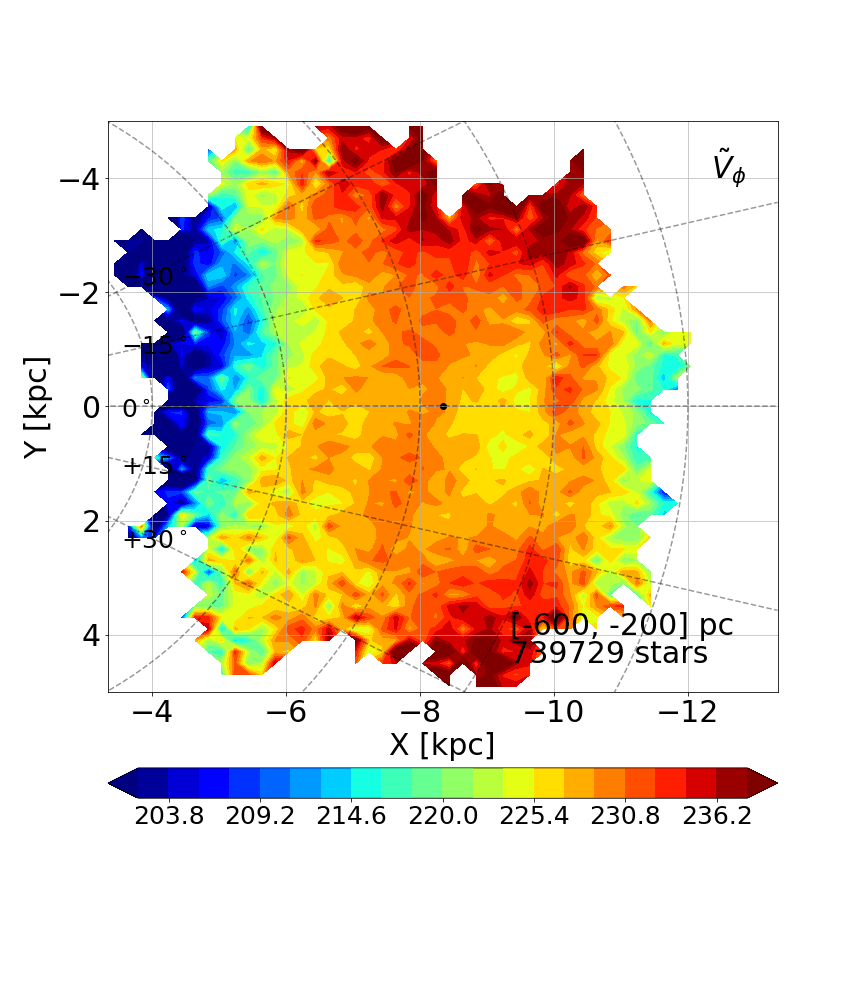}
\includegraphics[clip=true, trim = 10mm 40mm 20mm 40mm, width=0.3 \hsize]{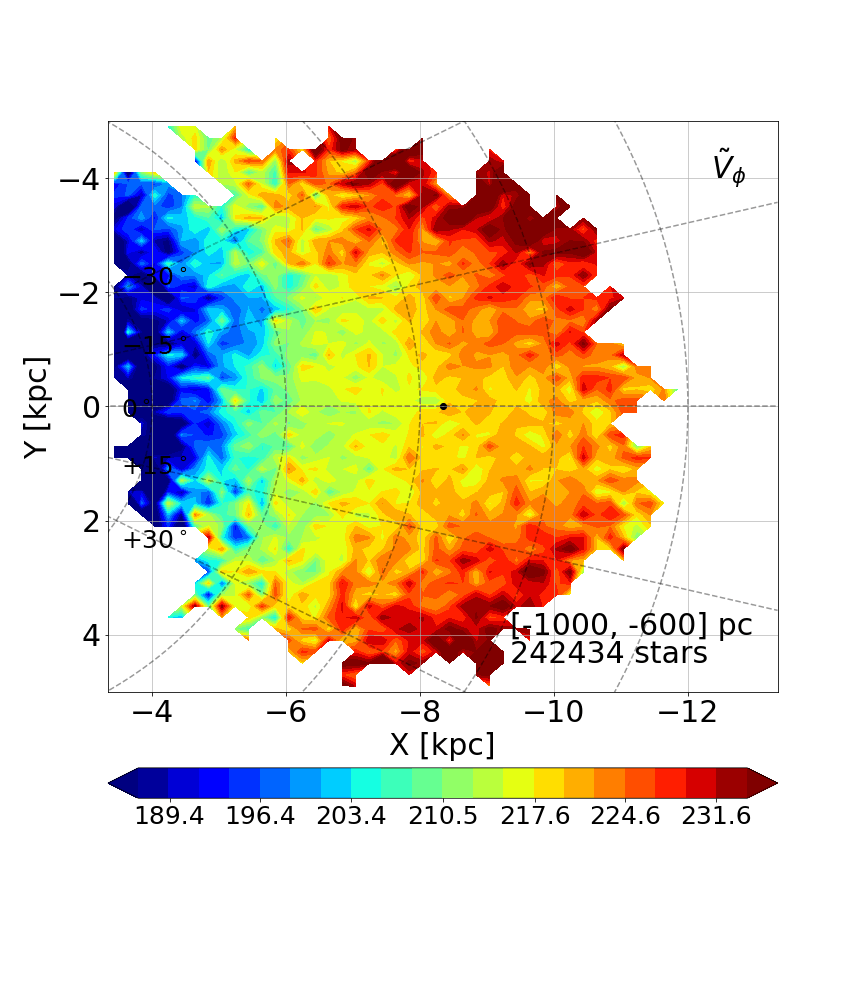}
\includegraphics[clip=true, trim = 10mm 40mm 20mm 40mm, width=0.3 \hsize]{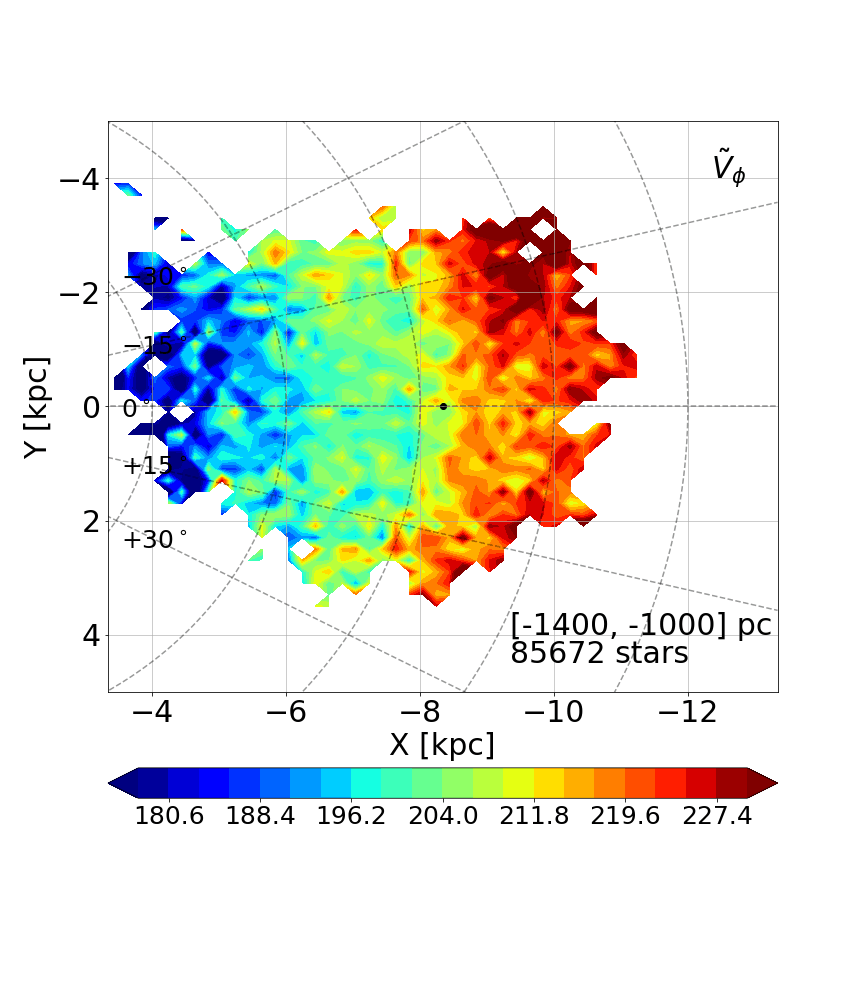}
\includegraphics[clip=true, trim = 10mm 40mm 20mm 40mm, width=0.3 \hsize]{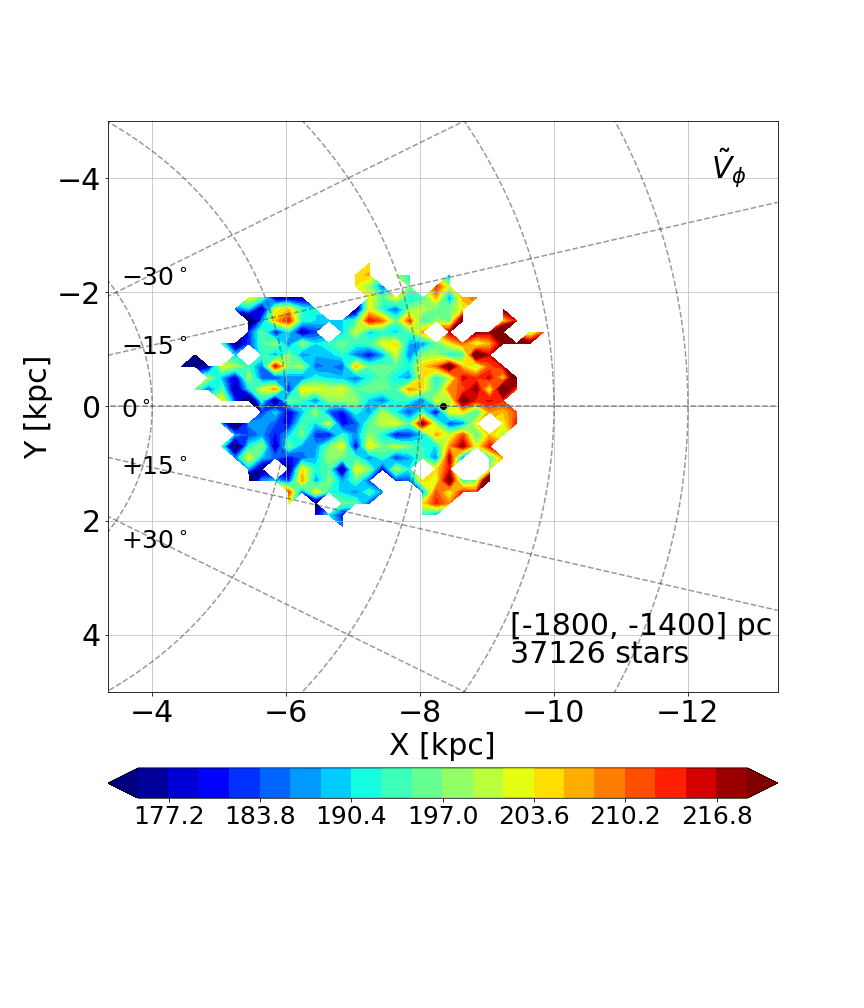}
\caption{Same as Fig.~\ref{fig:xyvrmed} for the median azimuthal velocity, $\tilde{V}_\phi$.}
\label{fig:xyvphimed}
\end{figure*}

\begin{figure*}[]
\centering
\includegraphics[clip=true, trim = 5mm 40mm 10mm 40mm, width=0.45 \hsize]{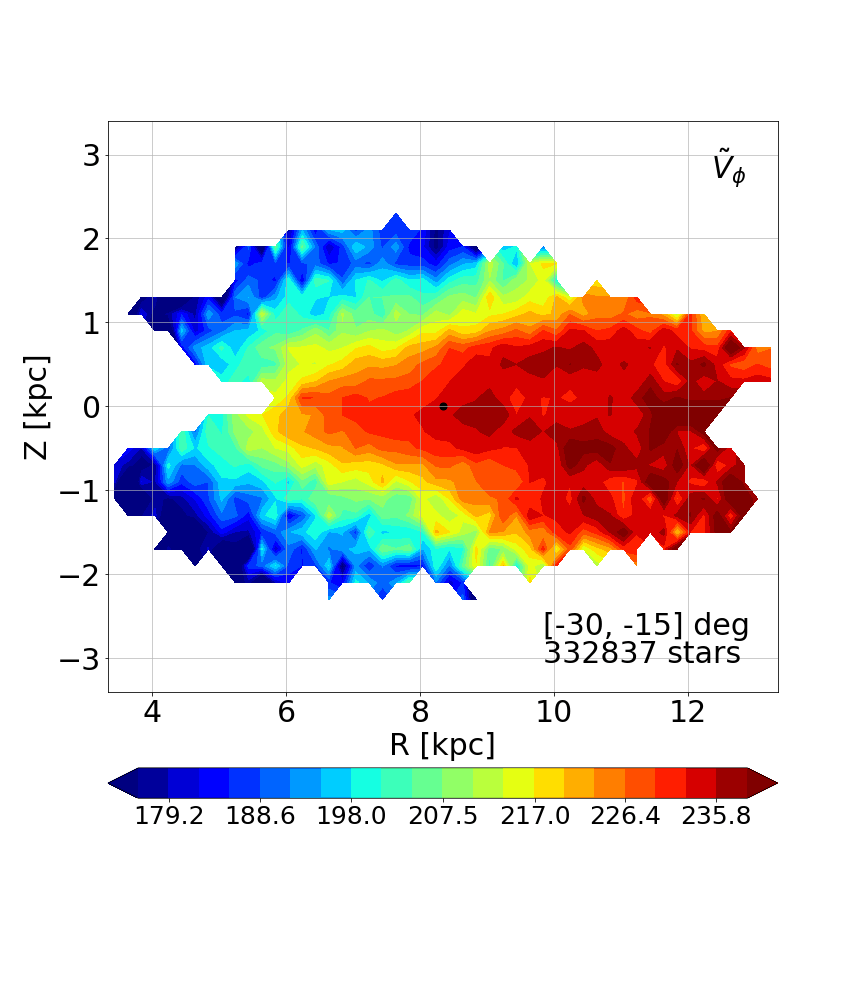}
\includegraphics[clip=true, trim = 5mm 40mm 10mm 40mm, width=0.45 \hsize]{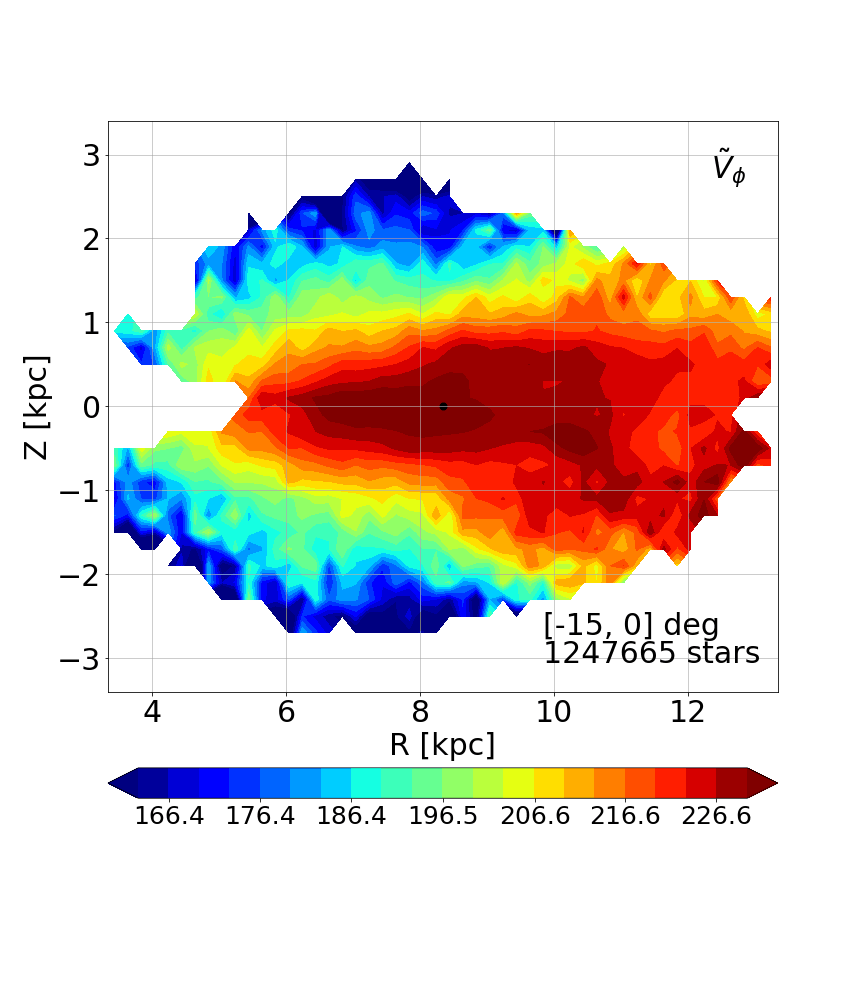}
\includegraphics[clip=true, trim = 5mm 40mm 10mm 40mm, width=0.45 \hsize]{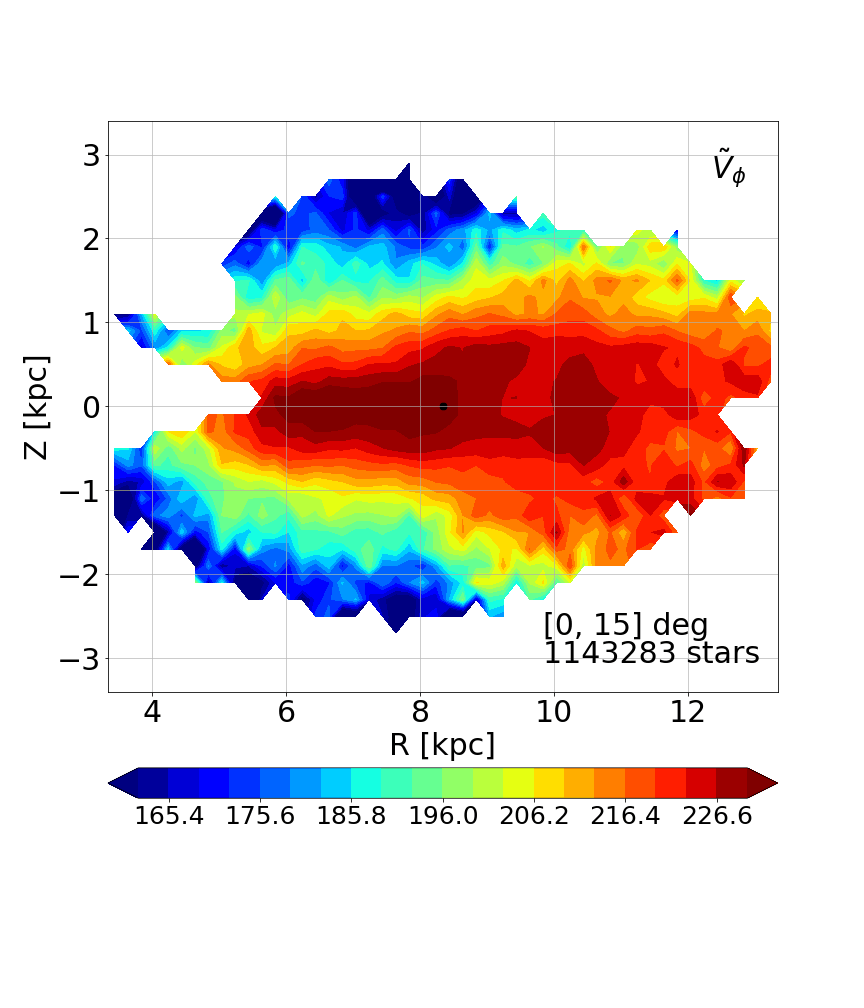}
\includegraphics[clip=true, trim = 5mm 40mm 10mm 40mm, width=0.45 \hsize]{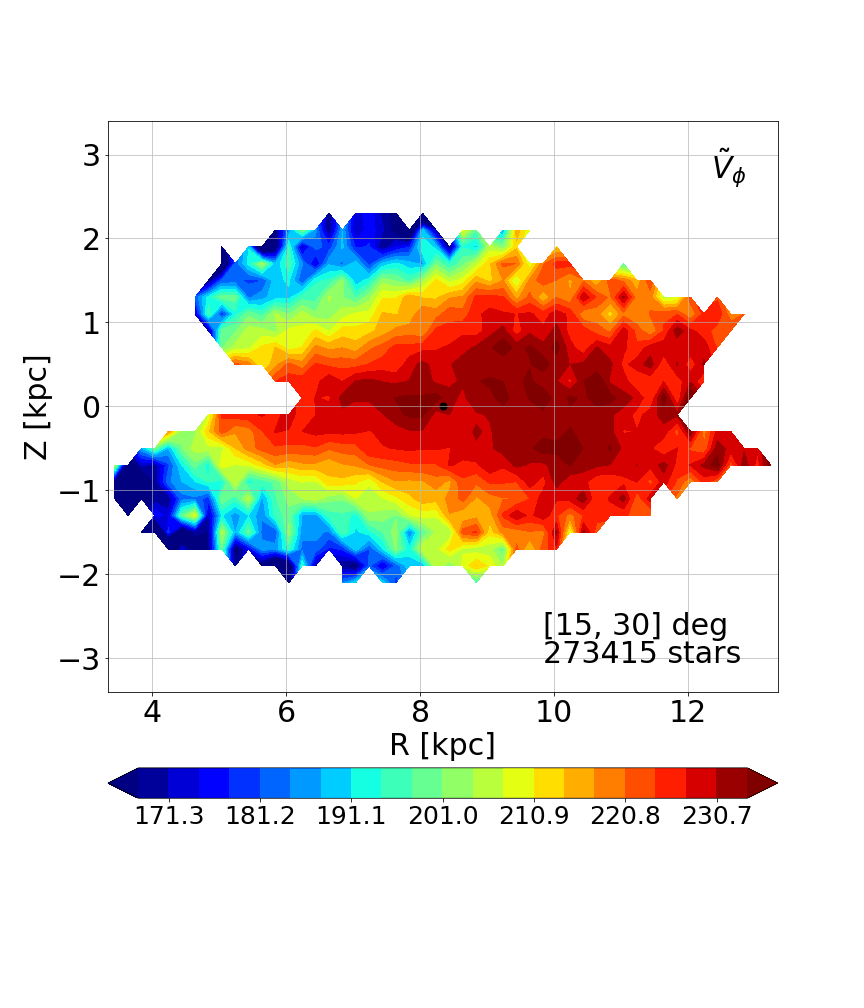}
\caption{Same as Fig.~\ref{fig:rzvrmed} for the median azimuthal velocity, $\tilde{V}_\phi$.}
\label{fig:rzvphimed}
\end{figure*}

\begin{figure*}[]
\centering
\includegraphics[clip=true, trim = 10mm 40mm 20mm 40mm, width=0.43 \hsize]{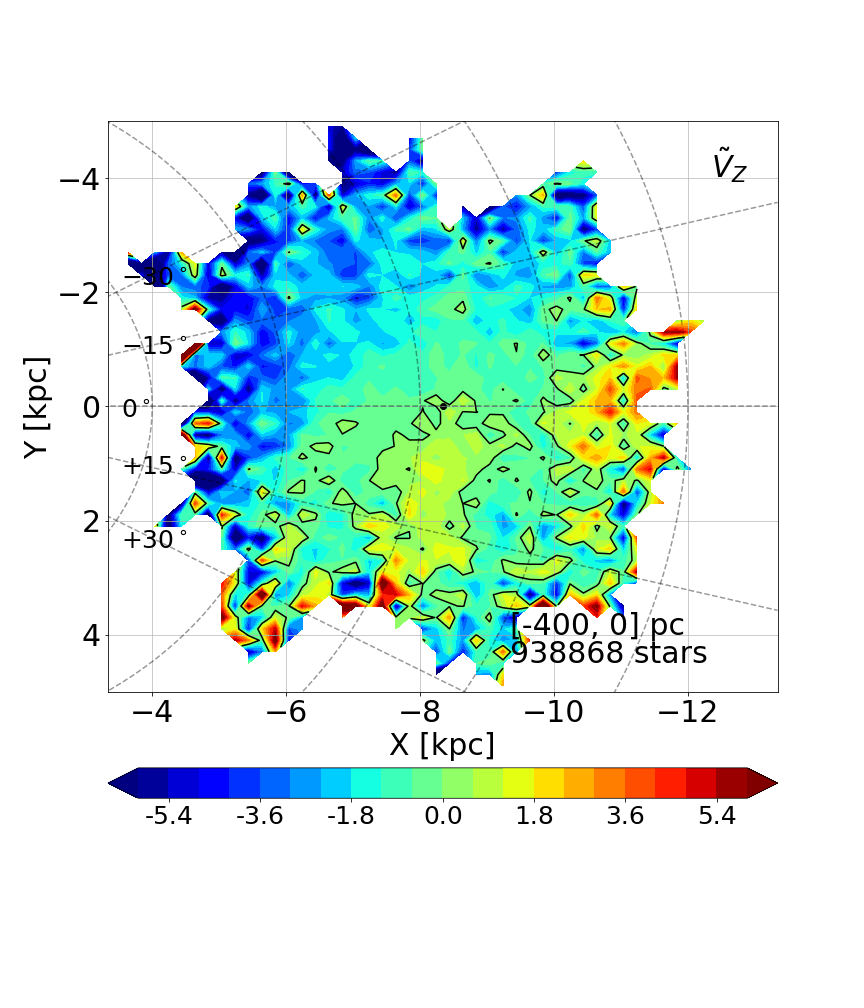}
\includegraphics[clip=true, trim = 10mm 40mm 20mm 40mm, width=0.43 \hsize]{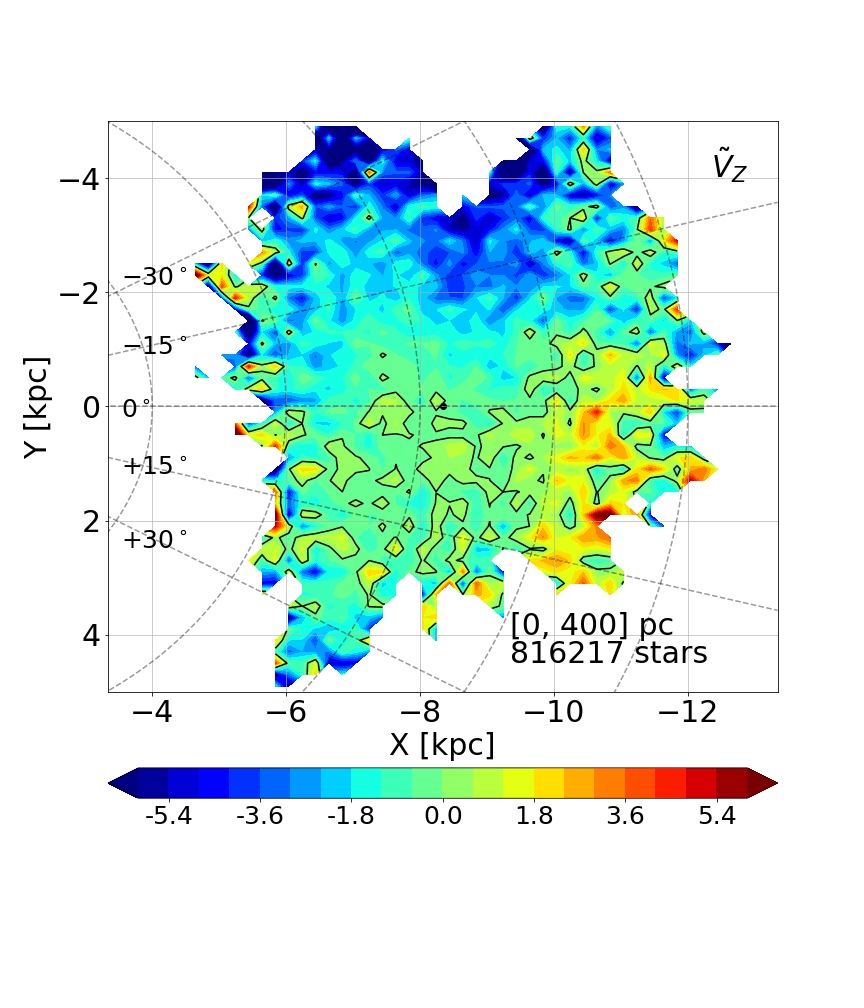}
\includegraphics[clip=true, trim = 10mm 40mm 20mm 40mm, width=0.43 \hsize]{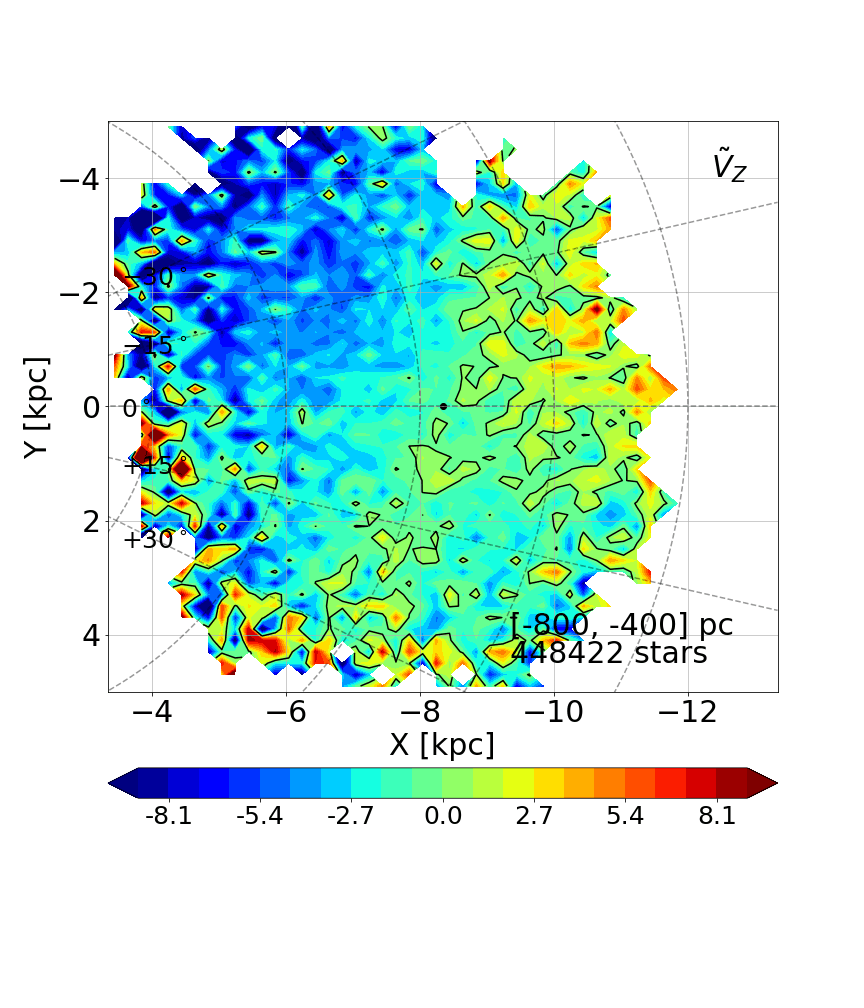}
\includegraphics[clip=true, trim = 10mm 40mm 20mm 40mm, width=0.43 \hsize]{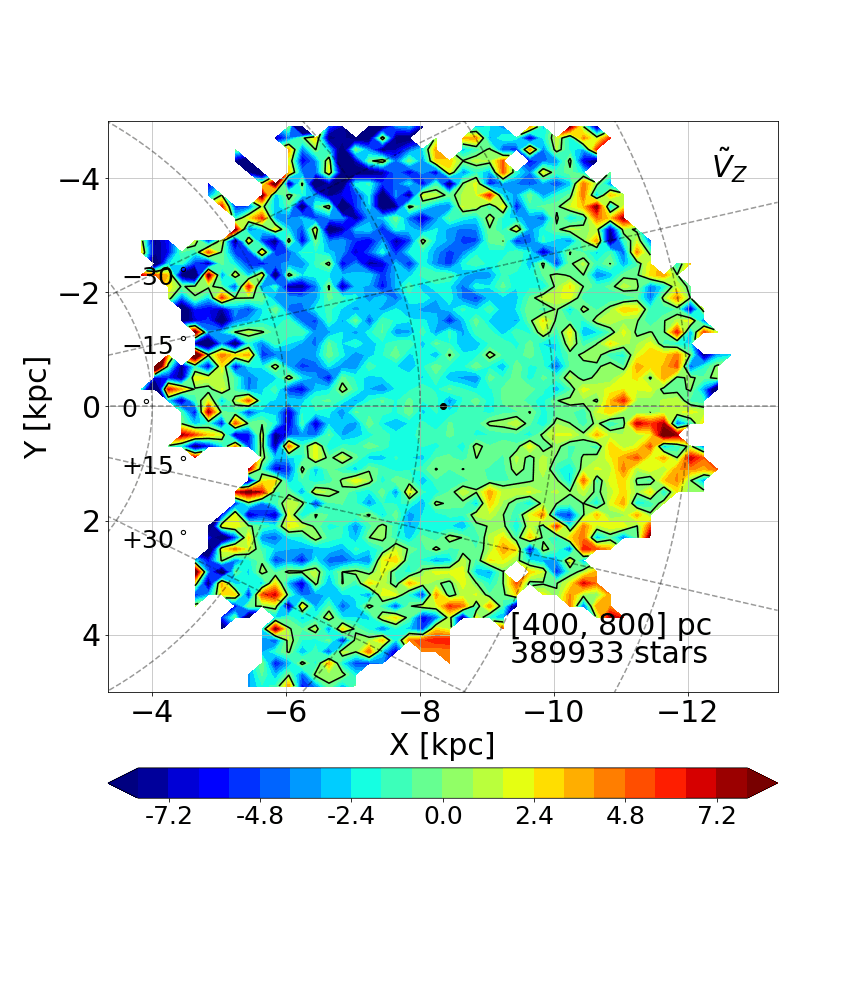}
\includegraphics[clip=true, trim = 10mm 40mm 20mm 40mm, width=0.43 \hsize]{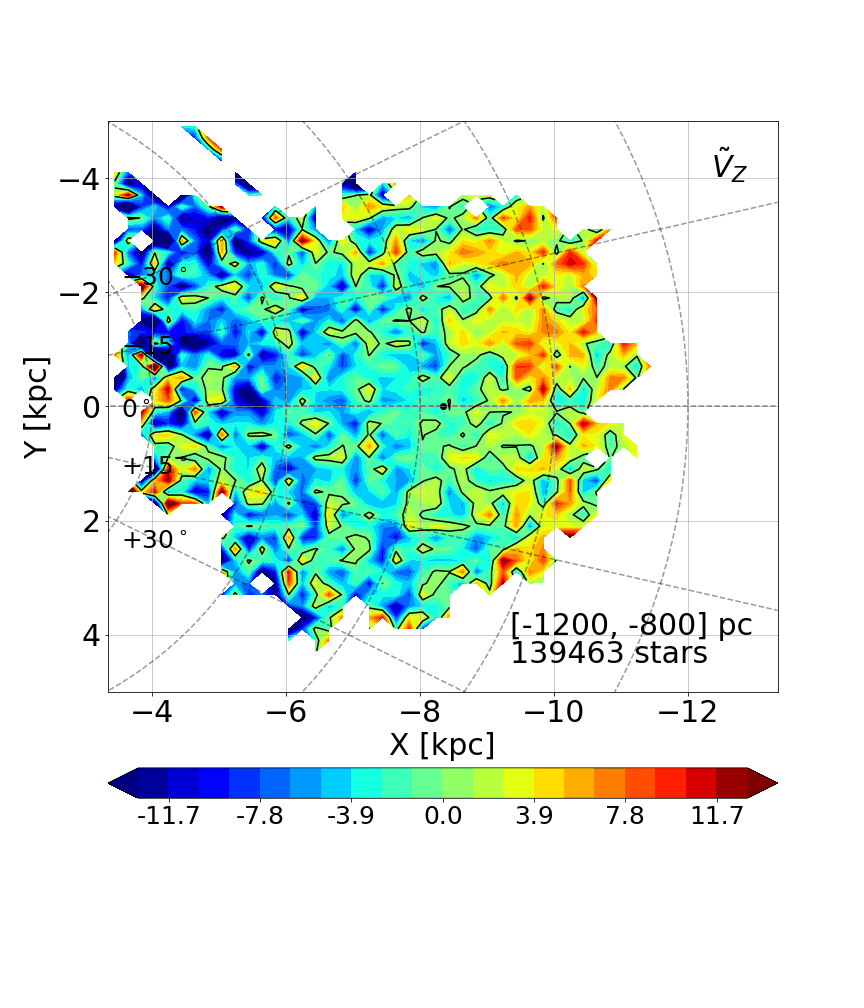}
\includegraphics[clip=true, trim = 10mm 40mm 20mm 40mm, width=0.43 \hsize]{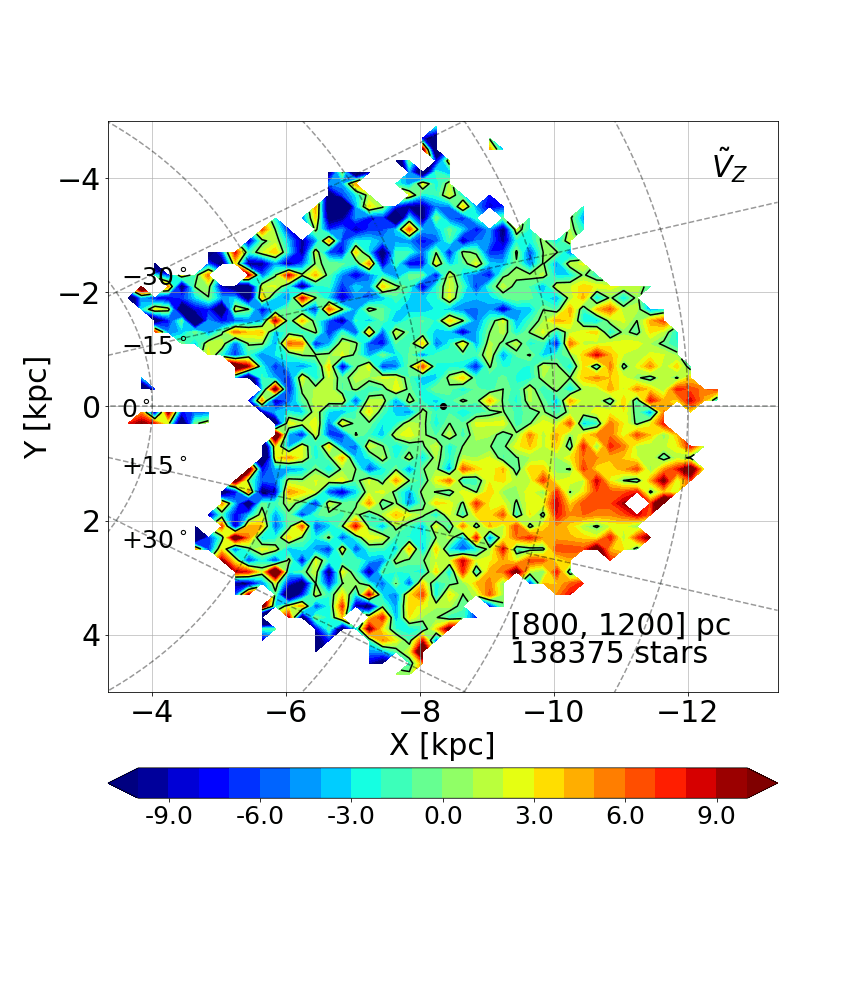}
\caption{Same as Fig.~\ref{fig:xyvrmed} for the median vertical velocity, $\tilde{V}_Z$. Here the disc has been divided into
six layers. {The southern Galactic hemisphere is on left and the northern is on the right. The distance to the Galactic mid-plane increases from top to bottom.}}
\label{fig:xyvzmed}
\end{figure*}

This appendix contains the mosaics of face-on and edge-on maps of the median velocities: $\tilde{V}_R$ (Fig.~\ref{fig:xyvrmed} and \ref{fig:rzvrmed}), $\tilde{V}_\phi$ (Fig.~\ref{fig:xyvphimed} and \ref{fig:rzvphimed}), and $\tilde{V}_Z$ (Fig.~\ref{fig:xyvzmed} and \ref{fig:rzvzmed}), and of the velocity dispersions: $\sigma_{V_R}$ (Fig.~\ref{fig:xyvrdisp} and \ref{fig:rzvrdisp}), $\sigma_{V_\phi}$ (Fig.~\ref{fig:xyvphidisp} and \ref{fig:rzvphidisp}), and $\sigma_{V_Z}$ (Fig.~\ref{fig:xyvzdisp} and \ref{fig:rzvzdisp}).

The face-on map mosaics are made of nine maps, each one corresponding to a 400~pc height $Z$ layer. The exception is the median vertical velocity, $\tilde{V}_Z$, whose mosaic is made of six maps: three above the disc mid-plane, and three below it. The edge-on map mosaics are made of four maps, each one corresponding to a 15~degrees slice in azimuth: $[-30, -15]$, $[-15, 0]$, $[0, +15]$ and $[+15, +30]$~degrees.

\begin{figure*}[]
\centering
\includegraphics[clip=true, trim = 10mm 40mm 20mm 40mm, width=0.42 \hsize]{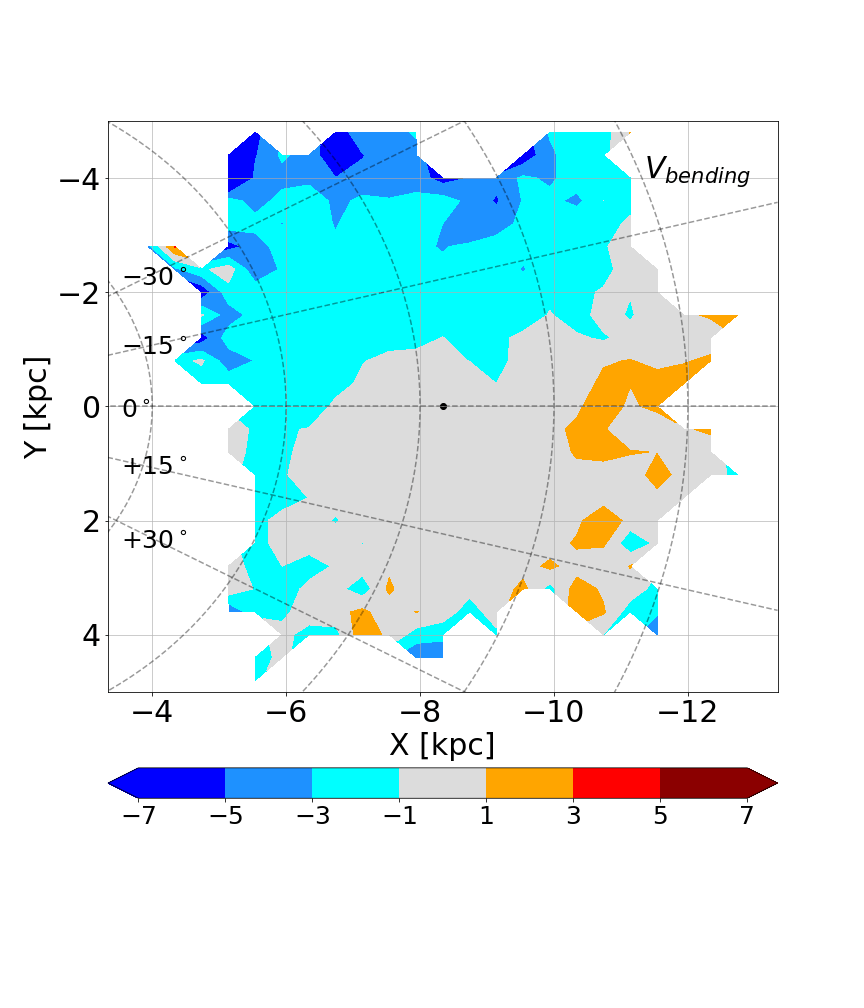}
\includegraphics[clip=true, trim = 10mm 40mm 20mm 40mm, width=0.42 \hsize]{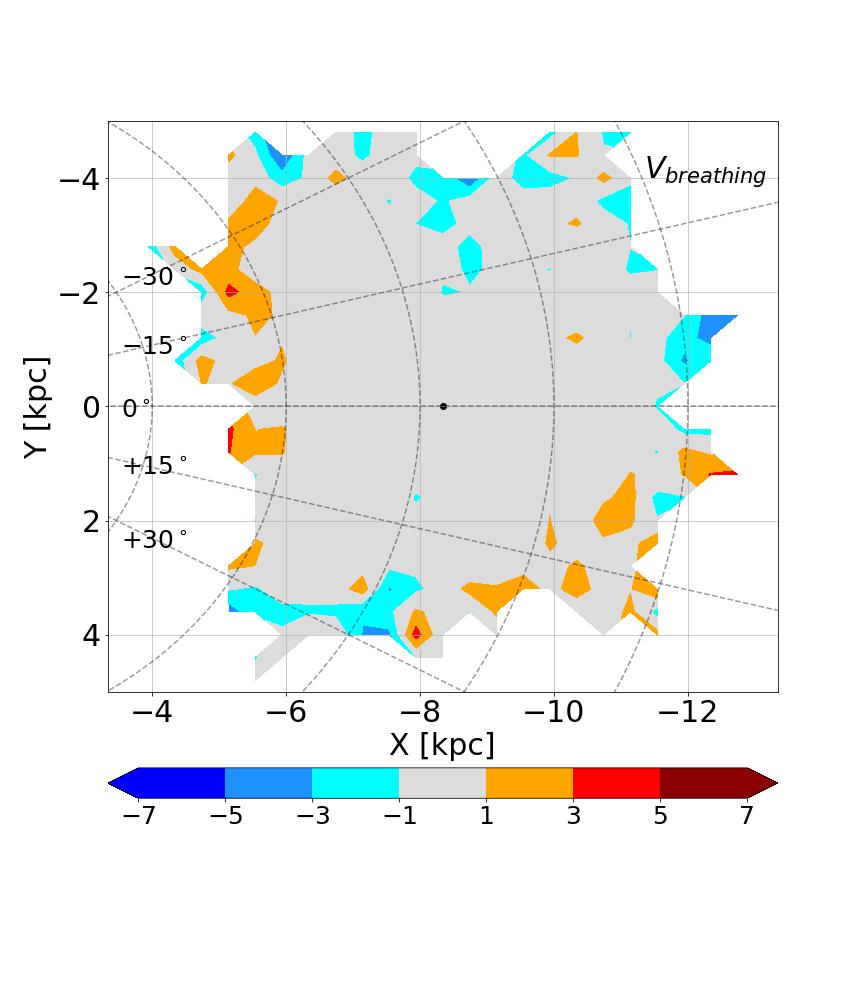}
\includegraphics[clip=true, trim = 10mm 40mm 20mm 40mm, width=0.42 \hsize]{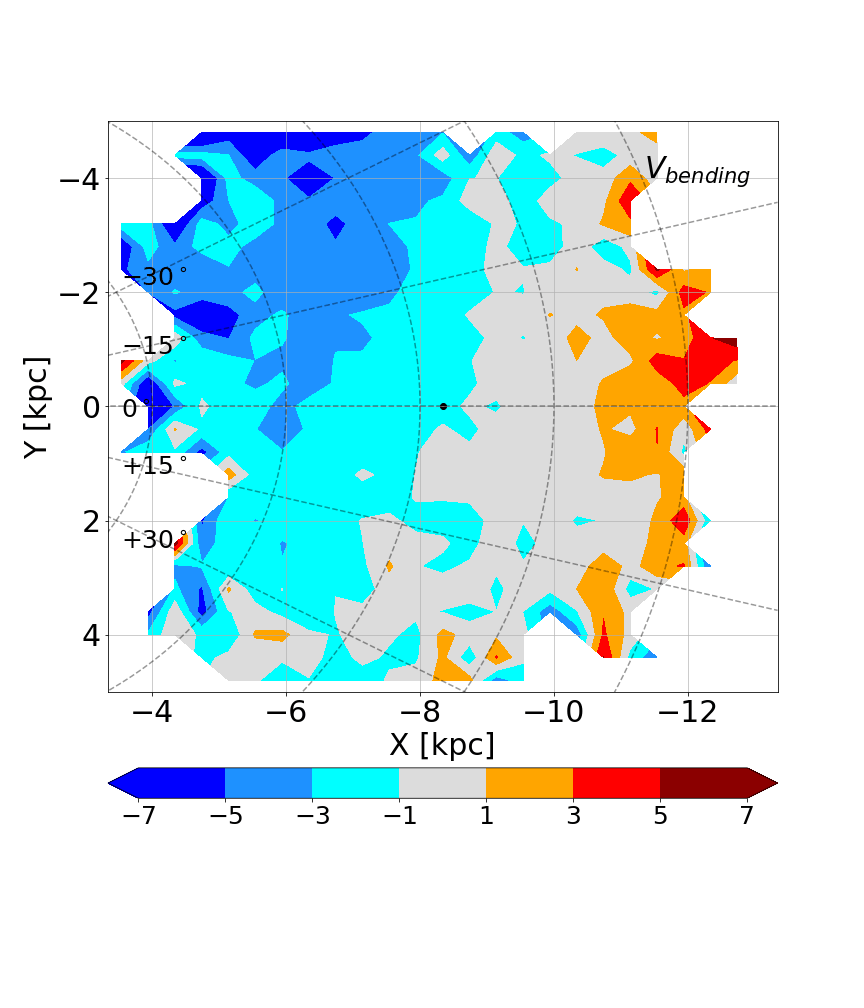}
\includegraphics[clip=true, trim = 10mm 40mm 20mm 40mm, width=0.42 \hsize]{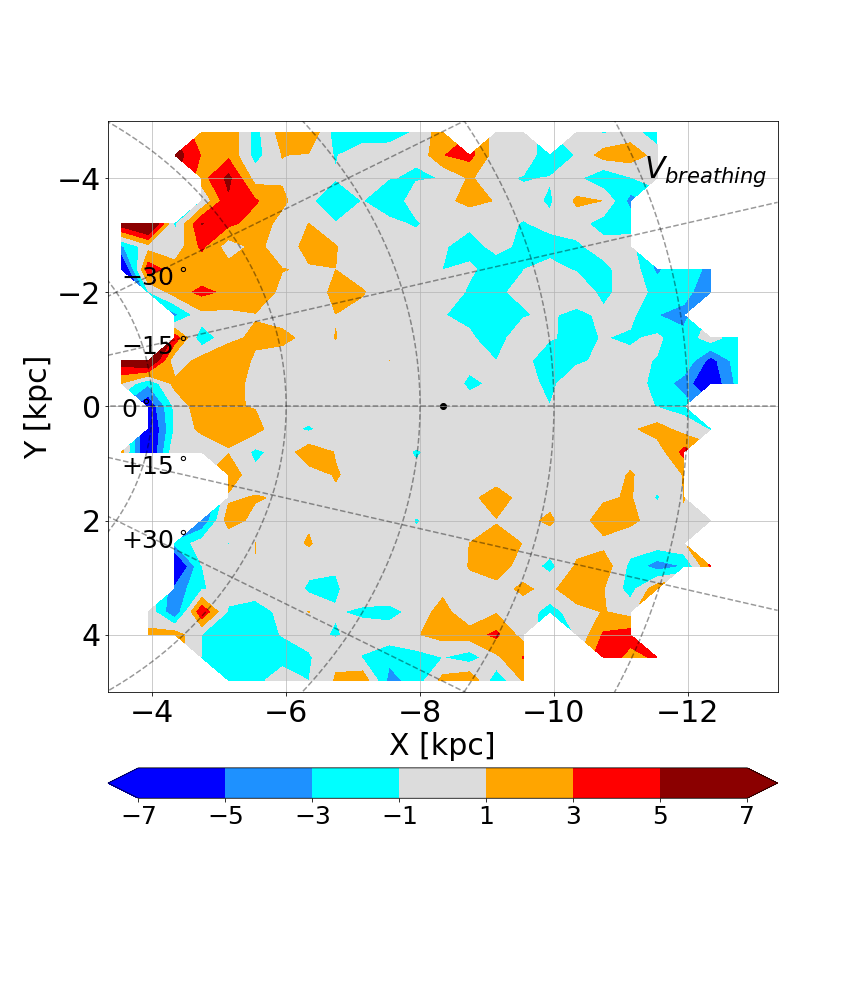}
\includegraphics[clip=true, trim = 10mm 40mm 20mm 40mm, width=0.42 \hsize]{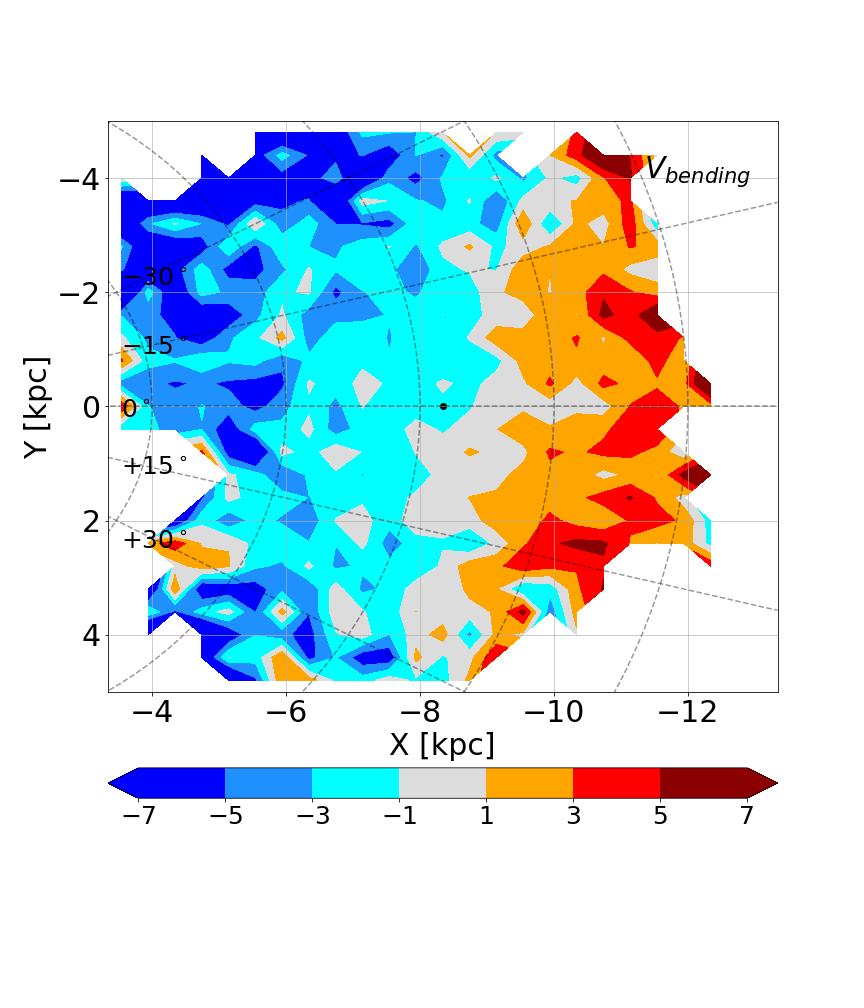}
\includegraphics[clip=true, trim = 10mm 40mm 20mm 40mm, width=0.42 \hsize]{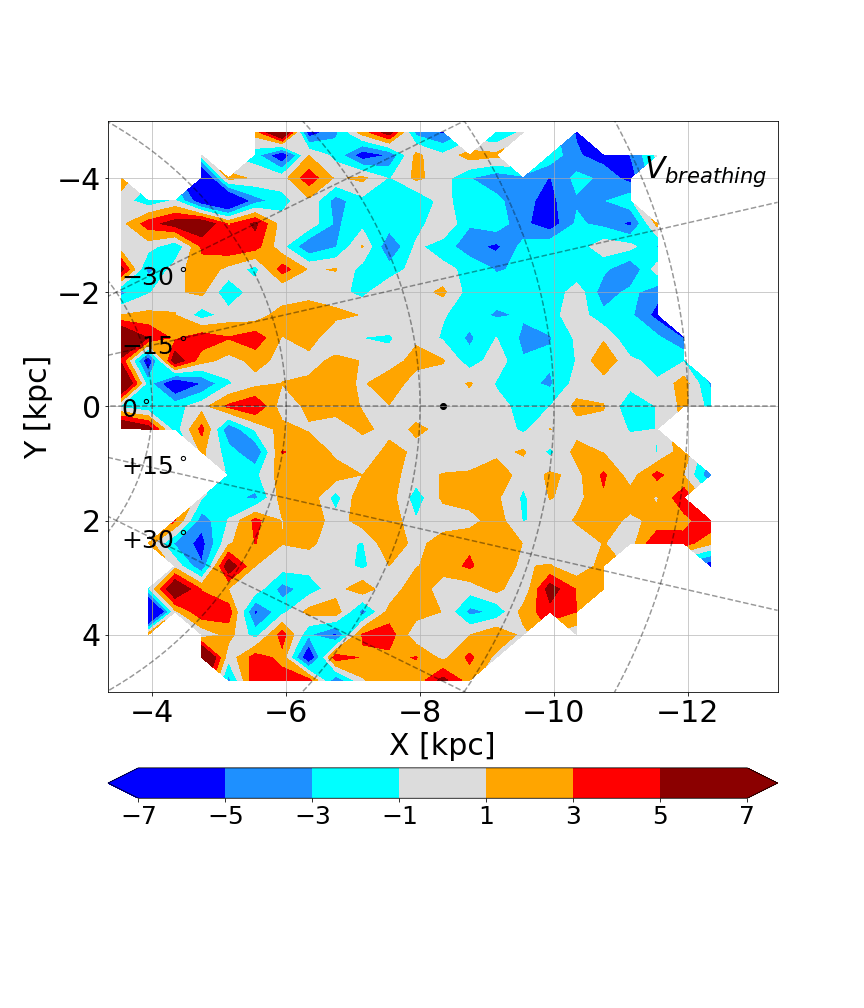}
\caption{{Same as Fig.~\ref{fig:xyvrmed} for the bending (left) and breathing (right) velocities. Here the disc has been divided into three groups of symmetric layers. The distance to the Galactic mid-plane increases from top to bottom: $[-400,0]$ and $[0,400]$~pc (top), $[-800,-400]$ and $[400,800]$~pc (middle), $[-1200,-800]$ and $[800,1200]$~pc (bottom). The bending and breathing velocities have been calculated using larger $(X,Y)$ cells than in the other maps of this appendix, i.e. 400~pc by 400~pc instead of 200~pc by 200~pc.}}
\label{fig:xyvbendbreath}
\end{figure*}

\begin{figure*}[]
\centering
\includegraphics[clip=true, trim = 5mm 40mm 10mm 40mm, width=0.45 \hsize]{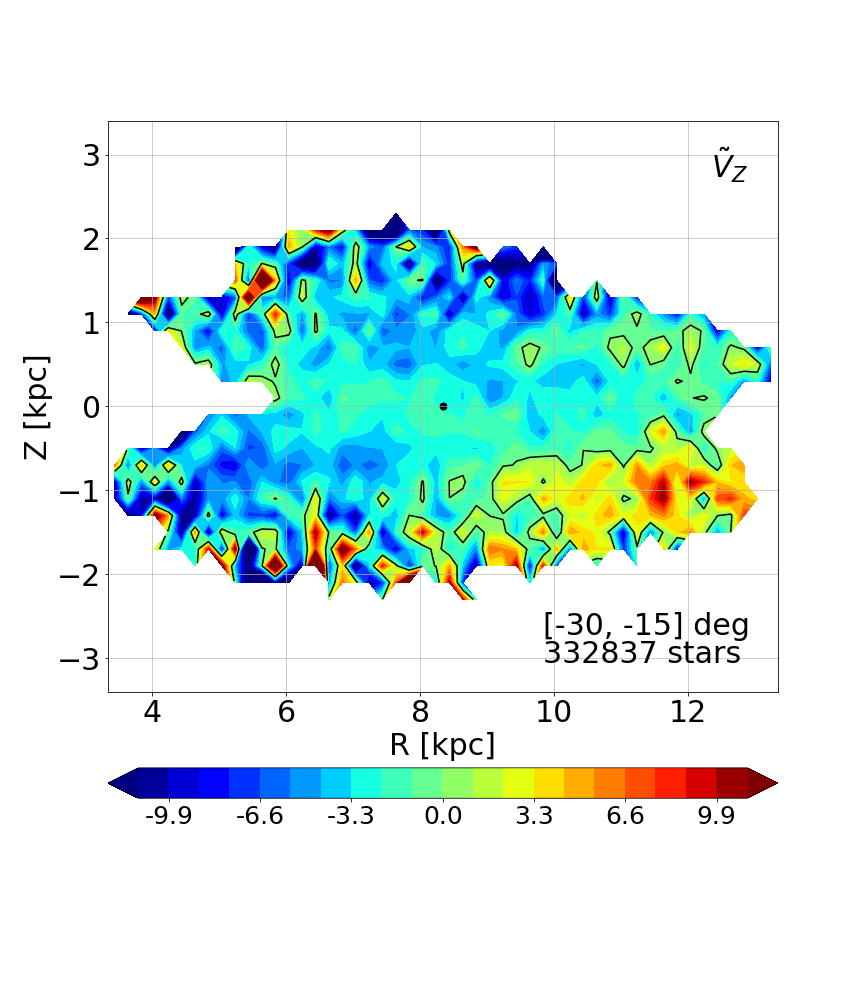}
\includegraphics[clip=true, trim = 5mm 40mm 10mm 40mm, width=0.45 \hsize]{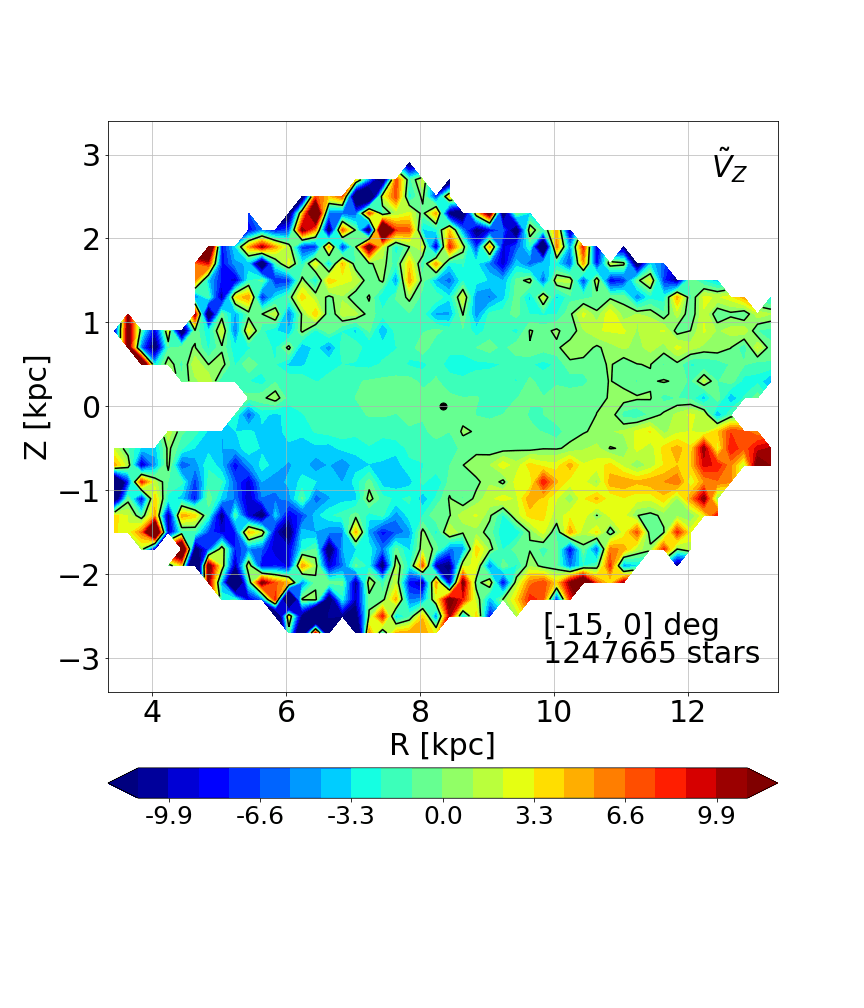}
\includegraphics[clip=true, trim = 5mm 40mm 10mm 40mm, width=0.45 \hsize]{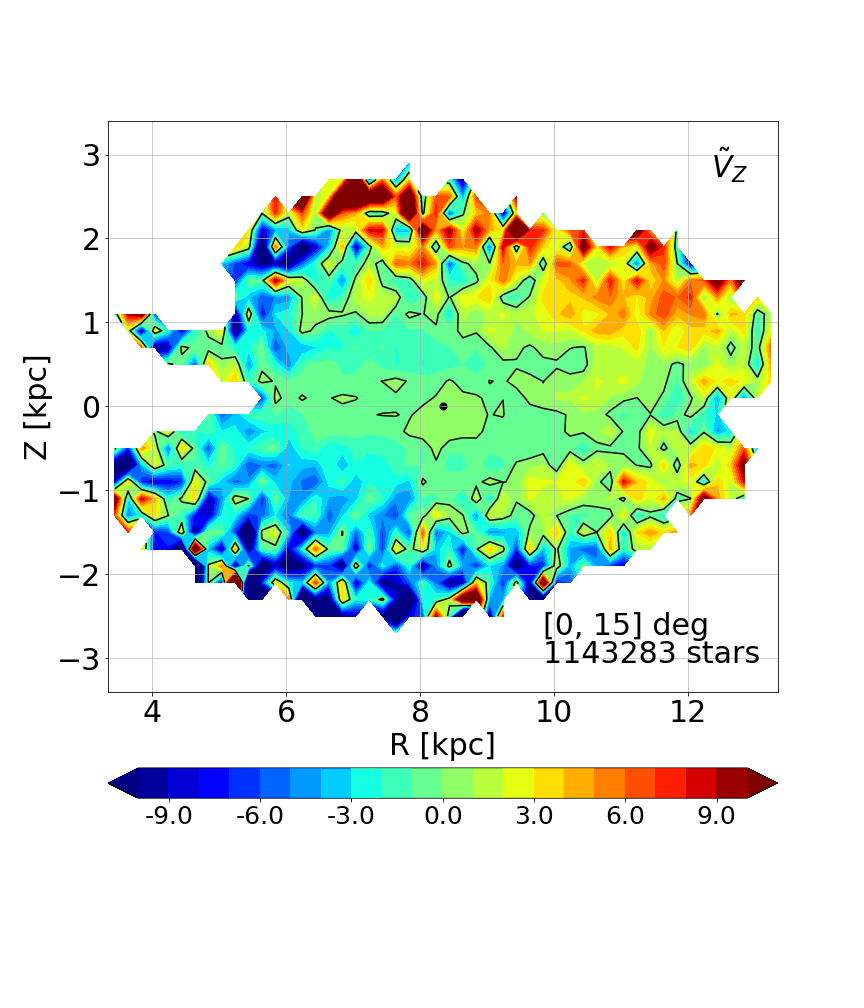}
\includegraphics[clip=true, trim = 5mm 40mm 10mm 40mm, width=0.45 \hsize]{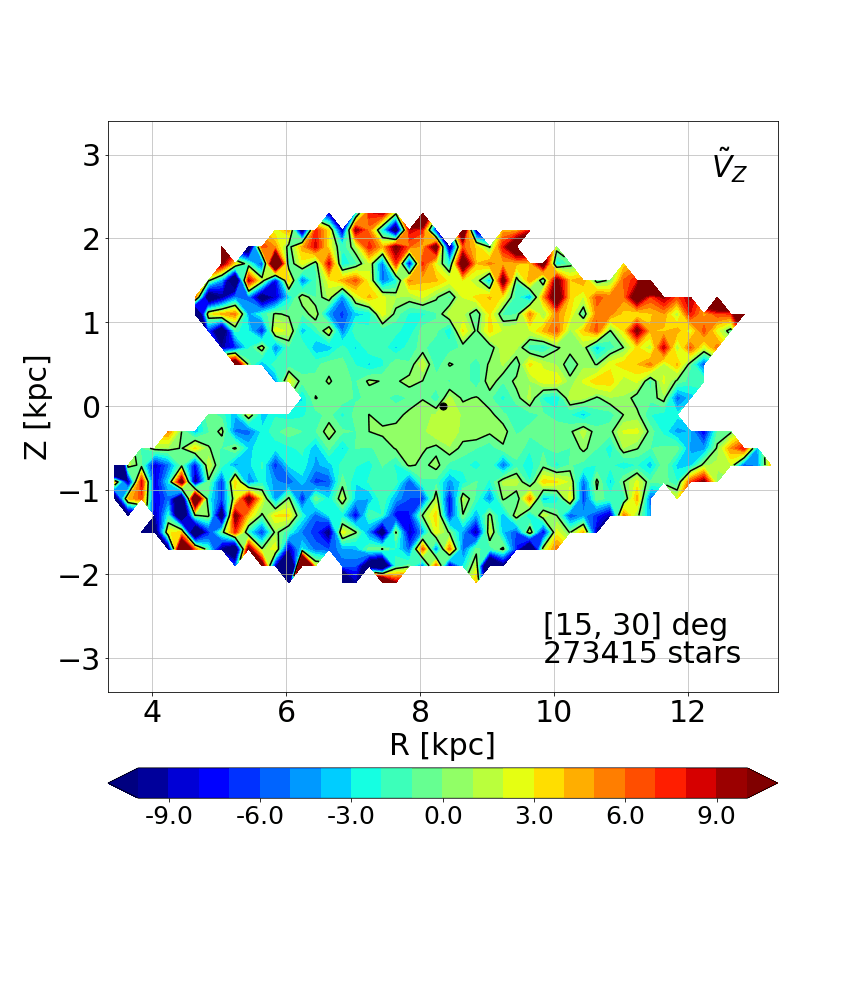}
\caption{Same as Fig.~\ref{fig:rzvrmed} for the median vertical velocity, $\tilde{V}_Z$.}
\label{fig:rzvzmed}
\end{figure*}

\begin{figure*}[]
\centering
\includegraphics[clip=true, trim = 10mm 40mm 20mm 40mm, width=0.3 \hsize]{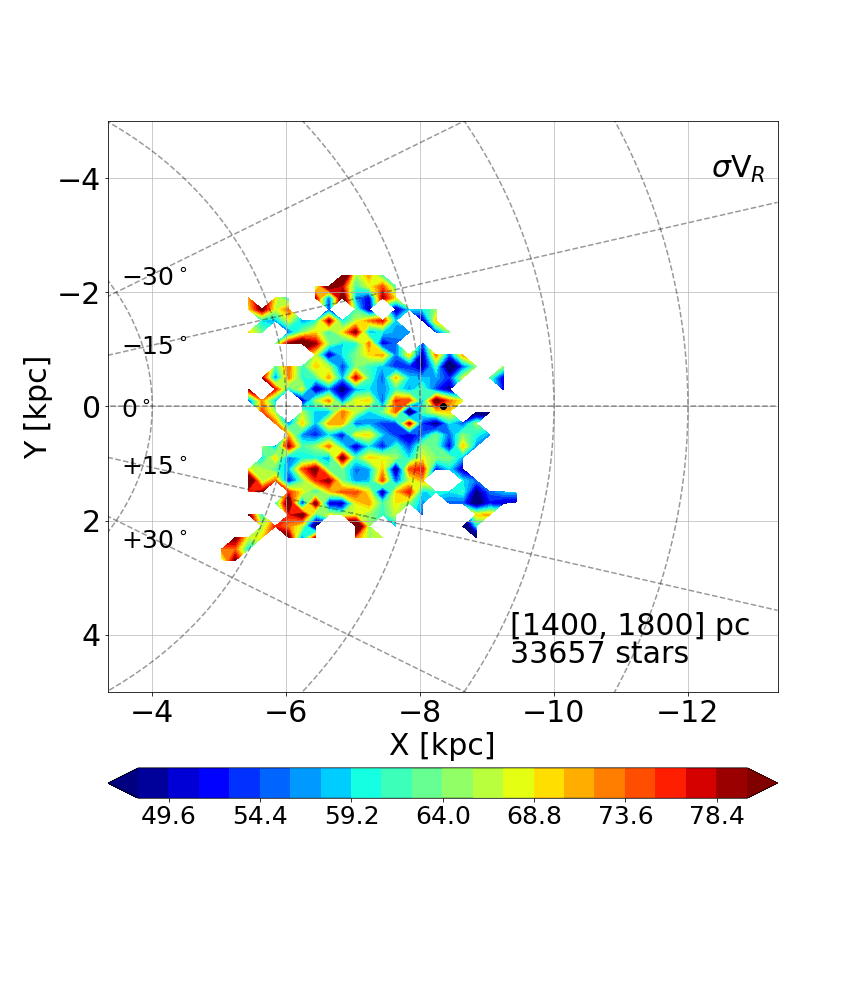}
\includegraphics[clip=true, trim = 10mm 40mm 20mm 40mm, width=0.3 \hsize]{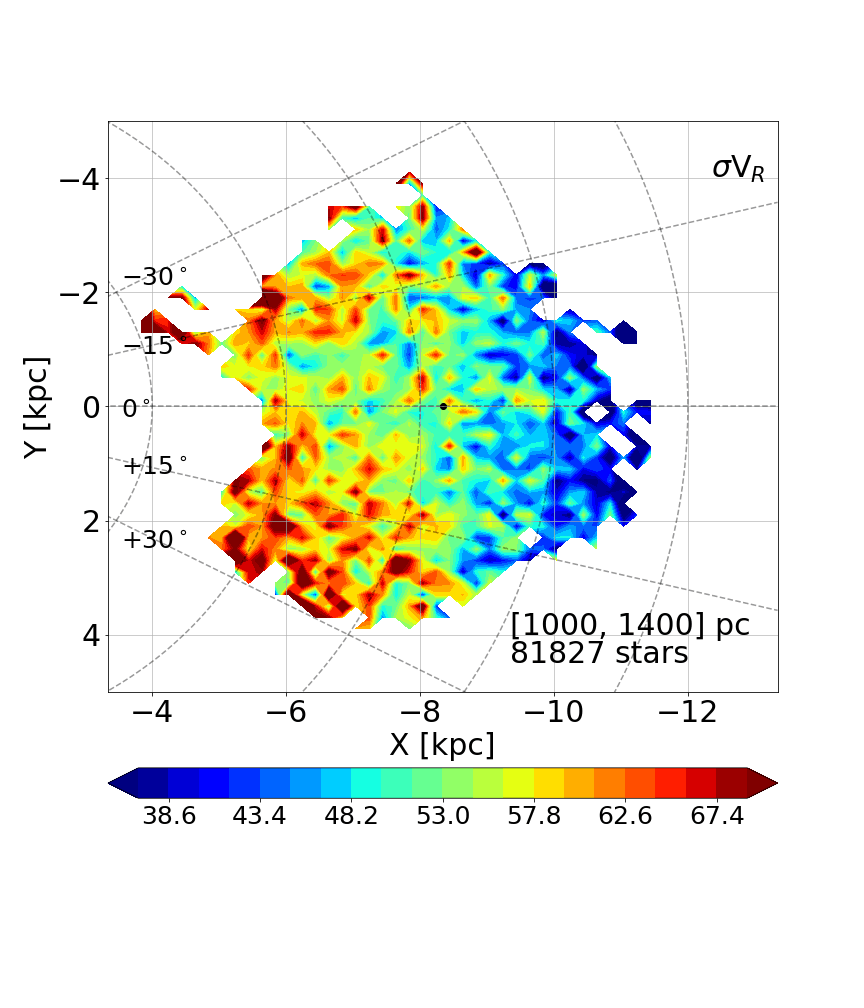}
\includegraphics[clip=true, trim = 10mm 40mm 20mm 40mm, width=0.3 \hsize]{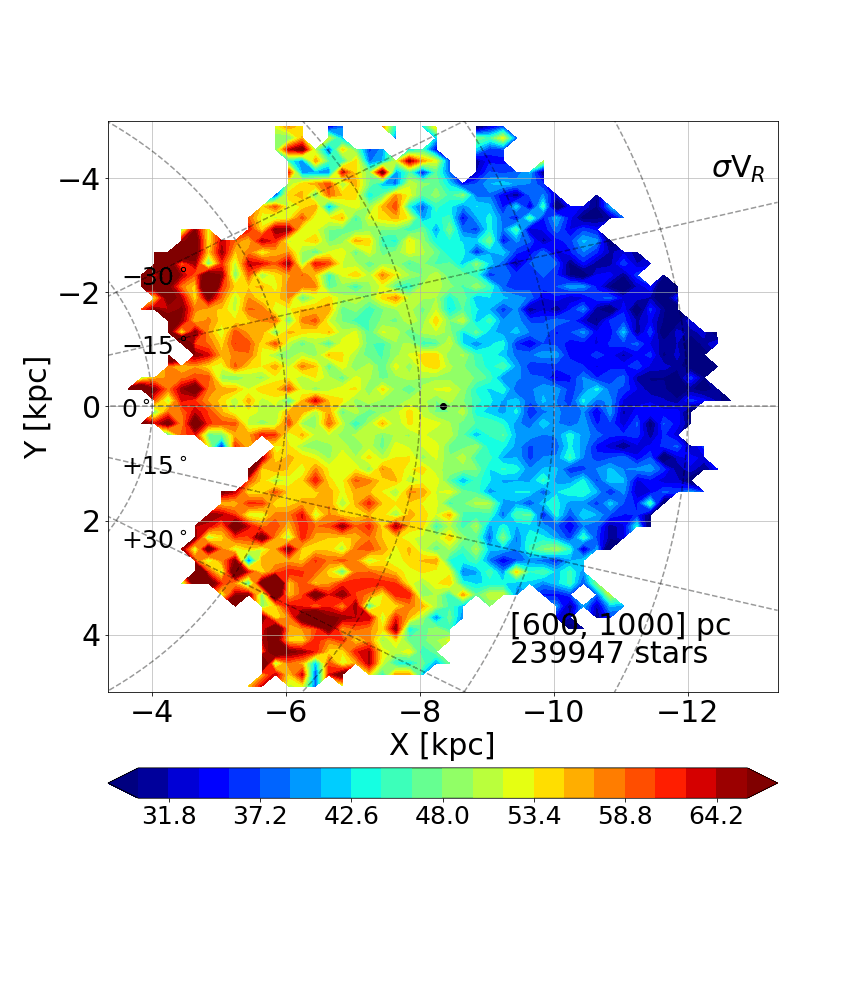}
\includegraphics[clip=true, trim = 10mm 40mm 20mm 40mm, width=0.3 \hsize]{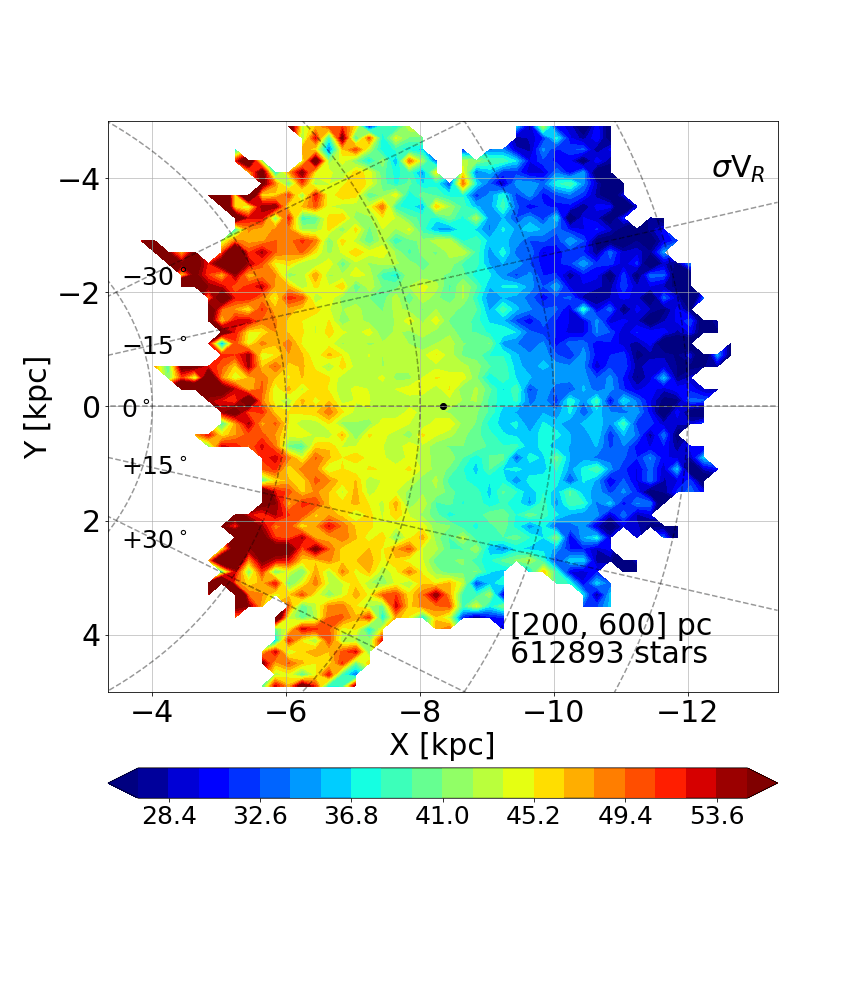}
\includegraphics[clip=true, trim = 10mm 40mm 20mm 40mm, width=0.3 \hsize]{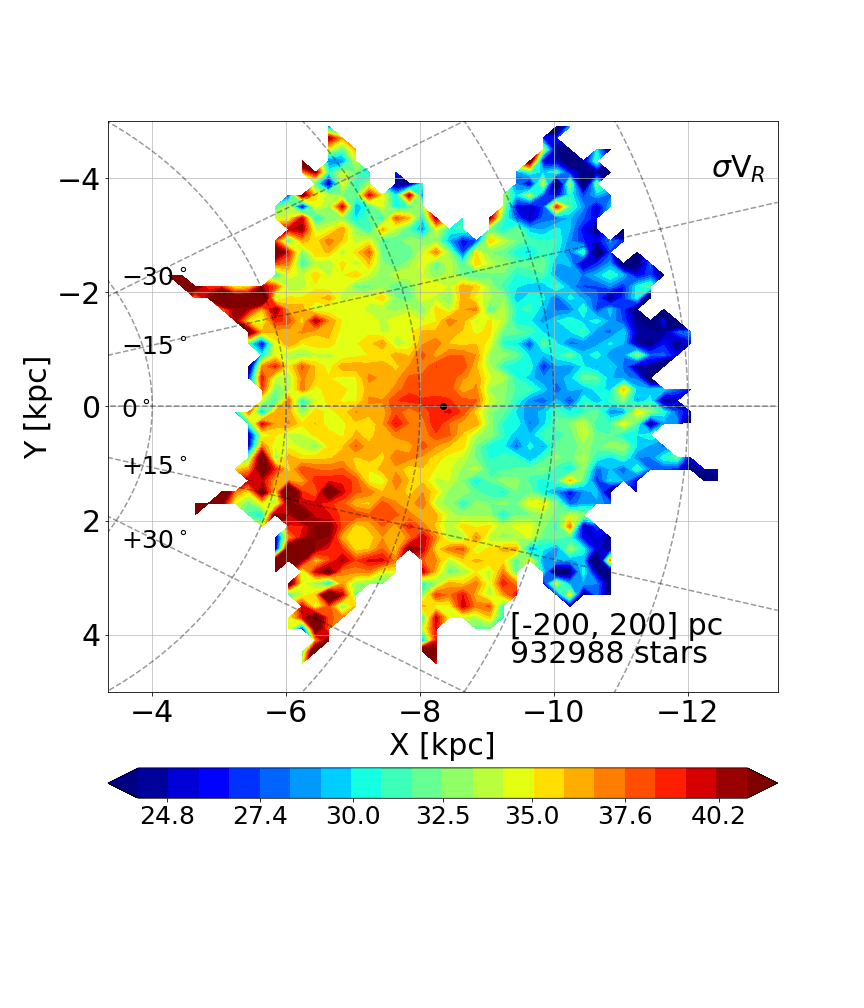}
\includegraphics[clip=true, trim = 10mm 40mm 20mm 40mm, width=0.3 \hsize]{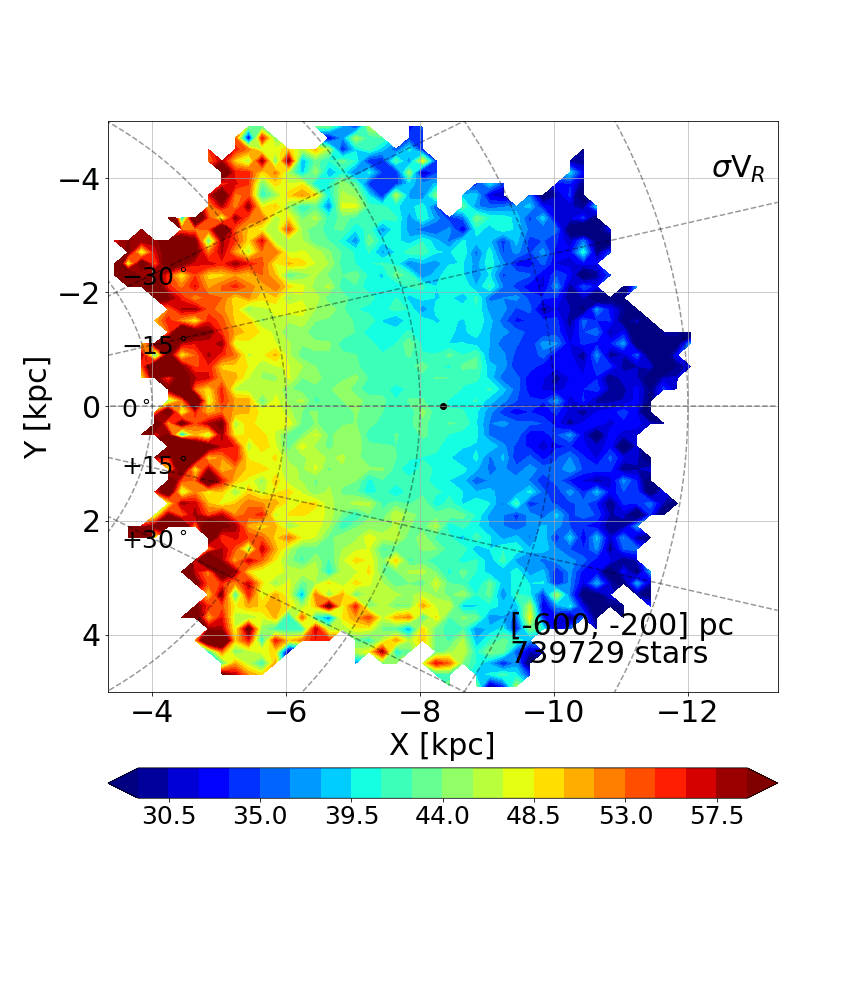}
\includegraphics[clip=true, trim = 10mm 40mm 20mm 40mm, width=0.3 \hsize]{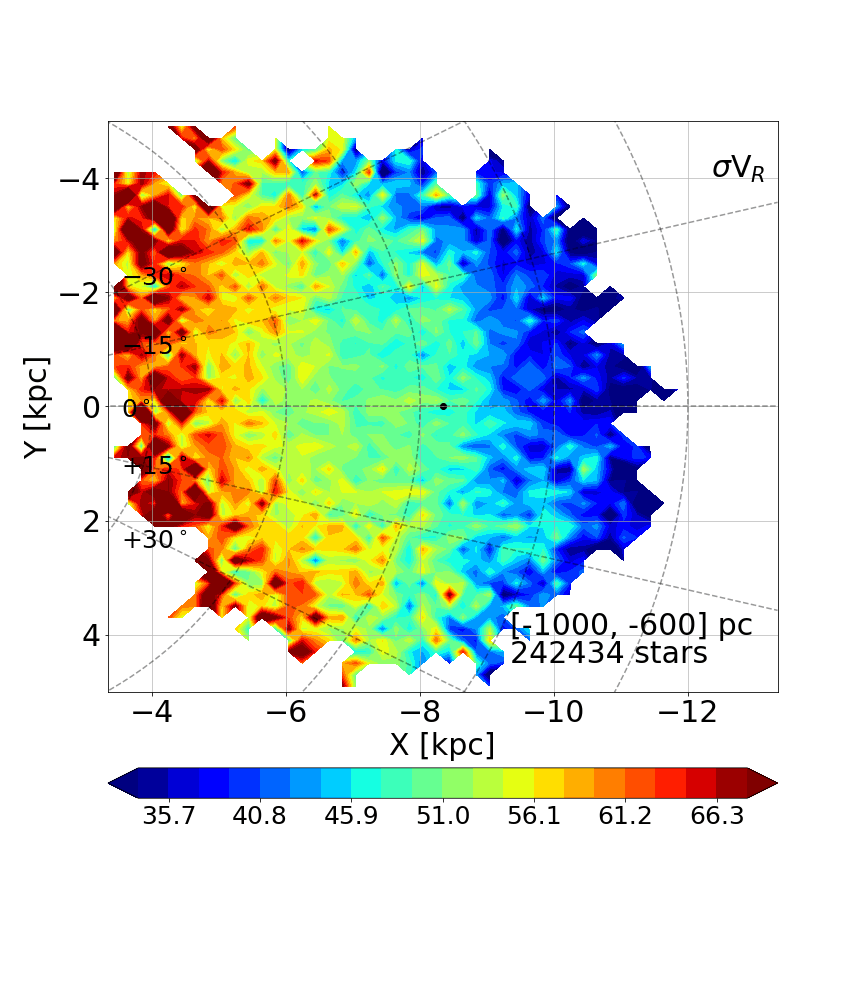}
\includegraphics[clip=true, trim = 10mm 40mm 20mm 40mm, width=0.3 \hsize]{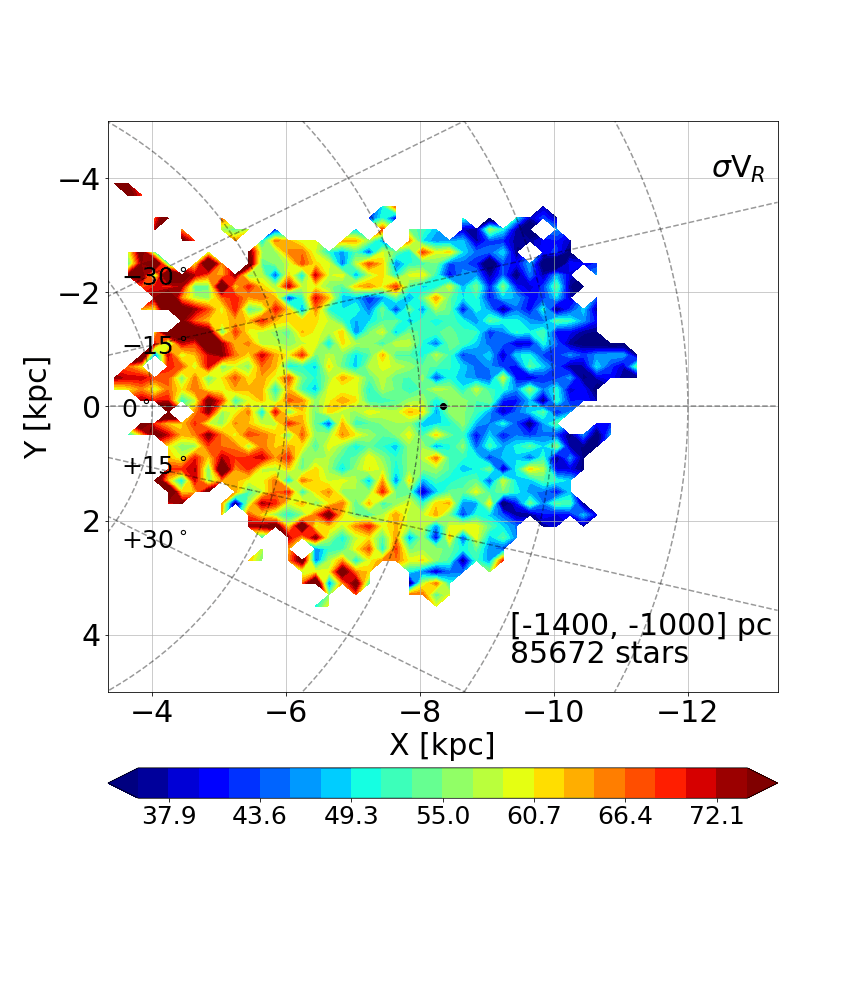}
\includegraphics[clip=true, trim = 10mm 40mm 20mm 40mm, width=0.3 \hsize]{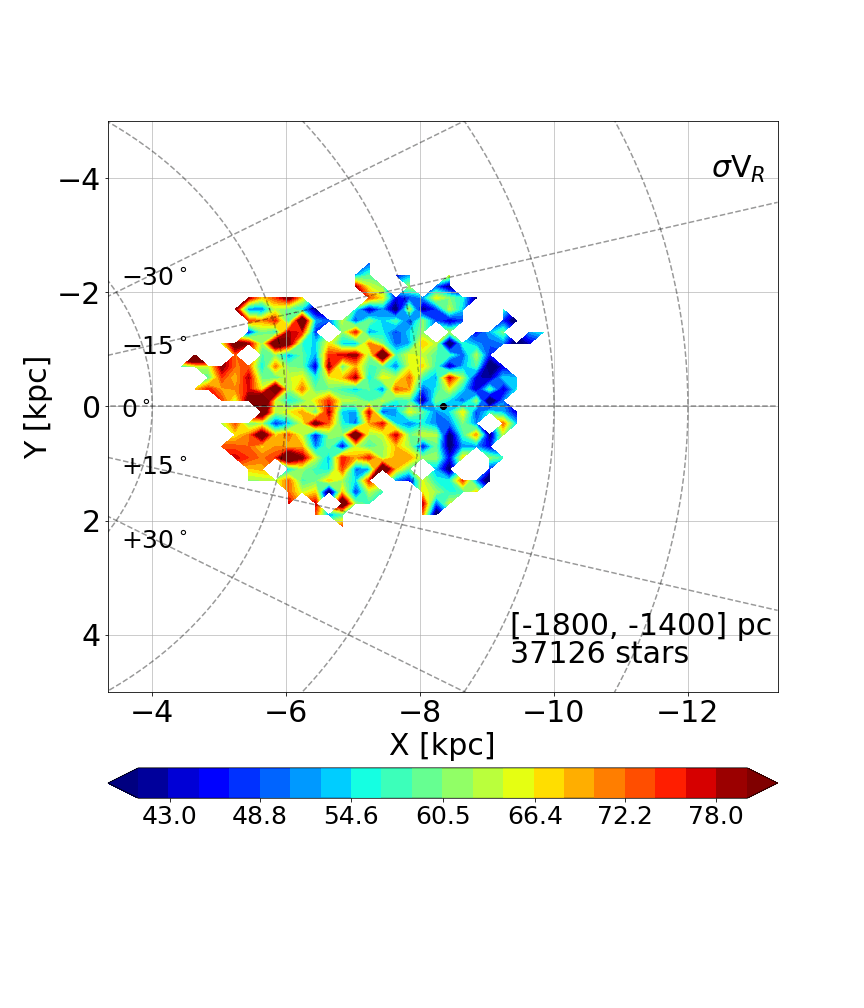}
\caption{Same as Fig.~\ref{fig:xyvrmed} for the radial velocity dispersion, $\sigma_{V_R}$.}
\label{fig:xyvrdisp}
\end{figure*}

\begin{figure*}[]
\centering
\includegraphics[clip=true, trim = 5mm 40mm 10mm 40mm, width=0.45 \hsize]{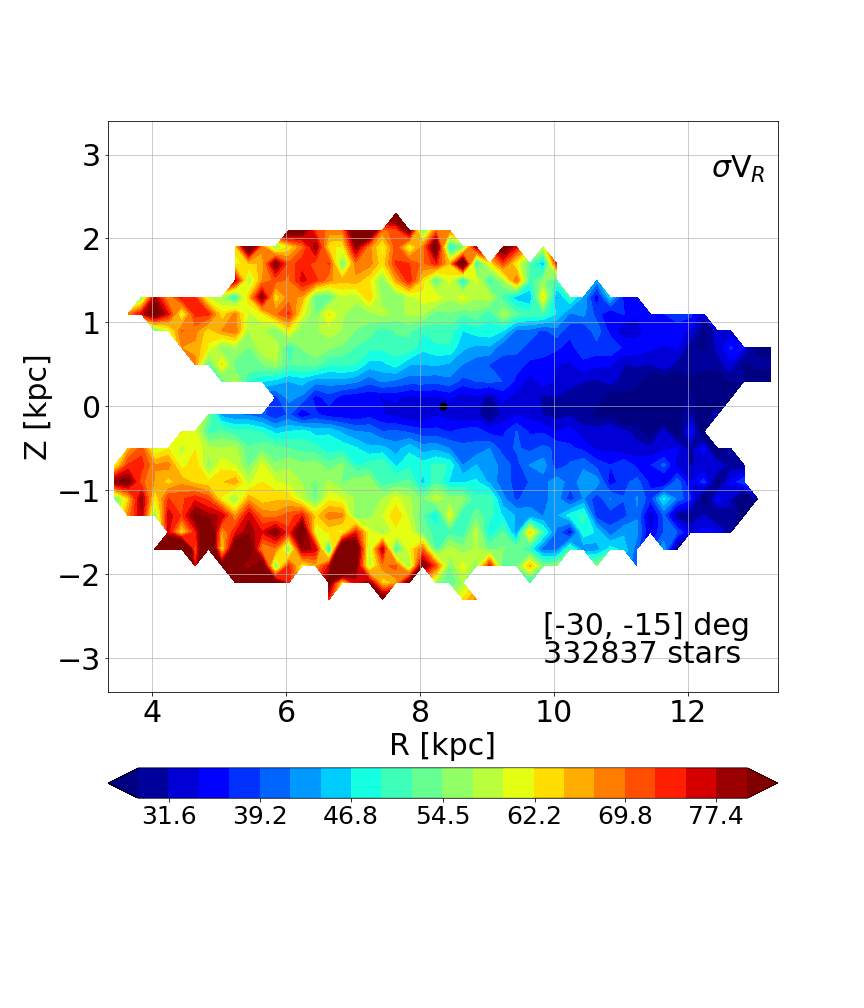}
\includegraphics[clip=true, trim = 5mm 40mm 10mm 40mm, width=0.45 \hsize]{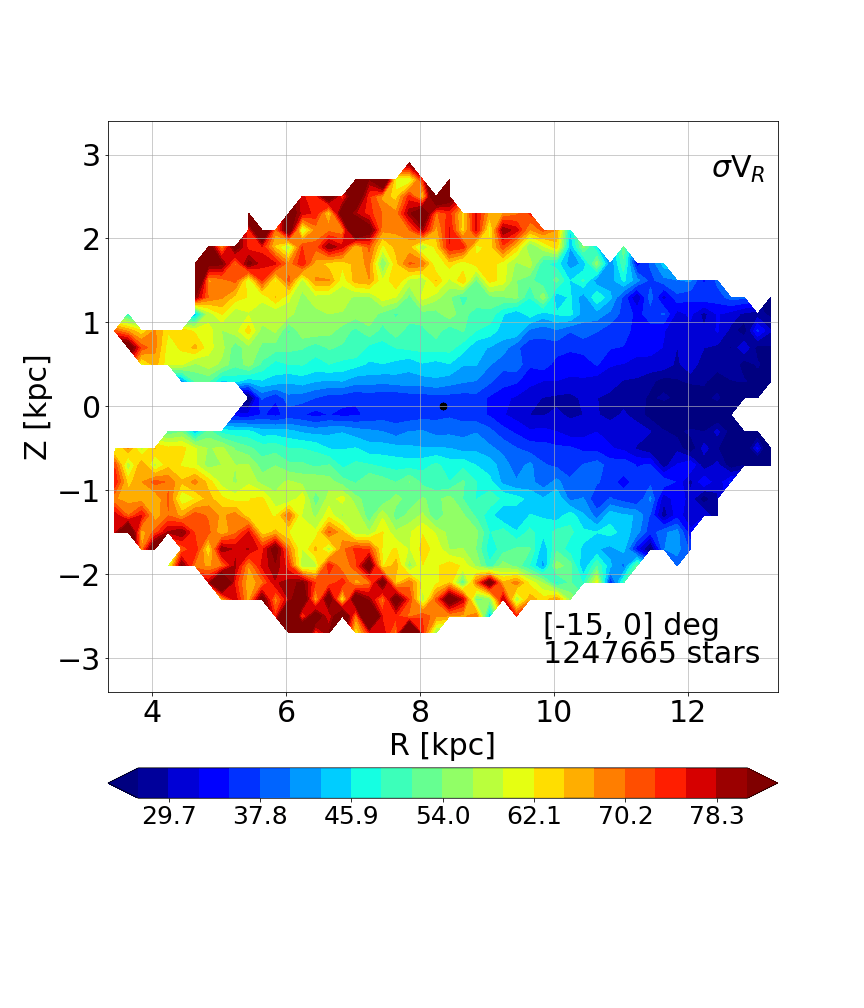}
\includegraphics[clip=true, trim = 5mm 40mm 10mm 40mm, width=0.45 \hsize]{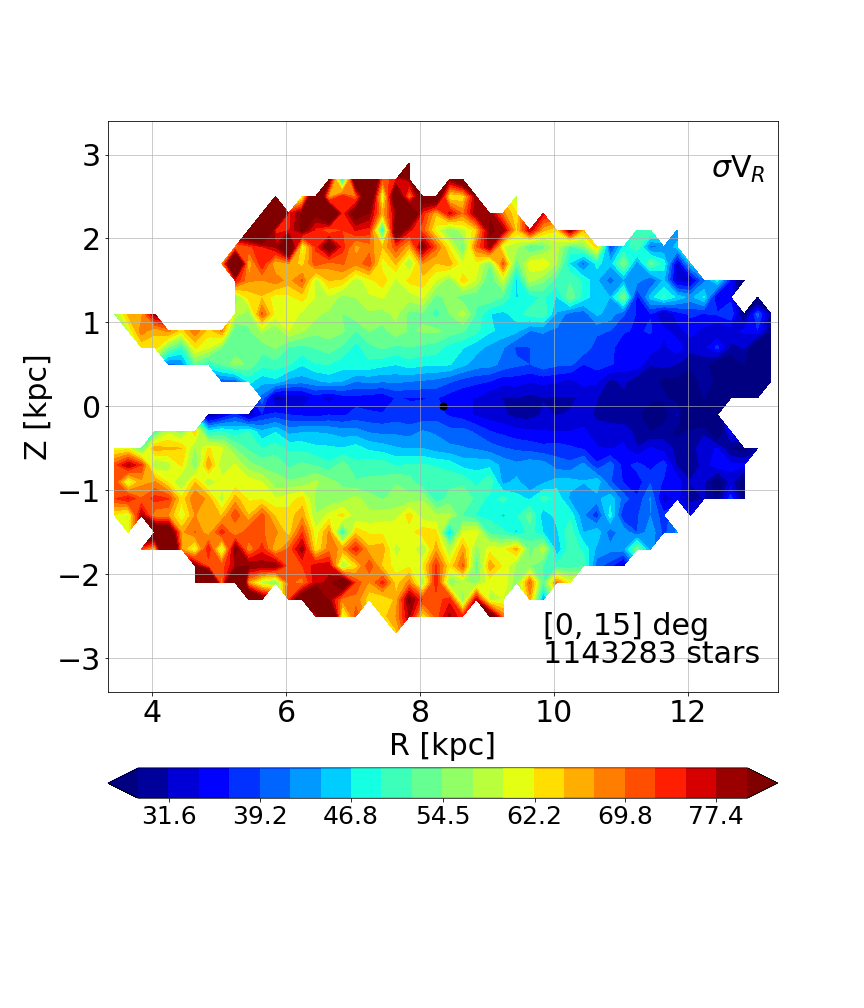}
\includegraphics[clip=true, trim = 5mm 40mm 10mm 40mm, width=0.45 \hsize]{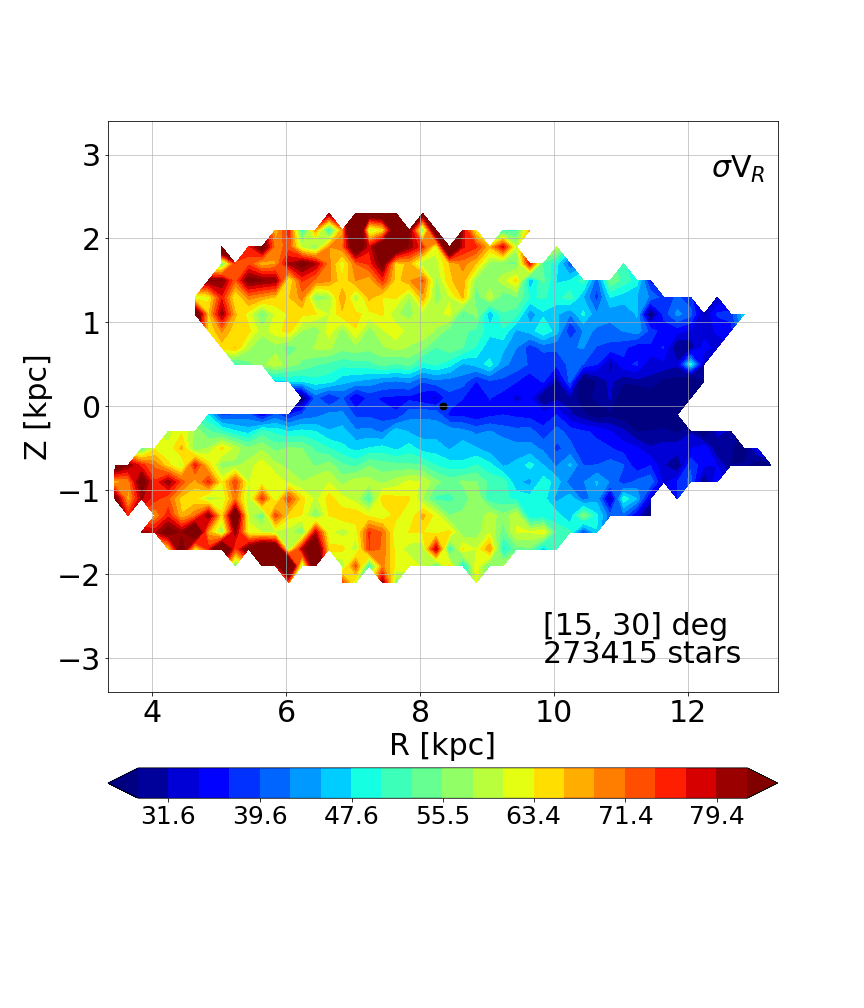}
\caption{Same as Fig.~\ref{fig:rzvrmed} for the radial velocity dispersion, $\sigma_{V_R}$.}
\label{fig:rzvrdisp}
\end{figure*}

\begin{figure*}[]
\centering
\includegraphics[clip=true, trim = 10mm 40mm 20mm 40mm, width=0.3 \hsize]{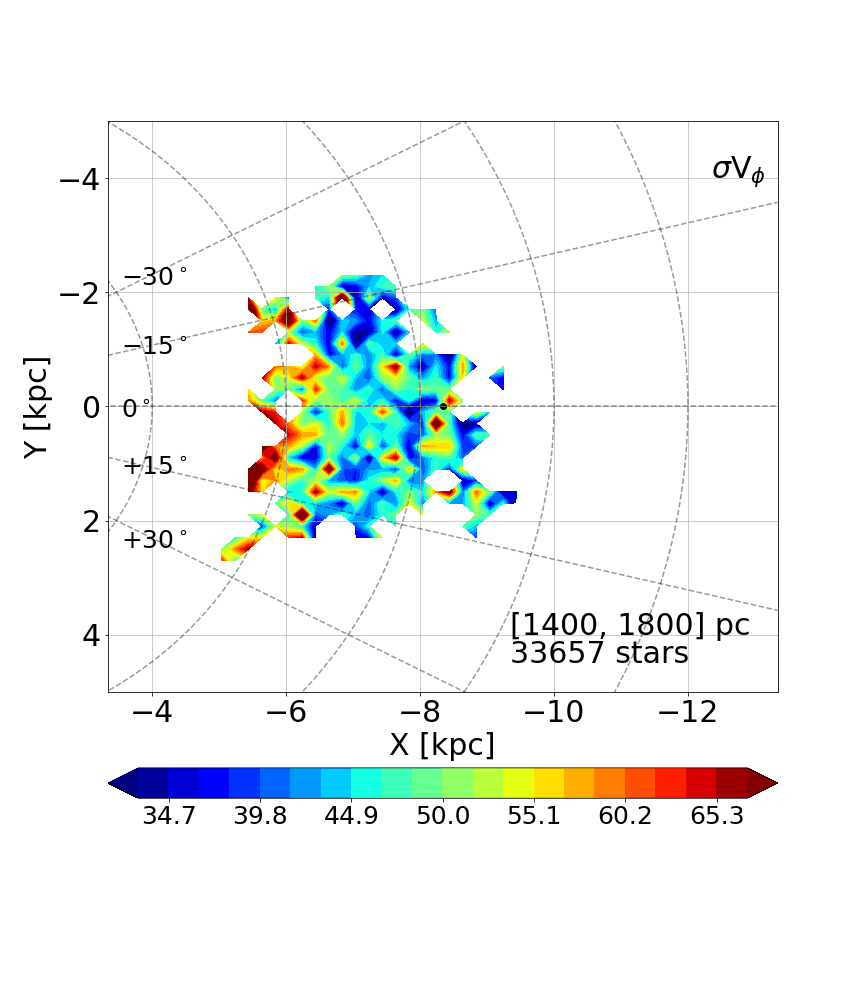}
\includegraphics[clip=true, trim = 10mm 40mm 20mm 40mm, width=0.3 \hsize]{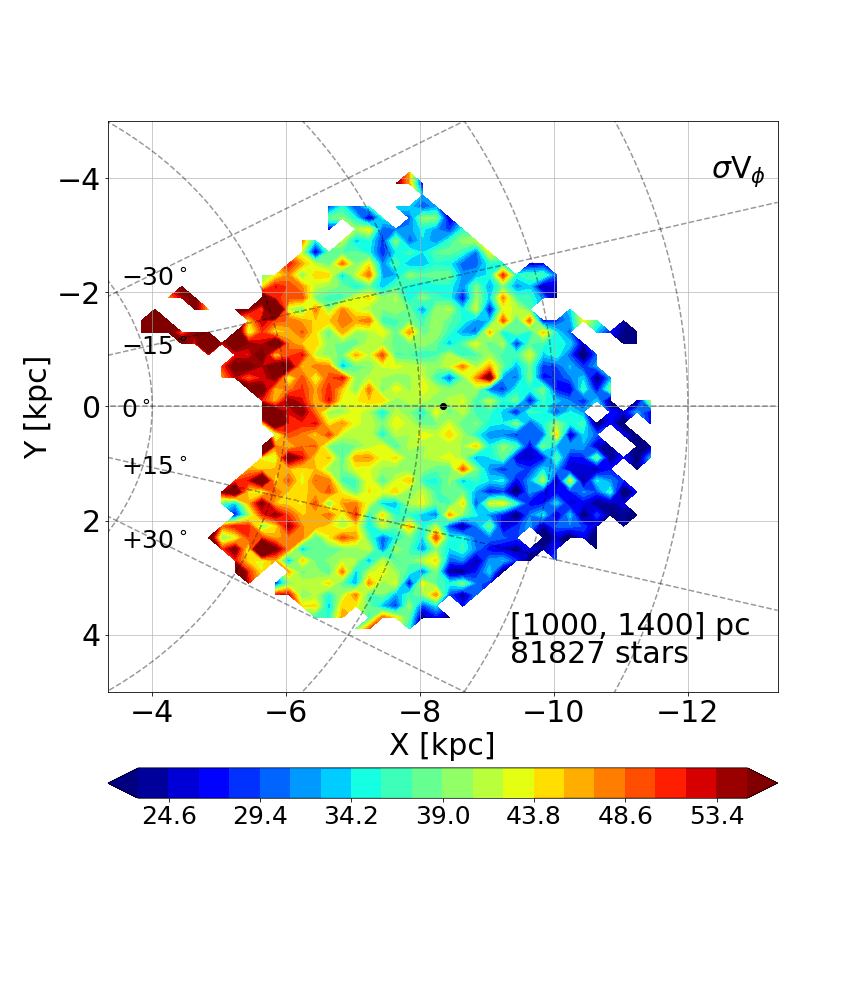}
\includegraphics[clip=true, trim = 10mm 40mm 20mm 40mm, width=0.3 \hsize]{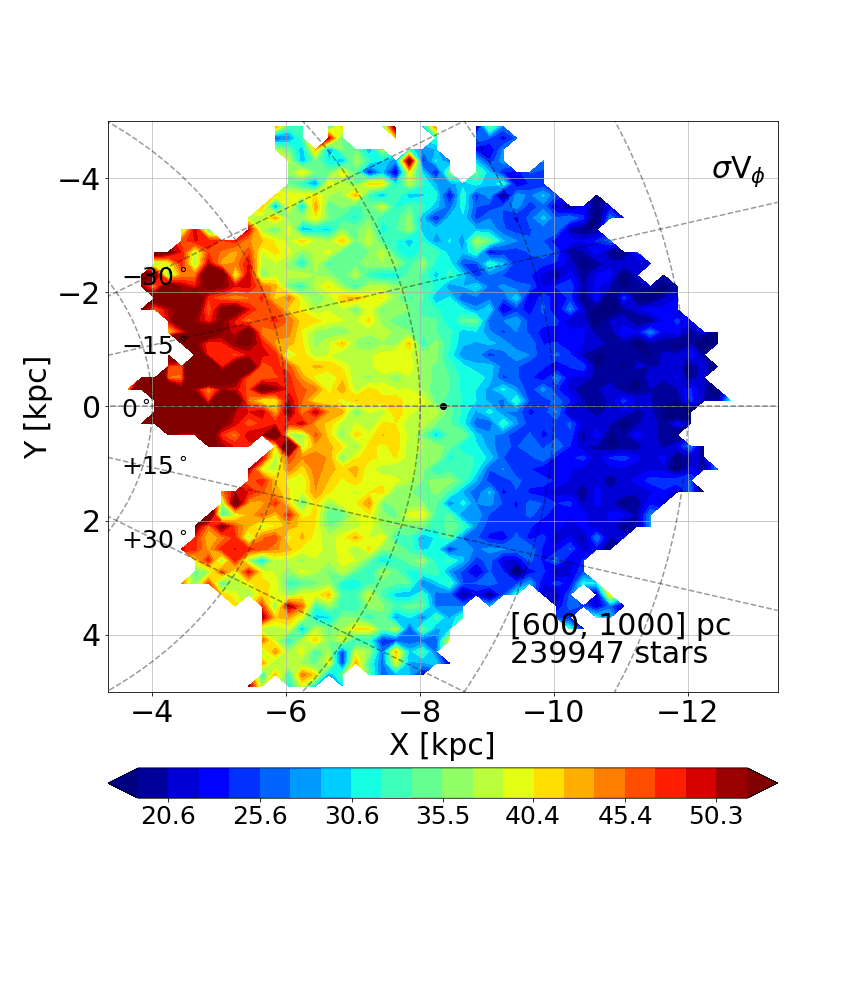}
\includegraphics[clip=true, trim = 10mm 40mm 20mm 40mm, width=0.3 \hsize]{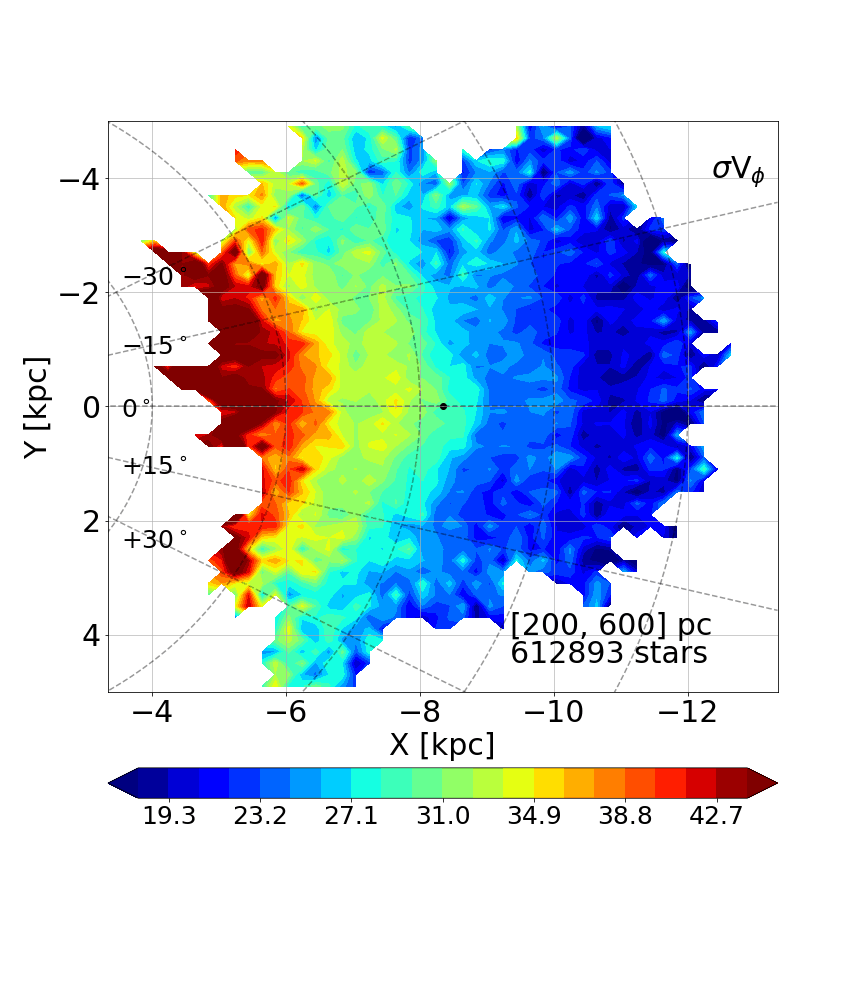}
\includegraphics[clip=true, trim = 10mm 40mm 20mm 40mm, width=0.3 \hsize]{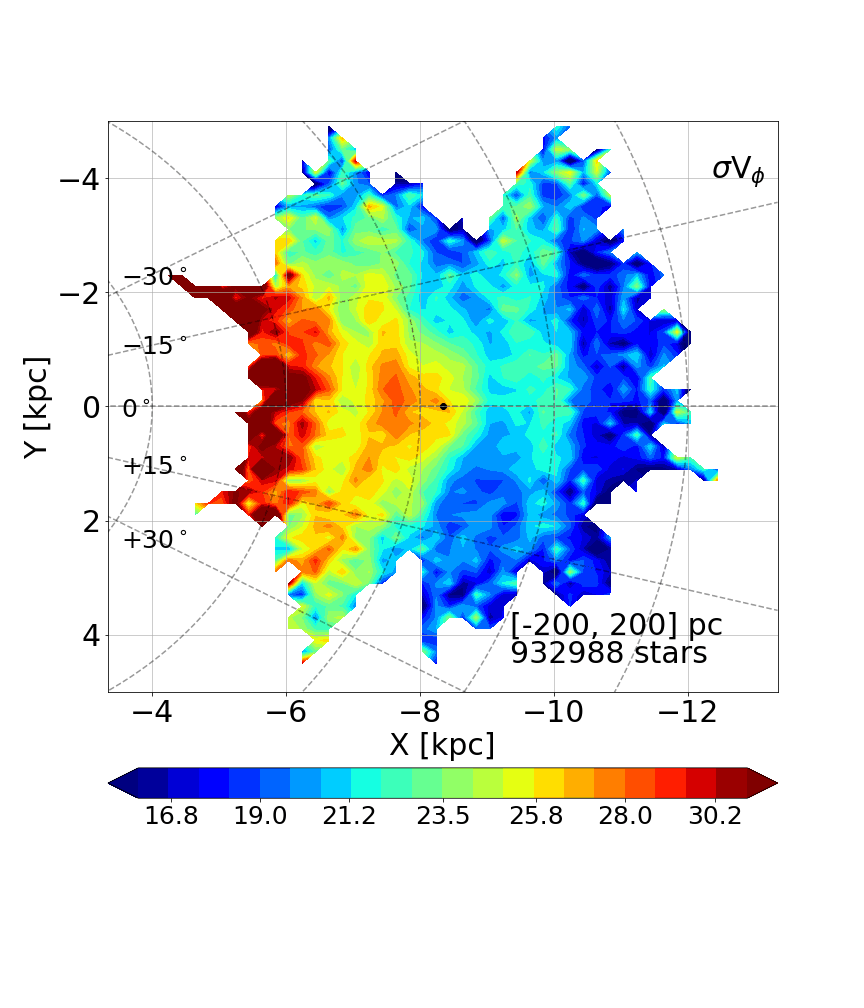}
\includegraphics[clip=true, trim = 10mm 40mm 20mm 40mm, width=0.3 \hsize]{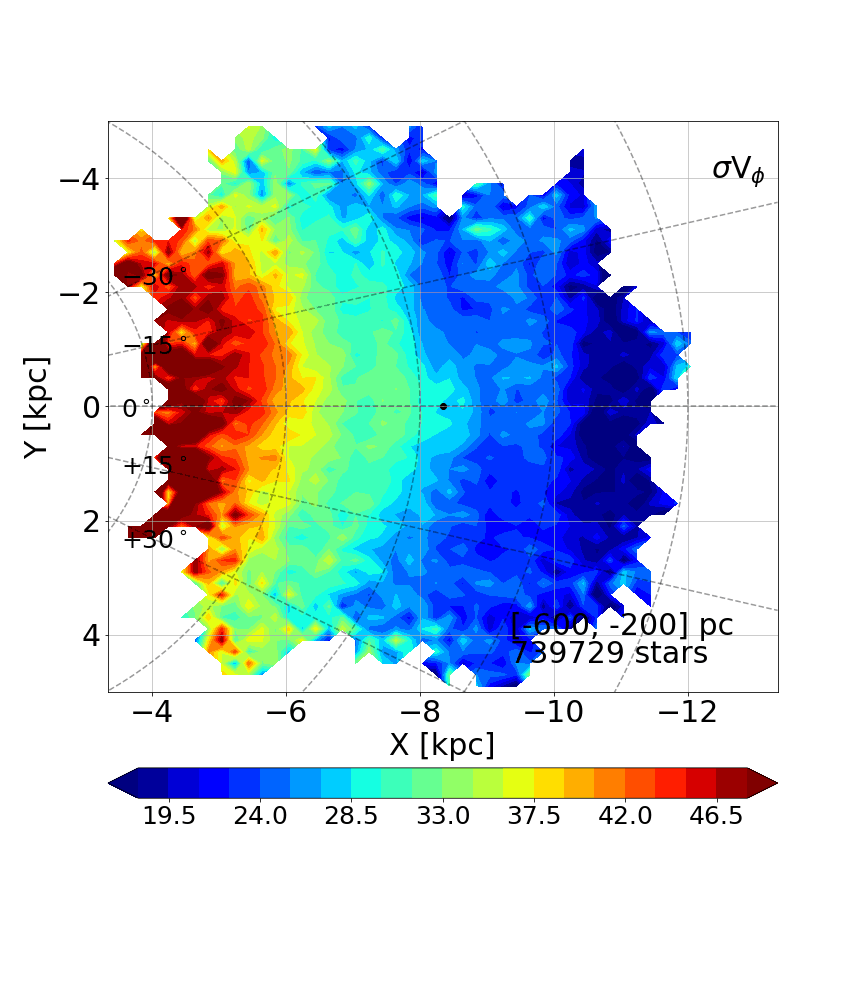}
\includegraphics[clip=true, trim = 10mm 40mm 20mm 40mm, width=0.3 \hsize]{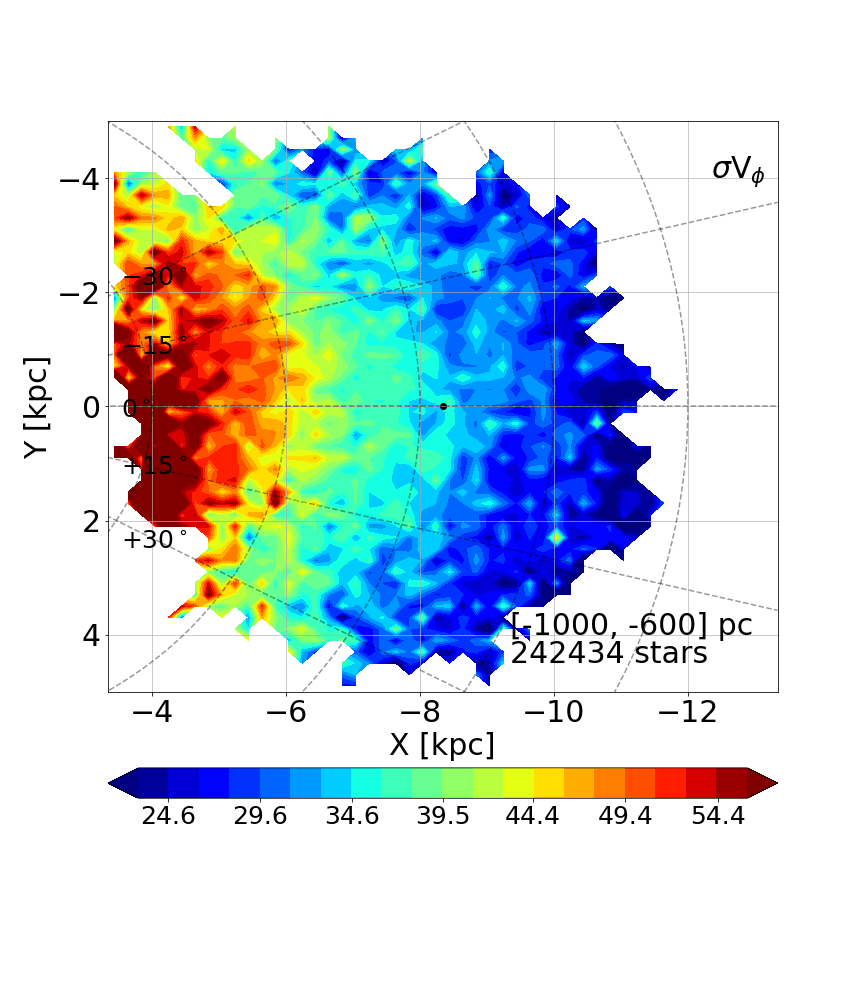}
\includegraphics[clip=true, trim = 10mm 40mm 20mm 40mm, width=0.3 \hsize]{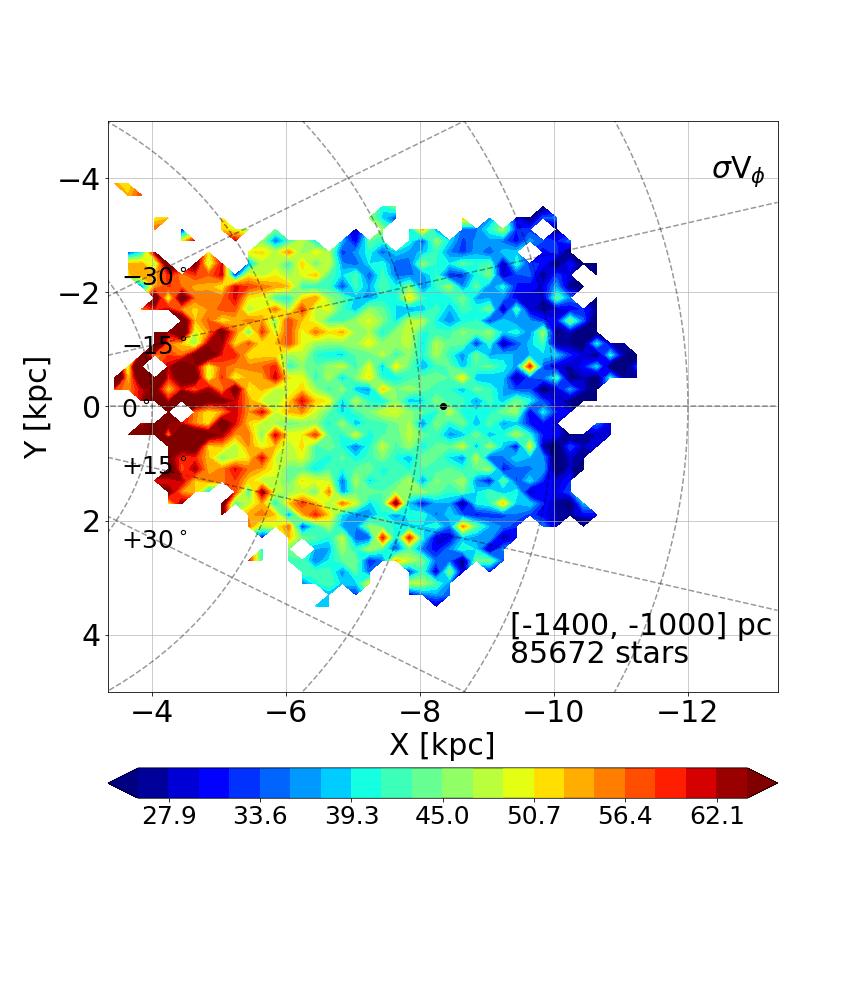}
\includegraphics[clip=true, trim = 10mm 40mm 20mm 40mm, width=0.3 \hsize]{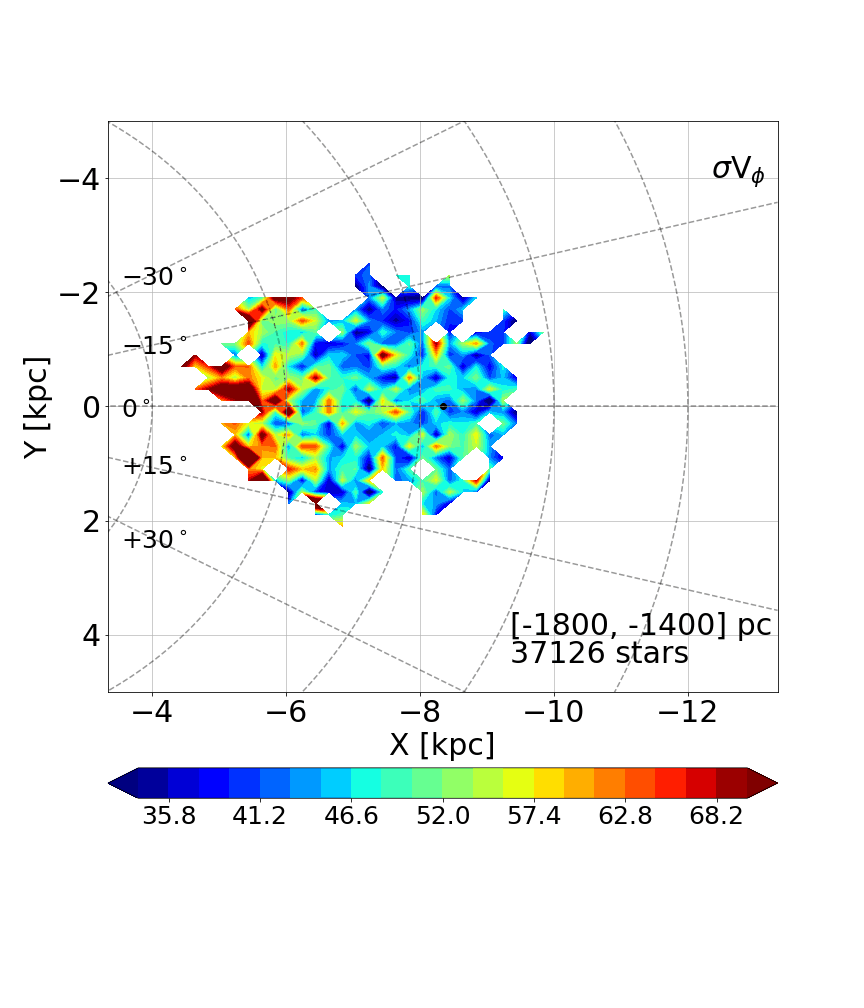}
\caption{Same as Fig.~\ref{fig:xyvrmed} for the azimuthal velocity dispersion, $\sigma_{V_\phi}$.}
\label{fig:xyvphidisp}
\end{figure*}

\begin{figure*}[]
\centering
\includegraphics[clip=true, trim = 5mm 40mm 10mm 40mm, width=0.45 \hsize]{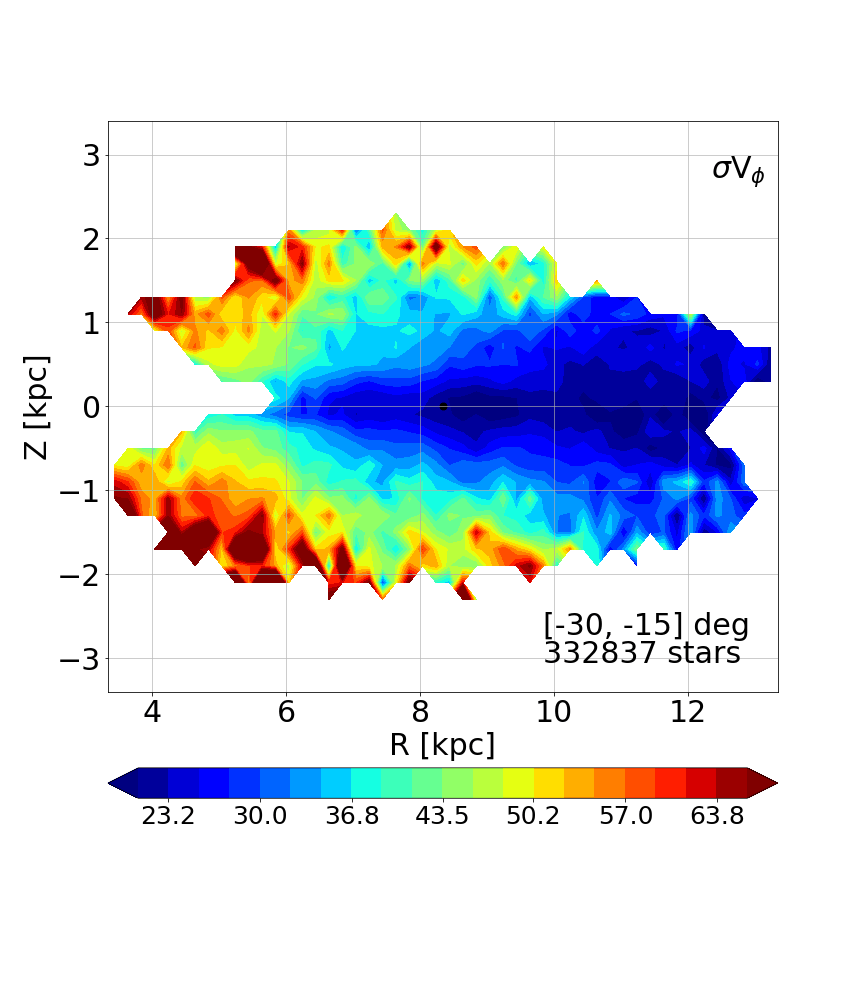}
\includegraphics[clip=true, trim = 5mm 40mm 10mm 40mm, width=0.45 \hsize]{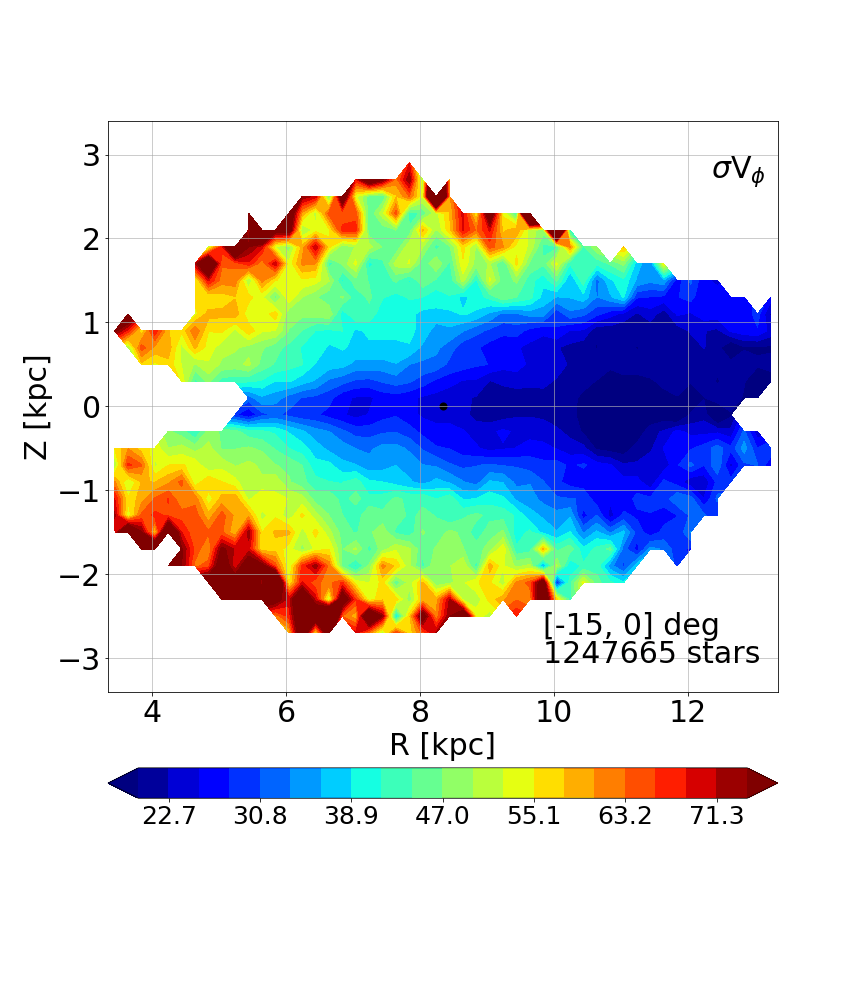}
\includegraphics[clip=true, trim = 5mm 40mm 10mm 40mm, width=0.45 \hsize]{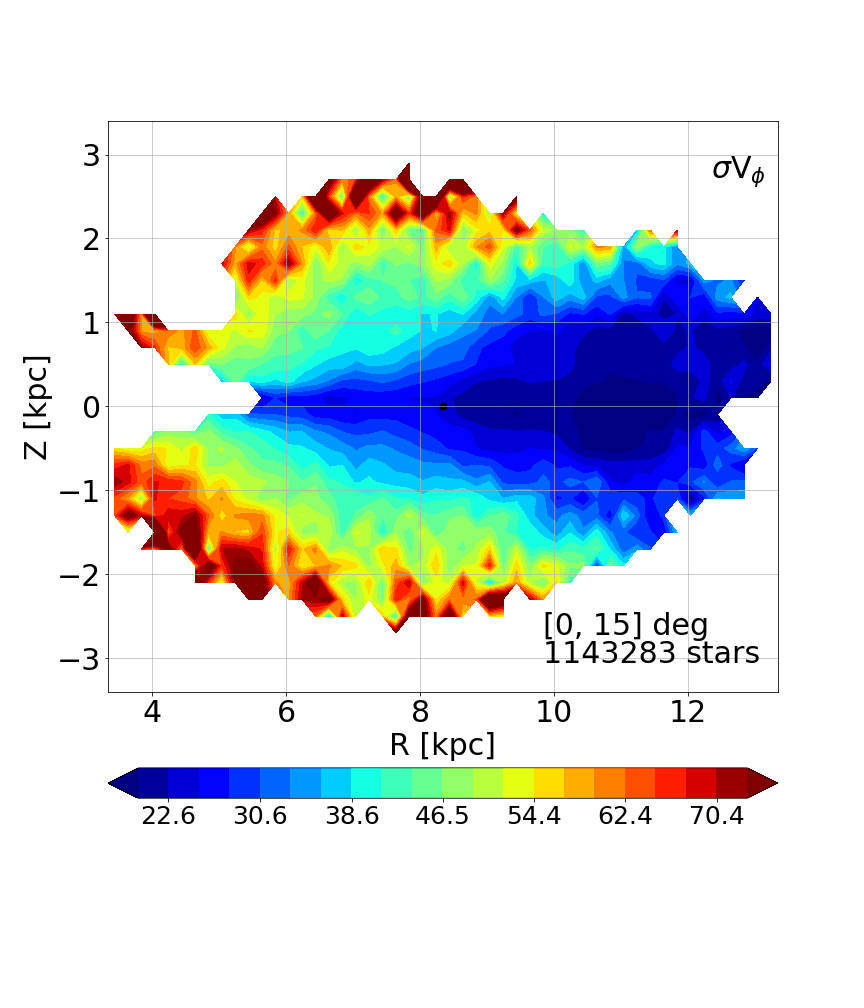}
\includegraphics[clip=true, trim = 5mm 40mm 10mm 40mm, width=0.45 \hsize]{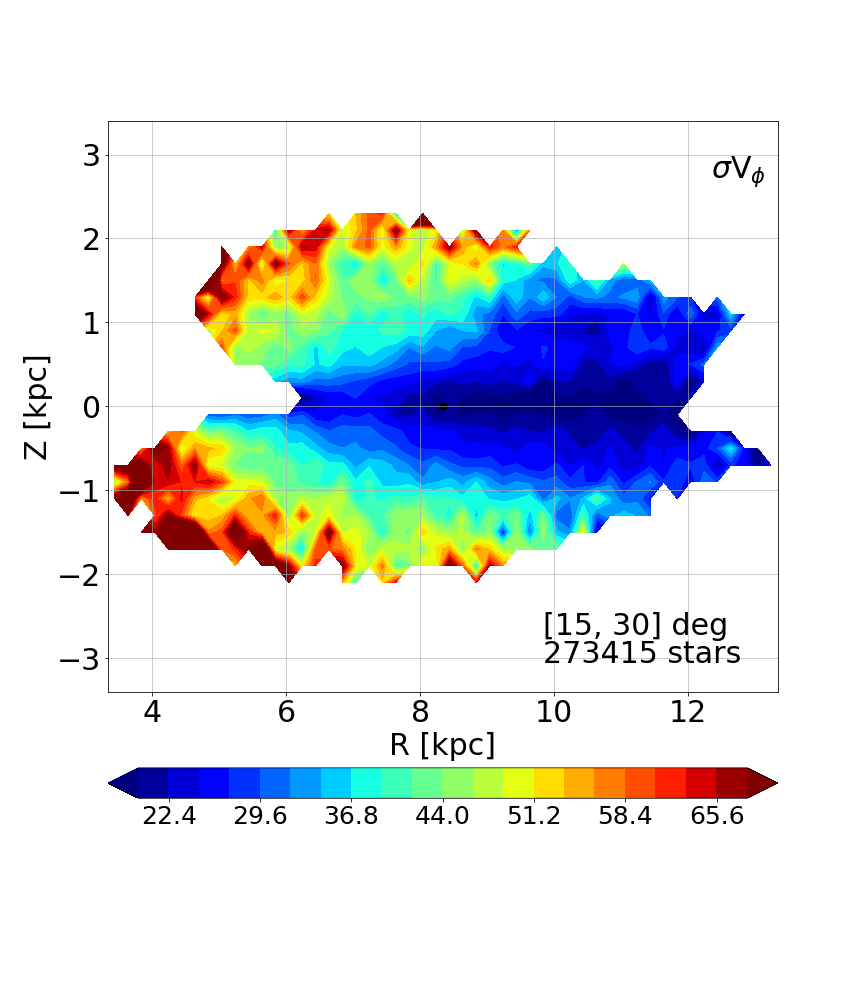}
\caption{Same as Fig.~\ref{fig:rzvrmed} for the azimuthal velocity dispersion, $\sigma_{V_\phi}$.}
\label{fig:rzvphidisp}
\end{figure*}

\begin{figure*}[]
\centering
\includegraphics[clip=true, trim = 10mm 40mm 20mm 40mm, width=0.3 \hsize]{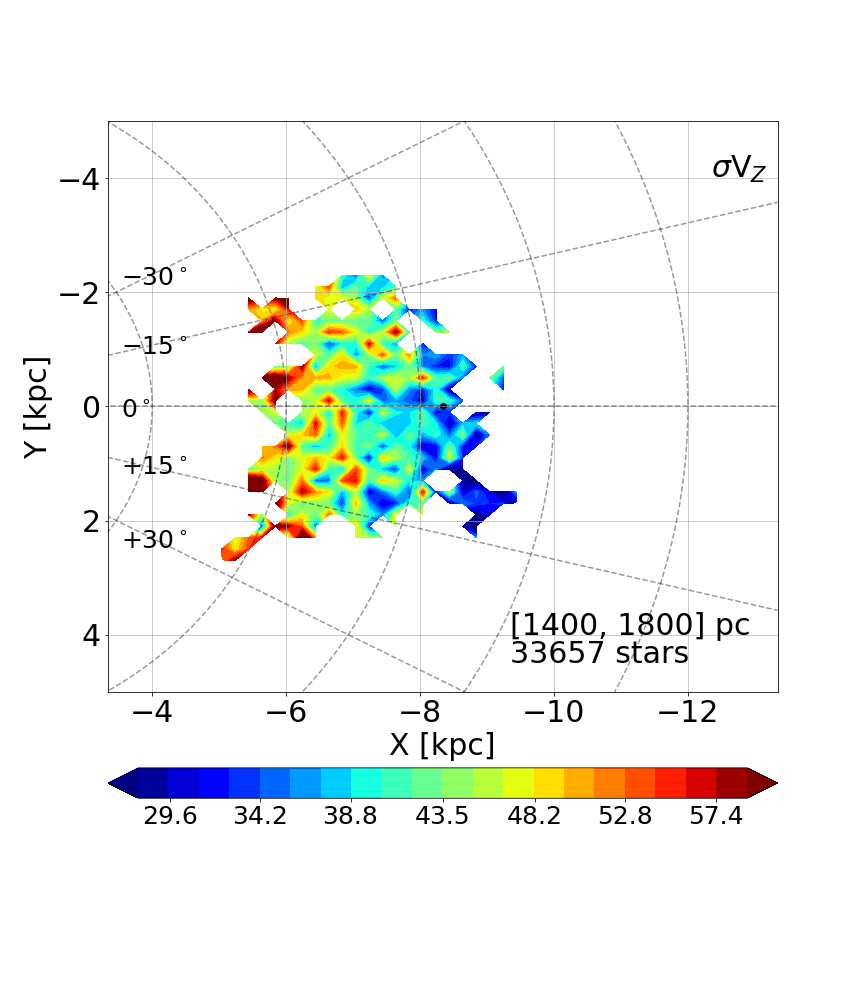}
\includegraphics[clip=true, trim = 10mm 40mm 20mm 40mm, width=0.3 \hsize]{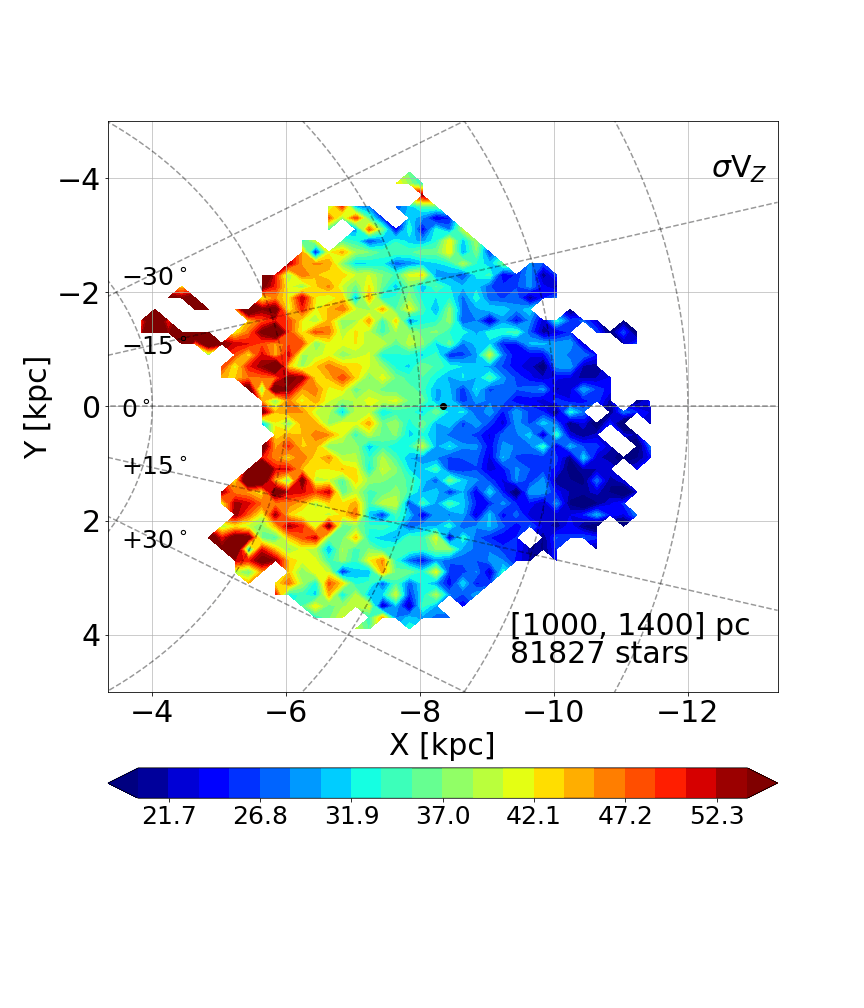}
\includegraphics[clip=true, trim = 10mm 40mm 20mm 40mm, width=0.3 \hsize]{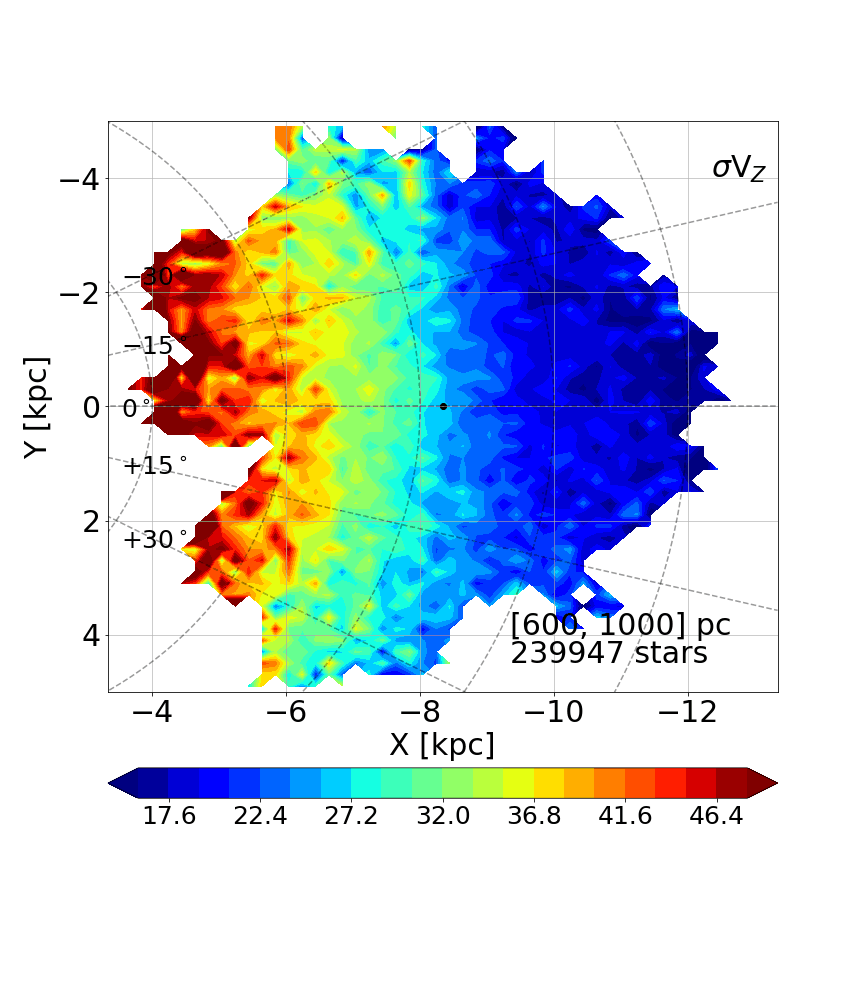}
\includegraphics[clip=true, trim = 10mm 40mm 20mm 40mm, width=0.3 \hsize]{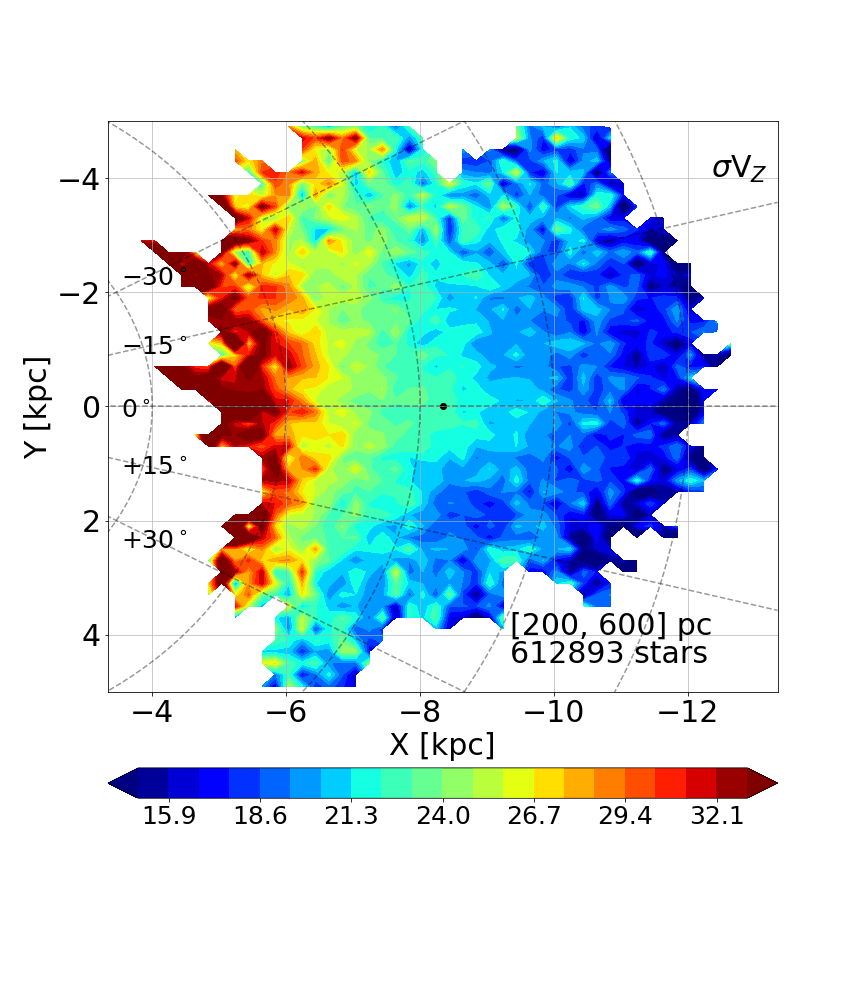}
\includegraphics[clip=true, trim = 10mm 40mm 20mm 40mm, width=0.3 \hsize]{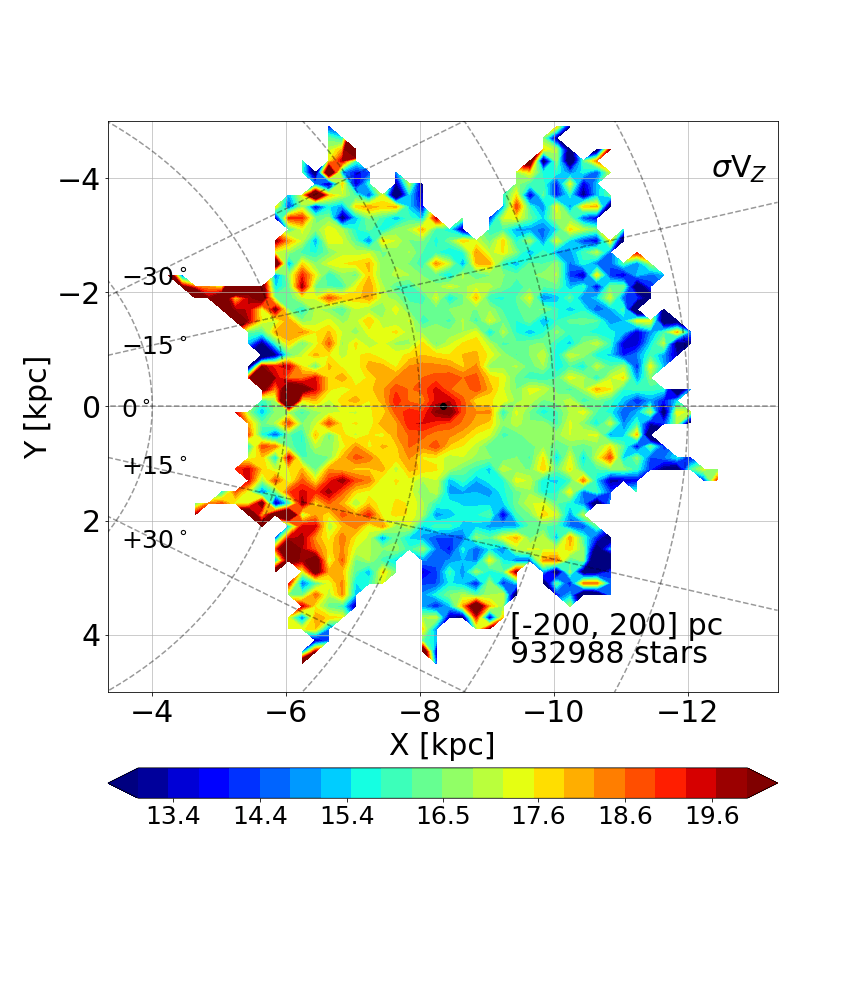}
\includegraphics[clip=true, trim = 10mm 40mm 20mm 40mm, width=0.3 \hsize]{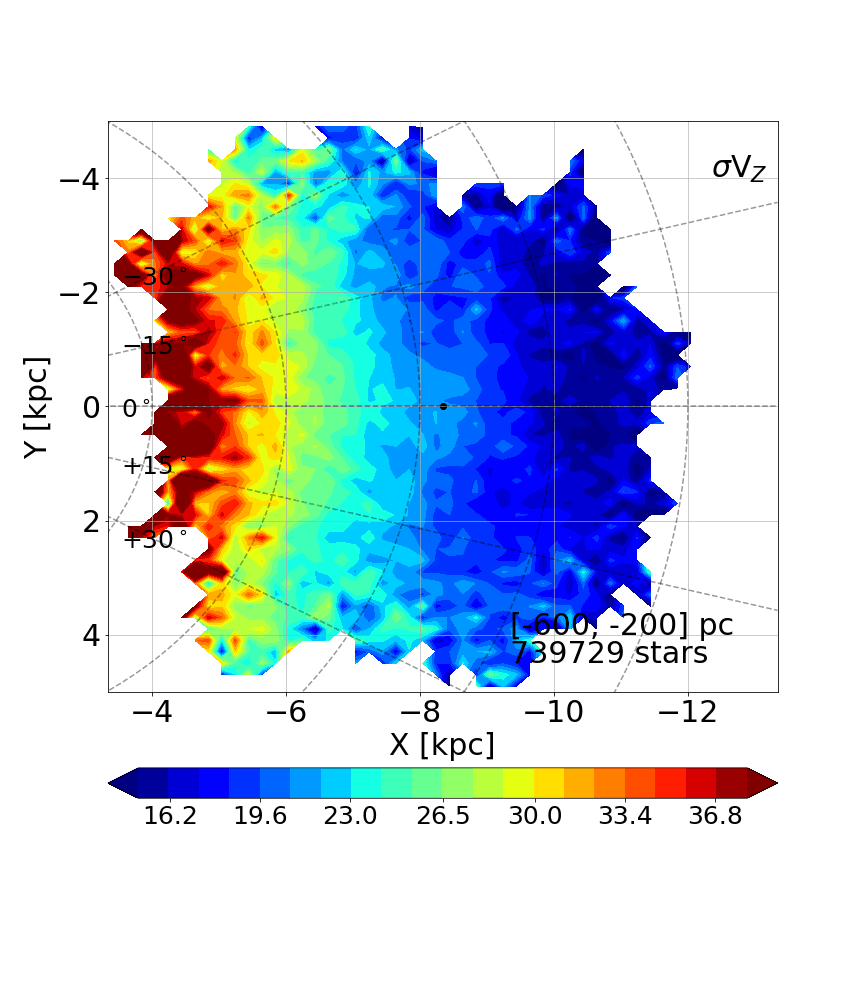}
\includegraphics[clip=true, trim = 10mm 40mm 20mm 40mm, width=0.3 \hsize]{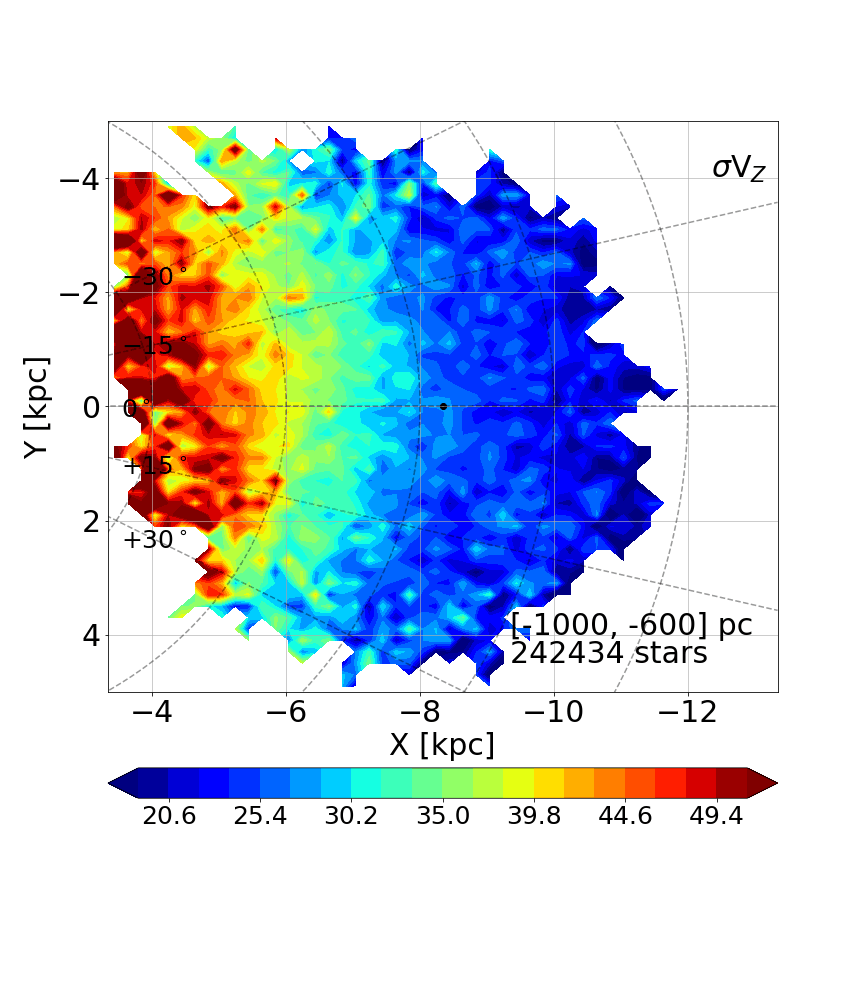}
\includegraphics[clip=true, trim = 10mm 40mm 20mm 40mm, width=0.3 \hsize]{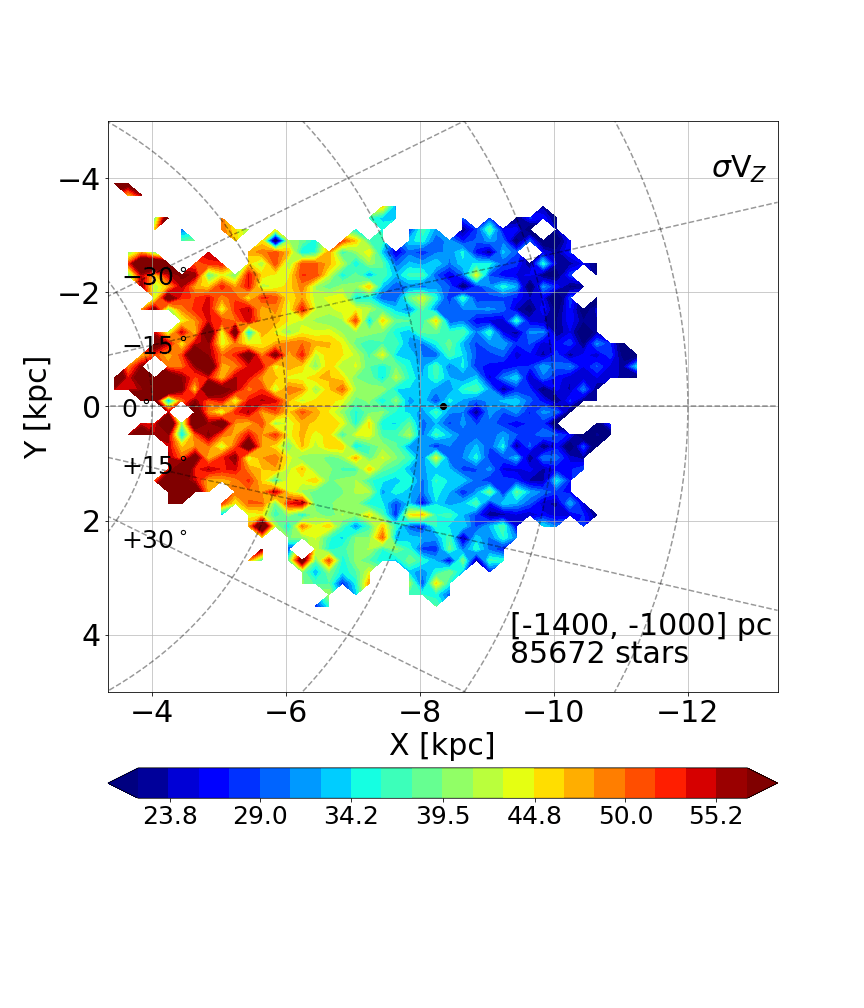}
\includegraphics[clip=true, trim = 10mm 40mm 20mm 40mm, width=0.3 \hsize]{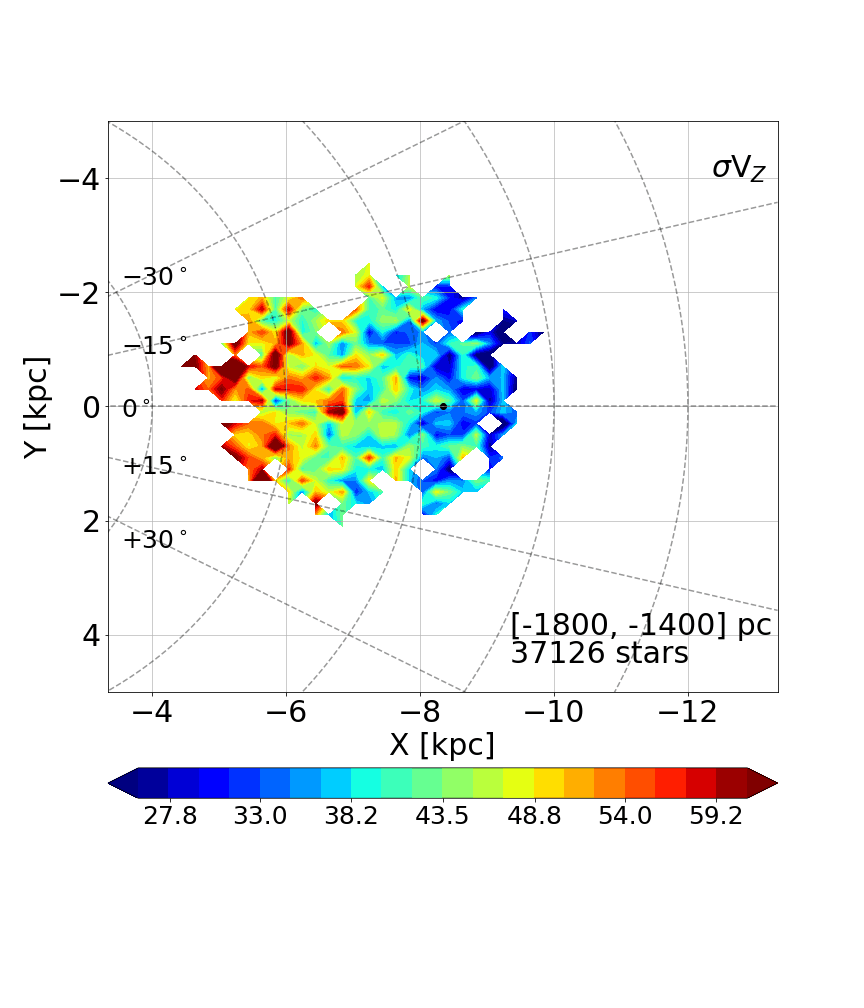}
\caption{Same as Fig.~\ref{fig:xyvrmed} for the vertical velocity dispersion, $\sigma_{V_Z}$.}
\label{fig:xyvzdisp}
\end{figure*}

\begin{figure*}[]
\centering
\includegraphics[clip=true, trim = 5mm 40mm 10mm 40mm, width=0.45 \hsize]{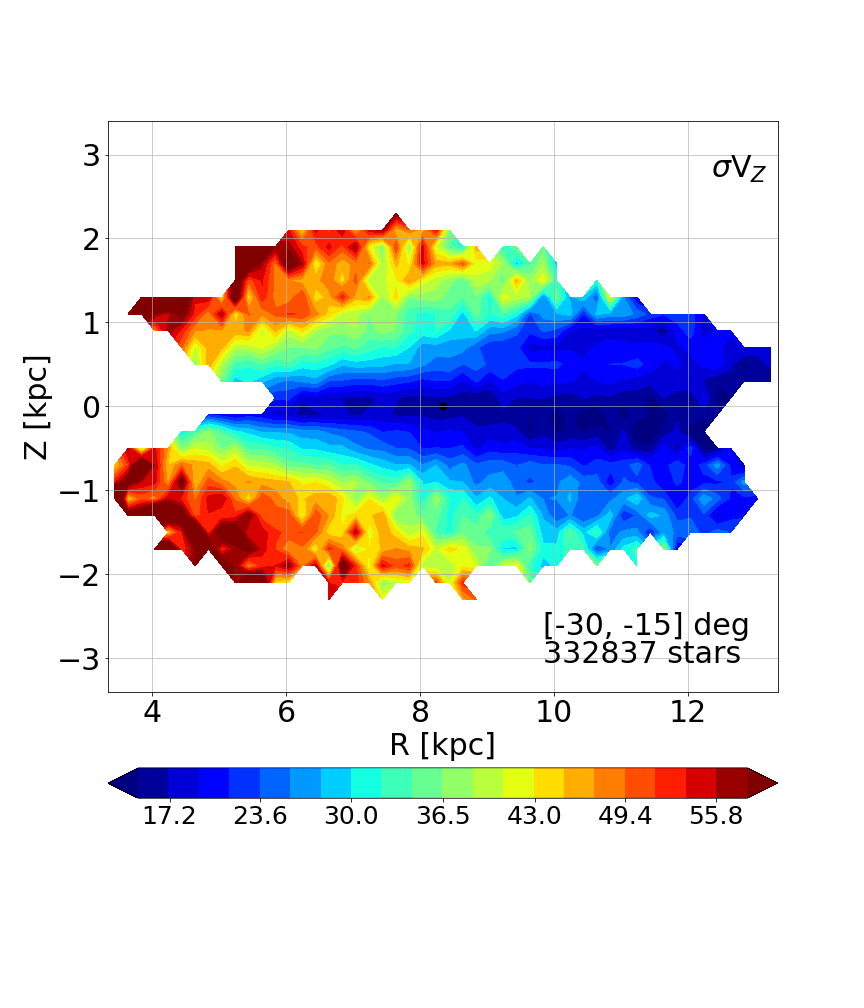}
\includegraphics[clip=true, trim = 5mm 40mm 10mm 40mm, width=0.45 \hsize]{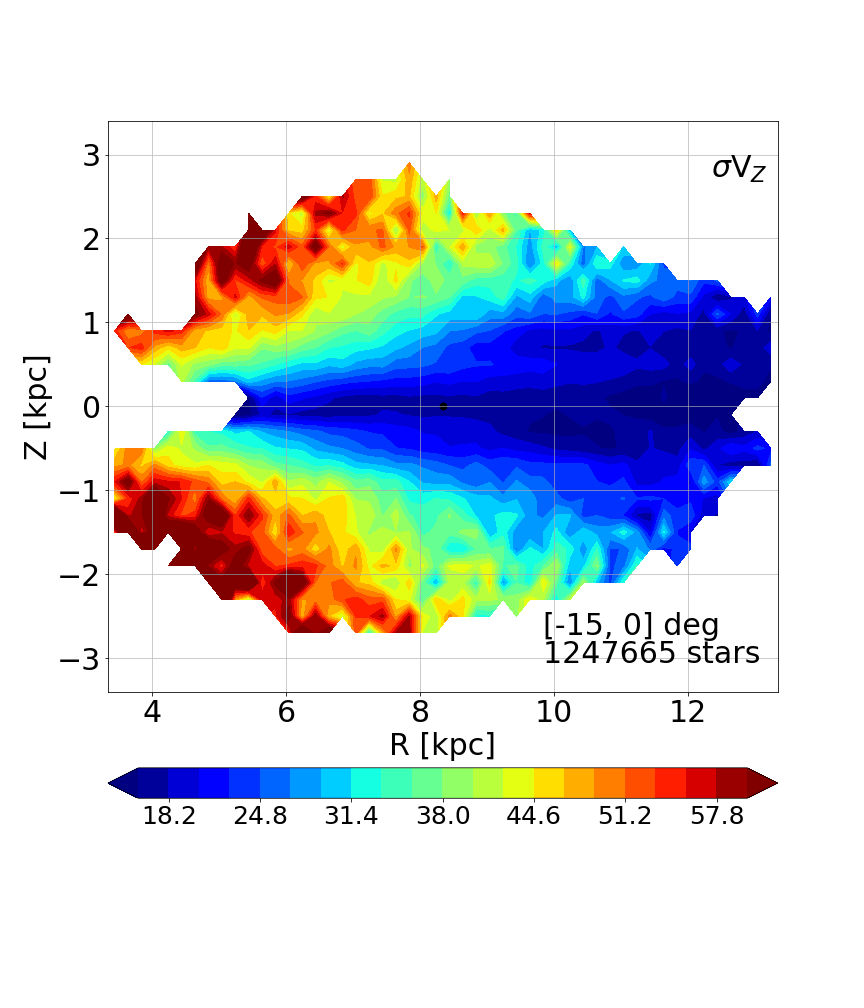}
\includegraphics[clip=true, trim = 5mm 40mm 10mm 40mm, width=0.45 \hsize]{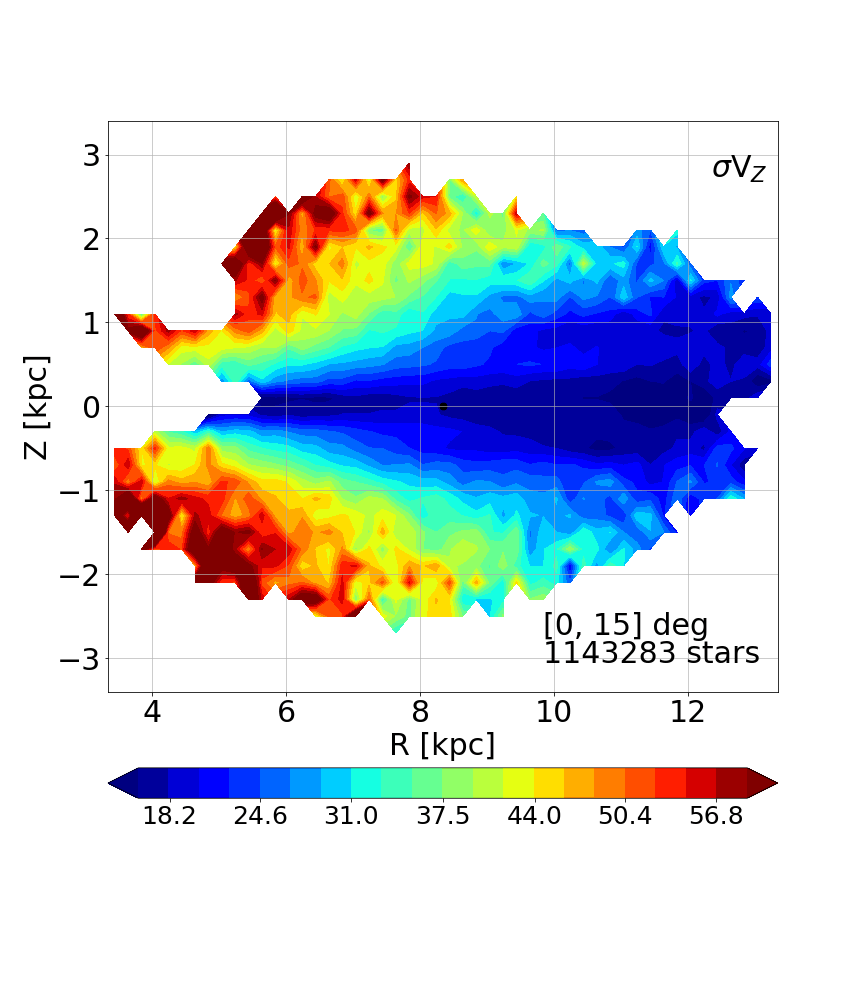}
\includegraphics[clip=true, trim = 5mm 40mm 10mm 40mm, width=0.45 \hsize]{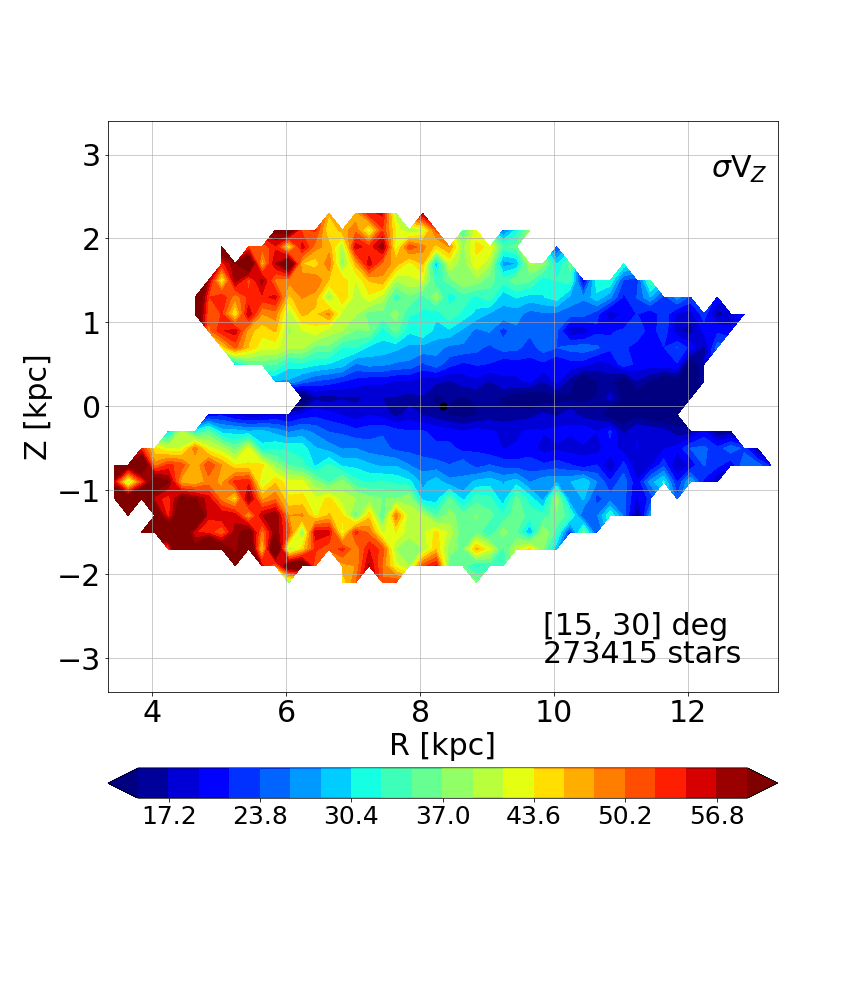}
\caption{Same as Fig.~\ref{fig:rzvrmed} for the vertical velocity dispersion, $\sigma_{V_Z}$.}
\label{fig:rzvzdisp}
\end{figure*}

{Figure~\ref{fig:xyvbendbreath} shows the face-on maps of the bending and breathing velocities (defined in Sect.~\ref{sec:methodo}).}

%
%

%
%

%
%

%
%

%
%

%
%

\end{appendix}

\end{document}